\documentclass[fleqn,useAMS]{Papers}
\usepackage[numbers]{natbib}
\usepackage{newtxtext,newtxmath}
\usepackage[T1]{fontenc}
\usepackage{makecell}

\usepackage{graphicx}	
\usepackage{lipsum}
\usepackage{caption}
\usepackage{comment}
\usepackage{multicol}

\usepackage{amsfonts}
\usepackage{amsmath}

\makeatletter
\newcommand*{\da@rightarrow}{\mathchar"0\hexnumber@\symAMSa 4B }
\newcommand*{\da@leftarrow}{\mathchar"0\hexnumber@\symAMSa 4C }
\newcommand*{\xdashrightarrow}[2][]{%
  \mathrel{%
    \mathpalette{\da@xarrow{#1}{#2}{}\da@rightarrow{\,}{}}{}%
  }%
}
\newcommand{\xdashleftarrow}[2][]{%
  \mathrel{%
    \mathpalette{\da@xarrow{#1}{#2}\da@leftarrow{}{}{\,}}{}%
  }%
}
\newcommand*{\da@xarrow}[7]{%
  \sbox0{$\ifx#7\scriptstyle\scriptscriptstyle\else\scriptstyle\fi#5#1#6\m@th$}%
  \sbox2{$\ifx#7\scriptstyle\scriptscriptstyle\else\scriptstyle\fi#5#2#6\m@th$}%
  \sbox4{$#7\dabar@\m@th$}%
  \dimen@=\wd0 %
  \ifdim\wd2 >\dimen@
    \dimen@=\wd2 %
  \fi
  \count@=2 %
  \def\da@bars{\dabar@\dabar@}%
  \@whiledim\count@\wd4<\dimen@\do{%
    \advance\count@\@ne
    \expandafter\def\expandafter\da@bars\expandafter{%
      \da@bars
      \dabar@ 
    }%
  }%
  \mathrel{#3}%
  \mathrel{%
    \mathop{\da@bars}\limits
    \ifx\\#1\\%
    \else
      _{\copy0}%
    \fi
    \ifx\\#2\\%
    \else
      ^{\copy2}%
    \fi
  }%
  \mathrel{#4}%
}
\makeatother
\usepackage{xcolor}  
\newcommand{\rev}[1]{{\color{black} #1}} 

\title[Superheavy dark matter and Hubble tension]{Cold freeze out of superheavy dark matter and Hubble tension}

\author[Z. Xu]{Zhijie (Jay) Xu,$^{1}$\thanks{E-mail: \href{mailto:zhijie.xu@pnnl.gov}{zhijie.xu@pnnl.gov};\href{mailto:zhijiexu@hotmail.com}{zhijiexu@hotmail.com}} 
\\
$^{1}$Physical and Computational Sciences Directorate, Pacific Northwest National Laboratory; Richland, WA 99354, USA}
\date{Accepted XXX. Received YYY; in original form ZZZ}

\pubyear{2025}

\begin{document}
\label{firstpage}
\pagerange{\pageref{firstpage}--\pageref{lastpage}}
\maketitle
\begin{abstract}
We present a unified framework, the "X miracle", in which dark matter consists of superheavy, nonthermal X particles whose relic abundance is determined not by the conventional weak-scale, semi-relativistic ("hot") freeze-out of WIMPs, but by annihilation or decay occurring within the smallest and earliest gravitationally bound objects. Unlike thermal WIMPs, which decouple at velocities of order 0.3$c$ with relic abundance $\rho_{\infty}$ set by weak-scale interactions, X particles are produced nonthermally with an initial overabundance $\rho_{ini}\gg \rho_{\infty}$. They become nonrelativistic extremely early, redshift to ultra-cold velocities, allowing collapse into compact bound structures characterized by a novel quantum-gravitational scale, $r_X=4\hbar^2/Gm_X^3=10^{-13}m\gg \hbar/m_Xc$, much larger than the Compton wavelength. The framework predicts a particle mass of $10^{12}$GeV and an enhanced cross section of $10^{-21}$m$^3$/s. Overlapping particle wavefunctions in these compact structures drive annihilation or decay into additional radiation, leading to a "cold" freeze-out that converts most of $\rho_{ini}$ into radiation while leaving a relic density $\rho_{\infty}$. Solutions to the Boltzmann equation indicate that an extreme ("big") depletion, with only one particle in a billion surviving, yields an additional radiation contribution $\Delta N_{eff}\approx$0.4, which could help alleviate the Hubble tension. For particles of $10^{12}$GeV, the scenario predicts a dark coupling constant $\alpha_X=0.09$ that is responsible for an instanton-induced decay process, consistent with current UHECR bounds. Early collapse at $10^{-6}$s may release binding energy as high-frequency (100kHz) gravitational waves or ultralight GUT-scale axions ($10^{-9}$eV). Superheavy sterile neutrinos provide a natural particle realization, linking dark matter to neutrino mass and baryogenesis. If gravitationally produced, this framework favors high-scale inflation and efficient reheating. The "X miracle" thus demonstrates that dark matter need not be weak-scale: gravitational dynamics can control freeze-out and evolution, producing multi-messenger observational signatures in UHECRs, axions, gravitational waves, and small-scale structures.

\end{abstract}



\vspace*{-50pt}
\setcounter{tocdepth}{1}
\begingroup
\let\clearpage\relax
\tableofcontents
\endgroup
\vspace*{-20pt}

\section{Introduction}
\label{sec:1}
Numerous astronomical observations support the existence of dark matter (DM). The most striking indications come from the dynamical motions of astronomical objects. The flat rotation curves of spiral galaxies suggest galactic dark matter halos with a total mass significantly greater than that of luminous matter \citep{Rubin:1970-Rotation-of-Andromeda-Nebula-f,Rubin:1980-Rotational-Properties-of-21-Sc}. Although the nature of dark matter remains unclear, the cold dark matter hypothesis (CDM) has quickly become a major component in the theory of cosmic structure formation and evolution \citep{Peebles:1982-Large-scale-background-temperature}. Incorporating CDM with the cosmological constant ($\Lambda$) representing the dark energy, $\Lambda$CDM theory is now generally considered the "standard" model of cosmology. Recent Planck measurements of the cosmic microwave background (CMB) anisotropies concluded that the amount of dark matter is 5.3 times that of baryonic matter \citep{Aghanim:2021-Planck-2018-results--VI--Cosmo}. 

In the standard $\Lambda$CDM paradigm, dark matter is cold (nonrelativistic), collisionless, dissipationless, nonbaryonic, and barely interacts with baryonic matter except through gravity \citep{Peebles:1984-Tests-of-cosmological-models,Spergel:2003-First-Year-Wilkinson-Microwave-Anisotropy,Komatsu:Seven-year-Wilkinson-Microwave-Anisotropy-Probe,Frenk:2012-Dark-matter-and-cosmic-structure}. 
Although the nature and properties of dark matter remain a great mystery, the leading cold dark matter candidate is often assumed to be a thermal relic, weakly interacting massive particles (WIMPs) that were in thermal equilibrium in the early universe \citep{Steigman:1985-Cosmological-Constraints-on-th}. These thermal relics freeze out as the reaction rate becomes comparable with the expansion rate of the Universe. The self-annihilation cross-section required by the right abundance of dark matter is of the same order as the typical electroweak cross-section, in alignment with the supersymmetric extensions of the standard model (WIMP miracle) \citep{Jungman:1996-Supersymmetric-dark-matter}. The mass of thermal WIMPs ranges from a few GeV to hundreds of GeV, with the unitarity argument giving an upper bound of $\sim$100TeV \citep{Griest:1990-Unitarity-Limits-on-the-Mass-a}, i.e., the Griest–Kamionkowski bound for thermal relics. However, no conclusive signals have been detected in this mass range, implying alternative theories are required beyond the WIMP paradigm. 

\begin{figure}
\includegraphics*[width=\columnwidth]{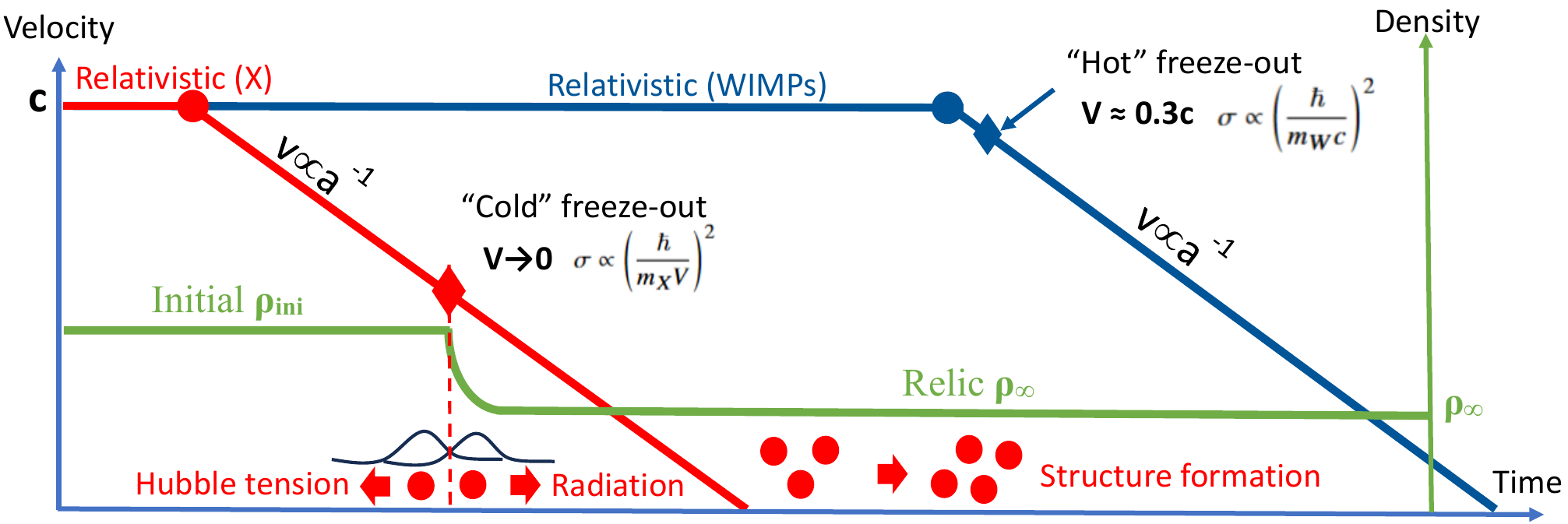}
\caption{The "hot" freeze-out for thermal WIMPs (blue) and the "cold" freeze-out for nonthermal heavy X particles (red). WIMPs are low-mass thermal relics that become nonrelativistic much later, with a thermal velocity around 0.3$c$ at freeze-out. The gravitational interaction between individual WIMP particles can be safely neglected due to fast free streaming. By contrast, nonthermal X particles are overproduced initially, become nonrelativistic much earlier, with thermal velocity redshifting rapidly and far below $c$. Due to their heavy mass, extreme coldness, and high initial density, these particles can form the smallest and the earliest gravitationally bound state. The overlapping particle wavefunctions within this bound state enable efficient annihilation or decay into extra radiation at extremely low velocities (i.e., a "cold" freeze-out). For X particles with an initial overabundance $\rho_{ini}\gg\rho_{\infty}$, this "cold" freeze-out also determines the relic abundance $\rho_{\infty}$. The density evolution of X particles is shown in green. The extra radiation produced during the "cold" freeze-out could alleviate the Hubble tension.} 
\label{fig:666}
\end{figure}

The WIMP miracle assumes low-mass thermal relics ($\sim$100GeV) that become nonrelativistic late in time, with a thermal velocity around 0.3$c$ before decoupling (Fig. \ref{fig:666}). In this scenario, while gravity still determines the background expansion, the gravitational interaction between individual particles can be safely neglected due to the fast free streaming, allowing us to derive the weak scale cross section. With gravity between particles neglected at freeze-out, the unitarity bound for thermal WIMPs requires that the annihilation cross section must be smaller than $m_X^{-2}$ with a relative collision velocity of the order of $c$. This relatively "hot" freeze-out scenario for thermal WIMPs, occurring close to 0.3$c$, leads to the relic density today.

However, dark matter particles (denoted as X) need not be thermally produced nor restricted to low masses. For nonthermal relics that never reach thermal equilibrium, their relic abundance is typically determined directly by the production mechanism—such as inflaton decay, gravitational particle creation, or freeze-in \citep{Moroi:2020-Light-Dark-Matter-from-Inflaton-Decay,Hall:2010-Freeze-inproduction-of-FIMP-dark-matter,Chung:2001-On-the-gravitational-production}. Beyond these scenarios, one may also consider nonthermal dark matter that is initially overproduced, with an initial abundance $\rho_{ini}$ exceeding the relic abundance $\rho_{\infty}$ (Fig. \ref{fig:666}). In this case, the relic density is still set by a subsequent annihilation or decay, just like thermal WIMPs, but the freeze-out process differs significantly from the “hot” freeze-out of thermal WIMPs.

As shown in Fig. \ref{fig:666}, because such particles can be heavy and become nonrelativistic much earlier, the thermal velocity of the particles redshifts rapidly to values far below the speed of light (Figs. \ref{fig:666} and \ref{fig:1111}). Due to their heavy mass, extreme coldness, and high initial density, these particles can form the smallest and the earliest gravitationally bound state, characterized by a length scale comparable to the particle's De Broglie wavelength. Within this bound state, the overlapping particle wavefunctions allow particles to quickly annihilate or decay into relativistic radiation, and ultimately set their relic abundance. This "cold" freeze-out process at low particle velocity -- referred to as the "X miracle", naturally predicts a much heavier particle mass $10^{12}$GeV and a much larger cross section $10^{-21}$m$^3$/s, as required to produce the relic density (Section \ref{sec:5-1-2}). The complete evolution of X particles can be obtained from exact solutions of the Boltzmann equation (Fig. \ref{fig:2222}). With extra radiation produced during the "cold" freeze-out, potential effects on Hubble tension and ultra-high energy cosmic rays (UHECRs) are also briefly discussed. 

\begin{figure}
\includegraphics*[width=\columnwidth]{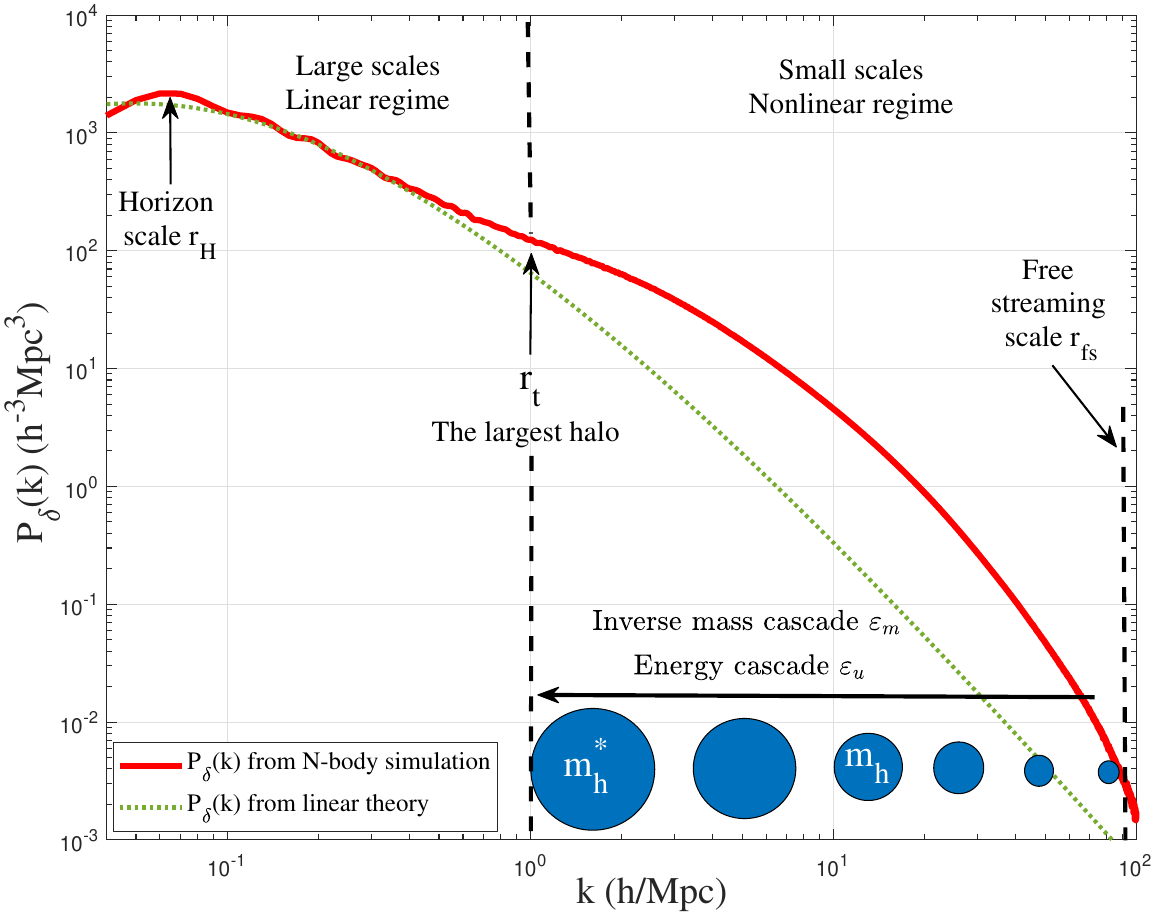}
\caption{The variation of density power spectrum $P_{\delta}(k)$ with comoving wavenumber $k$ at \textit{z}=0. Three important scales are identified. The pivot wavenumber denotes the size of the horizon $r_H$ at the matter-radiation equality. The scale $r_t\approx$1Mpc/h is roughly the size of the largest halo $m^*_h(z=0)$. The free streaming scale $r_{fs}$ represents the scale of the smallest nonlinear dark matter structure, below which dark matter forms a smooth background. The linear theory is in good agreement with N-body simulations on large scales ($r>r_t$), particularly in the linear regime. A universal transition spectrum with $n=-1$ can be identified around scale $r_t$ \citep{Xu:2021-Inverse-mass-cascade-mass-function}. On small scales $r<r_t$ (nonlinear regime), the linear theory is no longer valid and nonlinear theories are required. In this paper, we first focus on the free streaming scale $r_{fs}$ (Section \ref{sec:5-1}) and bound state formation (Section \ref{sec:5-1-1}) for mass and properties of dark matter particles, followed by an X miracle describing particle density evolution and "cold" freeze out (Section \ref{sec:5-1-2}). The appendix provides the structure formation in different eras for haloes on scales $r_{fs}<r<r_t$ (Section \ref{sec:5}). All suggest a consistent heavy particle mass of $10^{12}$ GeV.}
\label{fig:104}
\end{figure}

To facilitate the discussion, we start with three important scales in the matter density spectrum of dark matter. Figure \ref{fig:104} presents the variation of the density power spectrum with the comoving wavenumber $k$ from N-body simulations carried out by the Virgo consortium (SCDM) \citep{Frenk:2000-Public-Release-of-N-body-simul}. The pivot wavenumber denotes the size of the horizon $r_H$ at the matter-radiation equality. The scale $r_t$ roughly corresponds to the size of the characteristic halo that separates the small scales in a non-linear regime and the large scales in a linear regime. A universal transition spectrum with $n=-1$ can be identified around this scale $r_t$ \citep{Xu:2021-Inverse-mass-cascade-mass-function}. The free streaming scale $r_{fs}$ represents the scale of the smallest nonlinear halo structure that can be formed by dark matter. Linear theory predictions (green dotted line from \citep{Jenkins:1998-Evolution-of-structure-in-cold}) agree well with N-body simulations on large scales greater than $r_t$. However, linear theory cannot predict the structure evolution on small scales (i.e., $r<r_t$), which is highly nonlinear and can only be studied numerically or by simplified analytical tools. In particular, the free streaming scale $r_{fs}$ for the smallest structure is directly relevant to the mass and properties of dark matter particles. However, this scale is too small to be probed by observations or simulations.

In this paper, we aim to understand the mass and properties of cold dark matter. In particular, we focus on fermion nonthermal particle dark matter that is out of equilibrium in the early Universe and only interacts with the standard model (SM) and other dark matter (DM) particles via gravity. For dark matter satisfying these conditions, we can derive the exact particle mass and properties and postulate the nature of dark matter through two independent approaches: i) nonlinear structure formation at the free streaming scale $r_{fs}$ and ii) a WIMP-miracle-like analysis considering gravity at freeze-out ("X miracle"). As expected, both approaches suggest a superheavy scenario and a consistent particle mass of $10^{12}$GeV. 

The paper is structured as follows. Section \ref{sec:5-1} identifies a critical particle mass of $10^{12}$GeV from the free streaming mass calculations. Section \ref{sec:5-1-1} discusses the smallest and earliest gravitationally bound state that can be formed (a bottom-up approach). Section \ref{sec:5-1-2-2-2} presents the evolution of particle thermal and gravity-induced velocities, followed by Section \ref{sec:5-1-2} on the X miracle for nonthermal relics that also predicts a particle mass of $10^{12}$GeV. For this mass, Section \ref{sec:5-1-3} provides the relevant properties for dark matter particles, followed by Section \ref{sec:6-1-1} on potential dark matter candidates based on the properties we identified, especially superheavy sterile neutrinos. 

The first three sections describe a superheavy dark matter scenario from both free streaming and a "X miracle" for cold freeze-out. The impacts of this model are presented in the next several sections in chronological order. Section \ref{sec:6-1-11} explores the effect of an overabundant initial density on gravitational production. Section \ref{sec:6-1} discusses the nature of "radiation" during the formation of a gravitationally bound state, i.e., high-frequency ($\sim$100 kHz) gravitational waves (GW) or nano-eV GUT-scale axions. Section \ref{sec:5-1-2-33} focuses on the extra radiation produced during the cold freeze-out and connections to the Hubble tension (See \citep{Xu:2026-Superheavy-dark-matter-and-stepped-dark-radiation}). Section \ref{sec:5-1-2-34} presents the UHECRs from X particle annihilation or decay, and comparison with observational constraints. For completeness, the appendix provides the structure formation and evolution in different eras for haloes on scales $r_{fs}<r<r_t$. Scaling laws have been developed for dark matter halos. As expected, these halo-scale scaling laws also suggest a heavy particle mass $10^{12}$GeV (a top-down approach in Section \ref{sec:5}), consistent with the other two independent approaches (free-streaming and X miracle).

\begin{figure}
\centering
\includegraphics*[width=5cm, height=5cm]{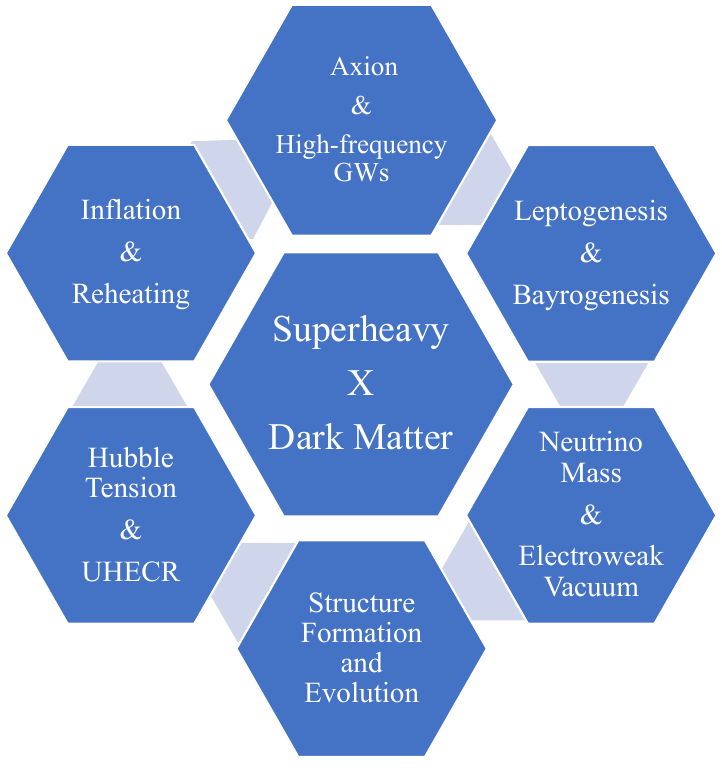}
\caption{Schematic overview of key cosmological and physical problems closely linked to the dark matter puzzle. The main analysis constrains the mass and properties of dark matter from structure formation at the free-streaming scale and from the proposed X-miracle mechanism involving gravity, while the appendix details the energy cascade and scaling laws governing structure formation and evolution across multiple scales. Collectively, these considerations favor a superheavy X dark matter candidate with mass $\sim10^{12}$GeV, though further work is required to fully elucidate the nature of dark matter and relations to related open problems.} 
\label{fig:S444}
\end{figure}

Based on the overview presented in this work, several outstanding problems in cosmology and particle physics may be closely related to the dark matter framework proposed here. These include the Hubble tension, leptogenesis/baryogenesis, ultra-high energy cosmic rays (UHECRs), and inflation and reheating for gravitational production. In scenarios where superheavy right-handed neutrinos constitute the dark matter, the origin of neutrino masses becomes particularly relevant. For models in which dark “radiation” emerges from the formation of the gravitationally bound state, GUT–scale nano-eV axions or high-frequency gravitational waves ($\sim$100 kHz) may be relevant. Figure \ref{fig:S444} summarizes the superheavy dark matter and its potential connections to these open problems. Nonetheless, the growing body of empirical and theoretical insights mainly highlights how incomplete our understanding remains. Substantial further work, well beyond the scope of this paper, is required to clarify the nature of dark matter and its relation to these unresolved issues.

\section{Free streaming and bound state formation}
We consider particle dark matter that is produced nonthermally. The early Universe conditions vary with the production mechanism. For dark matter produced gravitationally, we assume a standard inflationary epoch followed by rapid reheating to a high temperature (Section \ref{sec:6-1-11}). In this section, we focus on the post-production evolution after these particles have been created, involving free streaming and the formation of a gravitationally bound state.

\subsection{Critical particle mass from free streaming}
\label{sec:5-1}
The smallest scale of dark matter structures is the "free-streaming" scale ($r_{fs}$ in Fig. \ref{fig:104}). The thermal velocity of dark matter particles tends to erase the primordial perturbations below that scale. There are no coherent structures on scales smaller than the free streaming scale. Dark matter particles form a smooth background on scales $r<r_{fs}$. In this Section, we provide a brief analysis of the free-streaming mass $M_{fs}(m_X)$, i.e., the mass of the smallest dark matter structure as a function of particle mass $m_X$. A critical particle mass $m_{Xc}$ is then identified that exactly equals the free streaming mass, i.e., $M_{fs}(m_{Xc})=m_{Xc}$, which is the particle mass to form the smallest possible dark matter structure. 

The free streaming scale depends on both particle mass and the production mechanism \citep{Schneide:2013-Halo-mass-function-and-the-free-streaming-scale}. It is traditionally quantified by the comoving length a particle travels before primordial perturbations start to grow significantly, which is around the time $t_{eq}$ for matter-radiation equality. This assumes that the formation time $t_X$ of the first gravitationally bound structure is after the matter-radiation equality ($t_X>t_{eq}$). The comoving free streaming length $\lambda_{fs}$ is
\rev{
\begin{equation}
\begin{split}
\lambda_{fs}(t) &= \int^{t_{eq}}_0 \frac{V_{DM}(t)}{a(t)}dt \\
&= \int^{t_{1}}_0 \frac{V_{DM}}{a}dt + \int^{t_2}_{t_1} \frac{V_{DM}}{a}dt + \int^{t_{eq}}_{t_2} \frac{V_{DM}}{a}dt,
\end{split}
\label{eq:12-1}
\end{equation}
where $V_{DM}(t)$ is the particle thermal velocity, and $a(t)$ is the scale factor.} Two time scales, $t_1$ and $t_2$, represent either the time when particles become non-relativistic ($t_{NR}$) or the time when particles fall out of the thermal equilibrium and decouple (freeze-out) from the background ($t_{dec}$), depending on the nature of dark matter. For hot relics (hot dark matter), particles become non-relativistic after decoupling ($t_{NR}>t_{dec}$) and, therefore, $t_1=t_{dec}$ and $t_2=t_{NR}$. For cold relics (cold dark matter), particles become non-relativistic before decoupling ($t_{NR}<t_{dec}$), such that $t_1=t_{NR}$ and $t_2=t_{dec}$. For the radiation era before the matter-radiation equality, the scale factor $a(t)\propto t^{1/2}$. The free streaming mass $M_{fs}$ now reads
\begin{equation}
M_{fs} = \frac{\pi}{6} \rho_{DM0} \lambda_{fs}^3,
\label{eq:12-2}
\end{equation}
where $\rho_{DM0}$ is the dark matter density at $z=0$.

\rev{
An alternative way of understanding the free streaming scale is via the comoving Jeans length:
\begin{equation}
\begin{split}
\lambda_{J}(t) = \frac{V_{DM}(t)}{a(t)}\sqrt{\frac{\pi}{G\bar\rho(t)}},
\end{split}
\label{eq:12-3}
\end{equation}
where $\bar\rho(t)\propto a^{-4}$ is the background density. Only perturbations greater than the Jeans length will gravitationally collapse into haloes. Perturbations smaller than the Jeans length will be damped out. 

The thermal velocity $V_{DM}(t)$ also varies in different regimes: i) $V_{DM}(t)=c$ when the particles are still relativistic ($t<t_{NR}$). During this stage, the free streaming length $\lambda_{J}\propto a$; ii) $V_{DM}(t)\propto a^{-1/2}$ when the particle becomes non-relativistic, but still coupled ($t_{NR}<t<t_{dec}$). During this stage, the free streaming length $\lambda_{J}\propto a^{1/2}$; and iii) $V_{DM}(t)\propto a^{-1}$ when the particles become non-relativistic and also decoupled ($t>t_{dec}$ and $t>t_{NR}$). During this stage, $\lambda_{J}\propto a^0$ is frozen and independent of time. The variation of length $\lambda_{fs}(t)$ over time is similar to the Jeans length $\lambda_{J}(t)$.}  

\begin{figure}
\includegraphics*[width=\columnwidth]{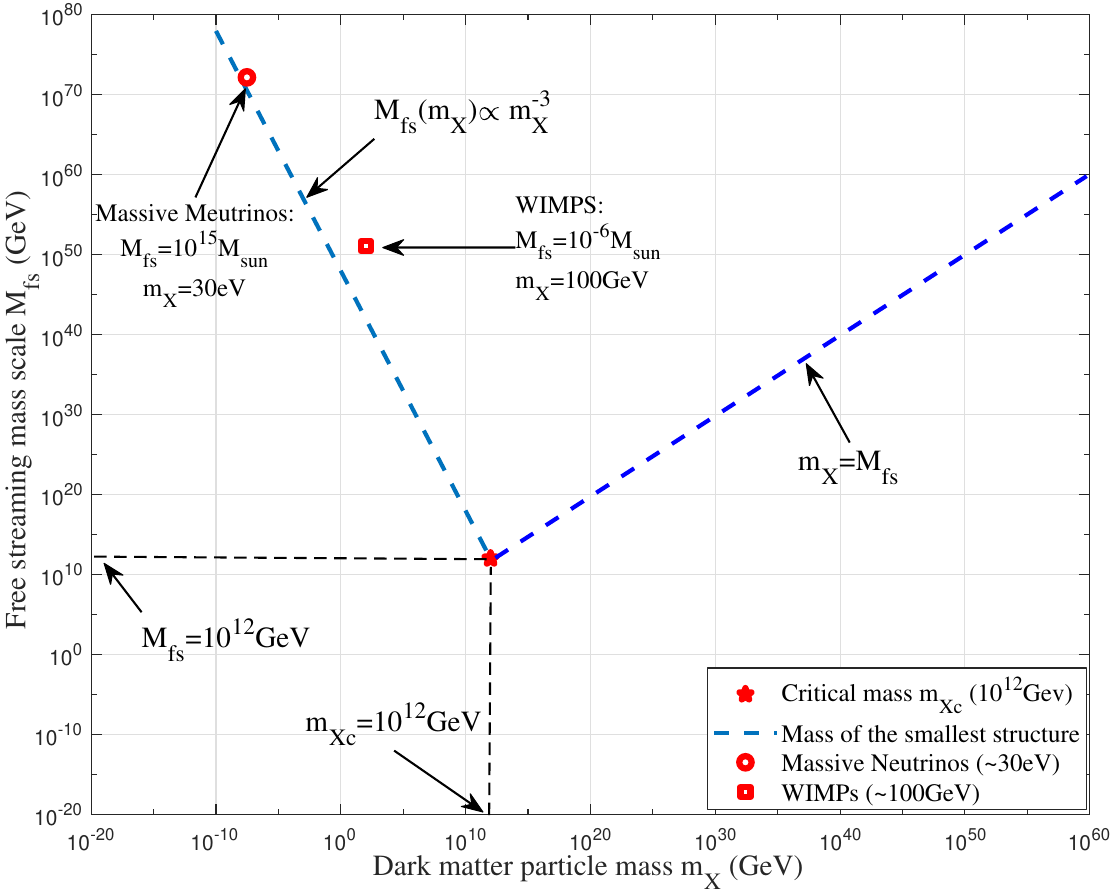}
\caption{The free streaming mass $M_{fs}$ decreases with the particle mass $m_X$. Hot massive neutrinos of mass 30eV may form the smallest structure of mass $M_{fs}\approx 10^{15}M_{\odot}$ (circle). For WIMP particles of mass 100GeV, $M_{fs}\approx 10^{-6}M_{\odot}$ (square). In this work, the calculated free streaming mass $M_{fs}$ for dark matter particles of mass $m_X$ follows the scaling $M_{fs}\propto m_X^{-3}$ (Eq. \eqref{eq:12-5-4}). A critical particle mass $m_{Xc}$ can be identified where $m_{Xc}=M_{fs}(m_{Xc})\approx 10^{12}$GeV (Eq. \eqref{eq:12-5-3}). The dashed line plots the mass of the smallest structure ($M_S$) that particles of mass $m_X$ can form. For particles with mass $m_X<m_{Xc}$, the mass of the smallest structure $M_S=M_{fs}$. For particles with mass $m_X>m_{Xc}$, $M_S=m_X$ because the free streaming mass is smaller than the particle mass, and the structure cannot be smaller than the particle itself. Therefore, $M_S$=max($M_{fs},m_X$). The critical mass $m_{Xc}$ represents the mass of the smallest structure that dark matter particles of any mass can form. Since smaller structures were formed earlier, particles of critical mass $m_{Xc}$ also form the earliest structure.} 
\label{fig:109}
\end{figure}

Based on these equations, we can estimate a free streaming mass $M_{fs}\approx 10^{15}M_{\odot}$ for hot massive neutrinos with a mass around 30eV (Fig. \ref{fig:109}). This is due to the smaller particle mass, so the particles become non-relativistic much later, resulting in a longer free streaming length and a larger mass. Therefore, for hot dark matter, the structure formation proceeds "top-down", which is inconsistent with observation. For WIMP particles with a much higher mass of 100GeV, the free streaming mass is around $M_{fs}\approx 10^{-6}M_{\odot}$ (the mass of the Earth) \citep{Bertschinger:2006-Effects-of-cold-dark-matter-decoupling} (Fig. \ref{fig:109}). Cold dark matter, such as WIMPs, undergoes hierarchical "bottom-up" structure formation, which is in better agreement with observations. Thus, the smallest structure that can form by particles of different masses is radically different. The larger the particle mass $m_{X}$, the smaller the free streaming mass $M_{fs}$. 

Figure \ref{fig:109} plots the decreasing free streaming mass $M_{fs}$ with the particle mass $m_X$. There should exist a critical particle mass $m_{X}$ such that $M_{fs}$ is exactly equal to the particle mass ($M_{fs}(m_{Xc}) \equiv m_{Xc}$). Particles with a mass $m_X<m_{Xc}$ should form the smallest structure of free streaming mass that can be much greater than the particle mass, i.e., $M_{fs}\gg m_X$. Particles with a mass $m_X>m_{Xc}$ have a free streaming mass smaller than the particle mass, such that the smallest structure formed by these particles should have a mass comparable to the particle mass. Therefore, the critical mass $m_{Xc}$ represents the mass of the smallest dark matter structure that can be formed by dark matter particles of any mass. 

\begin{figure}
\includegraphics*[width=\columnwidth]{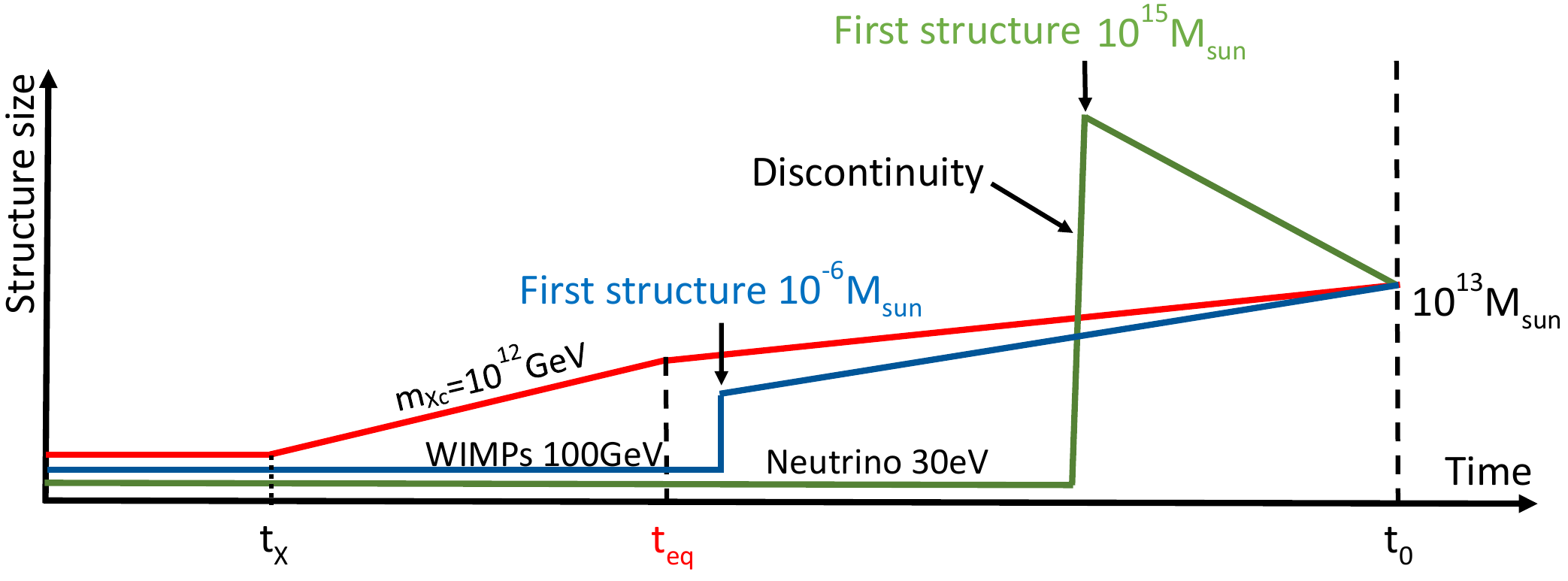}
\caption{A schematic plot of the structure size evolution for dark matter particles of different mass $m_X$. Today's typical haloes ($t=t_0$) are of mass around $10^{13}M_{\odot}$. Hot massive neutrinos of mass 30 eV form the first halo structure of mass $M_{fs}\approx 10^{15}M_{\odot}$ much later in the Universe (solid green), which conflicts with observations. Since no structures of any other mass may exist before the formation of the first structure, there is an enormous discontinuity in the structure size before and after the first structure formed by hot neutrinos. WIMP particles of mass 100 GeV form the first structure of mass $10^{-6}M_{\odot}$ much sooner with a relatively smaller discontinuity (solid blue). If we keep increasing the particle mass, particles of critical mass $m_{Xc}=10^{12}$GeV in Fig. \ref{fig:109} may form the smallest and the earliest structure with a smooth transition in structure size evolution (solid red, no discontinuity). We will identify the critical mass $m_{Xc}$ in this section.} 
\label{fig:S33}
\end{figure}

Figure \ref{fig:S33} presents a schematic plot of the structure size evolution for particles of different mass $m_X$. Today's observation confirms typical haloes of mass around $10^{13}M_{\odot}$. Haloes larger than this size are very rare. Hot massive neutrinos form the first halo of mass $10^{15}M_{\odot}$ much later in the Universe (green). This conflicts with observations, which exclude the hot dark matter. Since no structures of any other mass may exist before the formation of the first structure, there is an enormous discontinuity in the structure size evolution for hot neutrinos. On the other end, WIMP particles, as cold dark matter, form the first structure of mass $10^{-6}M_{\odot}$, which is still much greater than the mass of WIMP particles. The first structure was formed much sooner with a much smaller discontinuity (blue). Next, particles of critical mass $m_{Xc}=10^{12}$GeV in Fig. \ref{fig:109} form the smallest and the earliest structure with a smooth transition in structure evolution (red). For particles of this mass, there is no discontinuity in the evolution of structural size.

To calculate the critical particle mass $m_{Xc}$, we consider the free streaming until time $t_X$ when the first structure forms. The free streaming scale at this time determines the size of the first structure. This is different from Eq. \eqref{eq:12-1} because the first structure might be formed in the radiation era with $t_X<t_{eq}$, depending on the particle mass. Particles could be relativistic when produced and become nonrelativistic later at time $t_{NR}$ or $a_{NR}$. This includes: 
\begin{enumerate}
\item \noindent thermal relics that are produced with a particle energy E comparable to the universe temperature T, but much greater than the particle mass, i.e., $E\sim T\gg m$. They become nonrelativistic when particle energy redshifts along with temperature $E\sim T\propto a^{-1}$ to the scale of particle mass or $E\sim T\sim m$. 
\item \noindent nonthermal relics that are produced with an energy $E\gg m$ at $a_{prod}$ such that the kinetic energy is dominant over the rest mass energy. Due to their nonthermal nature, particles are out of thermal equilibrium. We can still define a ratio $\alpha_{prod}=E/T$ at production to relate the particle energy to the universe temperature. The particle momentum and energy redshift at the same rate as T. Nonthermal relics become nonrelativistic when their momentum redshifts below the particle mass, and the rest mass becomes dominant ($E\sim m$). 
\end{enumerate}

We consider the nonthermal relics produced with a particle energy comparable to the universe's temperature (or $\alpha_{prod}=E/T\sim 1$). These particles have a sufficiently large thermal velocity leading to a finite non-zero free streaming length before $t_X$. They are out of thermal equilibrium, which significantly simplifies the calculation, as the comoving Jeans length remains constant between $t_{NR}<t<t_{X}$ for nonthermal relics (Eq. \eqref{eq:12-3}). By contrast, the coupling of thermal relics (WIMPs, etc.) to the thermal bath leads to a period $t_{NR}<t<t_{dec}$ with an increasing Jeans length $\lambda_J\propto a^{1/2}$ (Eq. \eqref{eq:12-3}) and a larger free streaming mass. This was clearly shown in Fig. \ref{fig:109}, where WIMP is slightly above the model for nonthermal relics (dashed blue). 

For nonthermal relics with $E\sim T$, we only need to consider the free streaming length $\lambda_{fs}$ from the first term in Eq. \eqref{eq:12-1} (free streaming before $t_{NR}$), as Jeans length remains constant for $t_{NR}<t<t_X$: 
\begin{equation}
\begin{split}
\lambda_{fs} = \int^{t_{NR}}_0 \frac{c}{a(t)}dt=\frac{2ct_{NR}}{a_{NR}}=\frac{c}{H_0}\frac{a_{NR}}{\sqrt{\Omega_{rad}}},
\end{split}
\label{eq:12-4}
\end{equation}
where $a_{NR}$ is the scale factor when particles become non-relativistic. Here, we use the relation in the radiation era ($t\propto a^2$), 
\begin{equation}
\begin{split}
t = \frac{a^2}{2H_0\sqrt{\Omega_{rad}}},
\end{split}
\label{eq:12-4-1}
\end{equation}
where $\Omega_{rad}$ is the mass fraction of radiation at $z=0$. Next, the free streaming mass $M_{fs}$ reads (from Eqs. \eqref{eq:12-4} and \eqref{eq:12-2}),
\begin{equation}
\begin{split}
M_{fs}=\frac{\pi}{6}\Omega_{DM}\bar\rho_0\left(\frac{c}{H_0}\frac{a_{NR}}{\sqrt{\Omega_{rad}}}\right)^3,
\end{split}
\label{eq:12-4-2}
\end{equation}
where $\Omega_{DM}$ is the mass fraction of dark matter at $z=0$. The critical density of the Universe is $\bar\rho_0=3H_0^2/8\pi G$. 

The scale factor $a_{NR}$ also depends on the mass of the particles $m_X$. Since particle energy redshifts at the same rate as temperature, particles become non-relativistic when the radiation temperature $T_{\gamma}$ is comparable to the particle mass,
\begin{equation}
\begin{split}
3k_B T_{\gamma}(a_{NR}) = m_{X}c^2 \quad \textrm{and} \quad T_{\gamma}(a) = T_{\gamma0} a^{-1},
\end{split}
\label{eq:12-5-1}
\end{equation}
where the radiation temperature redshifts with time as $T_{\gamma}\propto a^{-1}$. Here, $T_{\gamma0}=2.7K$ is the radiation temperature (CMB) at $z=0$. Solving Eq. \eqref{eq:12-5-1}, the scale factor $a_{NR}$ reads
\begin{equation}
\begin{split}
a_{NR}=\frac{3k_BT_{\gamma_0}}{m_{X}c^2}.
\end{split}
\label{eq:12-5-2}
\end{equation}

Inserting $a_{NR}$ into Eq. \eqref{eq:12-4-2} leads to the final free streaming mass
\begin{equation}
\begin{split}
M_{fs}=\frac{1}{16}\left[\frac{\Omega_{DM}}{\Omega_{rad}^{3/2}}\frac{(3k_B T_{\gamma0})^3}{GH_0c^3}\right]m_X^{-3}\propto m_{X}^{-3}.
\end{split}
\label{eq:12-5-4}
\end{equation}
Let $M_{fs}=m_X$, we found the critical particle mass $m_{Xc}$:
\begin{equation}
\begin{split}
m_{Xc}=\frac{1}{2}\left[\frac{\Omega_{DM}}{\Omega_{rad}^{{3}/{2}}}\frac{(3k_B T_{\gamma0})^3}{GH_0c^3}\right]^{\frac{1}{4}}\approx \frac{3}{2}\times 10^{-15}kg\approx 10^{12}GeV.
\end{split}
\label{eq:12-5-3}
\end{equation}
The calculation is straightforward and yields a very important mass scale, which we will frequently revisit. 

We may also relate the critical mass $m_{Xc}$ to Planck quantities. Radiation density decreases as $\propto a^{-4}$ such that  
\begin{equation}
\begin{split}
&\Omega_{rad}\bar\rho_0a_{p}^{-4} = \alpha_p\rho_p \quad \textrm{and} \quad \rho_p = {c^5}/{(\hbar G^2)}, \\
&T_{\gamma0}a_p^{-1} = \beta_p T_p \quad \textrm{and} \quad T_p = \sqrt{{\hbar c^5}/{(Gk_B^2)}},
\end{split}
\label{eq:12-6}
\end{equation}
where $a_p$ is defined as the scale factor corresponding to the Planck time $t_p=\sqrt{\hbar G/c^5}=5.4\times 10^{-44}$s. Here, $\rho_p$ is the Planck density, $T_p$ is the Planck temperature, and $\alpha_p$ and $\beta_p$ are two numerical factors. The scale factor $a_p$ is related to the Planck time by Eq. \eqref{eq:12-4-1}, 
\begin{equation}
\begin{split}
&a_p=\left(2H_0t_p\sqrt{\Omega_{rad}}\right)^{1/2}\approx 4.76\times 10^{-32}.
\end{split}
\label{eq:12-7}
\end{equation}
Substituting Eq. \eqref{eq:12-7} into Eq. \eqref{eq:12-6}, we have 
\begin{equation}
\begin{split}
&T_{\gamma0} = \beta_p T_p \left(\frac{\Omega_{rad}\bar\rho_0}{\alpha_p \rho_p}\right)^{1/4}, \quad \alpha_p=\frac{3}{32\pi}, \quad \beta_p \approx \frac{2}{5}.
\end{split}
\label{eq:12-8}
\end{equation}
Now, substitute Eq. \eqref{eq:12-8} into Eqs. \eqref{eq:12-5-2} and \eqref{eq:12-4-2}, the critical particle mass $m_{Xc}$ can be obtained as
\begin{equation}
\begin{split}
m_{Xc} &= \frac{1}{2} \left(3\beta_p\right)^{\frac{3}{4}}\left(\frac{3}{8\pi\alpha_p}\right)^{\frac{3}{16}}\left(\frac{\Omega_{DM}^2}{\Omega_{rad}^{3/2}}\right)^{\frac{1}{8}}M_{pl}(H_0t_p)^{\frac{1}{8}} \\
&\approx 1.5\times 10^{-15}kg=0.9\times 10^{12}\textrm{GeV}.
\end{split}
\label{eq:12-9}
\end{equation}
Here, $M_{pl}=2.2\times 10^{-8}kg$ is the Planck mass. The dark matter fraction $\Omega_{DM}=0.12/h^2$, the radiation fraction $\Omega_{rad}=4.2\times 10^{-5}/h^2$, and $h=H_0$/(100km/s/Mpc) is the dimensionless Hubble constant. The critical mass $m_{Xc}$ is related to $M_{pl}$ by a factor of $(H_0t_p)^{1/8}$.

From this calculation, X particles of critical mass $m_{Xc}=10^{12}$GeV are relativistic when produced, leading to a free streaming length on the order of $10^4$m (Eq. \eqref{eq:12-4}). They become nonrelativistic and out-of-equilibrium at a time around $t_{NR}=10^{-29}$s, $a_{NR}=10^{-24}$, or temperature $T_{NR}=10^{11}$GeV. They can form the smallest bound structure among particles of any mass with a smooth structure size evolution in cosmic time (Fig. \ref{fig:S33}). Since smaller structures were usually formed earlier, they also formed the earliest structure among all particles of different masses. 

For more general scenarios, where nonthermal relics are produced at time $a_{Prod}$ with an arbitrary ratio $\alpha_{prod}=E/T$, the free streaming mass can be derived similarly, 
\begin{equation}
\begin{split}
M_{fs}=\frac{\pi}{6}\Omega_{DM}\bar\rho_0\left[\frac{c}{H_0\sqrt{\Omega_{rad}}}\left(\alpha_{prod}\frac{3k_BT_{\gamma0}}{m_Xc^2}-a_{prod}\right)\right],
\end{split}
\label{eq:12-9-9}
\end{equation}
where a smaller ratio $\alpha_{prod}$ leads to a smaller $M_{fs}$. The results reduce to Eq. \eqref{eq:12-4-2} with $\alpha_{prod}=1$ and will not be repeated here.

\subsection{Formation of gravitationally bound state}
\label{sec:5-1-1}
\begin{figure}
\includegraphics*[width=\columnwidth]{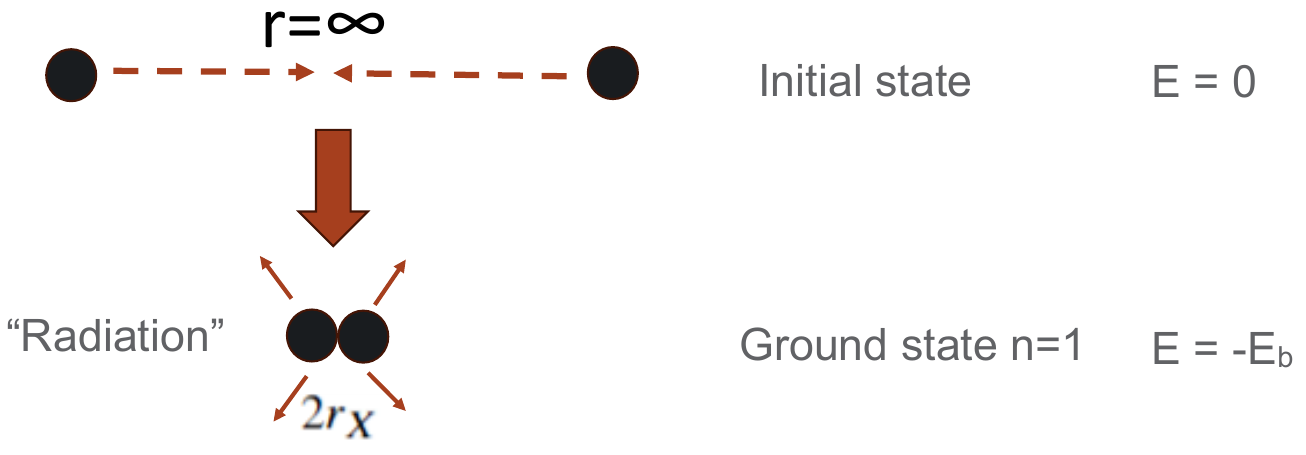}
\caption{Two X particles forming the smallest bound state X$\cdot$X. Two particles have a total energy E=0 in the initial state with infinite separation $r$. The final state is a gravitationally bound state with a separation of $l_X=2r_x$ and a binding energy of $E_b$. A form of "radiation" (with double quotes) of energy $E_b$ is postulated to be produced during the formation of the gravitationally bound state. Section \ref{sec:6-1} discusses possible nature of this dark "radiation".} 
\label{fig:S44}
\end{figure}

In this section, we consider the smallest gravitationally bound state formed by and only by two X particles (two-body bound state). This is possible for particles with a mass $m_X\ge m_{Xc}$ such that the free streaming mass $M_{fs}$ is smaller than the particle mass $m_X$ (Fig. \ref{fig:109}). Now consider the bound state with two X particles of mass $m_X$ and separation $l_X=2r_X$ (Fig. \ref{fig:S44}) formed at time $t_X$. From the virial equilibrium, we have a relation between the kinetic energy KE and the gravitational potential PE (2KE+PE=0) such that
\begin{equation}
\begin{split}
&2v_X^2=\frac{Gm_X}{2r_X},\\
\end{split}
\label{eq:5-2-7}
\end{equation}
where $r_X$ is particle size and $v_X$ is a characteristic particle velocity.

Next, we consider fermion dark matter, which is governed by the Pauli exclusion principle. In this scenario, the gravitational pressure should be balanced by the quantum pressure. This mimics the electron or neutron degeneracy pressure that also arises from the Pauli exclusion principle, resisting the gravitational collapse of stars. For a bound structure of fermion dark matter, the pressure balance is
\begin{equation}
\begin{split}
m_{X} v_{X}^2 n_X \sim \frac{\hbar^2}{m_X}n_X^{5/3}, \quad n_X=\frac{2}{\frac{4}{3}\pi r_X^3},
\end{split}
\label{eq:5-2-9-2}
\end{equation}
where $\hbar$ is the Planck constant and $n_X$ is the particle number density in this bound state. The gravitational pressure (left) balances the degeneracy pressure (right). This equation is equivalent to 
\begin{equation}
\begin{split}
m_{X} v_{X} \cdot {r_X} = n\hbar,
\end{split}
\label{eq:5-2-9}
\end{equation}
where $n$ is the quantum number with $n=1$ for the ground state. Equation \eqref{eq:5-2-9} requires that the angular momentum of the particle be a multiple of the Planck constant. The same idea was also employed in Bohr's atomic model, where the angular momentum of electrons is also a multiple of the Planck constant. 

Equations \eqref{eq:5-2-7} and \eqref{eq:5-2-9} represent a simple model for the smallest gravitationally bound state in the framework of Newtonian (non-relativistic) quantum gravity, which involves coupled gravity and quantum effects. Combining two equations, all other relevant physical quantities can be obtained as a function of the particle mass $m_X$. The particle size and velocity for the ground state $n=1$ read
\begin{equation}
\begin{split}
&r_X=\frac{l_X}{2}=\frac{4\hbar^2}{G(m_{X})^3} \quad \textrm{and} \quad v_X=\frac{G(m_{X})^2}{4\hbar}. 
\end{split}
\label{eq:5-2-10-2}
\end{equation}
The particle size $r_X\gg \hbar/m_Xc$ represents a novel quantum-gravitational scale due to gravity that allows for a much heavier particle mass and a larger cross section (Section \ref{sec:5-1-2}). The density of the bound state reads
\begin{equation}
\begin{split}
&\rho_{X}=n_Xm_X=\frac{2m_X}{\frac{4}{3}\pi r_X^3} = \frac{3}{2^{7}\pi}\frac{G^3m_X^{10}}{\hbar^6}. \\
\end{split}
\label{eq:5-2-8}
\end{equation}

The formation time $t_X$ of the smallest bound structure reads
\begin{equation}
\begin{split}
&t_X(m_X)=\frac{\pi}{\sqrt{2}} \frac{r_X}{v_X}=\frac{2^{7/2}\pi\hbar^3}{G^2(m_{X})^5}. \\
\end{split}
\label{eq:5-2-11}
\end{equation}
In the radiation era, the Hubble parameter $H_Xt_X=1/2$. This leads to the background density $\bar\rho$ at the time of formation
\begin{equation}
\begin{split}
&\bar\rho(t_X) = \frac{3H_X^2}{8\pi G}=\frac{3}{32\pi Gt_X^2}=\frac{3}{2^{12}\pi^3} \frac{G^3m_X^{10}}{\hbar^6}.\\
\end{split}
\label{eq:5-2-8-2}
\end{equation}
An interesting finding is that the density of the smallest structure $\rho_X$ is always exactly $32\pi^2$ times the background density $\bar\rho (t_X)$ at the formation time $t_X$, regardless of the particle mass $m_X$, 
\begin{equation}
\begin{split}
&\rho_{X}=\xi \bar\rho(t_X), \quad \xi=32\pi^2. \\
\end{split}
\label{eq:5-2-8-3}
\end{equation}
This is a typical feature of the gravitationally bound structure satisfying the virial equilibrium in Eq. \eqref{eq:5-2-7}. According to the spherical collapse model \citep{Gunn:1972-Infall-of-Matter-into-Clusters}, the density of the gravitationally bound halo structure should be a fixed ratio of the background density, i.e., $\rho_X(t_X)=\xi\bar\rho(t_X)$. For the radiation era and the matter era, this density ratio can be analytically obtained as $\xi=32\pi^2$ or $18\pi^2$. Section \ref{sec:2-2-1} provides additional details on the derivation of these density ratios in different eras from the spherical collapse model. 

For the smallest gravitationally bound structure, there exists a critical Hubble parameter $H_X$ comparable to the interaction rate of $n_X\langle \sigma_Xv_X\rangle$ involving a thermally-averaged cross section $\langle \sigma_Xv_X\rangle$
\begin{equation}
\begin{split}
\langle \sigma_Xv_X\rangle= \pi r_X^2v_X \quad \textrm{and} \quad H_X\sim n_X\langle \sigma_Xv_X\rangle=n_X\cdot \pi r_X^2v_X. \\
\end{split}
\label{eq:5-2-8-33}
\end{equation}
The cross section $\langle \sigma_Xv_X\rangle$ represents the cross section in the smallest and earliest bound structure. For any time before $t_X$, the expansion exceeds the interaction rate ($H\gg n_X\langle \sigma_Xv_X\rangle\sim v_X/r_X$) so that gravitationally bound structures cannot be formed any time earlier than the formation time $t_X$, which prohibits any annihilation or decay before $t_X$. This is important for the X miracle in the derivation of the particle mass (Fig. \ref{fig:5555} in Section \ref{sec:5-1-2}). The formation time $t_X$ is closely related to the particle mass $m_X$ (Eq. \eqref{eq:5-2-11}).

Figure \ref{fig:S44} describes the formation of the smallest bound state of two X particles. The initial state is represented by two X particles with total energy E=0 and infinite separation $r$. The final state is a two-body bound state (X$\cdot$X) with a separation $l_X=2r_x$ and a binding energy of $E_b$. The energy difference between the initial and final states suggests some form of "radiation" of binding energy $E_b$ that is produced during the formation of the bound state. The properties and nature of this "radiation" were discussed in Section \ref{sec:6-1}. The gravitational binding energy $E_b$ reads
\begin{equation}
\begin{split}
&E_b=m_X v_X^2 = \frac{G^2m_X^5}{16\hbar^2} = \frac{\left(\hbar/r_X\right)^2}{m_X} =\frac{\pi\hbar}{\sqrt{2}t_X}. \\
\end{split}
\label{eq:5-33-11}
\end{equation}
Note that the binding energy and the formation time satisfy the "uncertainty principle" of $E_bt_X\sim \hbar$. 

Similarly, we can extend the same analysis to find the properties of the smallest and the earliest structure that can be formed by particles with a mass $m_X\ll m_{Xc}$ such that the mass of that structure (the free streaming mass) $M_{fs}\gg m_X$. On the free streaming scale, we write the virial equilibrium and the pressure balance (similar to Eqs. \eqref{eq:5-2-7} and \eqref{eq:5-2-9} for two-particle structures):
\begin{equation}
\begin{split}
&v_{fs}^2=\frac{GM_{fs}}{4r_{fs}}, \quad m_Xv_{fs} r_{DM} = \hbar, \quad r_{DM}={r_{fs}}\left(\frac{m_X}{M_{fs}}\right)^{\frac{1}{3}}
\end{split}
\label{eq:5-2-9-9}
\end{equation}
where $r_{DM}$ is the average particle spacing. Here, $r_{fs}$ and $v_{fs}$ are the size and characteristic particle velocity of the smallest and earliest gravitationally bound structure that these particles can form. Solving these equations yields
\begin{equation}
\begin{split}
&r_{fs}=\left(\frac{2m_X}{M_{fs}}\right)^{1/3}\frac{4\hbar^2}{G(m_{X})^3}, \quad v_{fs}=\left(\frac{2m_X}{M_{fs}}\right)^{-2/3}\frac{G(m_{X})^2}{4\hbar}, \\
&\rho_{fs}= \frac{M_{fs}}{\frac{4}{3}\pi r_{fs}^3}=\frac{3}{2^{7}\pi}\left(\frac{2m_X}{M_{fs}}\right)^{-2}\frac{G^3m_X^{10}}{\hbar^6}=32\pi^2 \bar\rho(t_{fs}), \\
&t_{fs}=\left(\frac{2m_X}{M_{fs}}\right)\frac{2^{7/2}\pi\hbar^3}{G^2(m_{X})^5}, \\
&E_{fs}=\frac{1}{2}M_{fs} v_{fs}^2 = \left(\frac{2m_X}{M_{fs}}\right)^{-7/3}\frac{G^2m_X^5}{16\hbar^2}.
\end{split}
\label{eq:5-2-10-22}
\end{equation}
Here, $t_{fs}$ is the formation time, $\rho_{fs}$ and $\bar{\rho}_{fs}$ are the density of the structure and the background at the time of formation, $E_{fs}$ is the binding energy from structure formation. With $M_{fs}=2m_X$, Eq. \eqref{eq:5-2-10-22} reduces to results for the two-particle bound structure (Eq. \eqref{eq:5-2-10-2}). Table \ref{tab:55} presents the relevant properties of the smallest structure formed by three dark matter candidates. Particles with critical mass $10^{12}$GeV can form the smallest and earliest structure among all candidates, as well as a smooth transition in structure size (Fig. \ref{fig:S33}).

In this section, we focus on the smallest gravitationally bound structure that fermion dark matter particles can form. The balance between degeneracy and gravitational pressure determines all relevant properties of these structures. In the next section, we revisit the WIMP miracle, followed by adapting the WIMP miracle to nonthermal particles involving gravity, i.e., an "X miracle", which predicts the same particle mass $10^{12}$GeV independently as that from the free streaming scale calculation (Eq. \eqref{eq:12-5-3}).

\begin{table}
    \begin{center}
    \caption{The smallest structure formed by different DM candidates}
    \label{tab:55}
    \begin{tabular}{lccccc} 
    \hline
    Quantity                           & Hot neutrino           & WIMPs                  & Superheavy X     \\
    \hline
    Particle mass  ($m_X$)             & 30 eV                  & 100 GeV                & $10^{12}$ GeV \\
    Structure mass ($M_{fs}$)          & $10^{15} M_{\odot}$    & $10^{-6} M_{\odot}$    & $10^{12}$ GeV              \\
    Mass ratio ($M_{fs}/m_X$)          & $10^{79}$              & $10^{49}$              & 1                \\
    Structure size ($r_{fs}$)          & 0.5 kpc                & 6.6 m                  & $10^{-13}$ m            \\
    Velocity ($v_{fs}$) & $10^7$ m/s   & $10^6$ m/s             &  $10^{-7}$ m/s   \\
    Formation time ($t_{fs}$)          & $4\times 10^4$ years   & $10^{-5}$ s            &  $10^{-6}$ s      \\
    \hline
    \end{tabular}
  \end{center}
\end{table}

\section{Cold freeze out and X miracle}

\subsection{Thermal and gravity-induced velocities}
\label{sec:5-1-2-2-2}
\begin{figure}
\includegraphics*[width=\columnwidth]{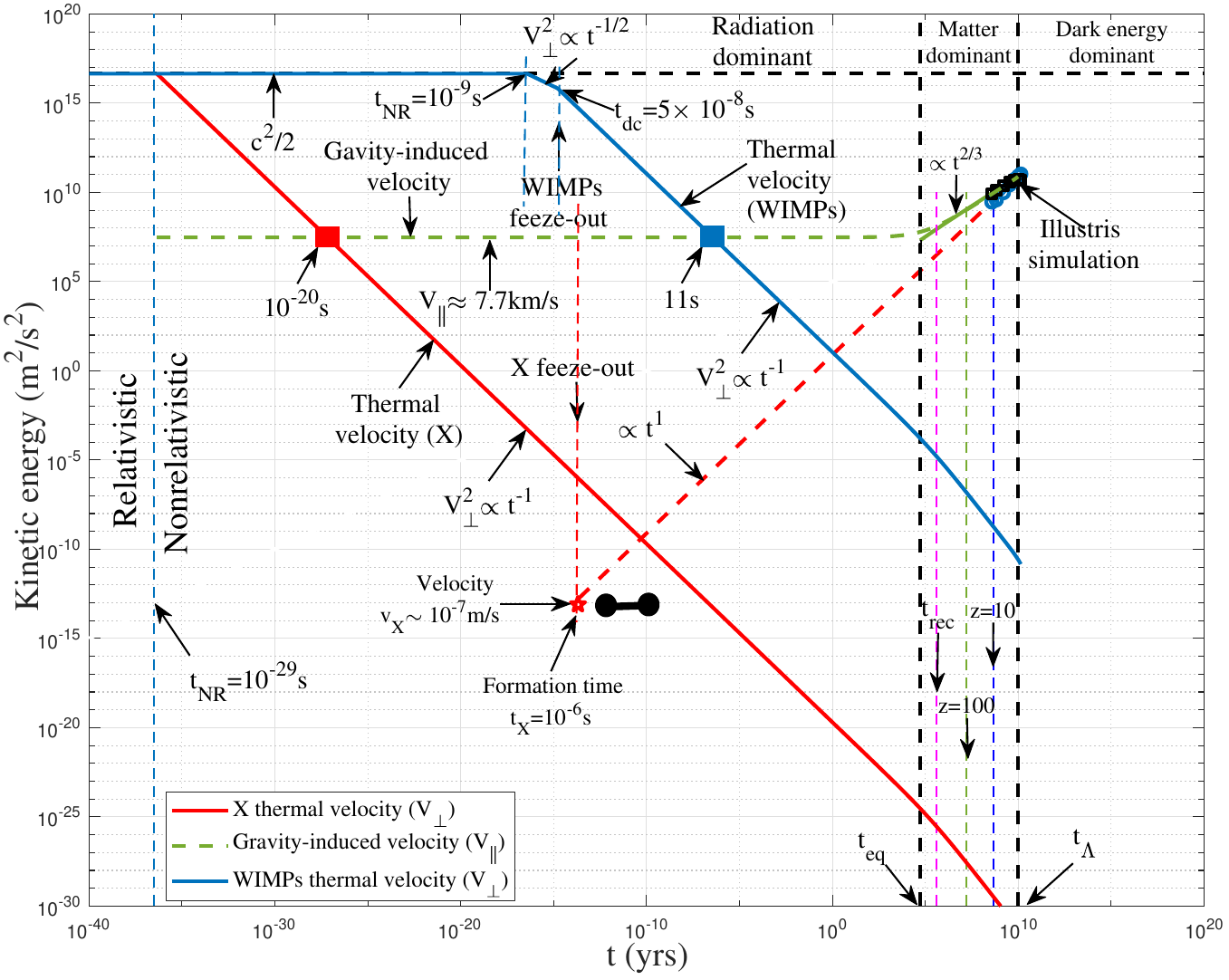}
\caption{Different velocity evolution in the early Universe for WIMP particles of 100 GeV (red) and superheavy X particles of $10^{12}$ GeV (blue), along with the constant gravity-induced velocity (dashed green). WIMP particles become nonrelativistic much later at $10^{-9}$s and are coupled to and in equilibrium with the thermal bath. They are decoupled and freeze out from the thermal bath at $5\times 10^{-8}$s (temperature 4 GeV). WIMP particles have a speed close to the speed of light $c$ before freeze-out, which is still dominated by their thermal velocity. Their speed becomes dominated by the gravity-induced velocity $V_{||}$ (dashed green) much later, only after 11s (blue square). For comparison, superheavy X particles become nonrelativistic much earlier at $10^{-29}$s and remain decoupled because of their lower number density and weak gravitational binding. Their speed becomes dominated by the gravity-induced velocity $V_{||}$ at $10^{-20}$s (red square), much earlier than WIMPs. X particles remain decoupled at this low velocity until the formation time $t_{X}$ of $10^{-6}$s (temperature 1 GeV) to form the smallest gravitationally bound state (red star).} 
\label{fig:1111}
\end{figure}

We first recognize the peculiar velocity of dark matter particles in the early Universe consisting of a parallel component $V_{||}$ due to the gravity of density perturbations on large scales and a perpendicular component $V_{\perp}$, i.e., the thermal velocity that is not induced by gravity. For the perpendicular component, the comoving velocity $\textbf{u}_{\perp}=V_{\perp}/a$ satisfies $\dot{\textbf{u}}_{\perp} +2H \textbf{u}_{\perp}=0$, so it decreases rapidly as $\textbf{u}_{\perp}\propto a^{-2}$ and $V_{\perp}\propto a^{-1}$. This is the thermal velocity that contributes to the vorticity. The thermal velocity $V_{\perp}$ decreases much more rapidly over time compared to the gravity-induced component $V_{||}$ (Fig. \ref{fig:1111}). Therefore, it is generally a good approximation to treat the large-scale dark matter velocity as being curl-free with a vanishing vorticity \citep{Xu:2023-On-the-statistical-theory-of-self-gravitating,Xu:2024-High-order-kinematic,Xu:2024-Scale-and-redshift-variation}. 

To compute the gravity-induced velocity $V_{||}$, we start with the comoving velocity $\textbf{u}_{||}=V_{||}/a$ that is coupled to dark matter density perturbations $\delta$. The linearized perturbation equations read
\begin{equation}
\begin{split}
&\dot \delta + \nabla\cdot \textbf{u}_{||}=0, \quad \textrm{(continuity equation)}\\
&\dot{\textbf{u}}_{||} +2H \textbf{u}_{||}=-\frac{1}{a^2}\nabla\delta\Phi, \quad \textrm{(momentum conservation)}\\
&\nabla^2 \delta\Phi = 4\pi G\rho_{DM}a^2\delta, \quad \textrm{(Poisson equation)} \\
\end{split}
\label{eq:5-2-2}
\end{equation}
where $\delta\Phi$ is the potential perturbation. The gravity-induced component can be expressed as the gradient of potential perturbation $\delta\Phi$, i.e., $\textbf{u}_{||}\propto -\nabla \delta\Phi$ (Zeldovich approximation). Substituting into Eq. \eqref{eq:5-2-2}, density perturbation $\delta$ and comoving velocity $\textbf{u}_{||}$ reads
\begin{equation}
\begin{split}
&\ddot{\delta} + 2H \dot{\delta} = 4\pi G\rho_{DM}\delta, \\
&\dot{\textbf{u}}_{||} +2H \textbf{u}_{||}=4\pi G\rho_{DM}\frac{\delta}{\dot\delta} \textbf{u}_{||}. \\
\end{split}
\label{eq:5-2-3}
\end{equation}
These equations are valid in both radiation and matter eras, where the Hubble parameter satisfies
\begin{equation}
\begin{split}
&H^2= \frac{8}{3}\pi G\rho_{DM}\left(1+\frac{a_{eq}}{a}\right),\quad  \dot{H}=-\frac{H^2}{2}\frac{3+4a_{eq}/a}{1+{a_{eq}}/{a}}.
\end{split}
\label{eq:5-2-3-2}
\end{equation}

If gravity is neglected on large scales (vanishing right-hand-side terms in Eq. \eqref{eq:5-2-3}), we find the logarithmic growth for density perturbation $\delta\propto \ln t$ and fast decaying of $\textbf{u}_{||}\propto a^{-2}$ in the radiation era, same as the thermal velocity $\textbf{u}_{\perp}\propto a^{-2}$. However, with gravity, a well-known exact solution of $\delta$ can be obtained from Eq. \eqref{eq:5-2-3} 
\begin{equation}
\begin{split}
&{\delta} = 2B/3+B(a/a_{eq}),
\end{split}
\label{eq:5-2-4}
\end{equation}
where $B$ is a constant. We therefore find that in the radiation era with $a\ll a_{eq}$, density perturbations are frozen and do not grow in the linear regime on large scales, i.e., the Meszaros effect \citep{Meszaros:1974-The-behaviour-of-point-masses-in-an-expanding-cosmological}. In the matter era with $a\gg a_{eq}$, perturbations grow as $\delta\propto a$. 

Combining Eqs. \eqref{eq:5-2-4} and \eqref{eq:5-2-3-2}, the right-hand-side of Eq. \eqref{eq:5-2-3} for $\textbf{u}_{||}$ reads
\begin{equation}
\begin{split}
&4\pi G\rho_{DM}\frac{\delta}{\dot{\delta}}=\frac{1}{2}\frac{2+3a/a_{eq}}{1+a/a_{eq}}H.\\
\end{split}
\label{eq:5-2-5-1}
\end{equation}
Substituting this into Eq. \eqref{eq:5-2-3} for $\textbf{u}_{||}$, the equation for $\textbf{u}_{||}$ reads
\begin{equation}
\begin{split}
&\dot {(\textbf{u}_{||}^2)}+H\textbf{u}_{||}^2\frac{2+a/a_{eq}}{1+a/a_{eq}}=0.
\end{split}
\label{eq:5-2-5-2}
\end{equation}
We obtain an exact solution for gravity-induced velocities
\begin{equation}
\begin{split}
&\textbf{u}_{||}^2 = C(1/a+1/a_{eq})/a, \\
&{V}_{||}^2 =\textbf{u}_{||}^2 a^2= C(1+a/a_{eq}), \\
\end{split}
\label{eq:5-2-5}
\end{equation}
where $C$ is a constant. Therefore, on large scales and in the radiation era with $a\ll a_{eq}$, there also exists a nonzero frozen constant velocity ${V}_{||}^2=C$ in parallel to the frozen density perturbations $\delta$=2B/3. This result is important because gravity-induced velocity ${V}_{||}$ is directly relevant to cross sections of nonthermal relics (Section \ref{sec:5-1-2}).

In the matter era, the velocity dispersion $V_{||}^2\propto a$ increases with time (Eq. \eqref{eq:5-2-5}). With $V_{||}$ on the order of several hundred km/s at the current epoch (450 km/s from the Virgo and Illustris simulations \citep{Xu:2023-Maximum-entropy-distributions-of-dark-matter,Xu:2023-Dark-matter-halo-mass-functions-and}) and the scale factor $a_{eq}\approx 1/3400$, we estimate the constant $C=6\times 10^7$m$^2$/s$^2$ such that a constant gravity-induced velocity ${V}_{||}\approx 7.7$ km/s is expected. The actual value may vary. However, a gravity-induced velocity on the km/s scale ${V}_{||}$ should be a robust estimate, which is also close to the sound speed (6 km/s) of baryons right after the time of recombination \citep{Tseliakhovich:2010-Relative-velocity-of-dark-matter-and-baryonic-fluids}. The gravity-induced velocity ${V}_{||}$ is independent of the particle mass $m_X$ and the time $t$ in the radiation era (Fig. \ref{fig:1111}). This velocity should be the same for different types of particle dark matter and be independent of the dark matter model. 

Figure \ref{fig:1111} plots the evolution of the velocity for WIMPs and superheavy X particles of mass $10^{12}$GeV. Low-mass particles like WIMPs become nonrelativistic much later, with their thermal velocity $V_{\perp}$ (solid blue) dominant over the gravity-induced velocity $V_{||}$ (dashed green) at freeze-out. In this case, the relative velocity of WIMPs is determined by the thermal component and is close to 0.3$c$, indicating a "hot" freeze-out at a high speed. However, heavy particles become nonrelativistic much earlier, with their thermal velocity (solid red) decreasing rapidly. Therefore, for heavy particles, the relative velocity is determined by $V_{||}$ (dashed green), which is much lower than the speed of light. The freeze-out at this low velocity is therefore "cold".

In the WIMP miracle, the gravitational interaction between individual WIMPs is neglected at freeze-out due to fast free streaming. By contrast, gravity between individual X particles is important: 1) on large scales, these particles become nonrelativistic much earlier, with their thermal velocity $V_{\perp}$ redshifted to a much smaller value than $V_{||}$ before freeze-out such that the gravity-induced velocity $V_{||}$ is dominant at freeze-out (Fig. \ref{fig:1111}); 2) on small scales, when two particles approach a scale comparable to the particle's size (de Broglie wavelength), $r_X=\hbar/(m_Xv_X)$, both gravity and quantum effects can be important (Eq. \eqref{eq:5-2-10-2}), leading to a different small-scale velocity, $v_X\ll V_{||}\ll c$. In this scenario, gravity between X particles introduces a new scale $r_X\gg \hbar/(m_Xc)$ and affects the cross section.

\subsection{The X miracle for nonthermal heavy relics}
\label{sec:5-1-2}
For completeness, Section \ref{sec:5-1-2-2} in the appendix revisits the WIMP miracle and assumptions. Similarly, in this section, we develop an X miracle for nonthermal relics with an initial overabundance $\rho_{ini}$ (Fig. \ref{fig:666}). The "cold" freeze-out results in the final relic abundance $\rho_{\infty}$. We will predict both the particle mass and the density ratio $\gamma=\rho_{\infty}/\rho_{ini}$. 

Similarly to WIMPs (Eq. \eqref{eq:5-22-112}), we focus on the reaction 
\begin{equation}
\begin{split}
&X+X \xrightarrow{1} X\cdot X \xrightarrow{2} \textrm{Dark Radiation} \xdashrightarrow {3} X+X,
\end{split}
\label{eq:5-22-111}
\end{equation}
where step 1 represents the formation of the smallest gravitationally bound state. Step 2 describes the annihilation (or decay) of X particles. The two-body bound states $X\cdot X$ have particle wavefunctions that overlap and typically decay (annihilate) much faster than the unbound pair. For particle annihilation or decay into dark radiation, it may require new interactions in the dark sector. Section \ref{sec:5-1-2-34} discusses an instanton-induced decay model, which requires some hidden gauge interaction. Both annihilation or decay could result in either dark sector particles (dark radiation) or SM particles (or both) that may potentially contribute to Hubble tension and Ultra-high Energy Cosmic Rays (UHECRs) (Section \ref{sec:5-1-2-33}). Finally, unlike WIMPs, there is no reverse reaction for nonthermal relics (step 3) due to Boltzmann-suppressed equilibrium density. Therefore, the annihilation/decay during the cold freeze-out is out of equilibrium. 

With velocities on large and small scales ($V_{||}$ in Eq. \eqref{eq:5-2-5} and $v_X$ in Eq. \eqref{eq:5-2-10-2}), three relevant cross sections can be defined:
\begin{equation}
\begin{split}
&\langle\sigma_{X} V_{X}\rangle = \pi r_{X}^2v_{X} = \left(\frac{\hbar}{m_Xv_{X}}\right)^2v_{X}=\frac{4\pi\hbar^3}{Gm_X^4}.\\
&\langle\sigma_{||} V_{||}\rangle = \pi r_{||}^2V_{||} = \left(\frac{\hbar}{m_XV_{||}}\right)^2V_{||} \quad \textrm{for $t<t_X$},\\
&\langle\sigma_{X} V_{||}\rangle = \pi r_{X}^2V_{||} = \left(\frac{\hbar}{m_Xv_{X}}\right)^2V_{||}=\frac{16\pi\hbar^4}{G^2m_X^6}V_{||}\quad \textrm{for $t\ge t_X$}.\\
\end{split}
\label{eq:5-33-10-111}
\end{equation}
All cross sections involve the gravitational constant $G$ due to gravity between individual X particles.

\begin{figure}
\includegraphics*[width=\columnwidth]{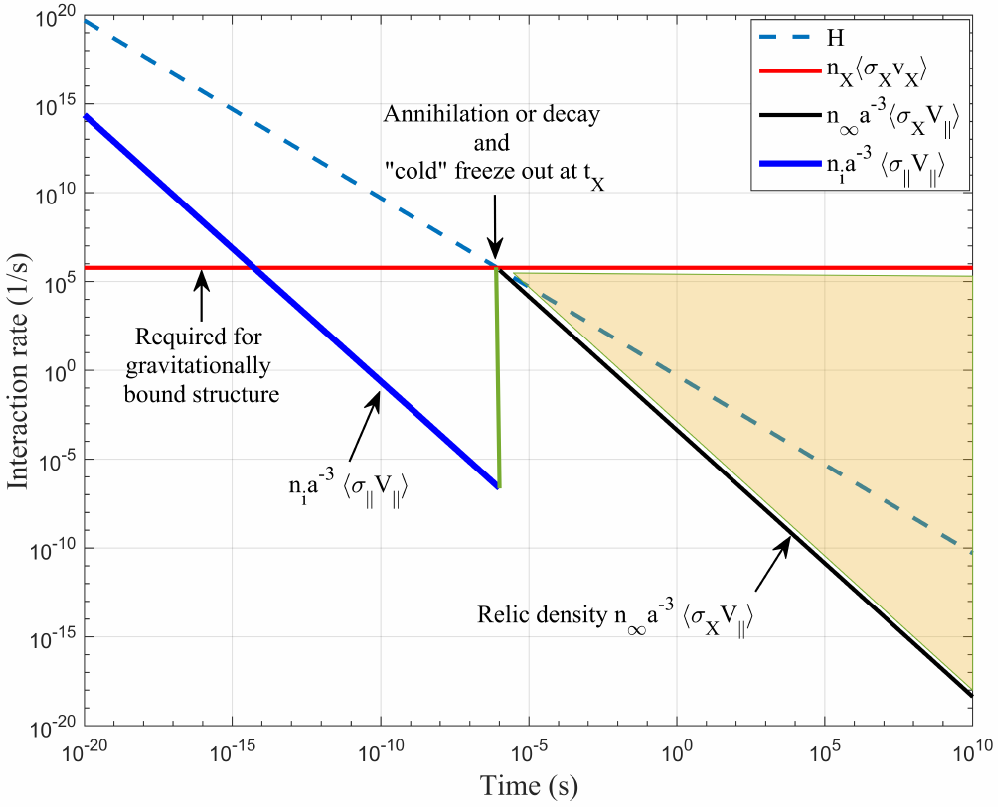}
\caption{The evolution of expansion rate $H$ (dashed blue) and interaction rates from three cross sections. For superheavy X particles with an initial overabundance $n_i$, the expansion rate $H$ is greater than both interaction rates $n_X\langle\sigma_X v_{X}\rangle$ (solid red) and $n_i\langle\sigma_{||} V_{||}\rangle$ (solid blue), such that both structure formation and particle annihilation/decay are suppressed until a critical time $t_X$. After $t_X$, the interaction rate with relic density $n_{\infty}\langle\sigma_X V_{||}\rangle$ (solid black) is below $H$ such that X particles are too dilute to find each other, i.e., X particles freeze out. Note that $n_X\langle\sigma_X v_{X}\rangle$ (red) is constant in time due to constant density $n_X$ for the smallest gravitationally bound structure (Eq. \eqref{eq:5-2-9-2}).} 
\label{fig:5555}
\end{figure}

Figure \ref{fig:5555} presents a plot for the evolution of the expansion rate $H$ (dashed blue) and interaction rates from three cross sections (solid lines). For "cold" freeze-out, the interplay between these interaction rates and expansion determines the evolution of particle density.
\begin{enumerate}
\item \noindent Initial stage before freeze-out. Interaction rates satisfy
\begin{equation}
\begin{split}
& H > n_{X}\langle\sigma_X v_X\rangle > n_ia^{-3} \langle\sigma_{||}V_{||}\rangle,
\end{split}
\label{eq:5-33-10-101}
\end{equation}
where $a$ is the scale factor, $n_i$ is the initial comoving number density, and $n_X$ is the constant number density in the smallest bound structure (Eq. \eqref{eq:5-2-8}). The cross section $\langle\sigma_X v_X\rangle$ represents the requirement to form a gravitationally bound state (solid red). The bound state cannot be formed before freeze-out due to fast expansion or $H\gg n_X\langle \sigma_Xv_X\rangle \sim v_X/r_X$ (Eq. \eqref{eq:5-2-8-33}). Before freeze-out, particle interaction is determined by the cross section $\langle\sigma_{||} V_{||}\rangle$ (solid blue), which depends on the gravity-induced velocity $V_{||}$ and associated de Broglie wavelength $r_{||}$ (Eq. \eqref{eq:5-33-10-111}). Therefore, particle annihilation/decay is also suppressed. 
\newline

\item \noindent Freeze-out stage. When the expansion is comparable to both interaction rates (around time $t_X$ or $a_X$), the formation of the smallest gravitationally bound state is permitted, as well as the particle annihilation/decay (solid green). Significant particle annihilation/decay occurs around this moment. Similar to the WIMP miracle (Eq. \eqref{eq:5-33-2}), the "cold" freeze-out condition reads \begin{equation}
\begin{split}
&n_{\infty}a_X^{-3}\langle\sigma_X V_{||}\rangle \sim H \sim n_{X}\langle\sigma_X v_X\rangle,
\end{split}
\label{eq:5-33-10-102}
\end{equation}
where $n_{\infty}$ is the (comoving) relic number density. 
\newline

\item \noindent Structure evolution stage. After cold freeze-out, the X particles approach a constant relic density $n_{\infty}$ when expansion is dominant over the interaction rates (solid black),
\begin{equation}
\begin{split}
& n_{\infty}a^{-3}\langle\sigma_X V_{||}\rangle <  H < n_X\langle\sigma_X v_X\rangle.
\end{split}
\label{eq:5-33-10-103}
\end{equation}
During this stage, gravitationally bound state formation is permitted. Here $\langle\sigma_X V_{||}\rangle$ (black line) represents the cross section on the largest scale, the average cross section required for particle annihilation/decay. Here, $\langle \sigma_Xv_X\rangle$ (solid red) represents the cross section on the smallest scale. The shaded yellow represents the cross sections on intermediate scales. However, X particles are too diluted to find each other, form the smallest bound structure, and annihilate/decay. Structures begin to form on different scales as haloes of various sizes, and the cross-section also becomes scale-dependent. 
\end{enumerate}

The entire "cold" freeze-out for X particles can be described by a nonequilibrium Boltzmann equation in analogy to the Boltzmann equation for thermal WIMPs (Eq. \eqref{eq:5-33-1}),
\begin{equation}
\begin{split}
&\frac{dn}{dt}=-3Hn-\langle\sigma_X V_{||}\rangle n^2, \quad n(t_{i})a_i^{3}=n_{i},
\end{split}
\label{eq:5-22-11}
\end{equation}
where $V_{||}$ is the gravity-induced velocity of km/s. On the right-hand side of Eq. \eqref{eq:5-22-11}, the first term accounts for the dilution from expansion, while the second term accounts for the annihilation/decay when two particles meet. The equilibrium number density $n_{eq}$ is neglected, reflecting the nonthermal nature. To solve Eq. \eqref{eq:5-22-11}, an initial number density $n_i$ and time $t_i$ (or scale factor $a_i$) are required.

Next, we present a simple analysis to derive the particle mass, similar to the WIMP miracle (Eq. \eqref{eq:5-33-2}), followed by a more rigorous analysis based on the exact solution of the Boltzmann equation (Eq. \eqref{eq:5-22-11}). When X particles freeze out, we should have (from Eq. \eqref{eq:5-33-10-102})
\begin{equation}
\begin{split}
&n_f\langle\sigma_X V_{||}\rangle \sim H_X \sim \frac{1}{t_X}\quad \textrm{and} \quad n_f = \frac{\bar\rho_{0}\Omega_{DM}}{m_Xa_X^3},\\
\end{split}
\label{eq:5-22-12}
\end{equation}
where $n_f=n_{\infty}a_X^{-3}$ is the freeze-out number density of $X$ particles at time $t_X$, $\bar\rho_{0}$ is the present critical density of the Universe, $\Omega_{DM}$ is the fraction of dark matter, and $a_X$ is the scale factor at time $t_X$. The relevant cross section becomes
\begin{equation}
\begin{split}
&\langle\sigma_X V_{||}\rangle \sim \frac{1}{n_ft_X}\sim \frac{m_X}{\bar\rho_{0}\Omega_{DM}}\frac{a_X^3}{t_X},
\end{split}
\label{eq:5-22-13}
\end{equation}
which depends on $\Omega_{DM}$ and the particle mass $m_X$. 

\begin{figure}
\includegraphics*[width=\columnwidth]{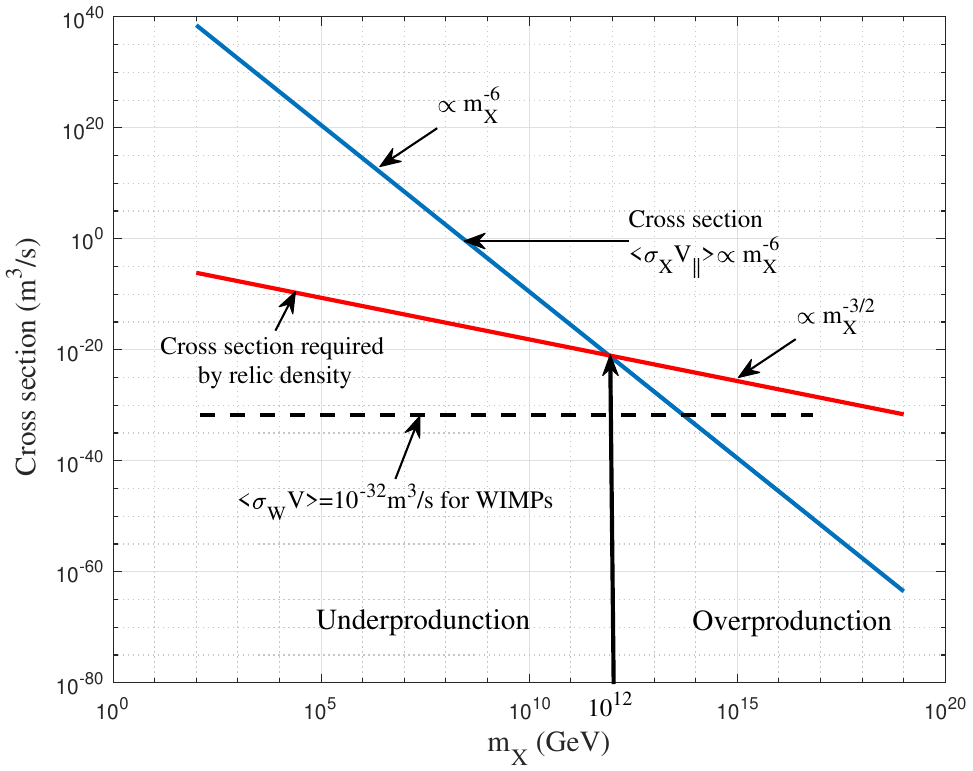}
\caption{The variation of cross section with particle mass $m_X$ for nonthermal relics (solid blue for LHS of Eq. \eqref{eq:5-22-13-2}). The red line plots the cross section required by the relic density (RHS of Eq. \eqref{eq:5-22-13-2}). By contrast, the cross section required for WIMPs (dashed line) is independent of the particle mass because of the constant $x_f$ (Eq. \eqref{eq:5-33-4-4}). By solving Eq. \eqref{eq:5-22-13-2}, particle mass $m_X=10^{12}$GeV is identified, which yields the right cross section required by the relic density, i.e., the X miracle. Particles lighter or heavier lead to under- or over-production of dark matter.}
\label{fig:4444-2}
\end{figure}

Similarly to the WIMP miracle for thermal relics (Eq. \eqref{eq:5-33-4-4}), we may relate the cross section to a dimensionless number $x_f^*$,
\begin{equation}
\begin{split}
&\langle\sigma_X V_{||}\rangle \sim {\frac{k_B T_{\gamma0}}{c^2M_{pl}}} \cdot {\frac{x_f^* k_B T_{\gamma0}}{c^2\bar\rho_{0}\Omega_{DM}}\cdot\frac{k_B T_{\gamma 0}}{\hbar}},\quad x_f^* = \frac{m_Xc^2}{k_BT_{X}},
\end{split}
\label{eq:5-22-13-2}
\end{equation}
where $T_X$ is the temperature at which X particles freeze out. Different from the WIMP miracle for thermal relics, the parameter $x_f^*$ is not a constant of particle mass for nonthermal relics. Instead, since $T_X\propto a_X^{-1}\propto t_X^{-1/2}\propto m_X^{5/2}$ from Eq. \eqref{eq:5-2-11}, we find $x_f^*\propto m_X^{-3/2}$ (Fig. \eqref{fig:4444-2}). We can naturally relate the cross-section $\langle\sigma_X V_{||}\rangle$ to the fundamental length scale $r_X$ and gravity-induced velocity $V_{||}$ (Eq. \eqref{eq:5-33-10-111}). Here, we provide expressions for all relevant quantities, 
\begin{equation}
\begin{split}
&\langle\sigma_X V_{||}\rangle=\pi r_X^2 V_{||}, \quad \bar\rho_{0}=\frac{3H_0^2}{8\pi G}, \quad t_X = \frac{a_X^2}{2H_0\sqrt{\Omega_{rad}}}.
\end{split}
\label{eq:5-22-14}
\end{equation}
The large-scale gravity-induced velocity $V_{||}$ is of km/s (Eq. \eqref{eq:5-2-5}). 

Figure \ref{fig:4444-2} plots the cross section variation with particle mass $m_X$ for nonthermal relics (solid blue line for LHS of Eq. \eqref{eq:5-22-13-2}). The cross section required by the relic density (RHS of Eq. \eqref{eq:5-22-13-2}) is presented as the solid red line. By contrast, the cross section required for thermal WIMPs is independent of the particle mass because of the constant $x_f$ (Eq. \eqref{eq:5-33-4-4}). By solving Eq. \eqref{eq:5-22-13-2}, mass $m_X=10^{12}$GeV is identified, which yields the right cross section required by the relic density. This is the X miracle: the cross section needed for the relic density happens to be the cross section because of the gravitational interaction between particles. Lighter or heavier particles lead to under- or over-production of dark matter. 

The explicit solution can be obtained by substituting the formation time $t_X$ and particle size $r_X$ (Eqs. \eqref{eq:5-2-11}, \eqref{eq:5-2-10-2}, and \eqref{eq:5-22-14}) into the cross section requirement (Eq. \eqref{eq:5-22-13}),
\begin{equation}
\begin{split}
&m_{X}=\left(\frac{9\cdot 2^{3/2}}{64\pi}\frac{H_0\Omega_{DM}^2}{\Omega_{rad}^{3/2}}\frac{V_{||}^2\hbar^5}{G^4}\right)^{1/9}\approx m_{Xc} \sim 10^{12}GeV. 
\end{split}
\label{eq:5-22-15}
\end{equation}
Here, the dark matter fraction $\Omega_{DM}=0.12/h^2$, the radiation fraction $\Omega_{rad}=4.2\times 10^{-5}/h^2$, and $h=H_0$/(100km/s/Mpc) is the dimensionless Hubble constant. The only uncertainty comes from the velocity $V_{||}$. However, this result is insensitive to the velocity value $V_{||}$ as $m_X\propto V_{||}^{2/9}$. For example, the difference in $m_X$ is only around 4.6 for a 1000 difference in velocity $V_{||}$. Interestingly, this simple analysis yields the same particle mass as that obtained from the structure formation at the free streaming scale (Eq. \eqref{eq:12-5-3}). Two independent approaches predict the same particle mass. 

Next, we develop a more rigorous and detailed analysis and determine the initial overabundance $n_i$. The exact solution of nonequilibrium Boltzmann Eq. \eqref{eq:5-22-11} for the number density $n(t)$ or the comoving density $na^3$ reads
\begin{equation}
\begin{split}
&n(t) = \frac{n_{\infty}\left(2H_0\sqrt{\Omega_{rad}}\right)^{-\frac{3}{2}}}{t^{\frac{3}{2}}-\left(1-\gamma\right)t_i^{\frac{1}{2}}t}\quad \textrm{or} \quad \frac{na^3}{n_{\infty}} = \frac{1}{1-(1-\gamma)\frac{a_i}{a}},
\end{split}
\label{eq:5-22-220}
\end{equation}
where $\gamma$ is a parameter for the ratio of the relic to the initial density,
\begin{equation}
\begin{split}
\gamma=\frac{n_{\infty}}{n_i}=\frac{1}{1+2\langle\sigma_X V_{||}\rangle n_it_ia_i^{-3}}=\frac{n_p-\bar n_p}{n_p+\bar n_p}.
\end{split}
\label{eq:5-22-230}
\end{equation}
In this solution, the initial time $t_i$ and the parameter $\gamma$ are the only two unknowns. The physical meaning of $\gamma$ could be related to the asymmetry in X particle annihilation. For fermion dark matter, let us consider X particles of number density $n_p$ and anti-X particles of number density $\bar n_p$, such that the total initial density $n_i=n_p+\bar n_p$. The annihilation leads to a relic density of $n_{\infty}=(n_p-\bar n_p)$. In this case, parameter $\gamma$ represents the asymmetry of annihilation (or decay).

\begin{figure}
\includegraphics*[width=\columnwidth]{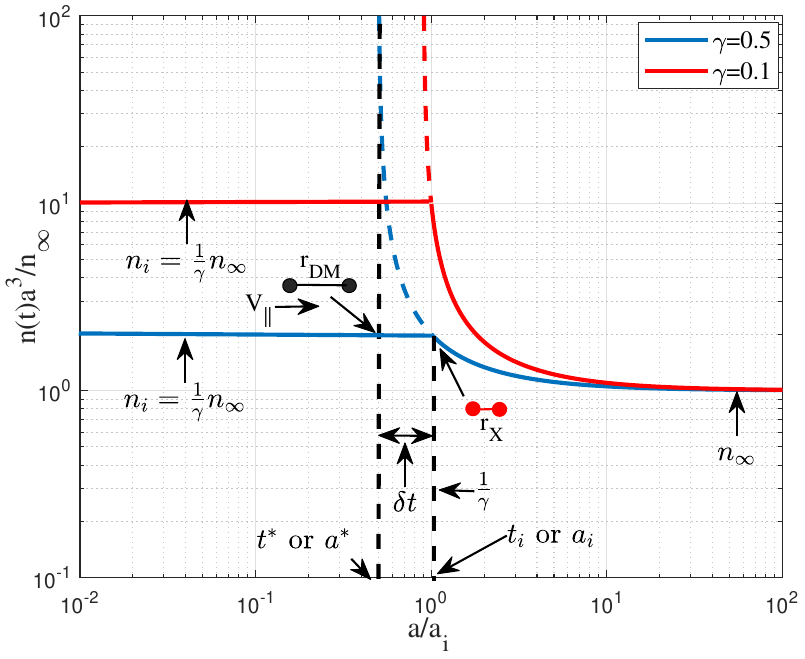}
\caption{The exact solution \eqref{eq:5-22-220} for the nonequilibrium Boltzmann equation \eqref{eq:5-22-11} for the evolution of X particle number density with scale factor $a$ (or time) for two different $\gamma$= 0.1 and 0.5. A smaller $\gamma$ leads to a higher initial density. There are two important times, $t^*$ and $t_i$, for density evolution. The time $t^*$ represents the critical time before which the smallest bound structures cannot be formed due to fast expansion, such that the comoving number density remains constant. After $t^*$, the smallest bound structures of size $r_X$ can be formed. However, there is a slight delay of a minimum time $\delta t$ due to the discrete nature of particles. The bound structures are actually formed, and particle annihilation begins at a slightly later time $t_i=t^*+\delta t$. The short delay $\delta t=r_{DM}/V_{||}$ is due to the time for particles traveling a distance of average spacing $r_{DM}\gg r_X$, from which we are able to calculate the initial number density $n_i$. Density evolution with $\gamma=10^{-9}$ is also plotted in Fig. \ref{fig:2222} for comparison to the evolution of WIMPs.}
\label{fig:4444}
\end{figure}

Figure \ref{fig:4444} plots the exact solution (Eq. \eqref{eq:5-22-220}) for two different $\gamma$. An important time $t^*$ naturally emerges from Eq. \eqref{eq:5-22-220} by requiring a vanishing denominator (using Eq. \eqref{eq:5-22-230}),
\begin{equation}
\begin{split}
&t^* = (1-\gamma)^2t_i=\frac{\langle\sigma_X V_{||}\rangle^2 n_{\infty}^2}{2(H_0\sqrt{\Omega_{rad}})^3}.
\end{split}
\label{eq:5-22-221}
\end{equation}
There was no physical solution before time $t^*$. The physical meaning of time $t^*$ can be related to the formation time $t_X$ (Eq. \eqref{eq:5-2-11}) of the smallest gravitationally bound structure. Before time $t^*$, structure formation and particle annihilation/decay are prohibited such that X particles have a constant comoving number density $n_i$. By setting two time scales $t^*=t_X$, the same particle mass in Eq. \eqref{eq:5-22-15} can be obtained and will not be repeated here. 

The time $t_i$ represents the actual formation time of the first bound structure and the beginning of particle annihilation. There is a slight delay between the two time scales, i.e., $t_i=t^*+\delta t$. In principle, the Boltzmann equation \eqref{eq:5-22-11} is a continuum description of the discrete particle system. The time delay $\delta t$ is due to the discrete nature of the particle system. This is the time required by particles traveling the average spacing $r_{DM}$ at time $t^*$ (or $a_X$). Since $\gamma\propto \delta t/t^*$ (Eq. \eqref{eq:5-22-221}), to determine the parameter $\gamma$, we need to calculate the time delay $\delta t$, 
\begin{equation}
\begin{split}
&\delta t= t_i-t^*\propto \frac{r_{DM}(t=t^*)}{V_{||}} \propto \frac{n_i^{-1/3}a_X}{V_{||}}, \\
&\gamma=\frac{n_{\infty}}{n_i}=1-\sqrt{\frac{t^*}{t_i}}\approx \frac{\delta t}{2t^*}\propto \frac{n_i^{-1/3}a_X}{2V_{||}t^*}=\frac{n_{\infty}^{-1/3}\gamma^{1/3}a_X}{2V_{||}t_X}.
\end{split}
\label{eq:5-22-222}
\end{equation}
Here, we use Eq. \eqref{eq:5-22-221} and assume $\gamma\ll 1$ for the approximation. 

Substituting Eqs. \eqref{eq:5-22-14} and \eqref{eq:5-2-11}, the final expression for $\gamma$ is
\begin{equation}
\begin{split}
&\gamma \propto \left(\frac{G^{16}m_X^{34}}{\hbar^{18}H_0^2V_{||}^{12}}\cdot\frac{\Omega_{rad}^3}{\Omega_{DM}^4}\right)^{1/8} \approx 10^{-9} \quad \textrm{to} \quad  10^{-8}.
\end{split}
\label{eq:5-22-2223}
\end{equation}
The exact value of $\gamma$ depends on the particle mass $m_X$ and the gravity-induced velocity as $\gamma\propto V_{||}^{-3/2}$. For $m_X=10^{12}$GeV and a km/s $V_{||}$, $\delta t\sim 10^{-15}$s and $\gamma$ is on the order of $10^{-9}$ to $10^{-8}$. Therefore, a significant amount of annihilation/decay takes place around time $t_X$, reducing the dark matter density from its initial value $n_i$ to its relic value of $n_{\infty}=\gamma n_i$, with only one particle in a billion surviving as relic dark matter ("big depletion"). Particle physics models may be developed to describe this process in a separate paper. The asymmetry parameter $\gamma$ for dark matter annihilation (or decay) could potentially be related to leptogenesis and baryogenesis. 

\begin{figure}
\includegraphics*[width=\columnwidth]{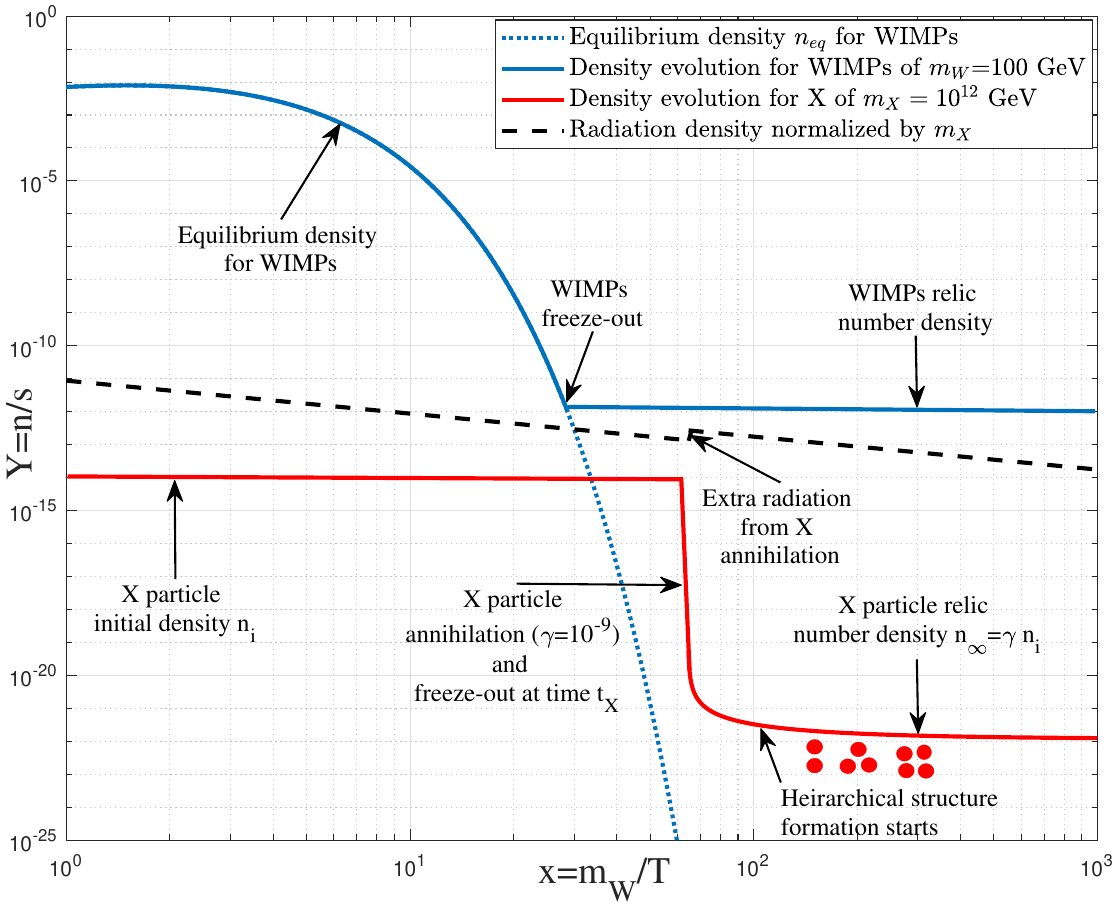}
\caption{The evolution of particle number density (normalized by entropy density $s$) with parameter $x$ (or time) for WIMPs of 100 GeV (solid blue) and superheavy X particles of $10^{12}$ GeV (solid red). WIMP particles are in equilibrium with the thermal bath. Their density follows the equilibrium density (dotted blue) before freeze-out and reaches a constant relic density after freeze-out. Superheavy X particles start with an initial density $n_{i}$ and remain decoupled and out of equilibrium. This is due to the fast expansion dominant over gravity to prevent structure formation and particle annihilation before time $t^*=t_X$. Shortly after time $t^*$, most X particles annihilate into (dark or SM) radiation in a very short time span $\sim1/n_i\langle\sigma v\rangle$ or $10^{-11}$s, followed by the beginning of the hierarchical structure formation (Section \ref{sec:5-3}). Long after time $t^*$, X particles become too dilute to find each other and reach a constant relic density $n_{\infty}=\gamma n_i$. The extra radiation from annihilation at time $t_X$ contributes to the radiation energy in the early Universe (dashed black), which may alleviate the Hubble tension (Section \ref{sec:5-1-2-33}). The SM particles from annihilation/decay today may contribute to the Ultra-high Energy Cosmic Rays (UHECRs) in Section \ref{sec:5-1-2-34}.} 
\label{fig:2222}
\end{figure}

Figure \ref{fig:2222} compares the evolution of the number density for WIMPs and X particles, normalized by the entropy density $s\propto a^{-3}$
\begin{equation}
\begin{split}
& s=\frac{2\pi^2}{45}g_*T^3,
\end{split}
\label{eq:5-33-10-99}
\end{equation}
where the effective number of relativistic degrees of freedom for entropy, $g_*\approx 100$ at T$\sim$100GeV, decreases with time to 15 at T$\sim$150MeV. For thermal WIMPs (solid blue), their initial density follows the equilibrium density (dotted blue). WIMPs freeze out at around $x\approx 25$ and approach a constant relic density. Superheavy X particles have an initial density $n_{i}$ and remain out of equilibrium. Fast expansion dominates gravity to prevent structure formation and particle annihilation before time $t^*=t_X$. Shortly after time $t^*$, most X particles annihilate/decay into (dark or SM) radiation in a very short time of $10^{-11}$s and freeze out. An extremely small fraction of X particles (one in every $10^9$) survive as relic dark matter and enable hierarchical structure formation to form larger and larger structures (Section \ref{sec:5-3}). Long after time $t^*$, X particles become too dilute to find each other, with their density reaching a constant relic density. The extra radiation from particle annihilation/decay at time $t_X$ contributes to the radiation energy in the early Universe, which may alleviate the Hubble tension problem (Section \ref{sec:5-1-2-33}). If SM particles are the annihilation/decay product, the annihilation today may contribute to the Ultra-high Energy Cosmic Rays (UHECRs). In this scenario, UHECR observations can be directly compared with our model prediction. The actual observations support this scenario, as discussed in Section \ref{sec:5-1-2-33}.  

To summarize, the WIMP miracle neglects the gravitational interactions between individual particles, assumes a weak-scale interaction, and predicts a weak-scale WIMP particle of 100GeV. WIMPs are in equilibrium with the thermal bath and annihilate into and are produced by SM particles before freeze-out. By contrast, the X miracle considers gravity between heavy particles, introducing a new fundamental scale $r_X$, two gravity-induced velocities on large and small scales ($V_{||}$ and $v_X$), and three relevant cross sections (Eq. \eqref{eq:5-33-10-111}), which enables the "cold" freeze-out of X particles. The competition between interaction rates and expansion governs the evolution of particle density. The X miracle predicts superheavy X particles with a mass of $10^{12}$GeV, which is surprisingly consistent with the critical particle mass $m_{Xc}$ that was obtained independently from the structure formation (Section \ref{sec:5-1}). 

Finally, instead of starting from the Boltzmann equation (Eq. \eqref{eq:5-22-11}), we calculate the direct collision time $\tau_{dc}$ for structure formation,
\begin{equation}
\begin{split}
&\tau_{dc}(m_X,t_X)=\frac{4}{3}\left(\frac{r_{DM}}{r_X}\right)^2\frac{r_{DM}}{V_{||}}\propto \frac{\left(r_{DM}\right)^3}{\langle\sigma_X v\rangle},
\end{split}
\label{eq:5-2-12-2}
\end{equation}
where $r_{DM}$ the mean spacing between the particles. The requirement for the direct collision time to be comparable to the formation time, i.e. $\tau_{dc}=t_X$, is equivalent to the cross section requirement in Eq. \eqref{eq:5-22-12}, as $n_f\propto 1/r_{DM}^3$ and $\sigma_X \propto r_X^2$. Therefore, this requirement yields the same mass constraint as in Eq. \eqref{eq:5-22-15}. Section \ref{sec:5-2} provides details for the direct collision time $\tau_{dc}$ for structure formation. Only X particles with critical mass $m_{Xc}$ have the right cross section and number density that allow the formation of the smallest structure. Particles heavier than $m_{Xc}$ have a much lower number density, such that particles cannot find each other to form these structures in a time comparable to the formation time $t_X$ in Eq. \eqref{eq:5-2-11}. On the other hand, for particles lighter than $m_{Xc}$, free streaming prevents the formation of these two-particle structures (Fig. \ref{fig:109}). Again, we identified a critical and exact particle mass $m_{Xc}=10^{12}$GeV, no more and no less. 

\section{The properties and nature of dark matter}
\subsection{Dark matter particle properties}
\label{sec:5-1-3}
Free streaming and the X miracle suggest a particle mass of $m_{Xc}=10^{12}$GeV. In this section, relevant properties can be obtained for this given mass. These properties are crucial for identifying the nature of dark matter. Examples are particle size (fundamental scale $r_X$), characteristic particle velocity and acceleration, formation time, and binding energy (using Eqs. \eqref{eq:5-2-10-2}, \eqref{eq:5-2-11}, and \eqref{eq:5-33-11}):
\begin{equation}
\begin{split}
&l_X=2r_X=\frac{8\hbar^2}{G(m_{Xc})^3}\approx 3\times 10^{-13}m, \\
&v_X=\frac{G(m_{Xc})^2}{4\hbar}  \approx 4\times 10^{-7}m/s, 
\end{split}
\label{eq:5-22-16}
\end{equation}
\begin{equation}
\begin{split}
&a_{X} = \frac{v_X^2}{r_X}=\frac{G^3m_{Xc}^7}{64\hbar^4} = 1{m/s^{2}},\\
&t_X=\frac{\pi}{\sqrt{2}} \frac{r_X}{v_X}=\frac{2^{7/2}\pi\hbar^3}{G^2(m_{Xc})^5}\approx 5.2\times 10^{-7}s, \\
&E_b = E_X = \frac{G^2m_{Xc}^5}{16\hbar^2} \approx 2.8\times 10^{-9}eV. 
\end{split}
\label{eq:5-22-17}
\end{equation}
The smallest gravitationally bound state formed by two X particles of mass $m_{Xc}=10^{12}$GeV has a size of $10^{-13}$m, a characteristic velocity of $10^{-7}$m/s and was formed at around $t_X\approx 10^{-6}$s or $a_X\approx 1.5\times 10^{-13}$ (or temperature 1GeV). We also identified a critical energy scale $E_X \approx 10^{-9}$eV that is the binding energy in the ground state, i.e., the amount of energy released as "radiation" when two X particles form the bound state. This "radiation" of $10^{-9}$eV, if it exists, is also produced during the formation of these structures around time $t_X\approx 10^{-6}$s. This "radiation" has a corresponding Compton wavelength of 1.4 km or a frequency of around 0.2MHz. The possible nature of this "radiation" is discussed in Section \ref{sec:6-1}.

The time scale $t_{X}$ is close to the characteristic time for weak interactions ($10^{-6}\sim10^{-10}$s). The fundamental length scale $l_{X}$ is greater than the characteristic range of strong interactions ($\sim 10^{-15}$m) and weak interactions ($\sim 10^{-18} m$), which are the Compton wavelength of respective force mediators (pions and W and Z bosons) for strong and weak interactions. This leads to a $10^{10}$ larger cross section than WIMPs (Eq. \eqref{eq:21-21}). The fundamental scale of $10^{-13}$m is equivalent to the Compton wavelength of MeV particles, suggesting possible $m_{\phi}\sim$MeV particles (scalar or gauge bosons) mediating the particle annihilation, with a mediator mass of (Eq. \eqref{eq:5-22-16})
\begin{equation}
\begin{split}
&m_{\phi} = \frac{\hbar}{r_Xc}=c\sqrt{m_XE_X}\sim \textrm{MeV}.
\end{split}
\label{eq:5-22-17-2}
\end{equation}

The cross section per unit mass $\sigma_X/m=\pi r_X^2/m_X$ is around $5 \times 10^{-11}$m$^2$/kg, which is extremely small, reflecting the collisionless nature of dark matter. Three relevant cross sections are (Eq. \eqref{eq:5-33-10-101})
\begin{equation} 
\label{eq:21-21} 
\begin{split}
&\langle\sigma_{||} V_{||}\rangle = \left(\frac{\hbar}{m_XV_{||}}\right)^2V_{||}=1.2\times 10^{-42}m^3/s,\\
&\langle \sigma_X V_{||} \rangle = 3\times 10^{-22} \textrm{m$^3$/s} \quad \textrm{and} \quad \langle \sigma_X v_X \rangle = 2\times 10^{-32} \textrm{m$^3$/s},
\end{split}
\end{equation} 
which determines the evolution of X particle density. Here, $\langle \sigma_X v_X \rangle$ represents the cross section of the smallest bound structures, while $\langle \sigma_X V_{||}\rangle$ is the large-scale cross section that can be potentially constrained by UHECR observations (Section \ref{sec:5-1-2-33}).

More importantly, a new physical constant $\mu _{X}$ (a constant for power or the rate of energy change) can be introduced,
\begin{equation} 
\label{eq:21-2-1} 
\begin{split}
\mu _{X}&=m_{Xc} a_{X}v_{X}=F_{X} v_{X} = \frac{G^4 m_{Xc}^{10}}{256\hbar^5} =3.5\times 10^{-22} kg \frac{m^{2}}{s^{3}},  
\end{split}
\end{equation} 
which is a different representation of mass $m_{Xc}$ that might reflect the uncertainty principle between momentum $m_Xv_X$ and acceleration $a_X$ \citep{Xu:2023-Universal-scaling-laws-and-density-slope}. Similar to the Planck constant representing the fundamental unit of quantized angular momentum, the constant $\mu_X$ represents a fundamental unit of power. Planck scale quantities are determined by three constants: $\hbar$, $G$, and $c$. Similarly, all other quantities on the energy scale of $m_X$ can be expressed by three constants: $\hbar$, $G$, and $\mu_{X}$. These three constants reflect the nonrelativistic coupling of gravity and quantum effects. The relevant quantities are
\begin{equation}
\label{eq:21-2-2} 
\begin{split}
&m_{Xc} \sim \left(\frac{\mu_X \hbar^5}{G^4}\right)^{1/10}=10^{12} \textrm{GeV}, \quad E_X \sim {\sqrt{\hbar\mu_{X}}} = 10^{-9} \textrm{eV},\\
&r_X\sim \left(\frac{\hbar^5G^2}{\mu_X^3}\right)^{1/10}, \quad v_X \sim \left(G\mu_X\right)^{1/5}, \quad t_X\sim \left(\frac{\hbar}{\mu_X}\right)^{1/2}. 
\end{split}
\end{equation}
The standard approach to determining critical energy or mass scales involves identifying the relevant physical constants associated with that scale. Since no collider experiment can reach such a high energy level to determine the constant $\mu_X$, we take an approach to determine the relevant constant from the mass or energy scale we identified (Eq. \eqref{eq:21-2-1}). The two mass/energy scales $m_{Xc}$ and $E_X$ are the most significant findings of this work. They represent the mass of dark matter particles and the energy of "radiation" associated with the bound state formation. They also determine the decay rate of X particles (Section \ref{sec:5-1-2-34}).

The density of the smallest bound structure reads (Eq. \eqref{eq:5-2-8})
\begin{equation}
\label{eq:22}
\begin{split}
&\rho_{X}=n_Xm_X=\frac{3}{2^{7}\pi}\frac{G^3m_X^{10}}{\hbar^6}=5.3\times 10^{23} \frac{kg}{m^3}, \\
\end{split}
\end{equation}
which is much higher than the nuclear density ($\sim 10^{17}$kg/m$^{3}$). The degeneracy pressure reads
\begin{equation}
\label{eq:22-2}
\begin{split}
P_X =m_{X} v_{X}^2 n_{X} &= \frac{3G^5m_{Xc}^{14}}{2^{11}\pi\hbar^8} \\
& =\left(\frac{4\pi^2}{9}\right)^{1/3}\frac{\hbar^2}{m_X}n_X^{5/3}=1.3\times 10^{11} P_{a}.
\end{split}
\end{equation}

With today's dark matter density around $2.2\times 10^{-27}$kg/m$^3$ and local density $7.2\times 10^{-22}$kg/m$^3$, the mean separation between \textbf{\textit{X}} particles of mass $10^{12}$ GeV is about $l_u \approx 10^4$m for the entire universe and $l_c \approx 130$m locally. This information is highly relevant to the direct detection of dark matter, as the event rate for interactions is directly proportional to the particle number density. 

Finally, a key parameter $\varepsilon_u$ can be introduced that significantly impacts the evolution of large-scale halo structures \citep{Xu:2023-Universal-scaling-laws-and-density-slope, Xu:2023-Dark-matter-halo-mass-functions-and,Xu:2024-Cosmic-quenching-and-scaling-laws,Xu:2021-Inverse-mass-cascade-mass-function},
\begin{equation}
\begin{split}
\varepsilon_u = \frac{v_X^2}{r_X/v_X}=\frac{G^4 m_{Xc}^9}{256\hbar^5}=\left(\frac{G^4\mu_X^9}{\hbar^5}\right)^{\frac{1}{10}}\approx 2.3\times 10^{-7} \frac{m^2}{s^3}. \\
\end{split}
\label{eq:5-22-9}
\end{equation}
We interpret the parameter $\varepsilon_u$ as the rate of kinetic energy $v_X^2$ produced in a turnaround time $t_X\sim r_X/v_X$. Due to the formation of a two-particle bound structure, the system's (specific) kinetic energy increases by $v_X^2$ during a time of $t_X$. This process can be extended to large scales due to the hierarchical formation of structures, i.e., $X+X\rightarrow XX$, $X+XX\rightarrow XXX$, and so on. Along with the formation of larger and larger structures, the kinetic energy of particles increases steadily as the structure size increases (see Eq. \eqref{eq:5-3-4-2} in Section \ref{sec:5-3}). The parameter $\varepsilon_u$ originating from the microphysics of DM particles is independent of the structure size. It affects the dynamics and evolution of halo structures on large scales. 

The same parameter can also be obtained from N-body simulations and observations \citep{Xu:2023-Universal-scaling-laws-and-density-slope,Xu:2023-Dark-matter-halo-mass-functions-and,Xu:2021-Inverse-mass-cascade-mass-function} and connected to the critical acceleration $a_0$ for galaxy dynamics \citep{Xu:2025-On-a-Critical-Acceleration-Scale}. For today's dark matter particles with a typical velocity of $v_0$=300 km/s and the age of the universe $t_0=13.7\times 10^9$years, a rough estimate of $\varepsilon_u = v_0^2/t_0=2\times 10^{-7}$m$^2$/s$^{3}$ can be obtained by following the same argument. Particle kinetic energy was increased by $v_0^2$ during a time of $t_0$ due to the hierarchical formation of structures. That estimated value is in surprising coincidence with the value obtained from the smallest bound structure in Eq. \eqref{eq:5-22-9}. This coincidence provides strong support for the proposed theory, reflecting the scale-independent nature of the parameter $\varepsilon_u$. It also strongly suggests the consistency between microscopic dark matter physics and the dynamics and evolution of large-scale dark matter haloes \citep{Xu:2023-Universal-scaling-laws-and-density-slope,Xu:2025-On-a-Critical-Acceleration-Scale}. For completeness, the Appendix provides details on how $\varepsilon_u$ impacts the structure formation and evolution.

\subsection{The nature of cold dark matter particles}
\label{sec:6-1-1}
Our analysis identifies a critical particle mass $m_{Xc}=10^{12}$GeV and size $r_X=10^{-13}$m, as well as other relevant properties. Although this particle mass is beyond the reach of the collider experiment, we can still briefly explore the nature of dark matter. Our results suggest these superheavy X particles 1) are out of equilibrium with the thermal bath (nonthermal); 2) have an initial density $n_i$ much larger than the relic density $n_{\infty}$ ($n_{\infty}=\gamma n_i$); 3) interact with SM and other dark matter particles primarily via gravity; 4) are fermions with quantum pressure counteracting gravity; 5) annihilate or decay into dark sector and/or SM particles. 

The predicted mass falls well beyond the mass range of standard thermal WIMPs but lies within the range of the so-called superheavy dark matter (SHDM). If gravity is the only interaction between the Standard Model and the dark sectors, these superheavy particles have also been dubbed Planckian-interacting massive particles (PIDM) \citep{Abreu:2023-Cosmological-implications-of-photon-flux-upper-limits}. Except for what we discussed in this work, there are other good motives for considering such superheavy dark matter particles. Since the instability of the electroweak Higgs vacuum can also be linked to an energy scale on the same order of $10^{12}$GeV \citep{Eichhorn:2015-The-Higgs-Mass-and-the-Scale, Goswami:2014-Higgs-instability-and}, our predicted mass approaches the energy scale at which new physics might emerge. This usually requires a new fundamental physical constant, similar to the Planck constant $\hbar$. An example is the constant $\mu_X$ introduced in Eq. \eqref{eq:21-2-1}. 

Based on the properties suggested for dark matter, a very interesting option is superheavy sterile neutrinos (or heavy neutral leptons) \citep{SHAPOSHNIKOV:2024-Sterile-neutrinos-as-dark-matter}. Such right-handed neutrinos are singlets that do not directly participate in strong, weak, or electromagnetic interactions. Their interaction with SM particles is heavily suppressed because of their superheavy mass. They can be Majorana fermions with a “Majorana mass” term that is often generated by a higher-dimensional operator and not related to the Higgs field. Majorana fermions are their own antiparticles, allowing two identical sterile neutrinos to annihilate or decay. The annihilation or decay of these superheavy sterile neutrinos naturally generates UHECRs (Ultra-high-energy cosmic rays) and explains the detection of extremely high-energy particles. The particle mass $10^{12}$GeV and the relevant cross section $10^{-21}$$m^3$/s (Table \ref{tab:555}) are also consistent with the UHECRs data \citep{Blasi:Ultra-high-energy-cosmic-rays-The-annihilation,Luce:2022-Observational-Constraints-on-Cosmic-Ray}. More studies are required to confirm these results and explore the underlying mechanisms of annihilation/decay.

Superheavy sterile neutrinos are also a very attractive dark matter candidate from the viewpoint of "Occam's razor": one tries to minimize the number of assumptions introduced while maximizing the number of problems that can be solved simultaneously. If superheavy sterile neutrinos of mass $10^{12}$GeV exist, then from the seesaw mechanism and without any fine-tuning, the Yukawa coupling of these superheavy sterile neutrinos should be of the order one to obtain the correct active sub-eV (left-handed) neutrino mass \citep{SHAPOSHNIKOV:2024-Sterile-neutrinos-as-dark-matter}. The Higgs quartic coupling $\lambda$ runs with energy. Due to the large negative contribution from the top quark, $\lambda$ can become negative at high energies, making the vacuum metastable. The new order 1 Yukawa coupling of superheavy sterile neutrinos might be sufficient to offset the negative top quark contribution and alleviate the instability. 

The X miracle predicts significant annihilation or decay during cold freeze-out in the early Universe. The asymmetry in X particle annihilation or decay can be related to the baryon asymmetry realized through thermal leptogenesis \citep{Fukugita:1986-Baryogenesis-Without-Grand-Unification} (CP-violating annihilation similar to CP-violating decay). The three necessary Sakharov conditions for baryon asymmetry are: 1) Baryon number violation; 2) C and CP-symmetry violation; and 3) Interactions out of thermal equilibrium. For superheavy X particles, annihilation/decay is naturally out of equilibrium because of the nonthermal nature and the Boltzmann suppression of particle abundance. If this annihilation/decay also violates C and CP, there will be an asymmetry between leptons and anti-leptons. The lepton asymmetry can be subsequently converted to a baryon asymmetry through sphalerons, a non-perturbative solution of the Standard Model. During particle annihilation/decay, the asymmetry in cross sections or decay rates producing particles ($\sigma$) and antiparticles ($\bar\sigma$) can be defined by $\varepsilon=(\sigma-\bar\sigma)/(\sigma+\bar\sigma)$. A rough estimate of cross section asymmetry is $\varepsilon\sim \gamma \Omega_{b}/\Omega_{DM}\sim 10^{-9}$, such that the baryon asymmetry number $Y_b=n_b/s\sim 10^{-10}$ can be related to the cross section asymmetry $\sigma$ and the asymmetry parameter $\gamma$. Here, $\Omega_b$ is the mass fraction of baryons with $\Omega_bh^2=0.02$. 

In principle, the existence of such superheavy sterile neutrinos with a mass of $10^{12}$GeV may simultaneously account for neutrino mass, dark matter, baryon asymmetry, and potentially stabilize the electroweak vacuum. Next, some important impacts of this superheavy dark matter scenario and cold freeze-out are presented in Sections \ref{sec:6-1-11} to \ref{sec:5-1-2-34} in chronological order.

\section{Gravitational production of X particles}
\label{sec:6-1-11}
Our results suggest superheavy X particles of mass $10^{12}$GeV. These particles: 1) are out of equilibrium with the thermal bath; 2) must be produced in overabundance with an initial density $n_i$ much larger than the relic density $n_{\infty}$ ($n_{\infty}=\gamma n_i$); 3) must be fermions following the Pauli exclusion principle. This information is useful in identifying and constraining the production mechanism of X particles. Note that our prediction is independent of the production mechanism. 

There is a wide range of possibilities. For nonthermal dark matter, one potential mechanism is the production of gravitational particles in quintessential inflation \citep{Parker:1969-Quantized-Fields-and-Particle-Creation,Ford:1987-Gravitational-particle-creation-and-inflation,Haro:2019-Gravitational-production-of-superheavy-baryonic-and-dark-matter}. These nonthermal relics from gravitational production do not have to be in the local equilibrium in the early universe or obey the unitarity bounds. To have the right abundance generated during inflation, a wide mass range between $10^{12}$ and $10^{16}$ GeV is estimated \citep{Chung:1999-Superheavy-dark-matter,Kolb:2017-Superheavy-dark-matter-through}, where an asymmetry parameter $\gamma=1$ is usually assumed, i.e., no particle annihilation such that $n_{\infty}=n_i$. The other possible candidate is the Crypton in the string or M theory with a mass around $10^{12}$GeV \citep{Ellis:1990-Confinement-of-fractional-charges-yields-integer-charged,Benakli:1999-Natural-candidates-for-superheavy-dark-matter}. 

Although similar analysis can be extended to other production mechanisms, we focus on the gravitational production of X particles in this section. The change in the spacetime metric at the end of inflation could create particles due to their coupling to the spacetime curvature and gravity. Gravitational particle production should be most efficient when the expansion rate is the largest, i.e., the inflation and reheating phases. The rate of production is dependent on two key parameters: inflationary scale, $H_I$, and reheating temperature $T_{RH}$. This section discusses relevant constraints on these two parameters. 

\rev{
We first relate the relic density to the particle density at the end of reheating $T_{RH}$. We also assume the energy density of X particles and the inflaton redshifts at the same rate from the end of inflation ($t_I\sim H_I^{-1}$) to the end of reheating ($T_{RH}$ or $t_{RH}\sim \Gamma_{\phi}^{-1}$) when most of the inflaton energy is converted into radiation \citep{Chung:1999-Superheavy-dark-matter}. In this way, the relic density can be related to the produced density at $H_I$ as} 
\begin{equation} 
\label{eq:24-1-2-33} 
\begin{split}
& \frac{\Omega_{DM}h^2}{\Omega_{rad}h^2}=\gamma \frac{T_{RH}}{T_{\gamma0}}\frac{\rho_X(T_{RH})}{\rho_{rad}(T_{RH})}\approx  \gamma \frac{T_{RH}}{T_{\gamma0}}\frac{\rho_X(H_I)}{\rho_{univ}(H_I)}.\\ 
\end{split}
\end{equation} 
Here, an asymmetry parameter $\gamma$ is involved to reflect the difference between the produced and relic densities ($n_{\infty=}\gamma n_i$). Most studies focus on the case $\gamma=1$. However, the exact solution of the Boltzmann equation suggests an extremely small $\gamma\sim 10^{-8}$ to $10^{-9}$ (Eq. \eqref{eq:5-22-2223}), which impacts the production mechanism and constraints. We use Einstein's equation to relate the universe density $\rho_{univ}$ to $H_I$. Two particle densities at $H_I$ are
\begin{equation} 
\label{eq:24-1-2-34} 
\begin{split}
& \rho_{univ}(H_I)=\frac{3H_I^2}{8\pi G} \quad \textrm{and} \quad \rho_X(H_I)=m_Xn_X \\ 
\end{split}
\end{equation}

The number density $n_X$ for gravitationally produced particles is usually related to the inflationary scale $H_I$, depending on the nature of the particles. Here, we adopt three different models for $n_X$,
\begin{equation} 
\label{eq:24-1-2-34-2} 
\begin{split}
& n_{X1} = \alpha_1 H_I^3, \quad \textrm{for} \quad \xi=0, \quad \alpha_1=1/12\pi,\\
& n_{X2} = \alpha_2 m_X H_I^2, \quad \textrm{for} \quad \xi=1/6, \quad \alpha_2 \approx 0.0018,\\
& n_{X3} = \alpha_3 m_X^{-1} H_I^4, \quad \textrm{for} \quad \xi=0, \quad \alpha_3 \approx 76 ,
\end{split}
\end{equation}
where the parameter $\xi$ reflects the coupling to gravity, with $\xi$=0 for the minimal coupling and $\xi$=1/6 for the conformal coupling. The first density $n_{X1}$ is for massless scalars with minimal coupling \citep{Ford:1987-Gravitational-particle-creation-and-inflation}. The other two models are taken from reference \citep{Kolb:2024-Cosmological-gravitational-particle-production}. The second density $n_{X2}$ applies to a conformally coupled massive scalar with $H_{RH}<m_X$ for late reheating. The gravitational production of massive fermion particles with spin-1/2 follows a trend similar to that predicted by this model. The third density $n_{X3}$ is for a minimally coupled massive scalar with $H_{RH}<m_X$ for late reheating. The two pre-factors $\alpha_2$ and $\alpha_3$ are chosen to match the data in Fig. 8 in the same reference. 

\begin{figure}
\includegraphics*[width=\columnwidth]{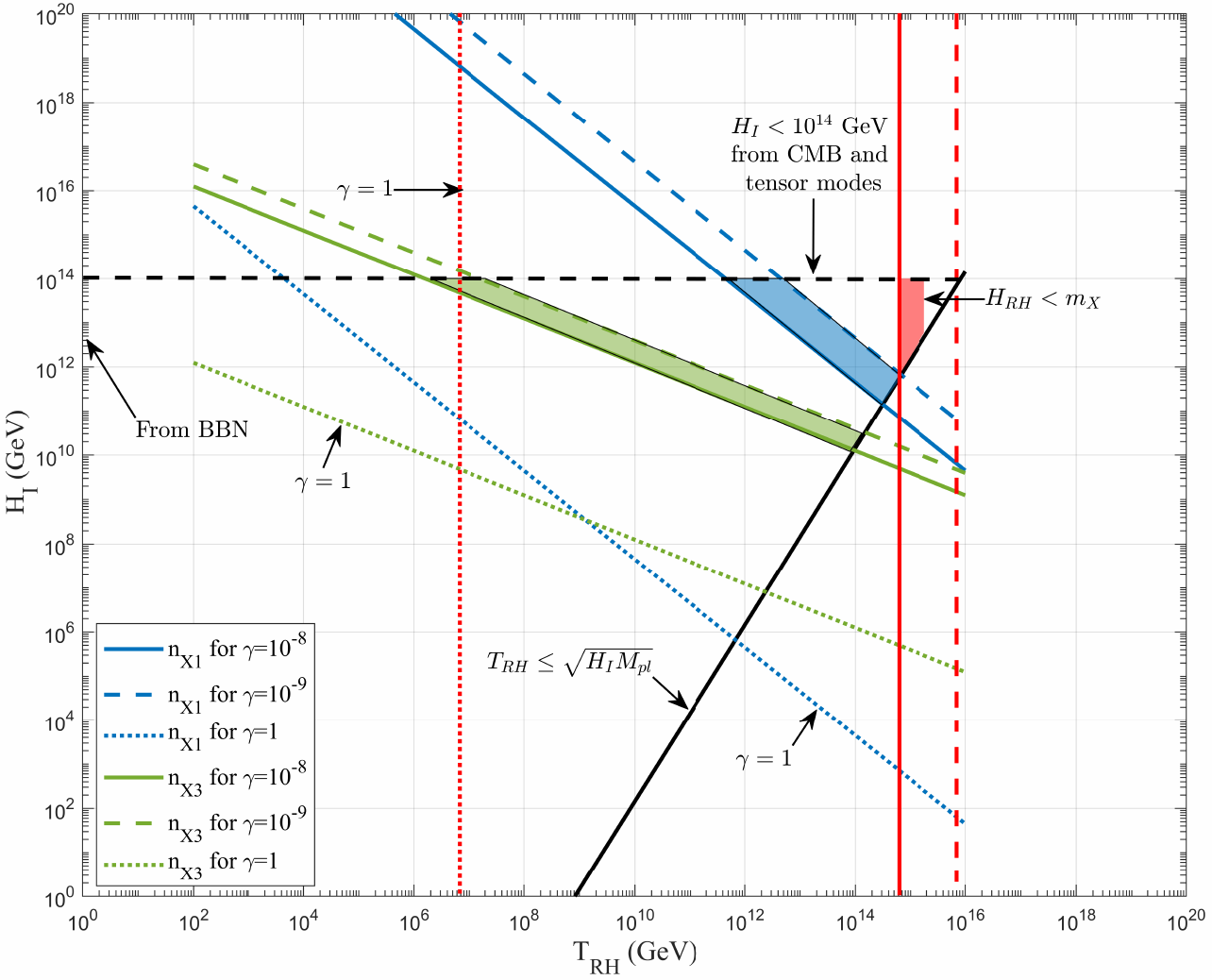}
\caption{The relations between $H_I$ and $T_{RH}$ for three different number density models in Eq. \eqref{eq:24-1-2-34-2} (blue lines for model 1, red lines for model 2, and green lines for model 3). Each model is presented with three different values of $\gamma$=1 (dotted), $10^{-8}$ (solid), and $10^{-9}$ (dashed). The solid black line presents the constraint from Eq. \eqref{eq:24-1-2-36} based on the energy conservation. The dashed black line presents the constraint from CMB observations. BBN sets a lower bound of 5MeV for $T_{RH}$. For X particles of $10^{12}$GeV, the late reheating with $H_{RH}<m_X$ is equivalent to $T_{RH}<10^{15}$GeV. Shaded regions represent the constraint region for an asymmetry parameter in the range of $10^{-9}<\gamma<10^{-8}$. The superheavy dark matter in this work employs model 2 for the production of fermion particles. It suggests high-scale inflation and efficient reheating, resulting in a high reheating temperature (red-shaded area). The $\gamma$=1 is only consistent with a much lower $T_{RH}\sim10^7$GeV.} 
\label{fig:6666}
\end{figure}

Substituting three number density models into Eq. \eqref{eq:24-1-2-33}, three relic densities can be related to the particle mass $m_X$, $H_I$, and $T_{RH}$ as
\begin{equation} 
\label{eq:24-1-2-35} 
\begin{split}
& \left(\frac{\Omega_{DM}h^2}{0.12}\right)_1=2.2\times 10^4 \gamma \frac{m_X}{10^{12}GeV}\cdot \frac{T_{RH}}{10^9GeV}\cdot \frac{H_I}{10^{13}GeV},\\ 
& \left(\frac{\Omega_{DM}h^2}{0.12}\right)_2=150 \gamma \left(\frac{m_X}{10^{12}GeV}\right)^2\cdot \frac{T_{RH}}{10^9GeV},\\ 
& \left(\frac{\Omega_{DM}h^2}{0.12}\right)_3=6.4\times 10^8 \gamma \frac{T_{RH}}{10^9GeV} \cdot \left(\frac{H_I}{10^{13}GeV}\right)^2.\\ 
\end{split}
\end{equation}
Taking our predicted mass $m_X=10^{12}$GeV and relic density $\Omega_{DM}h^2$ = 0.12, Fig. \ref{fig:6666} presents the relations between $H_I$ and $T_{RH}$ for three different models (blue lines for model 1, red lines for model 2 and green lines for model 3). Each model is presented with three different values of $\gamma$=1 (dotted), $10^{-8}$ (solid), and $10^{-9}$ (dashed). The shaded regions represent the constraint region for the asymmetry parameter in the range of $10^{-9}<\gamma<10^{-8}$. 

Next, we work on the other relevant constraints of $H_I$ and $T_{RH}$. The first constraint comes from energy conservation. The reheating temperature $T_{RH}$ could be constrained by the inflationary scale $H_I$ assuming an extremely efficient reheating \citep{kolb:1998-WIMPZILLAS!},
\begin{equation} 
\label{eq:24-1-2-36} 
\begin{split}
& T_{RH} \le 0.2\left(\frac{200}{g*}\right)^{1/4}\sqrt{H_IM_{pl}},\\ 
\end{split}
\end{equation}
where $g*\approx 100$ is the relativistic degree of freedom. This constraint is plotted as a solid black line in Fig. \ref{fig:6666}.

The CMB observations by the Planck satellite, combined with other relevant data, also lead to a limit on the tensor-to-scalar ratio, which constrains the inflationary scale $H_I<10^{14}$GeV \citep{Kolb:2024-Cosmological-gravitational-particle-production} (dashed black line). The uncertainty principle $E \cdot t_{prod}\sim \hbar/2$ requires that these particles be produced around a time $t_{prod}\sim 1/2H_I$. The lower limit of reheating temperature (5MeV) is set by BBN. For number density models with late reheating (models 2 and 3) in Eq. \eqref{eq:24-1-2-34-2}, the Hubble scale at reheating $H_{RH}<m_X=10^{12}$GeV, which is equivalent to a upper limit of reheating temperature $T_{RH}\le10^{15}$GeV (replacing $H_I$ by $H_{RH}$ in Eq. \eqref{eq:24-1-2-36}).

Figure \ref{fig:6666} presents a comprehensive comparison of different models and all relevant constraints. For our fermion superheavy dark matter, we use model 2 for particle number density relevant to fermions (red lines). We exclude $\gamma=1$, which leads to a significantly lower reheating temperature. The initial particle density, when produced, must be much larger than the relic density with $\gamma\ll 1$, as predicted by the X miracle (Eq. \eqref{eq:5-22-2223}). The red-shaded area of this work provides the most stringent constraints on $H_I$ and $T_{RH}$, suggesting high-scale inflation $H_I\sim 10^{13}$GeV and high reheating temperature $T_{RH}\sim 10^{15}$GeV for extremely efficient reheating. 

\section{Dark radiation and the Hubble tension}
In this section, we discuss two types of dark radiation that naturally emerge in our model. Section \ref{sec:6-1} discusses the first type of radiation that is produced during the formation of the smallest gravitationally bound state (Fig. \ref{fig:S44}). Section \ref{sec:5-1-2-33} focuses on the second type of radiation that is produced during the cold freeze-out (Fig. \ref{fig:2222}).

\subsection{Axion or gravitational wave from bound-state formation}
\label{sec:6-1}
The binding energy $E_X = 10^{-9} \textrm{eV}$ (Eq. \eqref{eq:5-22-17}) strongly suggests the existence of some "radiation" that is produced during the formation of bound states at time $t_X=10^{-6}s$ or temperature 1GeV (Fig. \ref{fig:S44}). The energy and time satisfy $E_Xt_X\sim \hbar$ (Eq. \eqref{eq:5-33-11}), a typical feature of gravitational production with $E_X\sim H_X$ (Section \ref{sec:6-1-11}). 

Although the nature of this dark radiation is not yet clear, possible options include dark photons, gravitational waves, or even nonrelativistic axions. Since gravity is the only interaction involved, the first natural option is gravitational waves (GWs). Based on the binding energy of $10^{-9}$eV, a high-frequency gravitational wave of 200 kHz (wavelength of 1.4 km) was produced during the formation of the bound state. This high-frequency gravitational wave is detectable by a Levitated Sensor Detector \citep{Aggarwal:2022-Searching-for-New-Physics} or other advanced approaches. GWs produced in the early Universe during the formation of the earliest bound state ($t_X=10^{-6}$s or $a_X=10^{-13}$) redshift to the nano-Hz frequency today, making them detectable by pulsar timing arrays.

The second option is axions. Axion is a strongly motivated dark matter candidate from the Peccei–Quinn (PQ) solution to the strong CP problem. Axions satisfy two conditions of cold dark matter: a sufficient non-relativistic amount and the effectively collisionless nature \citep{Duffy:2009-Axions-as-dark-matter-particles}. The axion mass could be generated by QCD instantons, which are temperature-dependent because of the suppression of tunneling at high temperatures. These instantons generate a non-zero potential that breaks the shift symmetry, and the axion gains a mass. The axion field is initially frozen, static, and massless. With the Universe expanding, the axion field becomes dynamic and begins damped harmonic motion when the axion mass is comparable to the expansion at a temperature $T_{dyn}$, time $t_{dyn}$, or scale factor $a_{dyn}$, i.e.,
\begin{equation}
\label{eq:24-22-2} 
m_a^{dyn}\equiv m_a(T_{dyn})\sim H(T_{dyn})\sim H_0\sqrt{\Omega_{rad}}\left(\frac{T_{dyn}}{T_{\gamma0}}\right)^2,
\end{equation}
where $T_{\gamma0}$ is the radiation temperature today. Here, $T_{dyn}>\Lambda_{QCD}$ is the critical temperature when the axion field becomes dynamic. This is the time that axion DM is produced. 

At high temperature, the temperature-dependent axion mass $m_a(t)$ is suppressed by the QCD scale $\Lambda_{QCD}$ and is usually modeled by a power law function with $n\approx 4$ \citep{KIM:2018-Invisible-QCD-axion-rolling}. At low temperature, the axion mass approaches a zero-temperature axion mass $m_a^0$ that is determined by the axion decay constant $f_a$ \citep{Chadha-Day:2022-Axion-dark-matter},
\begin{equation}
\label{eq:24-22} 
\begin{split}
& m_a(T) \sim m_{a}^0\left(\frac{\Lambda_{QCD}}{T}\right)^n \quad \textrm{for} \quad T\gg\Lambda_{QCD},\\
& m_{a}^0=5.7 \times 10^{-6} \textrm{eV} \left(\frac{10^{12} \textrm{GeV}}{f_a}\right) \quad \textrm{for} \quad T\ll\Lambda_{QCD}.
\end{split}
\end{equation}
Combining the axion mass with Eq. \eqref{eq:24-22-2}, we can determine $T_{dyn}$ and the zero-temperature mass $m_0^a$,
\begin{equation}
\label{eq:24-22-3} 
\begin{split}
& k_BT_{dyn}=\Lambda_{QCD}\left(\frac{2.24\times 10^{16} \textrm{GeV}^3}{f_a\Lambda_{QCD}^2}\right)^{\frac{1}{n+2}},\\
& m_{a}^0 = m_{a}^{dyn}\left(\frac{m_a^{dyn}\cdot 3.93\times 10^{18} \textrm{GeV}}{\Lambda_{QCD}^2}\right)^{\frac{n}{2}}. 
\end{split}
\end{equation}
For $\Lambda_{QCD}\sim$200 MeV, the zero-temperature axion mass is known only if we know the decay constant $f_a$, the critical temperature $T_{dyn}$, or the axion mass $m_{a}^{dyn}$ at $T_{dyn}$. Next, we will explore potential connections with our superheavy dark matter model. 

The massive axion is produced when its mass is comparable to the rate of expansion, i.e., $m_a(T)\sim H(T)$ or $m_a^{dyn}t_{dyn}=\hbar/2$, satisfying the uncertainty principle. This appears to be a typical feature associated with gravitational production. The formation of the smallest gravitationally bound state by two superheavy X particles at time $t_X$ breaks the translational symmetry, which is respected in the otherwise homogeneous Universe before time $t_X$. We demonstrate that a form of "radiation" is produced along with the bound state formation at time $t_X$ (Fig. \ref{fig:S44}). The energy $E_X$ of that "radiation" also satisfies the uncertainty principle, i.e., $E_Xt_X =\pi\hbar/\sqrt{2}$ (Eq. \eqref{eq:5-33-11}), similar to the mass of the axion in Eq. \eqref{eq:24-22-2}. The two relations are
\begin{equation}
\label{eq:24-22-6} 
\begin{split}
& m_a^{dyn}t_{dyn}=\frac{\hbar}{2} \quad \textrm{and} \quad E_Xt_X=\frac{\pi}{\sqrt{2}}\hbar.\\
\end{split}
\end{equation}
This suggests potential connections between the axion and "radiation" released from structure formation.

If the axion is the "radiation", the mass of axion should be on the order of the energy scale $E_X\sim 10^{-9} \textrm{eV}$ (Eq. \eqref{eq:5-22-16}). These axions should be produced around the time $t_X\approx 10^{-6}$s (Eq. \eqref{eq:5-22-16}), around the QCD phase transition time. This leads to a GUT (Grand Unified Theory) energy scale at which the Peccei-Quinn (PQ) symmetry is spontaneously broken, i.e., $f_a \approx 10^{16}$GeV (Eq. \eqref{eq:24-22-3}). The coupling of the QCD axion to SM particles is inversely proportional to $f_a$, leading to an extremely weak coupling to photons with an effective coupling constant $g_{a\gamma\gamma} \sim 10^{-18}\textrm{GeV}^{-1}$. This is beyond the range of the current direct search for the axion \citep{Yannis:2022-Axion-dark-matter} but might be covered within the next decade by DMRadio collaboration \citep{Brouwer:2022-Proposal-for-a-definitive-search}. 

It is well known that nano-eV axions with a Compton wavelength similar to the size of a black hole can be exponentially produced through superradiance, forming a “gravitational atom” and producing coherent monochromatic gravitational wave (GW) radiation that can be potentially detected on Earth \citep{Aggarwal:2022-Searching-for-New-Physics}. In principle, axions can be produced by a rapidly varying dynamic gravitational field that introduces a time-dependent perturbation to the spacetime metric. For an axion field in a dynamic gravitational field, the rapidly varying spacetime metric acts as a driving force, exciting the field, and producing axions. In our case, the rapidly varying local spacetime metric is induced by two superheavy X particles starting from infinity and forming a gravitationally bound state on a short distance of $r_X\sim 10^{-13}$m (Eq. \eqref{eq:5-22-16}). This rapidly varying spacetime may be sufficient to excite the axion field and produce the nano-eV axions. 

In this scenario, quantitative connections can be made between existing theories for QCD axions and superheavy X particles. Since we postulate that the axion is associated with the formation of the smallest bound state at $t_X$. The critical time $t_{dyn}$ in axion theory when axion DM is produced should satisfy $t_{dyn}=t_X$. Substituting of $t_X$ (Eq. \eqref{eq:5-2-11}) into Eq. \eqref{eq:24-22-3} from axion theory leads to the connection between axion decay constant $f_a$ and X particle mass $m_X$,
\begin{equation}
\label{eq:24-22-5} 
\begin{split}
& \frac{f_a}{10^{16}\textrm{GeV}} = 0.86\left(\frac{\Lambda_{QCD}}{1.6 \textrm{GeV}}\right)^n\left(\frac{10^{12}\textrm{GeV}}{m_X}\right)^{\frac{5}{2}(n+2)},\\
\end{split}
\end{equation}
Due to a large $n\approx4$, $f_a$ is highly sensitive to the particle mass $m_X$ such that $m_X$ is constrained to $\sim10^{12}$GeV for a reasonable $f_a$. 

Next, since the initial state deriving the binding energy $E_b$ is represented by two X particles separated by an infinite distance (Fig. \ref{fig:S44}), $E_b$ should be the zero-temperature axion mass, the largest converged axion mass such that $E_b=E_X=m_a^0=(\pi\sqrt{2}) m_a^{dyn}$, obtained from Eq. \eqref{eq:24-22-6} with $t_{dyn}=t_X$. Substituting this into Eqs. \eqref{eq:24-22} and \eqref{eq:24-22-3} leads to the axion mass and the X particle mass,
\begin{equation}
\label{eq:24-22-4} 
\begin{split}
&m_a^0 = \frac{\left(2\pi^2\right)^{\frac{n+2}{2n}}\Lambda_{QCD}^2}{3.9\times 10^{18}\textrm{GeV}} \quad\textrm{and}\quad  \frac{m_X}{10^{12}\textrm{GeV}} = \left(\frac{\Lambda_{QCD}}{0.36 \textrm{GeV}}\right)^{\frac{2}{5}}.\\
\end{split}
\end{equation}
Depending on the exact value of $\Lambda_{QCD}$, this leads to a GUT-scale axion. The axion mass is comparable to the energy scale $\Lambda_X$ for non-perturbative instanton-induced decay of X particles (Eq. \eqref{eq:24-1}). Future theories should unify these two energy scales, the hidden gauge interaction and the axion.  

Although the axions described here may not be dark matter, these properties suggest that the axion may be a very promising candidate for the "radiation" released during bound state formation, with a tiny mass and weak interactions with SM particles \citep{Mazumdar:2016-Nonthermal-axion-dark-radiation,Marsh:2016-Axion-cosmology}. Most axion models and experiments (ADMX etc.) assume axions make up 100\% of dark matter. For pre-inflationary axions produced by the misalignment mechanism, the initial misalignment angle $\theta_{ini}$ determines the axion relic density. However, the assumption that the axion makes up 100\% dark matter requires an extremely small initial angle $\theta_{ini}$, i.e., the axion fine-tuning problem for a large decay constant $f_a$.

In this work, since superheavy X particles make up most dark matter, the axion is only a very small fraction of dark matter. This naturally requires a small misalignment angle to be consistent with the small amount of axion. We first estimate the amount of "radiation" released at freeze-out, which is much less than the extra radiation from the cold freeze-out of X particles in Section {\ref{sec:5-1-2-33}}. If axions exist as the "radiation" from bound state formation, the axion fraction can be estimated as
\begin{equation}
\begin{split}
&\delta_{a}=\frac{\Omega_a}{\Omega_{DM}}=\frac{1}{\gamma}\cdot\frac{m_a}{m_X}\approx 10^{-21},\\
\end{split}
\label{eq:5-33-6-2-3}
\end{equation}
where $\gamma\sim 10^{-9}$ is the asymmetry parameter (Eq. \eqref{eq:5-22-2223}). For every particle annihilated (or decayed), there is one nonrelativistic axion particle produced. The mass fraction of axion is significantly smaller than the fraction of relic dark matter. Here, $m_a=10^{-9}$eV is the mass of "radiation" (Eq. \eqref{eq:5-22-17}), and $m_X=10^{12}$GeV is the mass of X particles. This motivates probing the parameter space of axions at nano-eV mass and a much lower axion abundance.

The initial misalignment angle for pre-inflationary axion can be related to the amount of axion as \citep{Kobayashi:2013-Isocurvature-constraints-and-anharmonic-effects},
\begin{equation}
\begin{split}
&\frac{\Omega_a}{\Omega_{DM}}\approx \frac{5}{3} \theta_{ini}^2\left(\frac{f_a}{10^{12}GeV}\right)^{7/6}.\\
\end{split}
\label{eq:5-33-6-2-33}
\end{equation}
From this relation and Eq. \eqref{eq:5-33-6-2-3}, we can estimate that the initial misalignment angle for pre-inflationary axion is $\theta_{ini}\approx 10^{-13}$. Such a small angle could be the direct result of the extremely small amount of axions produced. The total neutron electric dipole moment (EDM) is proportional to the total CP-violating angle $\bar\theta$, the sum of contributions from both strong and weak interactions, i.e., $|\textbf{d}|\sim 3.6\times 10^{-16}\bar\theta$ $e$ cm. The experiment measure of $|\textbf{d}|<1.8\times 10^{-26}$ $e\cdot$cm, implying a total effective angle $\bar\theta< 10^{-10}$ \citep{Chadha-Day:2022-Axion-dark-matter}. Since the weak contribution is usually small and negligible due to small quark masses and loop suppression, the experiment further implies a QCD effective angle $\bar\theta< 10^{-10}$. Such a small angle is unnatural and motivates solutions such as the Peccei-Quinn mechanism, introducing an oscillating axion field to dynamically set the effective angle $\bar\theta\approx 0$ with an oscillating magnitude $\delta\theta$ or a fluctuating EDM $|\textbf{d}|$. The predicted initial angle should set the maximum magnitude of oscillation $\delta\theta\propto \theta_{ini}a_{dyn}^{3/2} \ll\theta_{ini} \sim10^{-13}$, if experimentally observable.

For pre-inflationary axions, the initial misalignment angle must be constrained by observations. This is related to the tension arising from the interplay between the axion model and the constraints imposed by the CMB isocurvature perturbations. These constraints can be written as (Eq. (73) of \citep{PlanckCollaboration:2014-Constraints-on-inflation}),
\begin{equation}
\begin{split}
&\alpha_{iso}\approx \frac{5}{3\pi^2}\frac{\left(\Omega_a/\Omega_{DM}\right)}{A_s}\left(\frac{H_I}{f_a}\right)^2\left(\frac{f_a}{10^{12}GeV}\right)^{7/6}<0.038.\\
\end{split}
\label{eq:5-33-6-2-34}
\end{equation}
where $A_s\approx 2.1\times 10^{-9}$ is the adiabatic scalar amplitude and $H_I$ is the inflationary Hubble scale. For GUT scale axions ($f_a\sim10^{16}$ GeV), as predicted by this work, this constraint is translated into
\begin{equation}
\begin{split}
&\frac{\Omega_a}{\Omega_{DM}}<10^{-14}\left(\frac{f_a}{H_I}\right)^2.\\
\end{split}
\label{eq:5-33-6-2-35}
\end{equation}
For high-scale inflationary models with detectable tensor-to-scale ratio, the Hubble scale $H_I\sim 10^{13}$ GeV leads to an extremely small fraction of relic pre-inflationary axions. The predicted small axion fraction \eqref{eq:5-33-6-2-3} naturally solves the tension between high-scale inflation and isocurvature perturbations from axions.

\subsection{Dark radiation from freeze-out for Hubble tension}
\label{sec:5-1-2-33}
This section presents a brief discussion of the dark radiation produced from cold freeze-out, which could be highly relevant to the Hubble tension. The Hubble tension refers to the discrepancy in the current expansion rate $H_0$ between the prediction $\Lambda\textrm{CDM}$ based on CMB measurements and direct measurements in the local universe. Various solutions have been proposed to resolve or reconcile this discrepancy. The reader may find a comprehensive review of these solutions \citep{Valentino:2021-In-the-realm-of-the-Hubble-tension}. The most relevant solution to this work is additional relativistic degrees of freedom in addition to the SM neutrinos that would increase $H_0$ and thereby alleviate the Hubble tension \citep{Pandey:2020-Alleviating-the-H0-and, Chen-Constraints-on-dark-matter-to-dark-radiation}. 

In the early universe around time $t_X$, a significant amount of X particles annihilate/decay into dark radiation during the cold freeze-out (Fig. \ref{fig:2222}), leaving a relic density of $n_{\infty}=\gamma n_i$ after freeze-out. Here, the parameter $\gamma\approx 10^{-9}$ to $10^{-8}$ (Eq. \eqref{eq:5-22-2223}) reflects the asymmetry in dark matter annihilation/decay leading to the relic density $n_{\infty}$. The particle annihilation/decay enhances the radiation fraction and relativistic degrees of freedom, modifies the expansion history of the early Universe, deviates from the $\Lambda$CDM prediction, and potentially alleviates the Hubble tension. The fraction of extra radiation relative to the standard $\Lambda$CDM model is 
\begin{equation}
\begin{split}
&\delta_{r}=\frac{1}{\gamma} \cdot\frac{\Omega_{DM} a_X^{-3}}{\Omega_{rad} a_X^{-4}}, \quad \Delta N_{eff} =  \frac{8}{7}\left(\frac{11}{4}\right)^\frac{4}{3}\delta_r \frac{\Omega_{rad}}{\Omega_{\gamma}}\approx 0.4.
\end{split}
\label{eq:5-33-6-2-2}
\end{equation}
For a scale factor at freeze-out $a_X\approx 1.5\times 10^{-13}$ and $\gamma=10^{-8}$, this extra radiation corresponds to an increase of 5\% of the $\Lambda$CDM radiation. This is equivalent to an increase in the relativistic degree of freedom parametrized by $\Delta N_{eff}\approx 0.4$. Here, the ratio of total radiation to photon is $\Omega_{rad}/\Omega_{\gamma}\approx 1.69$. 

Extra radiation was proposed to alleviate the Hubble tension. However, the amount of additional radiation in the early Universe is unavoidably constrained by BBN, CMB, and the large-scale structures of the Universe. Recent studies show that the Hubble tension can be improved for $0.2\le\Delta N_{eff}\le 0.5$ \citep{seto:2021-Comparing-early-dark-energy-and-extra-radiation}. With a predicted $\Delta N_{eff}=0.4$, the annihilation/decay of superheavy X particles may provide a possible mechanism to alleviate Hubble tension. For a complete study, a separate paper will discuss the nature of dark radiation (free-streaming or self-interacting), its interactions with dark matter, effects on the Hubble tension, and relevant particle physics.

\section{UHECR from superheavy X particles}
\label{sec:5-1-2-34}
\rev{
The ultra-high energy cosmic rays (UHECRs) involve the most energetic particles that are accelerated to extremely high energies by unknown sources, with energies exceeding $10^{18}$eV (1EeV). The origin of UHECRs remains a significant mystery in astrophysics and particle physics. Readers may also find a review of potential sources of UHECRs \citep{Anchordoqui2019:Ultra-high-energy-cosmic-rays}. Current observations favor a primarily astrophysical origin, supported by their nuclear composition, near-isotropic arrival directions, and stringent limits from secondary gamma-ray. Nevertheless, a subdominant contribution from the annihilation or decay of superheavy dark matter (SHDM) remains a viable option to connect UHECR observations to the dark sector.}

\subsection{UHECR from annihilation of X particles}
Particle annihilation/decay occurs mainly during cold freeze-out, very early around the formation time $t_X\sim 10^{-6}$, producing significant extra radiation that may alleviate the Hubble tension. Particles may continue to annihilate/decay after freeze-out at a much lower rate, producing UHECRs as indirect signatures of dark matter. With an exact solution for number density (Eq. \eqref{eq:5-22-220}), the particle annihilation rate $\Gamma(t)$ ($m^{-3}s^{-1}$) in radiation and matter eras reads
\begin{equation}
\begin{split}
&\Gamma(t) = n(t)^2\langle\sigma_X V_{||}\rangle=\frac{n_{\infty}^2\left(2tH_0\sqrt{\Omega_{rad}}\right)^{-3}}{(1-(1-\gamma)(t_i/t)^{1/2})^2}\langle\sigma_X v_{||}\rangle,\\
&\Gamma(t) = n_{\infty}^2a^{-6}\langle\sigma_X V_{||}\rangle.
\end{split}
\label{eq:5-22-223}
\end{equation}

In a radiation era with constant velocity $V_{||}$ on the order of km/s, we find the annihilation rate $\Gamma(t)\propto t^{-3}$. In the era of matter with particle velocity $V_{||}\propto (a/a_{eq})^{1/2}$ (Fig. \ref{fig:1111}), the annihilation rate $\Gamma(t)\propto t^{-11/3}$. Since both particle number density and particle velocity are scale-dependent (dependent on halo mass), the cross section and annihilation rate are also scale-dependent. Small haloes with lower density and lower particle velocity tend to have a smaller cross section and a lower annihilation rate. The cross section $\langle \sigma_X v_X \rangle \sim 10^{-32}$ m$^3$/s represents the cross section in the smallest and earliest two-body bound state (Eq. \eqref{eq:21-21}). On large scales, the annihilation cross section $\langle \sigma_X V_{||} \rangle$ is of $10^{-23}$ m$^3$/s (radiation era with $V_{||}=10^3$m/s) and  of $10^{-21}$ m$^3$/s (present epoch with particle velocity $10^5$m/s). 

The average energy production rate density describes the UHECR energy produced per unit of volume and time. Taking into account that each annihilation produces two UHECR jets, each of energy $m_X$, the density of the UHECR energy production rate due to the X particle annihilation rate should be $J=m_Xc^2\cdot \Gamma(t)$. Based on the predicted X particle mass $10^{12}$GeV, table \ref{tab:555} presents the predicted cross section, particle number density, annihilation rate, and energy production rate density at different times and locations (scales). Note that haloes have a density $\sim$100 times the average dark matter density. The local density is $\sim 10^{5}$ times the average density. 

These predictions can be directly compared with UHECR observations. By matching the observed UHECR spectra, constraints on the annihilation cross section were derived. In order to provide enough events to explain the observed UHECRs, for particle mass of $10^{12}$GeV, the cross section $\langle\sigma v\rangle$ must be in the range of $3\times10^{-15}$cm$^3$/s to $3\times10^{-17}$cm$^3$/s \citep{Blasi:Ultra-high-energy-cosmic-rays-The-annihilation}, in good agreement with the predicted cross section in table \ref{tab:555}. Another constraint comes from the total observed UHECR energy density. To provide that total energy density, the density of the production rate of UHECRs is constrained to be $10^{45}$ erg Mpc$^{-3}$ Yr$^{-1}$ by integrating energy spectra above 0.63EeV \citep{Luce:2022-Observational-Constraints-on-Cosmic-Ray}. If we assume a subdominant contribution from decay of X particles to UHERCs, the UHECR production rate density provides an upper limit for the production rate density from X particle decay, i.e., $J<$$10^{45}$ erg Mpc$^{-3}$ Yr$^{-1}$. Table \ref{tab:555} presents the average $J$ on different scales. The average production rate density $J$ on large scales is $J\approx 10^{40}<$$10^{45}$ erg Mpc$^{-3}$ Yr$^{-1}$, which is within the constraint of the total UHECR energy density. 

However, particle annihilation models face the challenge of the observed isotropy of UHECRs. The annihilation rate $\propto \rho_{DM}^{2}$ should lead to a certain level of anisotropy of the observed UHECRs. In addition to the annihilation model, the decay of superheavy X particles into SM particles and the generation of UHECRs are also considered by different models. The particle decay models are less severely constrained by the observed isotropy of the UHECRs. In the next section, we will briefly discuss the decay of superheavy X particles.

\begin{table}
    \begin{center}
    \caption{Cross section, annihilation rate, and UHECR production rate}
    \label{tab:555}
    \begin{tabular}{lccccc} 
    \hline
    Quantity            & \makecell{Cross\\ section}    & \makecell{Number\\density}      & \makecell{Annihilation\\rate $\Gamma$}     & \makecell{Production\\rate density J}\\
    \hline
    Unit                & m$^3$s$^{-1}$ & m$^{-3}$      & m$^{-3}$s$^{-1}$      & erg/Mpc$^{3}$Yr$^{1}$ \\
    Freeze-out $t_X$    & $10^{-23}$    & $10^{34}$     & $10^{45}$             & $10^{130}$ \\
    Present Universe    & $10^{-21}$    & $10^{-12}$    & $10^{-44}$            & $10^{40}$   \\
    Milk way halo       & $10^{-21}$    & $10^{-10}$    & $10^{-40}$            & $10^{45}$    \\
    Local               & $10^{-21}$    & $10^{-7}$     & $10^{-33}$            & $10^{51}$     \\
    \hline
    \end{tabular}
  \end{center}
\end{table}

\subsection{UHECR from decay of X particles}
The longevity of superheavy X particles can be effectively maintained by introducing a new conserved quantum number (new symmetry) within the dark sector, which prevents these particles from decaying under normal interactions. However, even perturbatively stable particles are generally not immune to decay through non-perturbative processes in non-Abelian gauge theories. One such process involves instantons, which describe the quantum tunneling between distinct vacuum configurations of the gauge field. During these transitions, fermion fields undergo nontrivial changes, resulting in the decay of otherwise stable X particles. 

Decay processes of superheavy X particles triggered by instantons arise from non-perturbative phenomena through quantum tunneling between vacua with different topological properties. These effects are significant in gauge theories, where instantons can facilitate interactions that break global symmetries that are otherwise preserved in perturbative processes. In instanton physics, the transition probability between vacua is exponentially suppressed by the classical instanton action. This leads to a "non-perturbative mass scale $\Lambda_X$" — the scale at which the symmetry-violating processes become dynamically relevant, albeit rare. It is instructive to compare the instanton-induced decay with Quantum Chromodynamics (QCD). Both arise from non-perturbative effects and involve exponential suppression. 

In Quantum Chromodynamics (QCD), the gauge coupling $g_s$ and the coupling constant $\alpha_s=g_s^2/4\pi$ are not constants. They run with the energy scale $m_Z$ due to quantum corrections (asymptotic freedom) that lead to the generation of a dynamically determined mass scale $\Lambda_{QCD}$ through dimensional transmutation (1-loop solution) \citep{DEUR:2016-The-QCD-running-coupling}: 
\begin{equation} 
\label{eq:24-2} 
\alpha_s =\frac{2\pi}{\beta_0\ln\left(\frac{m_Z}{\Lambda_{QCD}}\right)} \quad \textrm{and} \quad  \Lambda_{QCD} = m_Z \exp\left(-\frac{2\pi}{\beta_0\alpha_s} \right).       
\end{equation} 
Here, $\Lambda_{QCD}$ is the scale where QCD becomes strongly coupled and perturbation theory breaks down. With parameter $\beta_0\approx 9$ and at a relevant energy scale $m_Z\sim 100$ GeV, the coupling constant $\alpha_s=0.12$ or $g_s=1.23$ and QCD scale $\Lambda_{QCD}\sim$200 MeV.

For instanton-induced decay of X particles, the gauge coupling $g_X$ and the coupling constant $\alpha_X=g_X^2/4\pi$ can be written similarly, 
\begin{equation} 
\label{eq:24-1} 
\begin{split}
&\alpha_X = \frac{2\pi}{\ln\left(\frac{m_X}{\Lambda_X}\right)} \quad \textrm{and} \quad  \Lambda_{X} = m_X \exp\left(-\frac{2\pi}{\alpha_X} \right),\\ 
\end{split}
\end{equation} 
where $\alpha _{X}$ is the reduced coupling constant of the hidden gauge interaction, $\Lambda_X$ is the effective energy scale of non-perturbative instanton effects. This scale marks the onset of rare instanton-induced decay events and sets the strength/suppression of symmetry-violating interactions, i.e., a scale below which non-perturbative phenomena dominate. Due to the long lifetime of X particles (much longer than the age of the Universe), we would expect $\Lambda_X\ll m_X$.

For energies above the QCD scale ($\Lambda_{QCD}$), QCD becomes asymptotically free. The strong coupling $\alpha_s$ decreases logarithmically, interactions become weaker, and quarks behave "free" at short distances. For energies around the QCD scale, the coupling becomes stronger, and the quarks become "confined" after the transition. Similarly, for instanton-induced decay, when particles are isolated and far apart, their wavefunctions barely overlap with the relevant energy $m_X\gg\Lambda_{X}$. For energies greater than $\Lambda_{X}$, the hidden gauge interactions become weaker, and particle decay is exponentially suppressed. For particles in a bound state, due to overlapping wavefunctions, shorter tunneling path, and less costly in action, the decay of X particles is not suppressed with the shortest particle lifetime or the widest decay width at energies around $\Lambda_{X}$. 

Based on Section \ref{sec:5-1-1}, X particles should have the shortest lifetime or the fastest decay in the smallest bound state, where the only relevant energy scale is the binding energy $E_X$ (Eq. \eqref{eq:5-33-11}). This is the potential or kinetic energy associated with the local temperature of X particles in a bound state. In this case, a reasonable estimate of the decay width is $\Lambda_X \sim E_X$ (Eq. \eqref{eq:21-2-2}) with a short particle lifetime $t_X\sim \hbar/E_X$ in a bound state. Therefore, the energy scale and the lifetime of the particles $\tau_X$ in an unbound state read (Eq. \eqref{eq:24-1}),
\begin{equation} 
\label{eq:24-1-2} 
\begin{split}
& \Lambda_X = E_X = m_X\exp\left(-\frac{2\pi}{\alpha_X}\right)\sim \sqrt{\hbar\mu_X},\\ 
&\tau_X = \frac{m_X}{E_X}t_X =  \frac{\hbar e^{{4\pi/\alpha _{X} } } }{m_{X} c^{2}}=\sqrt{\frac{\hbar}{\mu_X}}\exp\left(\frac{2\pi}{\alpha_X}\right)=\frac{c^2}{\varepsilon_u},\\
\end{split}
\end{equation} 
with $E_Xt_X\sim \hbar$ (Eq. \eqref{eq:5-33-11}). The particle lifetime $\tau_X$ shares the same expression as Eq. (14) in \citep{Abreu:2023-Cosmological-implications-of-photon-flux-upper-limits}. Plugging in the values of $m_X=10^{12}$GeV and $E_X=10^{-9}$eV (or constant $\mu_X$ in Eq. \eqref{eq:21-2-1}), we predict the coupling constant $\alpha_X$=0.09 or a hidden gauge coupling $g_X$=1.07, and a particle lifetime $\tau_X\sim 10^{24}s$ or $10^{17}$Yrs, much longer than the age of the Universe. 

From UHECR observations, the upper bounds of UHECR fluxes can be converted into tight constraints on the particle lifetime and the couplings governing the decay process. For X particles that decay into both dark sector and SM particles, less restrictive constraints may be obtained depending on the visible decay branching fractions $\alpha_{SM}$. A smaller SM branching fraction $\alpha_{SM}$ corresponds to a lower bound of the particle lifetime from observations ($\tau_X^{(min)}\propto \alpha_{SM}$). For instanton-induced decay, the constrained coupling constant $\alpha _{X} \le {4\pi/137}\approx 0.092$ for particles with mass $m_X=10^{12}$GeV \citep{Abreu:2023-Cosmological-implications-of-photon-flux-upper-limits}. Using Eq. \eqref{eq:24-1-2} and assuming $\alpha_{SM}=1$, this is equivalent to a particle lifetime $\tau_X\ge 1.4\times10^{23}$s. Our Predicted coupling $\alpha _{X}$=0.09 from Eq. \eqref{eq:24-1} and the lifetime are consistent with the corresponding constraints in \citep{Abreu:2023-Cosmological-implications-of-photon-flux-upper-limits}. In addition, the particle lifetime can also be related to the energy cascade rate $\varepsilon_u$ defined in Eq. \eqref{eq:5-22-9} (see 87i8o09Fig. \ref{fig:S1-3}). The cascade process can only increase the particle velocity to its maximum value, $c$, during the lifetime of the particle, $\tau_X$.

\section{Conclusions}
\label{sec:7}
In this paper, we consider nonthermal fermion dark matter particles. The free streaming scale calculation identifies a critical particle mass of $10^{12}$GeV. Dark matter particles of this mass have a free streaming mass comparable to the particle mass and can form the smallest and earliest bound state around a critical time $t_X\sim 10^{-6}$s. We also identify a "radiation" with an energy of $10^{-9}$eV released during the formation of this bound state. In this state, quantum pressure balances gravity, leading to a quantum-gravitational scale $r_X=10^{-13}$m. 

Alternatively, an X miracle, modified from the WIMP miracle, is adapted to the evolution of superheavy X particles, where gravity plays a significant role. When considering gravity, the fundamental scale $r_X$ emerges, leading to a much larger cross section than WIMPs, and enabling a "cold" freeze-out for nonthermal relics. The X miracle also predicts a superheavy particle mass of $10^{12}$GeV, consistent with the critical mass from free streaming calculation. The complete evolution of X particles can be found from the exact solution of the Boltzmann equation. These superheavy X particles are produced with an initial density $n_i$ and annihilate or decay into dark radiation (or SM) at a critical time $10^{-6}$s when the interaction rate is comparable to the expansion. A significant amount of annihilation/decay leads to a relic density $n_{\infty}=\gamma n_i$, where the asymmetry parameter $\gamma \sim 10^{-9}$. Roughly one in every billion X particles survived as relic dark matter, after which structure formation began.  

Based on this description, the annihilation/decay in the early Universe could be relevant to the Hubble tension and leptogenesis/baryogenesis. The annihilation/decay at the present epoch could be relevant to UHECRs (Ultra-high Energy Cosmic Rays). The model predicted a cross section $10^{-21}$m$^3$/s, an energy production rate density $10^{40}$erg Mpc$^{-3}$yr$^{-1}$, and a particle lifetime $10^{24}$s, which are in agreement with the observations of UHECR. For superheavy X particles, the right-handed neutrino is an excellent candidate that is also relevant to neutrino mass and Higgs vacuum stability. For "radiation" from bound state formation, underabundant GUT-scale nano-eV axions or high-frequency (200 kHz) gravitational waves could be very relevant. For hidden gauge interaction, which governs particle decay, a coupling constant $\alpha_X\approx$0.09 can be estimated. A more complete theory describing these interactions has yet to be developed.


To summarize, superheavy nonthermal X particles with these properties seem to prefer to have a mass $10^{12}$GeV and a cross section of $10^{-21}$m$^3$/s because i) particles of this mass can form the smallest and earliest bound state (Fig. \ref{fig:109}); ii) particles of this mass can have a smooth structure size evolution (Fig. \ref{fig:S33}); iii) particles of this mass and cross section satisfy the X miracle when gravity between individual particles is considered (Section \ref{sec:5-1-2}); iv) particles of this mass and cross section is consistent with UHECR observations; iv) particles of this mass have the shortest possible waiting time to allow for the formation of haloes observed today as large as $10^{13}M_{\odot}$ (Section \ref{sec:5-3}); v) particles of this mass give rise to the key parameter $\varepsilon_u$ consistent with the halo scale dynamics (Eq. \eqref{eq:5-22-9} and Section \ref{sec:5}).

\section*{Acknowledgment}
This research was supported by Laboratory Directed Research and Development at Pacific Northwest National Laboratory (PNNL). PNNL is a multiprogram national laboratory operated for the U.S. Department of Energy (DOE) by Battelle Memorial Institute under contract no. DE-AC05-76RL01830.

\section*{Data Availability}
Two datasets for this article, that is, halo-based and correlation-based statistics of dark matter flow, are available on Zenodo \citep{Xu:2022-Dark_matter-flow-dataset-part1,Xu:2022-Dark_matter-flow-dataset-part2}(http://dx.doi.org/10.5281/zenodo.6541230), along with the accompanying presentation "A comparative study of dark matter flow \& hydrodynamic turbulence and its applications" \citep{Xu:2022-Dark_matter-flow-and-hydrodynamic-turbulence-presentation} (http://dx.doi.org/10.5281/zenodo.6569901). All data are also available on GitHub \citep{Xu:Dark_matter_flow_dataset_2022_all_files} (http://dx.doi.org/10.5281/zenodo.6586212).

\bibliographystyle{Papers}
\bibliography{Papers}

\begin{thebibliography}{95}%
\makeatletter
\providecommand \@ifxundefined [1]{%
 \@ifx{#1\undefined}
}%
\providecommand \@ifnum [1]{%
 \ifnum #1\expandafter \@firstoftwo
 \else \expandafter \@secondoftwo
 \fi
}%
\providecommand \@ifx [1]{%
 \ifx #1\expandafter \@firstoftwo
 \else \expandafter \@secondoftwo
 \fi
}%
\providecommand \natexlab [1]{#1}%
\providecommand \enquote  [1]{``#1''}%
\providecommand \bibnamefont  [1]{#1}%
\providecommand \bibfnamefont [1]{#1}%
\providecommand \citenamefont [1]{#1}%
\providecommand \href@noop [0]{\@secondoftwo}%
\providecommand \href [0]{\begingroup \@sanitize@url \@href}%
\providecommand \@href[1]{\@@startlink{#1}\@@href}%
\providecommand \@@href[1]{\endgroup#1\@@endlink}%
\providecommand \@sanitize@url [0]{\catcode `\\12\catcode `\$12\catcode `\&12\catcode `\#12\catcode `\^12\catcode `\_12\catcode `\%12\relax}%
\providecommand \@@startlink[1]{}%
\providecommand \@@endlink[0]{}%
\providecommand \url  [0]{\begingroup\@sanitize@url \@url }%
\providecommand \@url [1]{\endgroup\@href {#1}{\urlprefix }}%
\providecommand \urlprefix  [0]{URL }%
\providecommand \Eprint [0]{\href }%
\providecommand \doibase [0]{http://dx.doi.org/}%
\providecommand \selectlanguage [0]{\@gobble}%
\providecommand \bibinfo  [0]{\@secondoftwo}%
\providecommand \bibfield  [0]{\@secondoftwo}%
\providecommand \translation [1]{[#1]}%
\providecommand \BibitemOpen [0]{}%
\providecommand \bibitemStop [0]{}%
\providecommand \bibitemNoStop [0]{.\EOS\space}%
\providecommand \EOS [0]{\spacefactor3000\relax}%
\providecommand \BibitemShut  [1]{\csname bibitem#1\endcsname}%
\let\auto@bib@innerbib\@empty
\bibitem [{\citenamefont {Rubin}\ and\ \citenamefont {Ford}(1970)}]{Rubin:1970-Rotation-of-Andromeda-Nebula-f}%
  \BibitemOpen
  \bibfield  {author} {\bibinfo {author} {\bibfnamefont {V.~C.}\ \bibnamefont {Rubin}}\ and\ \bibinfo {author} {\bibfnamefont {W.~K.}\ \bibnamefont {Ford}},\ }\href {\doibase 10.1086/150317} {\bibfield  {journal} {\bibinfo  {journal} {Astrophysical Journal}\ }\textbf {\bibinfo {volume} {159}},\ \bibinfo {pages} {379} (\bibinfo {year} {1970})}\BibitemShut {NoStop}%
\bibitem [{\citenamefont {Rubin}\ \emph {et~al.}(1980)\citenamefont {Rubin}, \citenamefont {Ford},\ and\ \citenamefont {Thonnard}}]{Rubin:1980-Rotational-Properties-of-21-Sc}%
  \BibitemOpen
  \bibfield  {author} {\bibinfo {author} {\bibfnamefont {V.~C.}\ \bibnamefont {Rubin}}, \bibinfo {author} {\bibfnamefont {W.~K.}\ \bibnamefont {Ford}}, \ and\ \bibinfo {author} {\bibfnamefont {N.}~\bibnamefont {Thonnard}},\ }\href {\doibase 10.1086/158003} {\bibfield  {journal} {\bibinfo  {journal} {Astrophysical Journal}\ }\textbf {\bibinfo {volume} {238}},\ \bibinfo {pages} {471} (\bibinfo {year} {1980})}\BibitemShut {NoStop}%
\bibitem [{\citenamefont {{Peebles}}(1982)}]{Peebles:1982-Large-scale-background-temperature}%
  \BibitemOpen
  \bibfield  {author} {\bibinfo {author} {\bibfnamefont {P.~J.~E.}\ \bibnamefont {{Peebles}}},\ }\href {\doibase 10.1086/183911} {\bibfield  {journal} {\bibinfo  {journal} {\apjl}\ }\textbf {\bibinfo {volume} {263}},\ \bibinfo {pages} {L1} (\bibinfo {year} {1982})}\BibitemShut {NoStop}%
\bibitem [{\citenamefont {Aghanim}\ \emph {et~al.}(2021)\citenamefont {Aghanim}, \citenamefont {Akrami}, \citenamefont {Ashdown}, \citenamefont {Aumont}, \citenamefont {Baccigalupi}, \citenamefont {Ballardini}, \citenamefont {Banday}, \citenamefont {Barreiro}, \citenamefont {Bartolo}, \citenamefont {Basak}, \citenamefont {Battye}, \citenamefont {Benabed}, \citenamefont {Bernard}, \citenamefont {Bersanelli}, \citenamefont {Bielewicz}, \citenamefont {Bock}, \citenamefont {Bond}, \citenamefont {Borrill}, \citenamefont {Bouchet}, \citenamefont {Boulanger}, \citenamefont {Bucher}, \citenamefont {Burigana}, \citenamefont {Butler}, \citenamefont {Calabrese}, \citenamefont {Cardoso}, \citenamefont {Carron}, \citenamefont {Challinor}, \citenamefont {Chiang}, \citenamefont {Chluba}, \citenamefont {Colombo}, \citenamefont {Combet}, \citenamefont {Contreras}, \citenamefont {Crill}, \citenamefont {Cuttaia}, \citenamefont {de~Bernardis}, \citenamefont {de~Zotti}, \citenamefont {Delabrouille}, \citenamefont {Delouis},
  \citenamefont {Di~Valentino}, \citenamefont {Diego}, \citenamefont {Dore}, \citenamefont {Douspis}, \citenamefont {Ducout}, \citenamefont {Dupac}, \citenamefont {Dusini}, \citenamefont {Efstathiou}, \citenamefont {Elsner}, \citenamefont {Ensslin}, \citenamefont {Eriksen}, \citenamefont {Fantaye}, \citenamefont {Farhang}, \citenamefont {Fergusson}, \citenamefont {Fernandez-Cobos}, \citenamefont {Finelli}, \citenamefont {Forastieri}, \citenamefont {Frailis}, \citenamefont {Fraisse}, \citenamefont {Franceschi}, \citenamefont {Frolov}, \citenamefont {Galeotta}, \citenamefont {Galli}, \citenamefont {Ganga}, \citenamefont {Genova-Santos}, \citenamefont {Gerbino}, \citenamefont {Ghosh}, \citenamefont {Gonzalez-Nuevo}, \citenamefont {Gorski}, \citenamefont {Gratton}, \citenamefont {Gruppuso}, \citenamefont {Gudmundsson}, \citenamefont {Hamann}, \citenamefont {Handley}, \citenamefont {Hansen}, \citenamefont {Herranz}, \citenamefont {Hildebrandt}, \citenamefont {Hivon}, \citenamefont {Huang}, \citenamefont {Jaffe},
  \citenamefont {Jones}, \citenamefont {Karakci}, \citenamefont {Keihanen}, \citenamefont {Keskitalo}, \citenamefont {Kiiveri}, \citenamefont {Kim}, \citenamefont {Kisner}, \citenamefont {Knox}, \citenamefont {Krachmalnicoff}, \citenamefont {Kunz}, \citenamefont {Kurki-Suonio}, \citenamefont {Lagache}, \citenamefont {Lamarre}, \citenamefont {Lasenby}, \citenamefont {Lattanzi}, \citenamefont {Lawrence}, \citenamefont {Le~Jeune}, \citenamefont {Lemos}, \citenamefont {Lesgourgues}, \citenamefont {Levrier}, \citenamefont {Lewis}, \citenamefont {Liguori} \emph {et~al.}}]{Aghanim:2021-Planck-2018-results--VI--Cosmo}%
  \BibitemOpen
  \bibfield  {author} {\bibinfo {author} {\bibfnamefont {N.}~\bibnamefont {Aghanim}}, \bibinfo {author} {\bibfnamefont {Y.}~\bibnamefont {Akrami}}, \bibinfo {author} {\bibfnamefont {M.}~\bibnamefont {Ashdown}}, \bibinfo {author} {\bibfnamefont {J.}~\bibnamefont {Aumont}}, \bibinfo {author} {\bibfnamefont {C.}~\bibnamefont {Baccigalupi}}, \bibinfo {author} {\bibfnamefont {M.}~\bibnamefont {Ballardini}}, \bibinfo {author} {\bibfnamefont {A.~J.}\ \bibnamefont {Banday}}, \bibinfo {author} {\bibfnamefont {R.~B.}\ \bibnamefont {Barreiro}}, \bibinfo {author} {\bibfnamefont {N.}~\bibnamefont {Bartolo}}, \bibinfo {author} {\bibfnamefont {S.}~\bibnamefont {Basak}}, \bibinfo {author} {\bibfnamefont {R.}~\bibnamefont {Battye}}, \bibinfo {author} {\bibfnamefont {K.}~\bibnamefont {Benabed}}, \bibinfo {author} {\bibfnamefont {J.~P.}\ \bibnamefont {Bernard}}, \bibinfo {author} {\bibfnamefont {M.}~\bibnamefont {Bersanelli}}, \bibinfo {author} {\bibfnamefont {P.}~\bibnamefont {Bielewicz}}, \bibinfo {author} {\bibfnamefont
  {J.~J.}\ \bibnamefont {Bock}}, \bibinfo {author} {\bibfnamefont {J.~R.}\ \bibnamefont {Bond}}, \bibinfo {author} {\bibfnamefont {J.}~\bibnamefont {Borrill}}, \bibinfo {author} {\bibfnamefont {F.~R.}\ \bibnamefont {Bouchet}}, \bibinfo {author} {\bibfnamefont {F.}~\bibnamefont {Boulanger}}, \bibinfo {author} {\bibfnamefont {M.}~\bibnamefont {Bucher}}, \bibinfo {author} {\bibfnamefont {C.}~\bibnamefont {Burigana}}, \bibinfo {author} {\bibfnamefont {R.~C.}\ \bibnamefont {Butler}}, \bibinfo {author} {\bibfnamefont {E.}~\bibnamefont {Calabrese}}, \bibinfo {author} {\bibfnamefont {J.~F.}\ \bibnamefont {Cardoso}}, \bibinfo {author} {\bibfnamefont {J.}~\bibnamefont {Carron}}, \bibinfo {author} {\bibfnamefont {A.}~\bibnamefont {Challinor}}, \bibinfo {author} {\bibfnamefont {H.~C.}\ \bibnamefont {Chiang}}, \bibinfo {author} {\bibfnamefont {J.}~\bibnamefont {Chluba}}, \bibinfo {author} {\bibfnamefont {L.~P.~L.}\ \bibnamefont {Colombo}}, \bibinfo {author} {\bibfnamefont {C.}~\bibnamefont {Combet}}, \bibinfo {author}
  {\bibfnamefont {D.}~\bibnamefont {Contreras}}, \bibinfo {author} {\bibfnamefont {B.~P.}\ \bibnamefont {Crill}}, \bibinfo {author} {\bibfnamefont {F.}~\bibnamefont {Cuttaia}}, \bibinfo {author} {\bibfnamefont {P.}~\bibnamefont {de~Bernardis}}, \bibinfo {author} {\bibfnamefont {G.}~\bibnamefont {de~Zotti}}, \bibinfo {author} {\bibfnamefont {J.}~\bibnamefont {Delabrouille}}, \bibinfo {author} {\bibfnamefont {J.~M.}\ \bibnamefont {Delouis}}, \bibinfo {author} {\bibfnamefont {E.}~\bibnamefont {Di~Valentino}}, \bibinfo {author} {\bibfnamefont {J.~M.}\ \bibnamefont {Diego}}, \bibinfo {author} {\bibfnamefont {O.}~\bibnamefont {Dore}}, \bibinfo {author} {\bibfnamefont {M.}~\bibnamefont {Douspis}}, \bibinfo {author} {\bibfnamefont {A.}~\bibnamefont {Ducout}}, \bibinfo {author} {\bibfnamefont {X.}~\bibnamefont {Dupac}}, \bibinfo {author} {\bibfnamefont {S.}~\bibnamefont {Dusini}}, \bibinfo {author} {\bibfnamefont {G.}~\bibnamefont {Efstathiou}}, \bibinfo {author} {\bibfnamefont {F.}~\bibnamefont {Elsner}}, \bibinfo
  {author} {\bibfnamefont {T.~A.}\ \bibnamefont {Ensslin}}, \bibinfo {author} {\bibfnamefont {H.~K.}\ \bibnamefont {Eriksen}}, \bibinfo {author} {\bibfnamefont {Y.}~\bibnamefont {Fantaye}}, \bibinfo {author} {\bibfnamefont {M.}~\bibnamefont {Farhang}}, \bibinfo {author} {\bibfnamefont {J.}~\bibnamefont {Fergusson}}, \bibinfo {author} {\bibfnamefont {R.}~\bibnamefont {Fernandez-Cobos}}, \bibinfo {author} {\bibfnamefont {F.}~\bibnamefont {Finelli}}, \bibinfo {author} {\bibfnamefont {F.}~\bibnamefont {Forastieri}}, \bibinfo {author} {\bibfnamefont {M.}~\bibnamefont {Frailis}}, \bibinfo {author} {\bibfnamefont {A.~A.}\ \bibnamefont {Fraisse}}, \bibinfo {author} {\bibfnamefont {E.}~\bibnamefont {Franceschi}}, \bibinfo {author} {\bibfnamefont {A.}~\bibnamefont {Frolov}}, \bibinfo {author} {\bibfnamefont {S.}~\bibnamefont {Galeotta}}, \bibinfo {author} {\bibfnamefont {S.}~\bibnamefont {Galli}}, \bibinfo {author} {\bibfnamefont {K.}~\bibnamefont {Ganga}}, \bibinfo {author} {\bibfnamefont {R.~T.}\ \bibnamefont
  {Genova-Santos}}, \bibinfo {author} {\bibfnamefont {M.}~\bibnamefont {Gerbino}}, \bibinfo {author} {\bibfnamefont {T.}~\bibnamefont {Ghosh}}, \bibinfo {author} {\bibfnamefont {J.}~\bibnamefont {Gonzalez-Nuevo}}, \bibinfo {author} {\bibfnamefont {K.~M.}\ \bibnamefont {Gorski}}, \bibinfo {author} {\bibfnamefont {S.}~\bibnamefont {Gratton}}, \bibinfo {author} {\bibfnamefont {A.}~\bibnamefont {Gruppuso}}, \bibinfo {author} {\bibfnamefont {J.~E.}\ \bibnamefont {Gudmundsson}}, \bibinfo {author} {\bibfnamefont {J.}~\bibnamefont {Hamann}}, \bibinfo {author} {\bibfnamefont {W.}~\bibnamefont {Handley}}, \bibinfo {author} {\bibfnamefont {F.~K.}\ \bibnamefont {Hansen}}, \bibinfo {author} {\bibfnamefont {D.}~\bibnamefont {Herranz}}, \bibinfo {author} {\bibfnamefont {S.~R.}\ \bibnamefont {Hildebrandt}}, \bibinfo {author} {\bibfnamefont {E.}~\bibnamefont {Hivon}}, \bibinfo {author} {\bibfnamefont {Z.}~\bibnamefont {Huang}}, \bibinfo {author} {\bibfnamefont {A.~H.}\ \bibnamefont {Jaffe}}, \bibinfo {author} {\bibfnamefont
  {W.~C.}\ \bibnamefont {Jones}}, \bibinfo {author} {\bibfnamefont {A.}~\bibnamefont {Karakci}}, \bibinfo {author} {\bibfnamefont {E.}~\bibnamefont {Keihanen}}, \bibinfo {author} {\bibfnamefont {R.}~\bibnamefont {Keskitalo}}, \bibinfo {author} {\bibfnamefont {K.}~\bibnamefont {Kiiveri}}, \bibinfo {author} {\bibfnamefont {J.}~\bibnamefont {Kim}}, \bibinfo {author} {\bibfnamefont {T.~S.}\ \bibnamefont {Kisner}}, \bibinfo {author} {\bibfnamefont {L.}~\bibnamefont {Knox}}, \bibinfo {author} {\bibfnamefont {N.}~\bibnamefont {Krachmalnicoff}}, \bibinfo {author} {\bibfnamefont {M.}~\bibnamefont {Kunz}}, \bibinfo {author} {\bibfnamefont {H.}~\bibnamefont {Kurki-Suonio}}, \bibinfo {author} {\bibfnamefont {G.}~\bibnamefont {Lagache}}, \bibinfo {author} {\bibfnamefont {J.~M.}\ \bibnamefont {Lamarre}}, \bibinfo {author} {\bibfnamefont {A.}~\bibnamefont {Lasenby}}, \bibinfo {author} {\bibfnamefont {M.}~\bibnamefont {Lattanzi}}, \bibinfo {author} {\bibfnamefont {C.~R.}\ \bibnamefont {Lawrence}}, \bibinfo {author}
  {\bibfnamefont {M.}~\bibnamefont {Le~Jeune}}, \bibinfo {author} {\bibfnamefont {P.}~\bibnamefont {Lemos}}, \bibinfo {author} {\bibfnamefont {J.}~\bibnamefont {Lesgourgues}}, \bibinfo {author} {\bibfnamefont {F.}~\bibnamefont {Levrier}}, \bibinfo {author} {\bibfnamefont {A.}~\bibnamefont {Lewis}}, \bibinfo {author} {\bibfnamefont {M.}~\bibnamefont {Liguori}},  \emph {et~al.},\ }\href {\doibase 10.1051/0004-6361/201833910} {\bibfield  {journal} {\bibinfo  {journal} {Astronomy \& Astrophysics}\ }\textbf {\bibinfo {volume} {652}} (\bibinfo {year} {2021}),\ 10.1051/0004-6361/201833910}\BibitemShut {NoStop}%
\bibitem [{\citenamefont {{Peebles}}(1984)}]{Peebles:1984-Tests-of-cosmological-models}%
  \BibitemOpen
  \bibfield  {author} {\bibinfo {author} {\bibfnamefont {P.~J.~E.}\ \bibnamefont {{Peebles}}},\ }\href {\doibase 10.1086/162425} {\bibfield  {journal} {\bibinfo  {journal} {\apj}\ }\textbf {\bibinfo {volume} {284}},\ \bibinfo {pages} {439} (\bibinfo {year} {1984})}\BibitemShut {NoStop}%
\bibitem [{\citenamefont {{Spergel}}\ \emph {et~al.}(2003)\citenamefont {{Spergel}}, \citenamefont {{Verde}}, \citenamefont {{Peiris}}, \citenamefont {{Komatsu}}, \citenamefont {{Nolta}}, \citenamefont {{Bennett}}, \citenamefont {{Halpern}}, \citenamefont {{Hinshaw}}, \citenamefont {{Jarosik}}, \citenamefont {{Kogut}}, \citenamefont {{Limon}}, \citenamefont {{Meyer}}, \citenamefont {{Page}}, \citenamefont {{Tucker}}, \citenamefont {{Weiland}}, \citenamefont {{Wollack}},\ and\ \citenamefont {{Wright}}}]{Spergel:2003-First-Year-Wilkinson-Microwave-Anisotropy}%
  \BibitemOpen
  \bibfield  {author} {\bibinfo {author} {\bibfnamefont {D.~N.}\ \bibnamefont {{Spergel}}}, \bibinfo {author} {\bibfnamefont {L.}~\bibnamefont {{Verde}}}, \bibinfo {author} {\bibfnamefont {H.~V.}\ \bibnamefont {{Peiris}}}, \bibinfo {author} {\bibfnamefont {E.}~\bibnamefont {{Komatsu}}}, \bibinfo {author} {\bibfnamefont {M.~R.}\ \bibnamefont {{Nolta}}}, \bibinfo {author} {\bibfnamefont {C.~L.}\ \bibnamefont {{Bennett}}}, \bibinfo {author} {\bibfnamefont {M.}~\bibnamefont {{Halpern}}}, \bibinfo {author} {\bibfnamefont {G.}~\bibnamefont {{Hinshaw}}}, \bibinfo {author} {\bibfnamefont {N.}~\bibnamefont {{Jarosik}}}, \bibinfo {author} {\bibfnamefont {A.}~\bibnamefont {{Kogut}}}, \bibinfo {author} {\bibfnamefont {M.}~\bibnamefont {{Limon}}}, \bibinfo {author} {\bibfnamefont {S.~S.}\ \bibnamefont {{Meyer}}}, \bibinfo {author} {\bibfnamefont {L.}~\bibnamefont {{Page}}}, \bibinfo {author} {\bibfnamefont {G.~S.}\ \bibnamefont {{Tucker}}}, \bibinfo {author} {\bibfnamefont {J.~L.}\ \bibnamefont {{Weiland}}}, \bibinfo
  {author} {\bibfnamefont {E.}~\bibnamefont {{Wollack}}}, \ and\ \bibinfo {author} {\bibfnamefont {E.~L.}\ \bibnamefont {{Wright}}},\ }\href {\doibase 10.1086/377226} {\bibfield  {journal} {\bibinfo  {journal} {\apjs}\ }\textbf {\bibinfo {volume} {148}},\ \bibinfo {pages} {175} (\bibinfo {year} {2003})},\ \Eprint {http://arxiv.org/abs/astro-ph/0302209} {arXiv:astro-ph/0302209 [astro-ph]} \BibitemShut {NoStop}%
\bibitem [{\citenamefont {{Komatsu}}\ \emph {et~al.}(2011)\citenamefont {{Komatsu}}, \citenamefont {{Smith}}, \citenamefont {{Dunkley}}, \citenamefont {{Bennett}}, \citenamefont {{Gold}}, \citenamefont {{Hinshaw}}, \citenamefont {{Jarosik}}, \citenamefont {{Larson}}, \citenamefont {{Nolta}}, \citenamefont {{Page}}, \citenamefont {{Spergel}}, \citenamefont {{Halpern}}, \citenamefont {{Hill}}, \citenamefont {{Kogut}}, \citenamefont {{Limon}}, \citenamefont {{Meyer}}, \citenamefont {{Odegard}}, \citenamefont {{Tucker}}, \citenamefont {{Weiland}}, \citenamefont {{Wollack}},\ and\ \citenamefont {{Wright}}}]{Komatsu:Seven-year-Wilkinson-Microwave-Anisotropy-Probe}%
  \BibitemOpen
  \bibfield  {author} {\bibinfo {author} {\bibfnamefont {E.}~\bibnamefont {{Komatsu}}}, \bibinfo {author} {\bibfnamefont {K.~M.}\ \bibnamefont {{Smith}}}, \bibinfo {author} {\bibfnamefont {J.}~\bibnamefont {{Dunkley}}}, \bibinfo {author} {\bibfnamefont {C.~L.}\ \bibnamefont {{Bennett}}}, \bibinfo {author} {\bibfnamefont {B.}~\bibnamefont {{Gold}}}, \bibinfo {author} {\bibfnamefont {G.}~\bibnamefont {{Hinshaw}}}, \bibinfo {author} {\bibfnamefont {N.}~\bibnamefont {{Jarosik}}}, \bibinfo {author} {\bibfnamefont {D.}~\bibnamefont {{Larson}}}, \bibinfo {author} {\bibfnamefont {M.~R.}\ \bibnamefont {{Nolta}}}, \bibinfo {author} {\bibfnamefont {L.}~\bibnamefont {{Page}}}, \bibinfo {author} {\bibfnamefont {D.~N.}\ \bibnamefont {{Spergel}}}, \bibinfo {author} {\bibfnamefont {M.}~\bibnamefont {{Halpern}}}, \bibinfo {author} {\bibfnamefont {R.~S.}\ \bibnamefont {{Hill}}}, \bibinfo {author} {\bibfnamefont {A.}~\bibnamefont {{Kogut}}}, \bibinfo {author} {\bibfnamefont {M.}~\bibnamefont {{Limon}}}, \bibinfo {author}
  {\bibfnamefont {S.~S.}\ \bibnamefont {{Meyer}}}, \bibinfo {author} {\bibfnamefont {N.}~\bibnamefont {{Odegard}}}, \bibinfo {author} {\bibfnamefont {G.~S.}\ \bibnamefont {{Tucker}}}, \bibinfo {author} {\bibfnamefont {J.~L.}\ \bibnamefont {{Weiland}}}, \bibinfo {author} {\bibfnamefont {E.}~\bibnamefont {{Wollack}}}, \ and\ \bibinfo {author} {\bibfnamefont {E.~L.}\ \bibnamefont {{Wright}}},\ }\href {\doibase 10.1088/0067-0049/192/2/18} {\bibfield  {journal} {\bibinfo  {journal} {\apjs}\ }\textbf {\bibinfo {volume} {192}},\ \bibinfo {eid} {18} (\bibinfo {year} {2011})},\ \Eprint {http://arxiv.org/abs/1001.4538} {arXiv:1001.4538 [astro-ph.CO]} \BibitemShut {NoStop}%
\bibitem [{\citenamefont {{Frenk}}\ and\ \citenamefont {{White}}(2012)}]{Frenk:2012-Dark-matter-and-cosmic-structure}%
  \BibitemOpen
  \bibfield  {author} {\bibinfo {author} {\bibfnamefont {C.~S.}\ \bibnamefont {{Frenk}}}\ and\ \bibinfo {author} {\bibfnamefont {S.~D.~M.}\ \bibnamefont {{White}}},\ }\href {\doibase 10.1002/andp.201200212} {\bibfield  {journal} {\bibinfo  {journal} {Annalen der Physik}\ }\textbf {\bibinfo {volume} {524}},\ \bibinfo {pages} {507} (\bibinfo {year} {2012})},\ \Eprint {http://arxiv.org/abs/1210.0544} {arXiv:1210.0544 [astro-ph.CO]} \BibitemShut {NoStop}%
\bibitem [{\citenamefont {Steigman}\ and\ \citenamefont {Turner}(1985)}]{Steigman:1985-Cosmological-Constraints-on-th}%
  \BibitemOpen
  \bibfield  {author} {\bibinfo {author} {\bibfnamefont {G.}~\bibnamefont {Steigman}}\ and\ \bibinfo {author} {\bibfnamefont {M.~S.}\ \bibnamefont {Turner}},\ }\href {\doibase 10.1016/0550-3213(85)90537-1} {\bibfield  {journal} {\bibinfo  {journal} {Nuclear Physics B}\ }\textbf {\bibinfo {volume} {253}},\ \bibinfo {pages} {375} (\bibinfo {year} {1985})}\BibitemShut {NoStop}%
\bibitem [{\citenamefont {Jungman}\ \emph {et~al.}(1996)\citenamefont {Jungman}, \citenamefont {Kamionkowski},\ and\ \citenamefont {Griest}}]{Jungman:1996-Supersymmetric-dark-matter}%
  \BibitemOpen
  \bibfield  {author} {\bibinfo {author} {\bibfnamefont {G.}~\bibnamefont {Jungman}}, \bibinfo {author} {\bibfnamefont {M.}~\bibnamefont {Kamionkowski}}, \ and\ \bibinfo {author} {\bibfnamefont {K.}~\bibnamefont {Griest}},\ }\href {\doibase 10.1016/0370-1573(95)00058-5} {\bibfield  {journal} {\bibinfo  {journal} {Physics Reports-Review Section of Physics Letters}\ }\textbf {\bibinfo {volume} {267}},\ \bibinfo {pages} {195} (\bibinfo {year} {1996})}\BibitemShut {NoStop}%
\bibitem [{\citenamefont {Griest}\ and\ \citenamefont {Kamionkowski}(1990)}]{Griest:1990-Unitarity-Limits-on-the-Mass-a}%
  \BibitemOpen
  \bibfield  {author} {\bibinfo {author} {\bibfnamefont {K.}~\bibnamefont {Griest}}\ and\ \bibinfo {author} {\bibfnamefont {M.}~\bibnamefont {Kamionkowski}},\ }\href {\doibase 10.1103/PhysRevLett.64.615} {\bibfield  {journal} {\bibinfo  {journal} {Physical Review Letters}\ }\textbf {\bibinfo {volume} {64}},\ \bibinfo {pages} {615} (\bibinfo {year} {1990})}\BibitemShut {NoStop}%
\bibitem [{\citenamefont {Moroi}\ and\ \citenamefont {Yin}(2021)}]{Moroi:2020-Light-Dark-Matter-from-Inflaton-Decay}%
  \BibitemOpen
  \bibfield  {author} {\bibinfo {author} {\bibfnamefont {T.}~\bibnamefont {Moroi}}\ and\ \bibinfo {author} {\bibfnamefont {W.}~\bibnamefont {Yin}},\ }\href {\doibase 10.1007/JHEP03(2021)301} {\bibfield  {journal} {\bibinfo  {journal} {JHEP}\ }\textbf {\bibinfo {volume} {03}},\ \bibinfo {pages} {301} (\bibinfo {year} {2021})},\ \Eprint {http://arxiv.org/abs/2011.09475} {arXiv:2011.09475 [hep-ph]} \BibitemShut {NoStop}%
\bibitem [{\citenamefont {{Hall}}\ \emph {et~al.}(2010)\citenamefont {{Hall}}, \citenamefont {{Jedamzik}}, \citenamefont {{March-Russell}},\ and\ \citenamefont {{West}}}]{Hall:2010-Freeze-inproduction-of-FIMP-dark-matter}%
  \BibitemOpen
  \bibfield  {author} {\bibinfo {author} {\bibfnamefont {L.~J.}\ \bibnamefont {{Hall}}}, \bibinfo {author} {\bibfnamefont {K.}~\bibnamefont {{Jedamzik}}}, \bibinfo {author} {\bibfnamefont {J.}~\bibnamefont {{March-Russell}}}, \ and\ \bibinfo {author} {\bibfnamefont {S.~M.}\ \bibnamefont {{West}}},\ }\href {\doibase 10.1007/JHEP03(2010)080} {\bibfield  {journal} {\bibinfo  {journal} {Journal of High Energy Physics}\ }\textbf {\bibinfo {volume} {2010}},\ \bibinfo {eid} {80} (\bibinfo {year} {2010})},\ \Eprint {http://arxiv.org/abs/0911.1120} {arXiv:0911.1120 [hep-ph]} \BibitemShut {NoStop}%
\bibitem [{\citenamefont {Chung}\ \emph {et~al.}(2001)\citenamefont {Chung}, \citenamefont {Crotty}, \citenamefont {Kolb},\ and\ \citenamefont {Riotto}}]{Chung:2001-On-the-gravitational-production}%
  \BibitemOpen
  \bibfield  {author} {\bibinfo {author} {\bibfnamefont {D.~J.~H.}\ \bibnamefont {Chung}}, \bibinfo {author} {\bibfnamefont {P.}~\bibnamefont {Crotty}}, \bibinfo {author} {\bibfnamefont {E.~W.}\ \bibnamefont {Kolb}}, \ and\ \bibinfo {author} {\bibfnamefont {A.}~\bibnamefont {Riotto}},\ }\href {\doibase 10.1103/PhysRevD.64.043503} {\bibfield  {journal} {\bibinfo  {journal} {Phys. Rev. D}\ }\textbf {\bibinfo {volume} {64}},\ \bibinfo {pages} {043503} (\bibinfo {year} {2001})}\BibitemShut {NoStop}%
\bibitem [{\citenamefont {Xu}(2021)}]{Xu:2021-Inverse-mass-cascade-mass-function}%
  \BibitemOpen
  \bibfield  {author} {\bibinfo {author} {\bibfnamefont {Z.}~\bibnamefont {Xu}},\ }\href {\doibase 10.48550/ARXIV.2109.09985} {\bibfield  {journal} {\bibinfo  {journal} {arXiv e-prints}\ ,\ \bibinfo {pages} {arXiv:2109.09985}} (\bibinfo {year} {2021})}\BibitemShut {NoStop}%
\bibitem [{\citenamefont {Frenk}\ \emph {et~al.}(2000)\citenamefont {Frenk}, \citenamefont {Colberg}, \citenamefont {Couchman}, \citenamefont {Efstathiou}, \citenamefont {Evrard}, \citenamefont {Jenkins}, \citenamefont {MacFarland}, \citenamefont {Moore}, \citenamefont {Peacock}, \citenamefont {Pearce}, \citenamefont {Thomas}, \citenamefont {White},\ and\ \citenamefont {Yoshida.}}]{Frenk:2000-Public-Release-of-N-body-simul}%
  \BibitemOpen
  \bibfield  {author} {\bibinfo {author} {\bibfnamefont {C.~S.}\ \bibnamefont {Frenk}}, \bibinfo {author} {\bibfnamefont {J.~M.}\ \bibnamefont {Colberg}}, \bibinfo {author} {\bibfnamefont {H.~M.~P.}\ \bibnamefont {Couchman}}, \bibinfo {author} {\bibfnamefont {G.}~\bibnamefont {Efstathiou}}, \bibinfo {author} {\bibfnamefont {A.~E.}\ \bibnamefont {Evrard}}, \bibinfo {author} {\bibfnamefont {A.}~\bibnamefont {Jenkins}}, \bibinfo {author} {\bibfnamefont {T.~J.}\ \bibnamefont {MacFarland}}, \bibinfo {author} {\bibfnamefont {B.}~\bibnamefont {Moore}}, \bibinfo {author} {\bibfnamefont {J.~A.}\ \bibnamefont {Peacock}}, \bibinfo {author} {\bibfnamefont {F.~R.}\ \bibnamefont {Pearce}}, \bibinfo {author} {\bibfnamefont {P.~A.}\ \bibnamefont {Thomas}}, \bibinfo {author} {\bibfnamefont {S.~D.~M.}\ \bibnamefont {White}}, \ and\ \bibinfo {author} {\bibfnamefont {N.}~\bibnamefont {Yoshida.}},\ }\href {\doibase 10.48550/arXiv.astro-ph/0007362} {\bibfield  {journal} {\bibinfo  {journal} {arXiv:astro-ph/0007362v1}\ } (\bibinfo
  {year} {2000}),\ 10.48550/arXiv.astro-ph/0007362}\BibitemShut {NoStop}%
\bibitem [{\citenamefont {Jenkins}\ \emph {et~al.}(1998)\citenamefont {Jenkins}, \citenamefont {Frenk}, \citenamefont {Pearce}, \citenamefont {Thomas}, \citenamefont {Colberg}, \citenamefont {White}, \citenamefont {Couchman}, \citenamefont {Peacock}, \citenamefont {Efstathiou},\ and\ \citenamefont {Nelson}}]{Jenkins:1998-Evolution-of-structure-in-cold}%
  \BibitemOpen
  \bibfield  {author} {\bibinfo {author} {\bibfnamefont {A.}~\bibnamefont {Jenkins}}, \bibinfo {author} {\bibfnamefont {C.~S.}\ \bibnamefont {Frenk}}, \bibinfo {author} {\bibfnamefont {F.~R.}\ \bibnamefont {Pearce}}, \bibinfo {author} {\bibfnamefont {P.~A.}\ \bibnamefont {Thomas}}, \bibinfo {author} {\bibfnamefont {J.~M.}\ \bibnamefont {Colberg}}, \bibinfo {author} {\bibfnamefont {S.~D.~M.}\ \bibnamefont {White}}, \bibinfo {author} {\bibfnamefont {H.~M.~P.}\ \bibnamefont {Couchman}}, \bibinfo {author} {\bibfnamefont {J.~A.}\ \bibnamefont {Peacock}}, \bibinfo {author} {\bibfnamefont {G.}~\bibnamefont {Efstathiou}}, \ and\ \bibinfo {author} {\bibfnamefont {A.~H.}\ \bibnamefont {Nelson}},\ }\href {\doibase 10.1086/305615} {\bibfield  {journal} {\bibinfo  {journal} {Astrophysical Journal}\ }\textbf {\bibinfo {volume} {499}},\ \bibinfo {pages} {20} (\bibinfo {year} {1998})}\BibitemShut {NoStop}%
\bibitem [{\citenamefont {Xu}(2026)}]{Xu:2026-Superheavy-dark-matter-and-stepped-dark-radiation}%
  \BibitemOpen
  \bibfield  {author} {\bibinfo {author} {\bibfnamefont {Z.}~\bibnamefont {Xu}},\ }\href {\doibase 10.21203/rs.3.rs-8570078/v1} {\bibfield  {journal} {\bibinfo  {journal} {Research Square e-prints}\ } (\bibinfo {year} {2026}),\ 10.21203/rs.3.rs-8570078/v1}\BibitemShut {NoStop}%
\bibitem [{\citenamefont {Schneider}\ \emph {et~al.}(2013)\citenamefont {Schneider}, \citenamefont {Smith},\ and\ \citenamefont {Reed}}]{Schneide:2013-Halo-mass-function-and-the-free-streaming-scale}%
  \BibitemOpen
  \bibfield  {author} {\bibinfo {author} {\bibfnamefont {A.}~\bibnamefont {Schneider}}, \bibinfo {author} {\bibfnamefont {R.~E.}\ \bibnamefont {Smith}}, \ and\ \bibinfo {author} {\bibfnamefont {D.}~\bibnamefont {Reed}},\ }\href {\doibase 10.1093/mnras/stt829} {\bibfield  {journal} {\bibinfo  {journal} {Monthly Notices of the Royal Astronomical Society}\ }\textbf {\bibinfo {volume} {433}},\ \bibinfo {pages} {1573} (\bibinfo {year} {2013})},\ \Eprint {http://arxiv.org/abs/https://academic.oup.com/mnras/article-pdf/433/2/1573/4926418/stt829.pdf} {https://academic.oup.com/mnras/article-pdf/433/2/1573/4926418/stt829.pdf} \BibitemShut {NoStop}%
\bibitem [{\citenamefont {Bertschinger}(2006)}]{Bertschinger:2006-Effects-of-cold-dark-matter-decoupling}%
  \BibitemOpen
  \bibfield  {author} {\bibinfo {author} {\bibfnamefont {E.}~\bibnamefont {Bertschinger}},\ }\href {\doibase 10.1103/PhysRevD.74.063509} {\bibfield  {journal} {\bibinfo  {journal} {Phys. Rev. D}\ }\textbf {\bibinfo {volume} {74}},\ \bibinfo {pages} {063509} (\bibinfo {year} {2006})}\BibitemShut {NoStop}%
\bibitem [{\citenamefont {Gunn}\ and\ \citenamefont {Gott}(1972)}]{Gunn:1972-Infall-of-Matter-into-Clusters}%
  \BibitemOpen
  \bibfield  {author} {\bibinfo {author} {\bibfnamefont {J.~E.}\ \bibnamefont {Gunn}}\ and\ \bibinfo {author} {\bibfnamefont {J.~R.}\ \bibnamefont {Gott}},\ }\href {\doibase 10.1086/151605} {\bibfield  {journal} {\bibinfo  {journal} {Astrophysical Journal}\ }\textbf {\bibinfo {volume} {176}},\ \bibinfo {pages} {1} (\bibinfo {year} {1972})}\BibitemShut {NoStop}%
\bibitem [{\citenamefont {Xu}(2023{\natexlab{a}})}]{Xu:2023-On-the-statistical-theory-of-self-gravitating}%
  \BibitemOpen
  \bibfield  {author} {\bibinfo {author} {\bibfnamefont {Z.}~\bibnamefont {Xu}},\ }\href {\doibase 10.1063/5.0151129} {\bibfield  {journal} {\bibinfo  {journal} {Physics of Fluids}\ }\textbf {\bibinfo {volume} {35}},\ \bibinfo {pages} {077105} (\bibinfo {year} {2023}{\natexlab{a}})},\ \Eprint {http://arxiv.org/abs/2202.00910} {arXiv:2202.00910 [astro-ph]} \BibitemShut {NoStop}%
\bibitem [{\citenamefont {Xu}(2024{\natexlab{a}})}]{Xu:2024-High-order-kinematic}%
  \BibitemOpen
  \bibfield  {author} {\bibinfo {author} {\bibfnamefont {Z.}~\bibnamefont {Xu}},\ }\href {\doibase 10.1063/5.0215026} {\bibfield  {journal} {\bibinfo  {journal} {Physics of Fluids}\ }\textbf {\bibinfo {volume} {36}},\ \bibinfo {pages} {075146} (\bibinfo {year} {2024}{\natexlab{a}})},\ \Eprint {http://arxiv.org/abs/2202.02991} {arXiv:2202.02991 [astro-ph]} \BibitemShut {NoStop}%
\bibitem [{\citenamefont {Xu}(2024{\natexlab{b}})}]{Xu:2024-Scale-and-redshift-variation}%
  \BibitemOpen
  \bibfield  {author} {\bibinfo {author} {\bibfnamefont {Z.}~\bibnamefont {Xu}},\ }\href {\doibase 10.1063/5.0236964} {\bibfield  {journal} {\bibinfo  {journal} {Physics of Fluids}\ }\textbf {\bibinfo {volume} {36}},\ \bibinfo {pages} {117158} (\bibinfo {year} {2024}{\natexlab{b}})},\ \Eprint {http://arxiv.org/abs/2202.06515} {arXiv:2202.06515 [astro-ph]} \BibitemShut {NoStop}%
\bibitem [{\citenamefont {{Meszaros}}(1974)}]{Meszaros:1974-The-behaviour-of-point-masses-in-an-expanding-cosmological}%
  \BibitemOpen
  \bibfield  {author} {\bibinfo {author} {\bibfnamefont {P.}~\bibnamefont {{Meszaros}}},\ }\href@noop {} {\bibfield  {journal} {\bibinfo  {journal} {\aap}\ }\textbf {\bibinfo {volume} {37}},\ \bibinfo {pages} {225} (\bibinfo {year} {1974})}\BibitemShut {NoStop}%
\bibitem [{\citenamefont {Xu}(2023{\natexlab{b}})}]{Xu:2023-Maximum-entropy-distributions-of-dark-matter}%
  \BibitemOpen
  \bibfield  {author} {\bibinfo {author} {\bibfnamefont {Z.}~\bibnamefont {Xu}},\ }\href {\doibase 10.1051/0004-6361/202346429} {\bibfield  {journal} {\bibinfo  {journal} {A\&A}\ }\textbf {\bibinfo {volume} {675}},\ \bibinfo {pages} {A92} (\bibinfo {year} {2023}{\natexlab{b}})},\ \Eprint {http://arxiv.org/abs/2110.03126} {arXiv:2110.03126 [astro-ph]} \BibitemShut {NoStop}%
\bibitem [{\citenamefont {Xu}(2023{\natexlab{c}})}]{Xu:2023-Dark-matter-halo-mass-functions-and}%
  \BibitemOpen
  \bibfield  {author} {\bibinfo {author} {\bibfnamefont {Z.}~\bibnamefont {Xu}},\ }\href {\doibase 10.1038/s41598-023-42958-6} {\bibfield  {journal} {\bibinfo  {journal} {Scientific Reports}\ }\textbf {\bibinfo {volume} {13}},\ \bibinfo {pages} {16531} (\bibinfo {year} {2023}{\natexlab{c}})},\ \Eprint {http://arxiv.org/abs/2210.01200} {arXiv:2210.01200 [astro-ph]} \BibitemShut {NoStop}%
\bibitem [{\citenamefont {Tseliakhovich}\ and\ \citenamefont {Hirata}(2010)}]{Tseliakhovich:2010-Relative-velocity-of-dark-matter-and-baryonic-fluids}%
  \BibitemOpen
  \bibfield  {author} {\bibinfo {author} {\bibfnamefont {D.}~\bibnamefont {Tseliakhovich}}\ and\ \bibinfo {author} {\bibfnamefont {C.}~\bibnamefont {Hirata}},\ }\href {\doibase 10.1103/PhysRevD.82.083520} {\bibfield  {journal} {\bibinfo  {journal} {Phys. Rev. D}\ }\textbf {\bibinfo {volume} {82}},\ \bibinfo {pages} {083520} (\bibinfo {year} {2010})}\BibitemShut {NoStop}%
\bibitem [{\citenamefont {Xu}(2023{\natexlab{d}})}]{Xu:2023-Universal-scaling-laws-and-density-slope}%
  \BibitemOpen
  \bibfield  {author} {\bibinfo {author} {\bibfnamefont {Z.}~\bibnamefont {Xu}},\ }\href {\doibase 10.1038/s41598-023-31083-z} {\bibfield  {journal} {\bibinfo  {journal} {Scientific Reports}\ }\textbf {\bibinfo {volume} {13}},\ \bibinfo {pages} {4165} (\bibinfo {year} {2023}{\natexlab{d}})},\ \Eprint {http://arxiv.org/abs/2209.03313} {arXiv:2209.03313 [astro-ph]} \BibitemShut {NoStop}%
\bibitem [{\citenamefont {Xu}(2024{\natexlab{c}})}]{Xu:2024-Cosmic-quenching-and-scaling-laws}%
  \BibitemOpen
  \bibfield  {author} {\bibinfo {author} {\bibfnamefont {Z.}~\bibnamefont {Xu}},\ }\href {\doibase 10.1093/mnras/stae2766} {\bibfield  {journal} {\bibinfo  {journal} {Monthly Notices of the Royal Astronomical Society}\ }\textbf {\bibinfo {volume} {536}},\ \bibinfo {pages} {3554} (\bibinfo {year} {2024}{\natexlab{c}})},\ \Eprint {http://arxiv.org/abs/2501.07608} {arXiv:2501.07608 [astro-ph]} \BibitemShut {NoStop}%
\bibitem [{\citenamefont {Xu}(2025)}]{Xu:2025-On-a-Critical-Acceleration-Scale}%
  \BibitemOpen
  \bibfield  {author} {\bibinfo {author} {\bibfnamefont {Z.}~\bibnamefont {Xu}},\ }\href {\doibase 10.3847/1538-4357/adaeb3} {\bibfield  {journal} {\bibinfo  {journal} {The Astrophysical Journal}\ }\textbf {\bibinfo {volume} {981}},\ \bibinfo {pages} {40} (\bibinfo {year} {2025})}\BibitemShut {NoStop}%
\bibitem [{\citenamefont {{Abreu}}\ \emph {et~al.}(2023)\citenamefont {{Abreu}}, \citenamefont {{Aglietta}}, \citenamefont {{Albury}}, \citenamefont {{Allekotte}}, \citenamefont {{Almeida Cheminant}}, \citenamefont {{Almela}}, \citenamefont {{Aloisio}}, \citenamefont {{Alvarez-Mu{\~n}iz}}, \citenamefont {{Alves Batista}}, \citenamefont {{Ammerman Yebra}}, \citenamefont {{Anastasi}}, \citenamefont {{Anchordoqui}}, \citenamefont {{Andrada}}, \citenamefont {{Andringa}}, \citenamefont {{Aramo}}, \citenamefont {{Ara{\'u}jo Ferreira}}, \citenamefont {{Arnone}}, \citenamefont {{Arteaga Vel{\'a}zquez}}, \citenamefont {{Asorey}}, \citenamefont {{Assis}}, \citenamefont {{Avila}}, \citenamefont {{Avocone}}, \citenamefont {{Badescu}}, \citenamefont {{Bakalova}}, \citenamefont {{Balaceanu}}, \citenamefont {{Barbato}}, \citenamefont {{Bellido}}, \citenamefont {{Berat}}, \citenamefont {{Bertaina}}, \citenamefont {{Bhatta}}, \citenamefont {{Biermann}}, \citenamefont {{Binet}}, \citenamefont {{Bismark}}, \citenamefont
  {{Bister}}, \citenamefont {{Biteau}}, \citenamefont {{Blazek}}, \citenamefont {{Bleve}}, \citenamefont {{Bl{\"u}mer}}, \citenamefont {{Boh{\'a}{\v{c}}ov{\'a}}}, \citenamefont {{Boncioli}}, \citenamefont {{Bonifazi}}, \citenamefont {{Bonneau Arbeletche}}, \citenamefont {{Borodai}}, \citenamefont {{Botti}}, \citenamefont {{Brack}}, \citenamefont {{Bretz}}, \citenamefont {{Brichetto Orchera}}, \citenamefont {{Briechle}}, \citenamefont {{Buchholz}}, \citenamefont {{Bueno}}, \citenamefont {{Buitink}}, \citenamefont {{Buscemi}}, \citenamefont {{B{\"u}sken}}, \citenamefont {{Caballero-Mora}}, \citenamefont {{Caccianiga}}, \citenamefont {{Canfora}}, \citenamefont {{Caracas}}, \citenamefont {{Caruso}}, \citenamefont {{Castellina}}, \citenamefont {{Catalani}}, \citenamefont {{Cataldi}}, \citenamefont {{Cazon}}, \citenamefont {{Cerda}}, \citenamefont {{Chinellato}}, \citenamefont {{Chudoba}}, \citenamefont {{Chytka}}, \citenamefont {{Clay}}, \citenamefont {{Cobos Cerutti}}, \citenamefont {{Colalillo}}, \citenamefont
  {{Coleman}}, \citenamefont {{Coluccia}}, \citenamefont {{Concei{\c{c}}{\~a}o}}, \citenamefont {{Condorelli}}, \citenamefont {{Consolati}}, \citenamefont {{Contreras}}, \citenamefont {{Convenga}}, \citenamefont {{Correia dos Santos}}, \citenamefont {{Covault}}, \citenamefont {{Dasso}}, \citenamefont {{Daumiller}}, \citenamefont {{Dawson}}, \citenamefont {{Day}}, \citenamefont {{de Almeida}}, \citenamefont {{de Jes{\'u}s}}, \citenamefont {{de Jong}}, \citenamefont {{de Mello Neto}}, \citenamefont {{De Mitri}}, \citenamefont {{de Oliveira}}, \citenamefont {{de Oliveira Franco}}, \citenamefont {{de Palma}}, \citenamefont {{de Souza}}, \citenamefont {{De Vito}}, \citenamefont {{Del Popolo}}, \citenamefont {{del R{\'\i}o}}, \citenamefont {{Deligny}}, \citenamefont {{Deval}}, \citenamefont {{di Matteo}}, \citenamefont {{Dobre}}, \citenamefont {{Dobrigkeit}}, \citenamefont {{D'Olivo}}, \citenamefont {{Domingues Mendes}}, \citenamefont {{dos Anjos}}, \citenamefont {{Dova}}, \citenamefont {{Ebr}}, \citenamefont
  {{Engel}}, \citenamefont {{Epicoco}}, \citenamefont {{Erdmann}}, \citenamefont {{Escobar}}, \citenamefont {{Etchegoyen}}, \citenamefont {{Falcke}}, \citenamefont {{Farmer}}, \citenamefont {{Farrar}}, \citenamefont {{Fauth}}, \citenamefont {{Fazzini}}, \citenamefont {{Feldbusch}}, \citenamefont {{Fenu}}, \citenamefont {{Fick}}, \citenamefont {{Figueira}}, \citenamefont {{Filip{\v{c}}i{\v{c}}}}, \citenamefont {{Fitoussi}}, \citenamefont {{Fodran}}, \citenamefont {{Fujii}}, \citenamefont {{Fuster}}, \citenamefont {{Galea}}, \citenamefont {{Galelli}}, \citenamefont {{Garc{\'\i}a}}, \citenamefont {{Garcia Vegas}}, \citenamefont {{Gemmeke}}, \citenamefont {{Gesualdi}}, \citenamefont {{Gherghel-Lascu}}, \citenamefont {{Ghia}}, \citenamefont {{Giaccari}}, \citenamefont {{Giammarchi}}, \citenamefont {{Glombitza}}, \citenamefont {{Gobbi}}, \citenamefont {{Gollan}}, \citenamefont {{Golup}}, \citenamefont {{G{\'o}mez Berisso}}, \citenamefont {{G{\'o}mez Vitale}}, \citenamefont {{Gongora}}, \citenamefont
  {{Gonz{\'a}lez}}, \citenamefont {{Gonz{\'a}lez}}, \citenamefont {{Goos}}, \citenamefont {{G{\'o}ra}}, \citenamefont {{Gorgi}}, \citenamefont {{Gottowik}}, \citenamefont {{Grubb}}, \citenamefont {{Guarino}}, \citenamefont {{Guedes}}, \citenamefont {{Guido}}, \citenamefont {{Hahn}}, \citenamefont {{Hamal}}, \citenamefont {{Hampel}}, \citenamefont {{Hansen}}, \citenamefont {{Harari}}, \citenamefont {{Harvey}}, \citenamefont {{Haungs}}, \citenamefont {{Hebbeker}}, \citenamefont {{Heck}}, \citenamefont {{Hill}}, \citenamefont {{Hojvat}}, \citenamefont {{H{\"o}randel}}, \citenamefont {{Horvath}}, \citenamefont {{Hrabovsk{\'y}}}, \citenamefont {{Huege}}, \citenamefont {{Insolia}}, \citenamefont {{Isar}}, \citenamefont {{Janecek}}, \citenamefont {{Johnsen}}, \citenamefont {{Jurysek}}, \citenamefont {{K{\"a}{\"a}p{\"a}}}, \citenamefont {{Kampert}}, \citenamefont {{Keilhauer}}, \citenamefont {{Khakurdikar}}, \citenamefont {{Kizakke Covilakam}}, \citenamefont {{Klages}}, \citenamefont {{Kleifges}}, \citenamefont
  {{Kleinfeller}}, \citenamefont {{Knapp}}, \citenamefont {{Kunka}}, \citenamefont {{Lago}}, \citenamefont {{Langner}}, \citenamefont {{Leigui de Oliveira}}, \citenamefont {{Lenok}}, \citenamefont {{Letessier-Selvon}}, \citenamefont {{Lhenry-Yvon}}, \citenamefont {{Lo Presti}}, \citenamefont {{Lopes}}, \citenamefont {{L{\'o}pez}}, \citenamefont {{Lu}}, \citenamefont {{Luce}}, \citenamefont {{Lundquist}}, \citenamefont {{Machado Payeras}}, \citenamefont {{Mancarella}}, \citenamefont {{Mandat}}, \citenamefont {{Manning}}, \citenamefont {{Manshanden}}, \citenamefont {{Mantsch}}, \citenamefont {{Marafico}}, \citenamefont {{Mariani}}, \citenamefont {{Mariazzi}}, \citenamefont {{Mari{\c{s}}}}, \citenamefont {{Marsella}}, \citenamefont {{Martello}}, \citenamefont {{Martinelli}}, \citenamefont {{Mart{\'\i}nez Bravo}}, \citenamefont {{Mastrodicasa}}, \citenamefont {{Mathes}}, \citenamefont {{Matthews}}, \citenamefont {{Matthiae}}, \citenamefont {{Mayotte}}, \citenamefont {{Mayotte}}, \citenamefont {{Mazur}},
  \citenamefont {{Medina-Tanco}}, \citenamefont {{Melo}}, \citenamefont {{Menshikov}}, \citenamefont {{Michal}}, \citenamefont {{Micheletti}}, \citenamefont {{Miramonti}}, \citenamefont {{Mollerach}}, \citenamefont {{Montanet}}, \citenamefont {{Morejon}}, \citenamefont {{Morello}}, \citenamefont {{Mostaf{\'a}}}, \citenamefont {{M{\"u}ller}}, \citenamefont {{Muller}}, \citenamefont {{Mulrey}}, \citenamefont {{Mussa}}, \citenamefont {{Muzio}}, \citenamefont {{Namasaka}}, \citenamefont {{Nasr-Esfahani}}, \citenamefont {{Nellen}}, \citenamefont {{Nicora}}, \citenamefont {{Niculescu-Oglinzanu}}, \citenamefont {{Niechciol}}, \citenamefont {{Nitz}}, \citenamefont {{Norwood}}, \citenamefont {{Nosek}}, \citenamefont {{Novotny}}, \citenamefont {{No{\v{z}}ka}}, \citenamefont {{Nucita}}, \citenamefont {{N{\'u}{\~n}ez}}, \citenamefont {{Oliveira}}, \citenamefont {{Palatka}}, \citenamefont {{Pallotta}}, \citenamefont {{Papenbreer}}, \citenamefont {{Parente}}, \citenamefont {{Parra}}, \citenamefont {{Pawlowsky}},
  \citenamefont {{Pech}}, \citenamefont {{Pekala}}, \citenamefont {{Pelayo}}, \citenamefont {{Pe{\~n}a-Rodriguez}}, \citenamefont {{Pereira Martins}}, \citenamefont {{Perez Armand}}, \citenamefont {{P{\'e}rez Bertolli}}, \citenamefont {{Perrone}}, \citenamefont {{Petrera}}, \citenamefont {{Petrucci}}, \citenamefont {{Pierog}}, \citenamefont {{Pimenta}}, \citenamefont {{Pirronello}}, \citenamefont {{Platino}}, \citenamefont {{Pont}}, \citenamefont {{Pothast}}, \citenamefont {{Privitera}}, \citenamefont {{Prouza}}, \citenamefont {{Puyleart}}, \citenamefont {{Querchfeld}}, \citenamefont {{Rautenberg}}, \citenamefont {{Ravignani}}, \citenamefont {{Reininghaus}}, \citenamefont {{Ridky}}, \citenamefont {{Riehn}}, \citenamefont {{Risse}}, \citenamefont {{Rizi}}, \citenamefont {{Rodrigues de Carvalho}}, \citenamefont {{Rodriguez Rojo}}, \citenamefont {{Roncoroni}}, \citenamefont {{Rossoni}}, \citenamefont {{Roth}}, \citenamefont {{Roulet}}, \citenamefont {{Rovero}}, \citenamefont {{Ruehl}}, \citenamefont {{Saftoiu}},
  \citenamefont {{Saharan}}, \citenamefont {{Salamida}}, \citenamefont {{Salazar}}, \citenamefont {{Salina}}, \citenamefont {{Sanabria Gomez}}, \citenamefont {{S{\'a}nchez}}, \citenamefont {{Santos}}, \citenamefont {{Santos}}, \citenamefont {{Sarazin}}, \citenamefont {{Sarmento}}, \citenamefont {{Sarmiento-Cano}}, \citenamefont {{Sato}}, \citenamefont {{Savina}}, \citenamefont {{Sch{\"a}fer}}, \citenamefont {{Scherini}}, \citenamefont {{Schieler}}, \citenamefont {{Schimassek}}, \citenamefont {{Schimp}}, \citenamefont {{Schl{\"u}ter}}, \citenamefont {{Schmidt}}, \citenamefont {{Scholten}}, \citenamefont {{Schoorlemmer}}, \citenamefont {{Schov{\'a}nek}}, \citenamefont {{Schr{\"o}der}}, \citenamefont {{Schulte}}, \citenamefont {{Schulz}}, \citenamefont {{Sciutto}}, \citenamefont {{Scornavacche}}, \citenamefont {{Segreto}}, \citenamefont {{Sehgal}}, \citenamefont {{Shellard}}, \citenamefont {{Sigl}}, \citenamefont {{Silli}}, \citenamefont {{Sima}}, \citenamefont {{Smau}}, \citenamefont {{{\v{S}}m{\'\i}da}},
  \citenamefont {{Sommers}}, \citenamefont {{Soriano}}, \citenamefont {{Squartini}}, \citenamefont {{Stadelmaier}}, \citenamefont {{Stanca}}, \citenamefont {{Stani{\v{c}}}}, \citenamefont {{Stasielak}}, \citenamefont {{Stassi}}, \citenamefont {{Streich}}, \citenamefont {{Su{\'a}rez-Dur{\'a}n}}, \citenamefont {{Sudholz}}, \citenamefont {{Suomij{\"a}rvi}}, \citenamefont {{Supanitsky}}, \citenamefont {{Szadkowski}}, \citenamefont {{Tapia}}, \citenamefont {{Taricco}}, \citenamefont {{Timmermans}}, \citenamefont {{Tkachenko}}, \citenamefont {{Tobiska}}, \citenamefont {{Todero Peixoto}}, \citenamefont {{Tom{\'e}}}, \citenamefont {{Torr{\`e}s}}, \citenamefont {{Travaini}}, \citenamefont {{Travnicek}}, \citenamefont {{Trimarelli}}, \citenamefont {{Tueros}}, \citenamefont {{Ulrich}}, \citenamefont {{Unger}}, \citenamefont {{Vaclavek}}, \citenamefont {{Vacula}}, \citenamefont {{Vald{\'e}s Galicia}}, \citenamefont {{Valore}}, \citenamefont {{Varela}}, \citenamefont {{V{\'a}squez-Ram{\'\i}rez}}, \citenamefont
  {{Veberi{\v{c}}}}, \citenamefont {{Ventura}}, \citenamefont {{Vergara Quispe}}, \citenamefont {{Verzi}}, \citenamefont {{Vicha}}, \citenamefont {{Vink}}, \citenamefont {{Vorobiov}}, \citenamefont {{Wahlberg}}, \citenamefont {{Watanabe}}, \citenamefont {{Watson}}, \citenamefont {{Weindl}}, \citenamefont {{Wiencke}}, \citenamefont {{Wilczy{\'n}ski}}, \citenamefont {{Wittkowski}}, \citenamefont {{Wundheiler}}, \citenamefont {{Yushkov}}, \citenamefont {{Zapparrata}}, \citenamefont {{Zas}}, \citenamefont {{Zavrtanik}}, \citenamefont {{Zavrtanik}}, \citenamefont {{Zehrer}},\ and\ \citenamefont {{Pierre Auger Collaboration}}}]{Abreu:2023-Cosmological-implications-of-photon-flux-upper-limits}%
  \BibitemOpen
  \bibfield  {author} {\bibinfo {author} {\bibfnamefont {P.}~\bibnamefont {{Abreu}}}, \bibinfo {author} {\bibfnamefont {M.}~\bibnamefont {{Aglietta}}}, \bibinfo {author} {\bibfnamefont {J.~M.}\ \bibnamefont {{Albury}}}, \bibinfo {author} {\bibfnamefont {I.}~\bibnamefont {{Allekotte}}}, \bibinfo {author} {\bibfnamefont {K.}~\bibnamefont {{Almeida Cheminant}}}, \bibinfo {author} {\bibfnamefont {A.}~\bibnamefont {{Almela}}}, \bibinfo {author} {\bibfnamefont {R.}~\bibnamefont {{Aloisio}}}, \bibinfo {author} {\bibfnamefont {J.}~\bibnamefont {{Alvarez-Mu{\~n}iz}}}, \bibinfo {author} {\bibfnamefont {R.}~\bibnamefont {{Alves Batista}}}, \bibinfo {author} {\bibfnamefont {J.}~\bibnamefont {{Ammerman Yebra}}}, \bibinfo {author} {\bibfnamefont {G.~A.}\ \bibnamefont {{Anastasi}}}, \bibinfo {author} {\bibfnamefont {L.}~\bibnamefont {{Anchordoqui}}}, \bibinfo {author} {\bibfnamefont {B.}~\bibnamefont {{Andrada}}}, \bibinfo {author} {\bibfnamefont {S.}~\bibnamefont {{Andringa}}}, \bibinfo {author} {\bibfnamefont
  {C.}~\bibnamefont {{Aramo}}}, \bibinfo {author} {\bibfnamefont {P.~R.}\ \bibnamefont {{Ara{\'u}jo Ferreira}}}, \bibinfo {author} {\bibfnamefont {E.}~\bibnamefont {{Arnone}}}, \bibinfo {author} {\bibfnamefont {J.~C.}\ \bibnamefont {{Arteaga Vel{\'a}zquez}}}, \bibinfo {author} {\bibfnamefont {H.}~\bibnamefont {{Asorey}}}, \bibinfo {author} {\bibfnamefont {P.}~\bibnamefont {{Assis}}}, \bibinfo {author} {\bibfnamefont {G.}~\bibnamefont {{Avila}}}, \bibinfo {author} {\bibfnamefont {E.}~\bibnamefont {{Avocone}}}, \bibinfo {author} {\bibfnamefont {A.~M.}\ \bibnamefont {{Badescu}}}, \bibinfo {author} {\bibfnamefont {A.}~\bibnamefont {{Bakalova}}}, \bibinfo {author} {\bibfnamefont {A.}~\bibnamefont {{Balaceanu}}}, \bibinfo {author} {\bibfnamefont {F.}~\bibnamefont {{Barbato}}}, \bibinfo {author} {\bibfnamefont {J.~A.}\ \bibnamefont {{Bellido}}}, \bibinfo {author} {\bibfnamefont {C.}~\bibnamefont {{Berat}}}, \bibinfo {author} {\bibfnamefont {M.~E.}\ \bibnamefont {{Bertaina}}}, \bibinfo {author} {\bibfnamefont
  {G.}~\bibnamefont {{Bhatta}}}, \bibinfo {author} {\bibfnamefont {P.~L.}\ \bibnamefont {{Biermann}}}, \bibinfo {author} {\bibfnamefont {V.}~\bibnamefont {{Binet}}}, \bibinfo {author} {\bibfnamefont {K.}~\bibnamefont {{Bismark}}}, \bibinfo {author} {\bibfnamefont {T.}~\bibnamefont {{Bister}}}, \bibinfo {author} {\bibfnamefont {J.}~\bibnamefont {{Biteau}}}, \bibinfo {author} {\bibfnamefont {J.}~\bibnamefont {{Blazek}}}, \bibinfo {author} {\bibfnamefont {C.}~\bibnamefont {{Bleve}}}, \bibinfo {author} {\bibfnamefont {J.}~\bibnamefont {{Bl{\"u}mer}}}, \bibinfo {author} {\bibfnamefont {M.}~\bibnamefont {{Boh{\'a}{\v{c}}ov{\'a}}}}, \bibinfo {author} {\bibfnamefont {D.}~\bibnamefont {{Boncioli}}}, \bibinfo {author} {\bibfnamefont {C.}~\bibnamefont {{Bonifazi}}}, \bibinfo {author} {\bibfnamefont {L.}~\bibnamefont {{Bonneau Arbeletche}}}, \bibinfo {author} {\bibfnamefont {N.}~\bibnamefont {{Borodai}}}, \bibinfo {author} {\bibfnamefont {A.~M.}\ \bibnamefont {{Botti}}}, \bibinfo {author} {\bibfnamefont {J.}~\bibnamefont
  {{Brack}}}, \bibinfo {author} {\bibfnamefont {T.}~\bibnamefont {{Bretz}}}, \bibinfo {author} {\bibfnamefont {P.~G.}\ \bibnamefont {{Brichetto Orchera}}}, \bibinfo {author} {\bibfnamefont {F.~L.}\ \bibnamefont {{Briechle}}}, \bibinfo {author} {\bibfnamefont {P.}~\bibnamefont {{Buchholz}}}, \bibinfo {author} {\bibfnamefont {A.}~\bibnamefont {{Bueno}}}, \bibinfo {author} {\bibfnamefont {S.}~\bibnamefont {{Buitink}}}, \bibinfo {author} {\bibfnamefont {M.}~\bibnamefont {{Buscemi}}}, \bibinfo {author} {\bibfnamefont {M.}~\bibnamefont {{B{\"u}sken}}}, \bibinfo {author} {\bibfnamefont {K.~S.}\ \bibnamefont {{Caballero-Mora}}}, \bibinfo {author} {\bibfnamefont {L.}~\bibnamefont {{Caccianiga}}}, \bibinfo {author} {\bibfnamefont {F.}~\bibnamefont {{Canfora}}}, \bibinfo {author} {\bibfnamefont {I.}~\bibnamefont {{Caracas}}}, \bibinfo {author} {\bibfnamefont {R.}~\bibnamefont {{Caruso}}}, \bibinfo {author} {\bibfnamefont {A.}~\bibnamefont {{Castellina}}}, \bibinfo {author} {\bibfnamefont {F.}~\bibnamefont {{Catalani}}},
  \bibinfo {author} {\bibfnamefont {G.}~\bibnamefont {{Cataldi}}}, \bibinfo {author} {\bibfnamefont {L.}~\bibnamefont {{Cazon}}}, \bibinfo {author} {\bibfnamefont {M.}~\bibnamefont {{Cerda}}}, \bibinfo {author} {\bibfnamefont {J.~A.}\ \bibnamefont {{Chinellato}}}, \bibinfo {author} {\bibfnamefont {J.}~\bibnamefont {{Chudoba}}}, \bibinfo {author} {\bibfnamefont {L.}~\bibnamefont {{Chytka}}}, \bibinfo {author} {\bibfnamefont {R.~W.}\ \bibnamefont {{Clay}}}, \bibinfo {author} {\bibfnamefont {A.~C.}\ \bibnamefont {{Cobos Cerutti}}}, \bibinfo {author} {\bibfnamefont {R.}~\bibnamefont {{Colalillo}}}, \bibinfo {author} {\bibfnamefont {A.}~\bibnamefont {{Coleman}}}, \bibinfo {author} {\bibfnamefont {M.~R.}\ \bibnamefont {{Coluccia}}}, \bibinfo {author} {\bibfnamefont {R.}~\bibnamefont {{Concei{\c{c}}{\~a}o}}}, \bibinfo {author} {\bibfnamefont {A.}~\bibnamefont {{Condorelli}}}, \bibinfo {author} {\bibfnamefont {G.}~\bibnamefont {{Consolati}}}, \bibinfo {author} {\bibfnamefont {F.}~\bibnamefont {{Contreras}}}, \bibinfo
  {author} {\bibfnamefont {F.}~\bibnamefont {{Convenga}}}, \bibinfo {author} {\bibfnamefont {D.}~\bibnamefont {{Correia dos Santos}}}, \bibinfo {author} {\bibfnamefont {C.~E.}\ \bibnamefont {{Covault}}}, \bibinfo {author} {\bibfnamefont {S.}~\bibnamefont {{Dasso}}}, \bibinfo {author} {\bibfnamefont {K.}~\bibnamefont {{Daumiller}}}, \bibinfo {author} {\bibfnamefont {B.~R.}\ \bibnamefont {{Dawson}}}, \bibinfo {author} {\bibfnamefont {J.~A.}\ \bibnamefont {{Day}}}, \bibinfo {author} {\bibfnamefont {R.~M.}\ \bibnamefont {{de Almeida}}}, \bibinfo {author} {\bibfnamefont {J.}~\bibnamefont {{de Jes{\'u}s}}}, \bibinfo {author} {\bibfnamefont {S.~J.}\ \bibnamefont {{de Jong}}}, \bibinfo {author} {\bibfnamefont {J.~R.~T.}\ \bibnamefont {{de Mello Neto}}}, \bibinfo {author} {\bibfnamefont {I.}~\bibnamefont {{De Mitri}}}, \bibinfo {author} {\bibfnamefont {J.}~\bibnamefont {{de Oliveira}}}, \bibinfo {author} {\bibfnamefont {D.}~\bibnamefont {{de Oliveira Franco}}}, \bibinfo {author} {\bibfnamefont {F.}~\bibnamefont {{de
  Palma}}}, \bibinfo {author} {\bibfnamefont {V.}~\bibnamefont {{de Souza}}}, \bibinfo {author} {\bibfnamefont {E.}~\bibnamefont {{De Vito}}}, \bibinfo {author} {\bibfnamefont {A.}~\bibnamefont {{Del Popolo}}}, \bibinfo {author} {\bibfnamefont {M.}~\bibnamefont {{del R{\'\i}o}}}, \bibinfo {author} {\bibfnamefont {O.}~\bibnamefont {{Deligny}}}, \bibinfo {author} {\bibfnamefont {L.}~\bibnamefont {{Deval}}}, \bibinfo {author} {\bibfnamefont {A.}~\bibnamefont {{di Matteo}}}, \bibinfo {author} {\bibfnamefont {M.}~\bibnamefont {{Dobre}}}, \bibinfo {author} {\bibfnamefont {C.}~\bibnamefont {{Dobrigkeit}}}, \bibinfo {author} {\bibfnamefont {J.~C.}\ \bibnamefont {{D'Olivo}}}, \bibinfo {author} {\bibfnamefont {L.~M.}\ \bibnamefont {{Domingues Mendes}}}, \bibinfo {author} {\bibfnamefont {R.~C.}\ \bibnamefont {{dos Anjos}}}, \bibinfo {author} {\bibfnamefont {M.~T.}\ \bibnamefont {{Dova}}}, \bibinfo {author} {\bibfnamefont {J.}~\bibnamefont {{Ebr}}}, \bibinfo {author} {\bibfnamefont {R.}~\bibnamefont {{Engel}}}, \bibinfo
  {author} {\bibfnamefont {I.}~\bibnamefont {{Epicoco}}}, \bibinfo {author} {\bibfnamefont {M.}~\bibnamefont {{Erdmann}}}, \bibinfo {author} {\bibfnamefont {C.~O.}\ \bibnamefont {{Escobar}}}, \bibinfo {author} {\bibfnamefont {A.}~\bibnamefont {{Etchegoyen}}}, \bibinfo {author} {\bibfnamefont {H.}~\bibnamefont {{Falcke}}}, \bibinfo {author} {\bibfnamefont {J.}~\bibnamefont {{Farmer}}}, \bibinfo {author} {\bibfnamefont {G.}~\bibnamefont {{Farrar}}}, \bibinfo {author} {\bibfnamefont {A.~C.}\ \bibnamefont {{Fauth}}}, \bibinfo {author} {\bibfnamefont {N.}~\bibnamefont {{Fazzini}}}, \bibinfo {author} {\bibfnamefont {F.}~\bibnamefont {{Feldbusch}}}, \bibinfo {author} {\bibfnamefont {F.}~\bibnamefont {{Fenu}}}, \bibinfo {author} {\bibfnamefont {B.}~\bibnamefont {{Fick}}}, \bibinfo {author} {\bibfnamefont {J.~M.}\ \bibnamefont {{Figueira}}}, \bibinfo {author} {\bibfnamefont {A.}~\bibnamefont {{Filip{\v{c}}i{\v{c}}}}}, \bibinfo {author} {\bibfnamefont {T.}~\bibnamefont {{Fitoussi}}}, \bibinfo {author} {\bibfnamefont
  {T.}~\bibnamefont {{Fodran}}}, \bibinfo {author} {\bibfnamefont {T.}~\bibnamefont {{Fujii}}}, \bibinfo {author} {\bibfnamefont {A.}~\bibnamefont {{Fuster}}}, \bibinfo {author} {\bibfnamefont {C.}~\bibnamefont {{Galea}}}, \bibinfo {author} {\bibfnamefont {C.}~\bibnamefont {{Galelli}}}, \bibinfo {author} {\bibfnamefont {B.}~\bibnamefont {{Garc{\'\i}a}}}, \bibinfo {author} {\bibfnamefont {A.~L.}\ \bibnamefont {{Garcia Vegas}}}, \bibinfo {author} {\bibfnamefont {H.}~\bibnamefont {{Gemmeke}}}, \bibinfo {author} {\bibfnamefont {F.}~\bibnamefont {{Gesualdi}}}, \bibinfo {author} {\bibfnamefont {A.}~\bibnamefont {{Gherghel-Lascu}}}, \bibinfo {author} {\bibfnamefont {P.~L.}\ \bibnamefont {{Ghia}}}, \bibinfo {author} {\bibfnamefont {U.}~\bibnamefont {{Giaccari}}}, \bibinfo {author} {\bibfnamefont {M.}~\bibnamefont {{Giammarchi}}}, \bibinfo {author} {\bibfnamefont {J.}~\bibnamefont {{Glombitza}}}, \bibinfo {author} {\bibfnamefont {F.}~\bibnamefont {{Gobbi}}}, \bibinfo {author} {\bibfnamefont {F.}~\bibnamefont
  {{Gollan}}}, \bibinfo {author} {\bibfnamefont {G.}~\bibnamefont {{Golup}}}, \bibinfo {author} {\bibfnamefont {M.}~\bibnamefont {{G{\'o}mez Berisso}}}, \bibinfo {author} {\bibfnamefont {P.~F.}\ \bibnamefont {{G{\'o}mez Vitale}}}, \bibinfo {author} {\bibfnamefont {J.~P.}\ \bibnamefont {{Gongora}}}, \bibinfo {author} {\bibfnamefont {J.~M.}\ \bibnamefont {{Gonz{\'a}lez}}}, \bibinfo {author} {\bibfnamefont {N.}~\bibnamefont {{Gonz{\'a}lez}}}, \bibinfo {author} {\bibfnamefont {I.}~\bibnamefont {{Goos}}}, \bibinfo {author} {\bibfnamefont {D.}~\bibnamefont {{G{\'o}ra}}}, \bibinfo {author} {\bibfnamefont {A.}~\bibnamefont {{Gorgi}}}, \bibinfo {author} {\bibfnamefont {M.}~\bibnamefont {{Gottowik}}}, \bibinfo {author} {\bibfnamefont {T.~D.}\ \bibnamefont {{Grubb}}}, \bibinfo {author} {\bibfnamefont {F.}~\bibnamefont {{Guarino}}}, \bibinfo {author} {\bibfnamefont {G.~P.}\ \bibnamefont {{Guedes}}}, \bibinfo {author} {\bibfnamefont {E.}~\bibnamefont {{Guido}}}, \bibinfo {author} {\bibfnamefont {S.}~\bibnamefont
  {{Hahn}}}, \bibinfo {author} {\bibfnamefont {P.}~\bibnamefont {{Hamal}}}, \bibinfo {author} {\bibfnamefont {M.~R.}\ \bibnamefont {{Hampel}}}, \bibinfo {author} {\bibfnamefont {P.}~\bibnamefont {{Hansen}}}, \bibinfo {author} {\bibfnamefont {D.}~\bibnamefont {{Harari}}}, \bibinfo {author} {\bibfnamefont {V.~M.}\ \bibnamefont {{Harvey}}}, \bibinfo {author} {\bibfnamefont {A.}~\bibnamefont {{Haungs}}}, \bibinfo {author} {\bibfnamefont {T.}~\bibnamefont {{Hebbeker}}}, \bibinfo {author} {\bibfnamefont {D.}~\bibnamefont {{Heck}}}, \bibinfo {author} {\bibfnamefont {G.~C.}\ \bibnamefont {{Hill}}}, \bibinfo {author} {\bibfnamefont {C.}~\bibnamefont {{Hojvat}}}, \bibinfo {author} {\bibfnamefont {J.~R.}\ \bibnamefont {{H{\"o}randel}}}, \bibinfo {author} {\bibfnamefont {P.}~\bibnamefont {{Horvath}}}, \bibinfo {author} {\bibfnamefont {M.}~\bibnamefont {{Hrabovsk{\'y}}}}, \bibinfo {author} {\bibfnamefont {T.}~\bibnamefont {{Huege}}}, \bibinfo {author} {\bibfnamefont {A.}~\bibnamefont {{Insolia}}}, \bibinfo {author}
  {\bibfnamefont {P.~G.}\ \bibnamefont {{Isar}}}, \bibinfo {author} {\bibfnamefont {P.}~\bibnamefont {{Janecek}}}, \bibinfo {author} {\bibfnamefont {J.~A.}\ \bibnamefont {{Johnsen}}}, \bibinfo {author} {\bibfnamefont {J.}~\bibnamefont {{Jurysek}}}, \bibinfo {author} {\bibfnamefont {A.}~\bibnamefont {{K{\"a}{\"a}p{\"a}}}}, \bibinfo {author} {\bibfnamefont {K.~H.}\ \bibnamefont {{Kampert}}}, \bibinfo {author} {\bibfnamefont {B.}~\bibnamefont {{Keilhauer}}}, \bibinfo {author} {\bibfnamefont {A.}~\bibnamefont {{Khakurdikar}}}, \bibinfo {author} {\bibfnamefont {V.~V.}\ \bibnamefont {{Kizakke Covilakam}}}, \bibinfo {author} {\bibfnamefont {H.~O.}\ \bibnamefont {{Klages}}}, \bibinfo {author} {\bibfnamefont {M.}~\bibnamefont {{Kleifges}}}, \bibinfo {author} {\bibfnamefont {J.}~\bibnamefont {{Kleinfeller}}}, \bibinfo {author} {\bibfnamefont {F.}~\bibnamefont {{Knapp}}}, \bibinfo {author} {\bibfnamefont {N.}~\bibnamefont {{Kunka}}}, \bibinfo {author} {\bibfnamefont {B.~L.}\ \bibnamefont {{Lago}}}, \bibinfo {author}
  {\bibfnamefont {N.}~\bibnamefont {{Langner}}}, \bibinfo {author} {\bibfnamefont {M.~A.}\ \bibnamefont {{Leigui de Oliveira}}}, \bibinfo {author} {\bibfnamefont {V.}~\bibnamefont {{Lenok}}}, \bibinfo {author} {\bibfnamefont {A.}~\bibnamefont {{Letessier-Selvon}}}, \bibinfo {author} {\bibfnamefont {I.}~\bibnamefont {{Lhenry-Yvon}}}, \bibinfo {author} {\bibfnamefont {D.}~\bibnamefont {{Lo Presti}}}, \bibinfo {author} {\bibfnamefont {L.}~\bibnamefont {{Lopes}}}, \bibinfo {author} {\bibfnamefont {R.}~\bibnamefont {{L{\'o}pez}}}, \bibinfo {author} {\bibfnamefont {L.}~\bibnamefont {{Lu}}}, \bibinfo {author} {\bibfnamefont {Q.}~\bibnamefont {{Luce}}}, \bibinfo {author} {\bibfnamefont {J.~P.}\ \bibnamefont {{Lundquist}}}, \bibinfo {author} {\bibfnamefont {A.}~\bibnamefont {{Machado Payeras}}}, \bibinfo {author} {\bibfnamefont {G.}~\bibnamefont {{Mancarella}}}, \bibinfo {author} {\bibfnamefont {D.}~\bibnamefont {{Mandat}}}, \bibinfo {author} {\bibfnamefont {B.~C.}\ \bibnamefont {{Manning}}}, \bibinfo {author}
  {\bibfnamefont {J.}~\bibnamefont {{Manshanden}}}, \bibinfo {author} {\bibfnamefont {P.}~\bibnamefont {{Mantsch}}}, \bibinfo {author} {\bibfnamefont {S.}~\bibnamefont {{Marafico}}}, \bibinfo {author} {\bibfnamefont {F.~M.}\ \bibnamefont {{Mariani}}}, \bibinfo {author} {\bibfnamefont {A.~G.}\ \bibnamefont {{Mariazzi}}}, \bibinfo {author} {\bibfnamefont {I.~C.}\ \bibnamefont {{Mari{\c{s}}}}}, \bibinfo {author} {\bibfnamefont {G.}~\bibnamefont {{Marsella}}}, \bibinfo {author} {\bibfnamefont {D.}~\bibnamefont {{Martello}}}, \bibinfo {author} {\bibfnamefont {S.}~\bibnamefont {{Martinelli}}}, \bibinfo {author} {\bibfnamefont {O.}~\bibnamefont {{Mart{\'\i}nez Bravo}}}, \bibinfo {author} {\bibfnamefont {M.}~\bibnamefont {{Mastrodicasa}}}, \bibinfo {author} {\bibfnamefont {H.~J.}\ \bibnamefont {{Mathes}}}, \bibinfo {author} {\bibfnamefont {J.}~\bibnamefont {{Matthews}}}, \bibinfo {author} {\bibfnamefont {G.}~\bibnamefont {{Matthiae}}}, \bibinfo {author} {\bibfnamefont {E.}~\bibnamefont {{Mayotte}}}, \bibinfo {author}
  {\bibfnamefont {S.}~\bibnamefont {{Mayotte}}}, \bibinfo {author} {\bibfnamefont {P.~O.}\ \bibnamefont {{Mazur}}}, \bibinfo {author} {\bibfnamefont {G.}~\bibnamefont {{Medina-Tanco}}}, \bibinfo {author} {\bibfnamefont {D.}~\bibnamefont {{Melo}}}, \bibinfo {author} {\bibfnamefont {A.}~\bibnamefont {{Menshikov}}}, \bibinfo {author} {\bibfnamefont {S.}~\bibnamefont {{Michal}}}, \bibinfo {author} {\bibfnamefont {M.~I.}\ \bibnamefont {{Micheletti}}}, \bibinfo {author} {\bibfnamefont {L.}~\bibnamefont {{Miramonti}}}, \bibinfo {author} {\bibfnamefont {S.}~\bibnamefont {{Mollerach}}}, \bibinfo {author} {\bibfnamefont {F.}~\bibnamefont {{Montanet}}}, \bibinfo {author} {\bibfnamefont {L.}~\bibnamefont {{Morejon}}}, \bibinfo {author} {\bibfnamefont {C.}~\bibnamefont {{Morello}}}, \bibinfo {author} {\bibfnamefont {M.}~\bibnamefont {{Mostaf{\'a}}}}, \bibinfo {author} {\bibfnamefont {A.~L.}\ \bibnamefont {{M{\"u}ller}}}, \bibinfo {author} {\bibfnamefont {M.~A.}\ \bibnamefont {{Muller}}}, \bibinfo {author} {\bibfnamefont
  {K.}~\bibnamefont {{Mulrey}}}, \bibinfo {author} {\bibfnamefont {R.}~\bibnamefont {{Mussa}}}, \bibinfo {author} {\bibfnamefont {M.}~\bibnamefont {{Muzio}}}, \bibinfo {author} {\bibfnamefont {W.~M.}\ \bibnamefont {{Namasaka}}}, \bibinfo {author} {\bibfnamefont {A.}~\bibnamefont {{Nasr-Esfahani}}}, \bibinfo {author} {\bibfnamefont {L.}~\bibnamefont {{Nellen}}}, \bibinfo {author} {\bibfnamefont {G.}~\bibnamefont {{Nicora}}}, \bibinfo {author} {\bibfnamefont {M.}~\bibnamefont {{Niculescu-Oglinzanu}}}, \bibinfo {author} {\bibfnamefont {M.}~\bibnamefont {{Niechciol}}}, \bibinfo {author} {\bibfnamefont {D.}~\bibnamefont {{Nitz}}}, \bibinfo {author} {\bibfnamefont {I.}~\bibnamefont {{Norwood}}}, \bibinfo {author} {\bibfnamefont {D.}~\bibnamefont {{Nosek}}}, \bibinfo {author} {\bibfnamefont {V.}~\bibnamefont {{Novotny}}}, \bibinfo {author} {\bibfnamefont {L.}~\bibnamefont {{No{\v{z}}ka}}}, \bibinfo {author} {\bibfnamefont {A.}~\bibnamefont {{Nucita}}}, \bibinfo {author} {\bibfnamefont {L.~A.}\ \bibnamefont
  {{N{\'u}{\~n}ez}}}, \bibinfo {author} {\bibfnamefont {C.}~\bibnamefont {{Oliveira}}}, \bibinfo {author} {\bibfnamefont {M.}~\bibnamefont {{Palatka}}}, \bibinfo {author} {\bibfnamefont {J.}~\bibnamefont {{Pallotta}}}, \bibinfo {author} {\bibfnamefont {P.}~\bibnamefont {{Papenbreer}}}, \bibinfo {author} {\bibfnamefont {G.}~\bibnamefont {{Parente}}}, \bibinfo {author} {\bibfnamefont {A.}~\bibnamefont {{Parra}}}, \bibinfo {author} {\bibfnamefont {J.}~\bibnamefont {{Pawlowsky}}}, \bibinfo {author} {\bibfnamefont {M.}~\bibnamefont {{Pech}}}, \bibinfo {author} {\bibfnamefont {J.}~\bibnamefont {{Pekala}}}, \bibinfo {author} {\bibfnamefont {R.}~\bibnamefont {{Pelayo}}}, \bibinfo {author} {\bibfnamefont {J.}~\bibnamefont {{Pe{\~n}a-Rodriguez}}}, \bibinfo {author} {\bibfnamefont {E.~E.}\ \bibnamefont {{Pereira Martins}}}, \bibinfo {author} {\bibfnamefont {J.}~\bibnamefont {{Perez Armand}}}, \bibinfo {author} {\bibfnamefont {C.}~\bibnamefont {{P{\'e}rez Bertolli}}}, \bibinfo {author} {\bibfnamefont {L.}~\bibnamefont
  {{Perrone}}}, \bibinfo {author} {\bibfnamefont {S.}~\bibnamefont {{Petrera}}}, \bibinfo {author} {\bibfnamefont {C.}~\bibnamefont {{Petrucci}}}, \bibinfo {author} {\bibfnamefont {T.}~\bibnamefont {{Pierog}}}, \bibinfo {author} {\bibfnamefont {M.}~\bibnamefont {{Pimenta}}}, \bibinfo {author} {\bibfnamefont {V.}~\bibnamefont {{Pirronello}}}, \bibinfo {author} {\bibfnamefont {M.}~\bibnamefont {{Platino}}}, \bibinfo {author} {\bibfnamefont {B.}~\bibnamefont {{Pont}}}, \bibinfo {author} {\bibfnamefont {M.}~\bibnamefont {{Pothast}}}, \bibinfo {author} {\bibfnamefont {P.}~\bibnamefont {{Privitera}}}, \bibinfo {author} {\bibfnamefont {M.}~\bibnamefont {{Prouza}}}, \bibinfo {author} {\bibfnamefont {A.}~\bibnamefont {{Puyleart}}}, \bibinfo {author} {\bibfnamefont {S.}~\bibnamefont {{Querchfeld}}}, \bibinfo {author} {\bibfnamefont {J.}~\bibnamefont {{Rautenberg}}}, \bibinfo {author} {\bibfnamefont {D.}~\bibnamefont {{Ravignani}}}, \bibinfo {author} {\bibfnamefont {M.}~\bibnamefont {{Reininghaus}}}, \bibinfo {author}
  {\bibfnamefont {J.}~\bibnamefont {{Ridky}}}, \bibinfo {author} {\bibfnamefont {F.}~\bibnamefont {{Riehn}}}, \bibinfo {author} {\bibfnamefont {M.}~\bibnamefont {{Risse}}}, \bibinfo {author} {\bibfnamefont {V.}~\bibnamefont {{Rizi}}}, \bibinfo {author} {\bibfnamefont {W.}~\bibnamefont {{Rodrigues de Carvalho}}}, \bibinfo {author} {\bibfnamefont {J.}~\bibnamefont {{Rodriguez Rojo}}}, \bibinfo {author} {\bibfnamefont {M.~J.}\ \bibnamefont {{Roncoroni}}}, \bibinfo {author} {\bibfnamefont {S.}~\bibnamefont {{Rossoni}}}, \bibinfo {author} {\bibfnamefont {M.}~\bibnamefont {{Roth}}}, \bibinfo {author} {\bibfnamefont {E.}~\bibnamefont {{Roulet}}}, \bibinfo {author} {\bibfnamefont {A.~C.}\ \bibnamefont {{Rovero}}}, \bibinfo {author} {\bibfnamefont {P.}~\bibnamefont {{Ruehl}}}, \bibinfo {author} {\bibfnamefont {A.}~\bibnamefont {{Saftoiu}}}, \bibinfo {author} {\bibfnamefont {M.}~\bibnamefont {{Saharan}}}, \bibinfo {author} {\bibfnamefont {F.}~\bibnamefont {{Salamida}}}, \bibinfo {author} {\bibfnamefont
  {H.}~\bibnamefont {{Salazar}}}, \bibinfo {author} {\bibfnamefont {G.}~\bibnamefont {{Salina}}}, \bibinfo {author} {\bibfnamefont {J.~D.}\ \bibnamefont {{Sanabria Gomez}}}, \bibinfo {author} {\bibfnamefont {F.}~\bibnamefont {{S{\'a}nchez}}}, \bibinfo {author} {\bibfnamefont {E.~M.}\ \bibnamefont {{Santos}}}, \bibinfo {author} {\bibfnamefont {E.}~\bibnamefont {{Santos}}}, \bibinfo {author} {\bibfnamefont {F.}~\bibnamefont {{Sarazin}}}, \bibinfo {author} {\bibfnamefont {R.}~\bibnamefont {{Sarmento}}}, \bibinfo {author} {\bibfnamefont {C.}~\bibnamefont {{Sarmiento-Cano}}}, \bibinfo {author} {\bibfnamefont {R.}~\bibnamefont {{Sato}}}, \bibinfo {author} {\bibfnamefont {P.}~\bibnamefont {{Savina}}}, \bibinfo {author} {\bibfnamefont {C.~M.}\ \bibnamefont {{Sch{\"a}fer}}}, \bibinfo {author} {\bibfnamefont {V.}~\bibnamefont {{Scherini}}}, \bibinfo {author} {\bibfnamefont {H.}~\bibnamefont {{Schieler}}}, \bibinfo {author} {\bibfnamefont {M.}~\bibnamefont {{Schimassek}}}, \bibinfo {author} {\bibfnamefont
  {M.}~\bibnamefont {{Schimp}}}, \bibinfo {author} {\bibfnamefont {F.}~\bibnamefont {{Schl{\"u}ter}}}, \bibinfo {author} {\bibfnamefont {D.}~\bibnamefont {{Schmidt}}}, \bibinfo {author} {\bibfnamefont {O.}~\bibnamefont {{Scholten}}}, \bibinfo {author} {\bibfnamefont {H.}~\bibnamefont {{Schoorlemmer}}}, \bibinfo {author} {\bibfnamefont {P.}~\bibnamefont {{Schov{\'a}nek}}}, \bibinfo {author} {\bibfnamefont {F.~G.}\ \bibnamefont {{Schr{\"o}der}}}, \bibinfo {author} {\bibfnamefont {J.}~\bibnamefont {{Schulte}}}, \bibinfo {author} {\bibfnamefont {T.}~\bibnamefont {{Schulz}}}, \bibinfo {author} {\bibfnamefont {S.~J.}\ \bibnamefont {{Sciutto}}}, \bibinfo {author} {\bibfnamefont {M.}~\bibnamefont {{Scornavacche}}}, \bibinfo {author} {\bibfnamefont {A.}~\bibnamefont {{Segreto}}}, \bibinfo {author} {\bibfnamefont {S.}~\bibnamefont {{Sehgal}}}, \bibinfo {author} {\bibfnamefont {R.~C.}\ \bibnamefont {{Shellard}}}, \bibinfo {author} {\bibfnamefont {G.}~\bibnamefont {{Sigl}}}, \bibinfo {author} {\bibfnamefont
  {G.}~\bibnamefont {{Silli}}}, \bibinfo {author} {\bibfnamefont {O.}~\bibnamefont {{Sima}}}, \bibinfo {author} {\bibfnamefont {R.}~\bibnamefont {{Smau}}}, \bibinfo {author} {\bibfnamefont {R.}~\bibnamefont {{{\v{S}}m{\'\i}da}}}, \bibinfo {author} {\bibfnamefont {P.}~\bibnamefont {{Sommers}}}, \bibinfo {author} {\bibfnamefont {J.~F.}\ \bibnamefont {{Soriano}}}, \bibinfo {author} {\bibfnamefont {R.}~\bibnamefont {{Squartini}}}, \bibinfo {author} {\bibfnamefont {M.}~\bibnamefont {{Stadelmaier}}}, \bibinfo {author} {\bibfnamefont {D.}~\bibnamefont {{Stanca}}}, \bibinfo {author} {\bibfnamefont {S.}~\bibnamefont {{Stani{\v{c}}}}}, \bibinfo {author} {\bibfnamefont {J.}~\bibnamefont {{Stasielak}}}, \bibinfo {author} {\bibfnamefont {P.}~\bibnamefont {{Stassi}}}, \bibinfo {author} {\bibfnamefont {A.}~\bibnamefont {{Streich}}}, \bibinfo {author} {\bibfnamefont {M.}~\bibnamefont {{Su{\'a}rez-Dur{\'a}n}}}, \bibinfo {author} {\bibfnamefont {T.}~\bibnamefont {{Sudholz}}}, \bibinfo {author} {\bibfnamefont {T.}~\bibnamefont
  {{Suomij{\"a}rvi}}}, \bibinfo {author} {\bibfnamefont {A.~D.}\ \bibnamefont {{Supanitsky}}}, \bibinfo {author} {\bibfnamefont {Z.}~\bibnamefont {{Szadkowski}}}, \bibinfo {author} {\bibfnamefont {A.}~\bibnamefont {{Tapia}}}, \bibinfo {author} {\bibfnamefont {C.}~\bibnamefont {{Taricco}}}, \bibinfo {author} {\bibfnamefont {C.}~\bibnamefont {{Timmermans}}}, \bibinfo {author} {\bibfnamefont {O.}~\bibnamefont {{Tkachenko}}}, \bibinfo {author} {\bibfnamefont {P.}~\bibnamefont {{Tobiska}}}, \bibinfo {author} {\bibfnamefont {C.~J.}\ \bibnamefont {{Todero Peixoto}}}, \bibinfo {author} {\bibfnamefont {B.}~\bibnamefont {{Tom{\'e}}}}, \bibinfo {author} {\bibfnamefont {Z.}~\bibnamefont {{Torr{\`e}s}}}, \bibinfo {author} {\bibfnamefont {A.}~\bibnamefont {{Travaini}}}, \bibinfo {author} {\bibfnamefont {P.}~\bibnamefont {{Travnicek}}}, \bibinfo {author} {\bibfnamefont {C.}~\bibnamefont {{Trimarelli}}}, \bibinfo {author} {\bibfnamefont {M.}~\bibnamefont {{Tueros}}}, \bibinfo {author} {\bibfnamefont {R.}~\bibnamefont
  {{Ulrich}}}, \bibinfo {author} {\bibfnamefont {M.}~\bibnamefont {{Unger}}}, \bibinfo {author} {\bibfnamefont {L.}~\bibnamefont {{Vaclavek}}}, \bibinfo {author} {\bibfnamefont {M.}~\bibnamefont {{Vacula}}}, \bibinfo {author} {\bibfnamefont {J.~F.}\ \bibnamefont {{Vald{\'e}s Galicia}}}, \bibinfo {author} {\bibfnamefont {L.}~\bibnamefont {{Valore}}}, \bibinfo {author} {\bibfnamefont {E.}~\bibnamefont {{Varela}}}, \bibinfo {author} {\bibfnamefont {A.}~\bibnamefont {{V{\'a}squez-Ram{\'\i}rez}}}, \bibinfo {author} {\bibfnamefont {D.}~\bibnamefont {{Veberi{\v{c}}}}}, \bibinfo {author} {\bibfnamefont {C.}~\bibnamefont {{Ventura}}}, \bibinfo {author} {\bibfnamefont {I.~D.}\ \bibnamefont {{Vergara Quispe}}}, \bibinfo {author} {\bibfnamefont {V.}~\bibnamefont {{Verzi}}}, \bibinfo {author} {\bibfnamefont {J.}~\bibnamefont {{Vicha}}}, \bibinfo {author} {\bibfnamefont {J.}~\bibnamefont {{Vink}}}, \bibinfo {author} {\bibfnamefont {S.}~\bibnamefont {{Vorobiov}}}, \bibinfo {author} {\bibfnamefont {H.}~\bibnamefont
  {{Wahlberg}}}, \bibinfo {author} {\bibfnamefont {C.}~\bibnamefont {{Watanabe}}}, \bibinfo {author} {\bibfnamefont {A.~A.}\ \bibnamefont {{Watson}}}, \bibinfo {author} {\bibfnamefont {A.}~\bibnamefont {{Weindl}}}, \bibinfo {author} {\bibfnamefont {L.}~\bibnamefont {{Wiencke}}}, \bibinfo {author} {\bibfnamefont {H.}~\bibnamefont {{Wilczy{\'n}ski}}}, \bibinfo {author} {\bibfnamefont {D.}~\bibnamefont {{Wittkowski}}}, \bibinfo {author} {\bibfnamefont {B.}~\bibnamefont {{Wundheiler}}}, \bibinfo {author} {\bibfnamefont {A.}~\bibnamefont {{Yushkov}}}, \bibinfo {author} {\bibfnamefont {O.}~\bibnamefont {{Zapparrata}}}, \bibinfo {author} {\bibfnamefont {E.}~\bibnamefont {{Zas}}}, \bibinfo {author} {\bibfnamefont {D.}~\bibnamefont {{Zavrtanik}}}, \bibinfo {author} {\bibfnamefont {M.}~\bibnamefont {{Zavrtanik}}}, \bibinfo {author} {\bibfnamefont {L.}~\bibnamefont {{Zehrer}}}, \ and\ \bibinfo {author} {\bibnamefont {{Pierre Auger Collaboration}}},\ }\href {\doibase 10.1103/PhysRevD.107.042002} {\bibfield  {journal}
  {\bibinfo  {journal} {\prd}\ }\textbf {\bibinfo {volume} {107}},\ \bibinfo {eid} {042002} (\bibinfo {year} {2023})},\ \Eprint {http://arxiv.org/abs/2208.02353} {arXiv:2208.02353 [astro-ph.HE]} \BibitemShut {NoStop}%
\bibitem [{\citenamefont {Eichhorn}\ \emph {et~al.}(2015)\citenamefont {Eichhorn}, \citenamefont {Gies}, \citenamefont {Jaeckel}, \citenamefont {Plehn}, \citenamefont {Scherer},\ and\ \citenamefont {Sondenheimer}}]{Eichhorn:2015-The-Higgs-Mass-and-the-Scale}%
  \BibitemOpen
  \bibfield  {author} {\bibinfo {author} {\bibfnamefont {A.}~\bibnamefont {Eichhorn}}, \bibinfo {author} {\bibfnamefont {H.}~\bibnamefont {Gies}}, \bibinfo {author} {\bibfnamefont {J.}~\bibnamefont {Jaeckel}}, \bibinfo {author} {\bibfnamefont {T.}~\bibnamefont {Plehn}}, \bibinfo {author} {\bibfnamefont {M.~M.}\ \bibnamefont {Scherer}}, \ and\ \bibinfo {author} {\bibfnamefont {R.}~\bibnamefont {Sondenheimer}},\ }\href {\doibase 10.1007/JHEP04(2015)022} {\bibfield  {journal} {\bibinfo  {journal} {JHEP}\ }\textbf {\bibinfo {volume} {04}},\ \bibinfo {pages} {022} (\bibinfo {year} {2015})},\ \Eprint {http://arxiv.org/abs/1501.02812} {arXiv:1501.02812 [hep-ph]} \BibitemShut {NoStop}%
\bibitem [{\citenamefont {Goswami}\ and\ \citenamefont {Mohanty}(2015)}]{Goswami:2014-Higgs-instability-and}%
  \BibitemOpen
  \bibfield  {author} {\bibinfo {author} {\bibfnamefont {G.}~\bibnamefont {Goswami}}\ and\ \bibinfo {author} {\bibfnamefont {S.}~\bibnamefont {Mohanty}},\ }\href {\doibase 10.1016/j.physletb.2015.10.027} {\bibfield  {journal} {\bibinfo  {journal} {Phys. Lett. B}\ }\textbf {\bibinfo {volume} {751}},\ \bibinfo {pages} {113} (\bibinfo {year} {2015})},\ \Eprint {http://arxiv.org/abs/1406.5644} {arXiv:1406.5644 [hep-ph]} \BibitemShut {NoStop}%
\bibitem [{\citenamefont {Shaposhnikov}(2024)}]{SHAPOSHNIKOV:2024-Sterile-neutrinos-as-dark-matter}%
  \BibitemOpen
  \bibfield  {author} {\bibinfo {author} {\bibfnamefont {M.}~\bibnamefont {Shaposhnikov}},\ }\href {\doibase https://doi.org/10.1016/j.nuclphysb.2024.116496} {\bibfield  {journal} {\bibinfo  {journal} {Nuclear Physics B}\ }\textbf {\bibinfo {volume} {1003}},\ \bibinfo {pages} {116496} (\bibinfo {year} {2024})},\ \bibinfo {note} {special Issue of Nobel Symposium 182 on Dark Matter}\BibitemShut {NoStop}%
\bibitem [{\citenamefont {Blasi}\ \emph {et~al.}(2002)\citenamefont {Blasi}, \citenamefont {Dick},\ and\ \citenamefont {Kolb}}]{Blasi:Ultra-high-energy-cosmic-rays-The-annihilation}%
  \BibitemOpen
  \bibfield  {author} {\bibinfo {author} {\bibfnamefont {P.}~\bibnamefont {Blasi}}, \bibinfo {author} {\bibfnamefont {R.}~\bibnamefont {Dick}}, \ and\ \bibinfo {author} {\bibfnamefont {E.~W.}\ \bibnamefont {Kolb}},\ }\href {\doibase 10.1016/S0920-5632(02)01545-1} {\bibfield  {journal} {\bibinfo  {journal} {Nucl. Phys. B Proc. Suppl.}\ }\textbf {\bibinfo {volume} {110}},\ \bibinfo {pages} {494} (\bibinfo {year} {2002})},\ \Eprint {http://arxiv.org/abs/astro-ph/0111531} {arXiv:astro-ph/0111531} \BibitemShut {NoStop}%
\bibitem [{\citenamefont {Luce}\ \emph {et~al.}(2022)\citenamefont {Luce}, \citenamefont {Marafico}, \citenamefont {Biteau}, \citenamefont {Condorelli},\ and\ \citenamefont {Deligny}}]{Luce:2022-Observational-Constraints-on-Cosmic-Ray}%
  \BibitemOpen
  \bibfield  {author} {\bibinfo {author} {\bibfnamefont {Q.}~\bibnamefont {Luce}}, \bibinfo {author} {\bibfnamefont {S.}~\bibnamefont {Marafico}}, \bibinfo {author} {\bibfnamefont {J.}~\bibnamefont {Biteau}}, \bibinfo {author} {\bibfnamefont {A.}~\bibnamefont {Condorelli}}, \ and\ \bibinfo {author} {\bibfnamefont {O.}~\bibnamefont {Deligny}},\ }\href {\doibase 10.3847/1538-4357/ac81cc} {\bibfield  {journal} {\bibinfo  {journal} {The Astrophysical Journal}\ }\textbf {\bibinfo {volume} {936}},\ \bibinfo {pages} {62} (\bibinfo {year} {2022})}\BibitemShut {NoStop}%
\bibitem [{\citenamefont {Fukugita}\ and\ \citenamefont {Yanagida}(1986)}]{Fukugita:1986-Baryogenesis-Without-Grand-Unification}%
  \BibitemOpen
  \bibfield  {author} {\bibinfo {author} {\bibfnamefont {M.}~\bibnamefont {Fukugita}}\ and\ \bibinfo {author} {\bibfnamefont {T.}~\bibnamefont {Yanagida}},\ }\href {\doibase 10.1016/0370-2693(86)91126-3} {\bibfield  {journal} {\bibinfo  {journal} {Phys. Lett. B}\ }\textbf {\bibinfo {volume} {174}},\ \bibinfo {pages} {45} (\bibinfo {year} {1986})}\BibitemShut {NoStop}%
\bibitem [{\citenamefont {Parker}(1969)}]{Parker:1969-Quantized-Fields-and-Particle-Creation}%
  \BibitemOpen
  \bibfield  {author} {\bibinfo {author} {\bibfnamefont {L.}~\bibnamefont {Parker}},\ }\href {\doibase 10.1103/PhysRev.183.1057} {\bibfield  {journal} {\bibinfo  {journal} {Phys. Rev.}\ }\textbf {\bibinfo {volume} {183}},\ \bibinfo {pages} {1057} (\bibinfo {year} {1969})}\BibitemShut {NoStop}%
\bibitem [{\citenamefont {Ford}(1987)}]{Ford:1987-Gravitational-particle-creation-and-inflation}%
  \BibitemOpen
  \bibfield  {author} {\bibinfo {author} {\bibfnamefont {L.~H.}\ \bibnamefont {Ford}},\ }\href {\doibase 10.1103/PhysRevD.35.2955} {\bibfield  {journal} {\bibinfo  {journal} {Phys. Rev. D}\ }\textbf {\bibinfo {volume} {35}},\ \bibinfo {pages} {2955} (\bibinfo {year} {1987})}\BibitemShut {NoStop}%
\bibitem [{\citenamefont {Haro}\ and\ \citenamefont {Sal\'o}(2019)}]{Haro:2019-Gravitational-production-of-superheavy-baryonic-and-dark-matter}%
  \BibitemOpen
  \bibfield  {author} {\bibinfo {author} {\bibfnamefont {J.}~\bibnamefont {Haro}}\ and\ \bibinfo {author} {\bibfnamefont {L.~A.}\ \bibnamefont {Sal\'o}},\ }\href {\doibase 10.1103/PhysRevD.100.043519} {\bibfield  {journal} {\bibinfo  {journal} {Phys. Rev. D}\ }\textbf {\bibinfo {volume} {100}},\ \bibinfo {pages} {043519} (\bibinfo {year} {2019})}\BibitemShut {NoStop}%
\bibitem [{\citenamefont {Chung}\ \emph {et~al.}(1999)\citenamefont {Chung}, \citenamefont {Kolb},\ and\ \citenamefont {Riotto}}]{Chung:1999-Superheavy-dark-matter}%
  \BibitemOpen
  \bibfield  {author} {\bibinfo {author} {\bibfnamefont {D.~J.~H.}\ \bibnamefont {Chung}}, \bibinfo {author} {\bibfnamefont {E.~W.}\ \bibnamefont {Kolb}}, \ and\ \bibinfo {author} {\bibfnamefont {A.}~\bibnamefont {Riotto}},\ }\href {\doibase 10.1103/PhysRevD.59.023501} {\bibfield  {journal} {\bibinfo  {journal} {Physical Review D}\ }\textbf {\bibinfo {volume} {59}} (\bibinfo {year} {1999}),\ 10.1103/PhysRevD.59.023501}\BibitemShut {NoStop}%
\bibitem [{\citenamefont {Kolb}\ and\ \citenamefont {Long}(2017)}]{Kolb:2017-Superheavy-dark-matter-through}%
  \BibitemOpen
  \bibfield  {author} {\bibinfo {author} {\bibfnamefont {E.~W.}\ \bibnamefont {Kolb}}\ and\ \bibinfo {author} {\bibfnamefont {A.~J.}\ \bibnamefont {Long}},\ }\href {\doibase 10.1103/PhysRevD.96.103540} {\bibfield  {journal} {\bibinfo  {journal} {Physical Review D}\ }\textbf {\bibinfo {volume} {96}} (\bibinfo {year} {2017}),\ 10.1103/PhysRevD.96.103540}\BibitemShut {NoStop}%
\bibitem [{\citenamefont {{Ellis}}\ \emph {et~al.}(1990)\citenamefont {{Ellis}}, \citenamefont {{Lopez}},\ and\ \citenamefont {{Nanopoulos}}}]{Ellis:1990-Confinement-of-fractional-charges-yields-integer-charged}%
  \BibitemOpen
  \bibfield  {author} {\bibinfo {author} {\bibfnamefont {J.}~\bibnamefont {{Ellis}}}, \bibinfo {author} {\bibfnamefont {J.~L.}\ \bibnamefont {{Lopez}}}, \ and\ \bibinfo {author} {\bibfnamefont {D.~V.}\ \bibnamefont {{Nanopoulos}}},\ }\href {\doibase 10.1016/0370-2693(90)90893-B} {\bibfield  {journal} {\bibinfo  {journal} {Physics Letters B}\ }\textbf {\bibinfo {volume} {247}},\ \bibinfo {pages} {257} (\bibinfo {year} {1990})}\BibitemShut {NoStop}%
\bibitem [{\citenamefont {{Benakli}}\ \emph {et~al.}(1999)\citenamefont {{Benakli}}, \citenamefont {{Ellis}},\ and\ \citenamefont {{Nanopoulos}}}]{Benakli:1999-Natural-candidates-for-superheavy-dark-matter}%
  \BibitemOpen
  \bibfield  {author} {\bibinfo {author} {\bibfnamefont {K.}~\bibnamefont {{Benakli}}}, \bibinfo {author} {\bibfnamefont {J.}~\bibnamefont {{Ellis}}}, \ and\ \bibinfo {author} {\bibfnamefont {D.~V.}\ \bibnamefont {{Nanopoulos}}},\ }\href {\doibase 10.1103/PhysRevD.59.047301} {\bibfield  {journal} {\bibinfo  {journal} {\prd}\ }\textbf {\bibinfo {volume} {59}},\ \bibinfo {eid} {047301} (\bibinfo {year} {1999})},\ \Eprint {http://arxiv.org/abs/hep-ph/9803333} {arXiv:hep-ph/9803333 [hep-ph]} \BibitemShut {NoStop}%
\bibitem [{\citenamefont {Kolb}\ and\ \citenamefont {Long}(2024)}]{Kolb:2024-Cosmological-gravitational-particle-production}%
  \BibitemOpen
  \bibfield  {author} {\bibinfo {author} {\bibfnamefont {E.~W.}\ \bibnamefont {Kolb}}\ and\ \bibinfo {author} {\bibfnamefont {A.~J.}\ \bibnamefont {Long}},\ }\href {\doibase 10.1103/RevModPhys.96.045005} {\bibfield  {journal} {\bibinfo  {journal} {Rev. Mod. Phys.}\ }\textbf {\bibinfo {volume} {96}},\ \bibinfo {pages} {045005} (\bibinfo {year} {2024})}\BibitemShut {NoStop}%
\bibitem [{\citenamefont {Kolb}\ \emph {et~al.}(1998)\citenamefont {Kolb}, \citenamefont {Chung},\ and\ \citenamefont {Riotto}}]{kolb:1998-WIMPZILLAS!}%
  \BibitemOpen
  \bibfield  {author} {\bibinfo {author} {\bibfnamefont {E.~W.}\ \bibnamefont {Kolb}}, \bibinfo {author} {\bibfnamefont {D.~J.~H.}\ \bibnamefont {Chung}}, \ and\ \bibinfo {author} {\bibfnamefont {A.}~\bibnamefont {Riotto}},\ }\href {https://arxiv.org/abs/hep-ph/9810361} {\enquote {\bibinfo {title} {Wimpzillas!}}\ } (\bibinfo {year} {1998}),\ \Eprint {http://arxiv.org/abs/hep-ph/9810361} {arXiv:hep-ph/9810361 [hep-ph]} \BibitemShut {NoStop}%
\bibitem [{\citenamefont {Aggarwal}\ \emph {et~al.}(2022)\citenamefont {Aggarwal}, \citenamefont {Winstone}, \citenamefont {Teo}, \citenamefont {Baryakhtar}, \citenamefont {Larson}, \citenamefont {Kalogera},\ and\ \citenamefont {Geraci}}]{Aggarwal:2022-Searching-for-New-Physics}%
  \BibitemOpen
  \bibfield  {author} {\bibinfo {author} {\bibfnamefont {N.}~\bibnamefont {Aggarwal}}, \bibinfo {author} {\bibfnamefont {G.~P.}\ \bibnamefont {Winstone}}, \bibinfo {author} {\bibfnamefont {M.}~\bibnamefont {Teo}}, \bibinfo {author} {\bibfnamefont {M.}~\bibnamefont {Baryakhtar}}, \bibinfo {author} {\bibfnamefont {S.~L.}\ \bibnamefont {Larson}}, \bibinfo {author} {\bibfnamefont {V.}~\bibnamefont {Kalogera}}, \ and\ \bibinfo {author} {\bibfnamefont {A.~A.}\ \bibnamefont {Geraci}},\ }\href {\doibase 10.1103/PhysRevLett.128.111101} {\bibfield  {journal} {\bibinfo  {journal} {Phys. Rev. Lett.}\ }\textbf {\bibinfo {volume} {128}},\ \bibinfo {pages} {111101} (\bibinfo {year} {2022})}\BibitemShut {NoStop}%
\bibitem [{\citenamefont {Duffy}\ and\ \citenamefont {van Bibber}(2009)}]{Duffy:2009-Axions-as-dark-matter-particles}%
  \BibitemOpen
  \bibfield  {author} {\bibinfo {author} {\bibfnamefont {L.~D.}\ \bibnamefont {Duffy}}\ and\ \bibinfo {author} {\bibfnamefont {K.}~\bibnamefont {van Bibber}},\ }\href {\doibase 10.1088/1367-2630/11/10/105008} {\bibfield  {journal} {\bibinfo  {journal} {New Journal of Physics}\ }\textbf {\bibinfo {volume} {11}},\ \bibinfo {pages} {105008} (\bibinfo {year} {2009})}\BibitemShut {NoStop}%
\bibitem [{\citenamefont {Kim}\ and\ \citenamefont {Kim}(2018)}]{KIM:2018-Invisible-QCD-axion-rolling}%
  \BibitemOpen
  \bibfield  {author} {\bibinfo {author} {\bibfnamefont {J.~E.}\ \bibnamefont {Kim}}\ and\ \bibinfo {author} {\bibfnamefont {S.-J.}\ \bibnamefont {Kim}},\ }\href {\doibase https://doi.org/10.1016/j.physletb.2018.07.020} {\bibfield  {journal} {\bibinfo  {journal} {Physics Letters B}\ }\textbf {\bibinfo {volume} {783}},\ \bibinfo {pages} {357} (\bibinfo {year} {2018})}\BibitemShut {NoStop}%
\bibitem [{\citenamefont {Chadha-Day}\ \emph {et~al.}(2022)\citenamefont {Chadha-Day}, \citenamefont {Ellis},\ and\ \citenamefont {Marsh}}]{Chadha-Day:2022-Axion-dark-matter}%
  \BibitemOpen
  \bibfield  {author} {\bibinfo {author} {\bibfnamefont {F.}~\bibnamefont {Chadha-Day}}, \bibinfo {author} {\bibfnamefont {J.}~\bibnamefont {Ellis}}, \ and\ \bibinfo {author} {\bibfnamefont {D.~J.~E.}\ \bibnamefont {Marsh}},\ }\href {\doibase 10.1126/sciadv.abj3618} {\bibfield  {journal} {\bibinfo  {journal} {Science Advances}\ }\textbf {\bibinfo {volume} {8}},\ \bibinfo {pages} {eabj3618} (\bibinfo {year} {2022})},\ \Eprint {http://arxiv.org/abs/https://www.science.org/doi/pdf/10.1126/sciadv.abj3618} {https://www.science.org/doi/pdf/10.1126/sciadv.abj3618} \BibitemShut {NoStop}%
\bibitem [{\citenamefont {Semertzidis}\ and\ \citenamefont {Youn}(2022)}]{Yannis:2022-Axion-dark-matter}%
  \BibitemOpen
  \bibfield  {author} {\bibinfo {author} {\bibfnamefont {Y.~K.}\ \bibnamefont {Semertzidis}}\ and\ \bibinfo {author} {\bibfnamefont {S.}~\bibnamefont {Youn}},\ }\href {\doibase 10.1126/sciadv.abm9928} {\bibfield  {journal} {\bibinfo  {journal} {Science Advances}\ }\textbf {\bibinfo {volume} {8}},\ \bibinfo {pages} {eabm9928} (\bibinfo {year} {2022})},\ \Eprint {http://arxiv.org/abs/https://www.science.org/doi/pdf/10.1126/sciadv.abm9928} {https://www.science.org/doi/pdf/10.1126/sciadv.abm9928} \BibitemShut {NoStop}%
\bibitem [{\citenamefont {Brouwer}\ \emph {et~al.}(2022)\citenamefont {Brouwer}, \citenamefont {Chaudhuri}, \citenamefont {Cho}, \citenamefont {Corbin}, \citenamefont {Dawson}, \citenamefont {Droster}, \citenamefont {Foster}, \citenamefont {Fry}, \citenamefont {Graham}, \citenamefont {Henning}, \citenamefont {Irwin}, \citenamefont {Kadribasic}, \citenamefont {Kahn}, \citenamefont {Keller}, \citenamefont {Kolevatov}, \citenamefont {Kuenstner}, \citenamefont {Leder}, \citenamefont {Li}, \citenamefont {Ouellet}, \citenamefont {Pappas}, \citenamefont {Phipps}, \citenamefont {Rapidis}, \citenamefont {Safdi}, \citenamefont {Salemi}, \citenamefont {Simanovskaia}, \citenamefont {Singh}, \citenamefont {van Assendelft}, \citenamefont {van Bibber}, \citenamefont {Wells}, \citenamefont {Winslow}, \citenamefont {Wisniewski},\ and\ \citenamefont {Young}}]{Brouwer:2022-Proposal-for-a-definitive-search}%
  \BibitemOpen
  \bibfield  {author} {\bibinfo {author} {\bibfnamefont {L.}~\bibnamefont {Brouwer}}, \bibinfo {author} {\bibfnamefont {S.}~\bibnamefont {Chaudhuri}}, \bibinfo {author} {\bibfnamefont {H.-M.}\ \bibnamefont {Cho}}, \bibinfo {author} {\bibfnamefont {J.}~\bibnamefont {Corbin}}, \bibinfo {author} {\bibfnamefont {C.~S.}\ \bibnamefont {Dawson}}, \bibinfo {author} {\bibfnamefont {A.}~\bibnamefont {Droster}}, \bibinfo {author} {\bibfnamefont {J.~W.}\ \bibnamefont {Foster}}, \bibinfo {author} {\bibfnamefont {J.~T.}\ \bibnamefont {Fry}}, \bibinfo {author} {\bibfnamefont {P.~W.}\ \bibnamefont {Graham}}, \bibinfo {author} {\bibfnamefont {R.}~\bibnamefont {Henning}}, \bibinfo {author} {\bibfnamefont {K.~D.}\ \bibnamefont {Irwin}}, \bibinfo {author} {\bibfnamefont {F.}~\bibnamefont {Kadribasic}}, \bibinfo {author} {\bibfnamefont {Y.}~\bibnamefont {Kahn}}, \bibinfo {author} {\bibfnamefont {A.}~\bibnamefont {Keller}}, \bibinfo {author} {\bibfnamefont {R.}~\bibnamefont {Kolevatov}}, \bibinfo {author} {\bibfnamefont
  {S.}~\bibnamefont {Kuenstner}}, \bibinfo {author} {\bibfnamefont {A.~F.}\ \bibnamefont {Leder}}, \bibinfo {author} {\bibfnamefont {D.}~\bibnamefont {Li}}, \bibinfo {author} {\bibfnamefont {J.~L.}\ \bibnamefont {Ouellet}}, \bibinfo {author} {\bibfnamefont {K.~M.~W.}\ \bibnamefont {Pappas}}, \bibinfo {author} {\bibfnamefont {A.}~\bibnamefont {Phipps}}, \bibinfo {author} {\bibfnamefont {N.~M.}\ \bibnamefont {Rapidis}}, \bibinfo {author} {\bibfnamefont {B.~R.}\ \bibnamefont {Safdi}}, \bibinfo {author} {\bibfnamefont {C.~P.}\ \bibnamefont {Salemi}}, \bibinfo {author} {\bibfnamefont {M.}~\bibnamefont {Simanovskaia}}, \bibinfo {author} {\bibfnamefont {J.}~\bibnamefont {Singh}}, \bibinfo {author} {\bibfnamefont {E.~C.}\ \bibnamefont {van Assendelft}}, \bibinfo {author} {\bibfnamefont {K.}~\bibnamefont {van Bibber}}, \bibinfo {author} {\bibfnamefont {K.}~\bibnamefont {Wells}}, \bibinfo {author} {\bibfnamefont {L.}~\bibnamefont {Winslow}}, \bibinfo {author} {\bibfnamefont {W.~J.}\ \bibnamefont {Wisniewski}}, \ and\
  \bibinfo {author} {\bibfnamefont {B.~A.}\ \bibnamefont {Young}} (\bibinfo {collaboration} {DMRadio Collaboration}),\ }\href {\doibase 10.1103/PhysRevD.106.112003} {\bibfield  {journal} {\bibinfo  {journal} {Phys. Rev. D}\ }\textbf {\bibinfo {volume} {106}},\ \bibinfo {pages} {112003} (\bibinfo {year} {2022})}\BibitemShut {NoStop}%
\bibitem [{\citenamefont {Mazumdar}\ \emph {et~al.}(2016)\citenamefont {Mazumdar}, \citenamefont {Qutub},\ and\ \citenamefont {Saikawa}}]{Mazumdar:2016-Nonthermal-axion-dark-radiation}%
  \BibitemOpen
  \bibfield  {author} {\bibinfo {author} {\bibfnamefont {A.}~\bibnamefont {Mazumdar}}, \bibinfo {author} {\bibfnamefont {S.}~\bibnamefont {Qutub}}, \ and\ \bibinfo {author} {\bibfnamefont {K.}~\bibnamefont {Saikawa}},\ }\href {\doibase 10.1103/PhysRevD.94.065030} {\bibfield  {journal} {\bibinfo  {journal} {Phys. Rev. D}\ }\textbf {\bibinfo {volume} {94}},\ \bibinfo {pages} {065030} (\bibinfo {year} {2016})}\BibitemShut {NoStop}%
\bibitem [{\citenamefont {{Marsh}}(2016)}]{Marsh:2016-Axion-cosmology}%
  \BibitemOpen
  \bibfield  {author} {\bibinfo {author} {\bibfnamefont {D.~J.~E.}\ \bibnamefont {{Marsh}}},\ }\href {\doibase 10.1016/j.physrep.2016.06.005} {\bibfield  {journal} {\bibinfo  {journal} {\physrep}\ }\textbf {\bibinfo {volume} {643}},\ \bibinfo {pages} {1} (\bibinfo {year} {2016})},\ \Eprint {http://arxiv.org/abs/1510.07633} {arXiv:1510.07633 [astro-ph.CO]} \BibitemShut {NoStop}%
\bibitem [{\citenamefont {Kobayashi}\ \emph {et~al.}(2013)\citenamefont {Kobayashi}, \citenamefont {Kurematsu},\ and\ \citenamefont {Takahashi}}]{Kobayashi:2013-Isocurvature-constraints-and-anharmonic-effects}%
  \BibitemOpen
  \bibfield  {author} {\bibinfo {author} {\bibfnamefont {T.}~\bibnamefont {Kobayashi}}, \bibinfo {author} {\bibfnamefont {R.}~\bibnamefont {Kurematsu}}, \ and\ \bibinfo {author} {\bibfnamefont {F.}~\bibnamefont {Takahashi}},\ }\href {\doibase 10.1088/1475-7516/2013/09/032} {\bibfield  {journal} {\bibinfo  {journal} {Journal of Cosmology and Astroparticle Physics}\ }\textbf {\bibinfo {volume} {2013}},\ \bibinfo {pages} {032} (\bibinfo {year} {2013})}\BibitemShut {NoStop}%
\bibitem [{\citenamefont {{Planck Collaboration}}\ \emph {et~al.}(2014)\citenamefont {{Planck Collaboration}}, \citenamefont {{Ade}}, \citenamefont {{Aghanim}}, \citenamefont {{Armitage-Caplan}}, \citenamefont {{Arnaud}}, \citenamefont {{Ashdown}}, \citenamefont {{Atrio-Barandela}}, \citenamefont {{Aumont}}, \citenamefont {{Baccigalupi}}, \citenamefont {{Banday}}, \citenamefont {{Barreiro}}, \citenamefont {{Bartlett}}, \citenamefont {{Bartolo}}, \citenamefont {{Battaner}}, \citenamefont {{Benabed}}, \citenamefont {{Beno{\^\i}t}}, \citenamefont {{Benoit-L{\'e}vy}}, \citenamefont {{Bernard}}, \citenamefont {{Bersanelli}}, \citenamefont {{Bielewicz}}, \citenamefont {{Bobin}}, \citenamefont {{Bock}}, \citenamefont {{Bonaldi}}, \citenamefont {{Bond}}, \citenamefont {{Borrill}}, \citenamefont {{Bouchet}}, \citenamefont {{Bridges}}, \citenamefont {{Bucher}}, \citenamefont {{Burigana}}, \citenamefont {{Butler}}, \citenamefont {{Calabrese}}, \citenamefont {{Cardoso}}, \citenamefont {{Catalano}}, \citenamefont
  {{Challinor}}, \citenamefont {{Chamballu}}, \citenamefont {{Chiang}}, \citenamefont {{Chiang}}, \citenamefont {{Christensen}}, \citenamefont {{Church}}, \citenamefont {{Clements}}, \citenamefont {{Colombi}}, \citenamefont {{Colombo}}, \citenamefont {{Couchot}}, \citenamefont {{Coulais}}, \citenamefont {{Crill}}, \citenamefont {{Curto}}, \citenamefont {{Cuttaia}}, \citenamefont {{Danese}}, \citenamefont {{Davies}}, \citenamefont {{Davis}}, \citenamefont {{de Bernardis}}, \citenamefont {{de Rosa}}, \citenamefont {{de Zotti}}, \citenamefont {{Delabrouille}}, \citenamefont {{Delouis}}, \citenamefont {{D{\'e}sert}}, \citenamefont {{Dickinson}}, \citenamefont {{Diego}}, \citenamefont {{Dole}}, \citenamefont {{Donzelli}}, \citenamefont {{Dor{\'e}}}, \citenamefont {{Douspis}}, \citenamefont {{Dunkley}}, \citenamefont {{Dupac}}, \citenamefont {{Efstathiou}}, \citenamefont {{En{\ss}lin}}, \citenamefont {{Eriksen}}, \citenamefont {{Finelli}}, \citenamefont {{Forni}}, \citenamefont {{Frailis}}, \citenamefont
  {{Franceschi}}, \citenamefont {{Galeotta}}, \citenamefont {{Ganga}}, \citenamefont {{Gauthier}}, \citenamefont {{Giard}}, \citenamefont {{Giardino}}, \citenamefont {{Giraud-H{\'e}raud}}, \citenamefont {{Gonz{\'a}lez-Nuevo}}, \citenamefont {{G{\'o}rski}}, \citenamefont {{Gratton}}, \citenamefont {{Gregorio}}, \citenamefont {{Gruppuso}}, \citenamefont {{Hamann}}, \citenamefont {{Hansen}}, \citenamefont {{Hanson}}, \citenamefont {{Harrison}}, \citenamefont {{Henrot-Versill{\'e}}}, \citenamefont {{Hern{\'a}ndez-Monteagudo}}, \citenamefont {{Herranz}}, \citenamefont {{Hildebrandt}}, \citenamefont {{Hivon}}, \citenamefont {{Hobson}}, \citenamefont {{Holmes}}, \citenamefont {{Hornstrup}}, \citenamefont {{Hovest}}, \citenamefont {{Huffenberger}}, \citenamefont {{Jaffe}}, \citenamefont {{Jaffe}}, \citenamefont {{Jones}}, \citenamefont {{Juvela}}, \citenamefont {{Keih{\"a}nen}}, \citenamefont {{Keskitalo}}, \citenamefont {{Kisner}}, \citenamefont {{Kneissl}}, \citenamefont {{Knoche}}, \citenamefont {{Knox}},
  \citenamefont {{Kunz}}, \citenamefont {{Kurki-Suonio}}, \citenamefont {{Lagache}}, \citenamefont {{L{\"a}hteenm{\"a}ki}}, \citenamefont {{Lamarre}}, \citenamefont {{Lasenby}}, \citenamefont {{Laureijs}}, \citenamefont {{Lawrence}}, \citenamefont {{Leach}}, \citenamefont {{Leahy}}, \citenamefont {{Leonardi}}, \citenamefont {{Lesgourgues}}, \citenamefont {{Lewis}}, \citenamefont {{Liguori}}, \citenamefont {{Lilje}}, \citenamefont {{Linden-V{\o}rnle}}, \citenamefont {{L{\'o}pez-Caniego}}, \citenamefont {{Lubin}}, \citenamefont {{Mac{\'\i}as-P{\'e}rez}}, \citenamefont {{Maffei}}, \citenamefont {{Maino}}, \citenamefont {{Mandolesi}}, \citenamefont {{Maris}}, \citenamefont {{Marshall}}, \citenamefont {{Martin}}, \citenamefont {{Mart{\'\i}nez-Gonz{\'a}lez}}, \citenamefont {{Masi}}, \citenamefont {{Massardi}}, \citenamefont {{Matarrese}}, \citenamefont {{Matthai}}, \citenamefont {{Mazzotta}}, \citenamefont {{Meinhold}}, \citenamefont {{Melchiorri}}, \citenamefont {{Mendes}}, \citenamefont {{Mennella}},
  \citenamefont {{Migliaccio}}, \citenamefont {{Mitra}}, \citenamefont {{Miville-Desch{\^e}nes}}, \citenamefont {{Moneti}}, \citenamefont {{Montier}}, \citenamefont {{Morgante}}, \citenamefont {{Mortlock}}, \citenamefont {{Moss}}, \citenamefont {{Munshi}}, \citenamefont {{Murphy}}, \citenamefont {{Naselsky}}, \citenamefont {{Nati}}, \citenamefont {{Natoli}}, \citenamefont {{Netterfield}}, \citenamefont {{N{\o}rgaard-Nielsen}}, \citenamefont {{Noviello}}, \citenamefont {{Novikov}}, \citenamefont {{Novikov}}, \citenamefont {{O'Dwyer}}, \citenamefont {{Osborne}}, \citenamefont {{Oxborrow}}, \citenamefont {{Paci}}, \citenamefont {{Pagano}}, \citenamefont {{Pajot}}, \citenamefont {{Paladini}}, \citenamefont {{Pandolfi}}, \citenamefont {{Paoletti}}, \citenamefont {{Partridge}}, \citenamefont {{Pasian}}, \citenamefont {{Patanchon}}, \citenamefont {{Peiris}}, \citenamefont {{Perdereau}}, \citenamefont {{Perotto}}, \citenamefont {{Perrotta}}, \citenamefont {{Piacentini}}, \citenamefont {{Piat}}, \citenamefont
  {{Pierpaoli}}, \citenamefont {{Pietrobon}}, \citenamefont {{Plaszczynski}}, \citenamefont {{Pointecouteau}}, \citenamefont {{Polenta}}, \citenamefont {{Ponthieu}}, \citenamefont {{Popa}}, \citenamefont {{Poutanen}}, \citenamefont {{Pratt}}, \citenamefont {{Pr{\'e}zeau}}, \citenamefont {{Prunet}}, \citenamefont {{Puget}}, \citenamefont {{Rachen}}, \citenamefont {{Rebolo}}, \citenamefont {{Reinecke}}, \citenamefont {{Remazeilles}}, \citenamefont {{Renault}}, \citenamefont {{Ricciardi}}, \citenamefont {{Riller}}, \citenamefont {{Ristorcelli}}, \citenamefont {{Rocha}}, \citenamefont {{Rosset}},\ and\ \citenamefont {{Roudier}}}]{PlanckCollaboration:2014-Constraints-on-inflation}%
  \BibitemOpen
  \bibfield  {author} {\bibinfo {author} {\bibnamefont {{Planck Collaboration}}}, \bibinfo {author} {\bibfnamefont {P.~A.~R.}\ \bibnamefont {{Ade}}}, \bibinfo {author} {\bibfnamefont {N.}~\bibnamefont {{Aghanim}}}, \bibinfo {author} {\bibfnamefont {C.}~\bibnamefont {{Armitage-Caplan}}}, \bibinfo {author} {\bibfnamefont {M.}~\bibnamefont {{Arnaud}}}, \bibinfo {author} {\bibfnamefont {M.}~\bibnamefont {{Ashdown}}}, \bibinfo {author} {\bibfnamefont {F.}~\bibnamefont {{Atrio-Barandela}}}, \bibinfo {author} {\bibfnamefont {J.}~\bibnamefont {{Aumont}}}, \bibinfo {author} {\bibfnamefont {C.}~\bibnamefont {{Baccigalupi}}}, \bibinfo {author} {\bibfnamefont {A.~J.}\ \bibnamefont {{Banday}}}, \bibinfo {author} {\bibfnamefont {R.~B.}\ \bibnamefont {{Barreiro}}}, \bibinfo {author} {\bibfnamefont {J.~G.}\ \bibnamefont {{Bartlett}}}, \bibinfo {author} {\bibfnamefont {N.}~\bibnamefont {{Bartolo}}}, \bibinfo {author} {\bibfnamefont {E.}~\bibnamefont {{Battaner}}}, \bibinfo {author} {\bibfnamefont {K.}~\bibnamefont
  {{Benabed}}}, \bibinfo {author} {\bibfnamefont {A.}~\bibnamefont {{Beno{\^\i}t}}}, \bibinfo {author} {\bibfnamefont {A.}~\bibnamefont {{Benoit-L{\'e}vy}}}, \bibinfo {author} {\bibfnamefont {J.~P.}\ \bibnamefont {{Bernard}}}, \bibinfo {author} {\bibfnamefont {M.}~\bibnamefont {{Bersanelli}}}, \bibinfo {author} {\bibfnamefont {P.}~\bibnamefont {{Bielewicz}}}, \bibinfo {author} {\bibfnamefont {J.}~\bibnamefont {{Bobin}}}, \bibinfo {author} {\bibfnamefont {J.~J.}\ \bibnamefont {{Bock}}}, \bibinfo {author} {\bibfnamefont {A.}~\bibnamefont {{Bonaldi}}}, \bibinfo {author} {\bibfnamefont {J.~R.}\ \bibnamefont {{Bond}}}, \bibinfo {author} {\bibfnamefont {J.}~\bibnamefont {{Borrill}}}, \bibinfo {author} {\bibfnamefont {F.~R.}\ \bibnamefont {{Bouchet}}}, \bibinfo {author} {\bibfnamefont {M.}~\bibnamefont {{Bridges}}}, \bibinfo {author} {\bibfnamefont {M.}~\bibnamefont {{Bucher}}}, \bibinfo {author} {\bibfnamefont {C.}~\bibnamefont {{Burigana}}}, \bibinfo {author} {\bibfnamefont {R.~C.}\ \bibnamefont {{Butler}}},
  \bibinfo {author} {\bibfnamefont {E.}~\bibnamefont {{Calabrese}}}, \bibinfo {author} {\bibfnamefont {J.~F.}\ \bibnamefont {{Cardoso}}}, \bibinfo {author} {\bibfnamefont {A.}~\bibnamefont {{Catalano}}}, \bibinfo {author} {\bibfnamefont {A.}~\bibnamefont {{Challinor}}}, \bibinfo {author} {\bibfnamefont {A.}~\bibnamefont {{Chamballu}}}, \bibinfo {author} {\bibfnamefont {H.~C.}\ \bibnamefont {{Chiang}}}, \bibinfo {author} {\bibfnamefont {L.~Y.}\ \bibnamefont {{Chiang}}}, \bibinfo {author} {\bibfnamefont {P.~R.}\ \bibnamefont {{Christensen}}}, \bibinfo {author} {\bibfnamefont {S.}~\bibnamefont {{Church}}}, \bibinfo {author} {\bibfnamefont {D.~L.}\ \bibnamefont {{Clements}}}, \bibinfo {author} {\bibfnamefont {S.}~\bibnamefont {{Colombi}}}, \bibinfo {author} {\bibfnamefont {L.~P.~L.}\ \bibnamefont {{Colombo}}}, \bibinfo {author} {\bibfnamefont {F.}~\bibnamefont {{Couchot}}}, \bibinfo {author} {\bibfnamefont {A.}~\bibnamefont {{Coulais}}}, \bibinfo {author} {\bibfnamefont {B.~P.}\ \bibnamefont {{Crill}}}, \bibinfo
  {author} {\bibfnamefont {A.}~\bibnamefont {{Curto}}}, \bibinfo {author} {\bibfnamefont {F.}~\bibnamefont {{Cuttaia}}}, \bibinfo {author} {\bibfnamefont {L.}~\bibnamefont {{Danese}}}, \bibinfo {author} {\bibfnamefont {R.~D.}\ \bibnamefont {{Davies}}}, \bibinfo {author} {\bibfnamefont {R.~J.}\ \bibnamefont {{Davis}}}, \bibinfo {author} {\bibfnamefont {P.}~\bibnamefont {{de Bernardis}}}, \bibinfo {author} {\bibfnamefont {A.}~\bibnamefont {{de Rosa}}}, \bibinfo {author} {\bibfnamefont {G.}~\bibnamefont {{de Zotti}}}, \bibinfo {author} {\bibfnamefont {J.}~\bibnamefont {{Delabrouille}}}, \bibinfo {author} {\bibfnamefont {J.~M.}\ \bibnamefont {{Delouis}}}, \bibinfo {author} {\bibfnamefont {F.~X.}\ \bibnamefont {{D{\'e}sert}}}, \bibinfo {author} {\bibfnamefont {C.}~\bibnamefont {{Dickinson}}}, \bibinfo {author} {\bibfnamefont {J.~M.}\ \bibnamefont {{Diego}}}, \bibinfo {author} {\bibfnamefont {H.}~\bibnamefont {{Dole}}}, \bibinfo {author} {\bibfnamefont {S.}~\bibnamefont {{Donzelli}}}, \bibinfo {author}
  {\bibfnamefont {O.}~\bibnamefont {{Dor{\'e}}}}, \bibinfo {author} {\bibfnamefont {M.}~\bibnamefont {{Douspis}}}, \bibinfo {author} {\bibfnamefont {J.}~\bibnamefont {{Dunkley}}}, \bibinfo {author} {\bibfnamefont {X.}~\bibnamefont {{Dupac}}}, \bibinfo {author} {\bibfnamefont {G.}~\bibnamefont {{Efstathiou}}}, \bibinfo {author} {\bibfnamefont {T.~A.}\ \bibnamefont {{En{\ss}lin}}}, \bibinfo {author} {\bibfnamefont {H.~K.}\ \bibnamefont {{Eriksen}}}, \bibinfo {author} {\bibfnamefont {F.}~\bibnamefont {{Finelli}}}, \bibinfo {author} {\bibfnamefont {O.}~\bibnamefont {{Forni}}}, \bibinfo {author} {\bibfnamefont {M.}~\bibnamefont {{Frailis}}}, \bibinfo {author} {\bibfnamefont {E.}~\bibnamefont {{Franceschi}}}, \bibinfo {author} {\bibfnamefont {S.}~\bibnamefont {{Galeotta}}}, \bibinfo {author} {\bibfnamefont {K.}~\bibnamefont {{Ganga}}}, \bibinfo {author} {\bibfnamefont {C.}~\bibnamefont {{Gauthier}}}, \bibinfo {author} {\bibfnamefont {M.}~\bibnamefont {{Giard}}}, \bibinfo {author} {\bibfnamefont {G.}~\bibnamefont
  {{Giardino}}}, \bibinfo {author} {\bibfnamefont {Y.}~\bibnamefont {{Giraud-H{\'e}raud}}}, \bibinfo {author} {\bibfnamefont {J.}~\bibnamefont {{Gonz{\'a}lez-Nuevo}}}, \bibinfo {author} {\bibfnamefont {K.~M.}\ \bibnamefont {{G{\'o}rski}}}, \bibinfo {author} {\bibfnamefont {S.}~\bibnamefont {{Gratton}}}, \bibinfo {author} {\bibfnamefont {A.}~\bibnamefont {{Gregorio}}}, \bibinfo {author} {\bibfnamefont {A.}~\bibnamefont {{Gruppuso}}}, \bibinfo {author} {\bibfnamefont {J.}~\bibnamefont {{Hamann}}}, \bibinfo {author} {\bibfnamefont {F.~K.}\ \bibnamefont {{Hansen}}}, \bibinfo {author} {\bibfnamefont {D.}~\bibnamefont {{Hanson}}}, \bibinfo {author} {\bibfnamefont {D.}~\bibnamefont {{Harrison}}}, \bibinfo {author} {\bibfnamefont {S.}~\bibnamefont {{Henrot-Versill{\'e}}}}, \bibinfo {author} {\bibfnamefont {C.}~\bibnamefont {{Hern{\'a}ndez-Monteagudo}}}, \bibinfo {author} {\bibfnamefont {D.}~\bibnamefont {{Herranz}}}, \bibinfo {author} {\bibfnamefont {S.~R.}\ \bibnamefont {{Hildebrandt}}}, \bibinfo {author}
  {\bibfnamefont {E.}~\bibnamefont {{Hivon}}}, \bibinfo {author} {\bibfnamefont {M.}~\bibnamefont {{Hobson}}}, \bibinfo {author} {\bibfnamefont {W.~A.}\ \bibnamefont {{Holmes}}}, \bibinfo {author} {\bibfnamefont {A.}~\bibnamefont {{Hornstrup}}}, \bibinfo {author} {\bibfnamefont {W.}~\bibnamefont {{Hovest}}}, \bibinfo {author} {\bibfnamefont {K.~M.}\ \bibnamefont {{Huffenberger}}}, \bibinfo {author} {\bibfnamefont {A.~H.}\ \bibnamefont {{Jaffe}}}, \bibinfo {author} {\bibfnamefont {T.~R.}\ \bibnamefont {{Jaffe}}}, \bibinfo {author} {\bibfnamefont {W.~C.}\ \bibnamefont {{Jones}}}, \bibinfo {author} {\bibfnamefont {M.}~\bibnamefont {{Juvela}}}, \bibinfo {author} {\bibfnamefont {E.}~\bibnamefont {{Keih{\"a}nen}}}, \bibinfo {author} {\bibfnamefont {R.}~\bibnamefont {{Keskitalo}}}, \bibinfo {author} {\bibfnamefont {T.~S.}\ \bibnamefont {{Kisner}}}, \bibinfo {author} {\bibfnamefont {R.}~\bibnamefont {{Kneissl}}}, \bibinfo {author} {\bibfnamefont {J.}~\bibnamefont {{Knoche}}}, \bibinfo {author} {\bibfnamefont
  {L.}~\bibnamefont {{Knox}}}, \bibinfo {author} {\bibfnamefont {M.}~\bibnamefont {{Kunz}}}, \bibinfo {author} {\bibfnamefont {H.}~\bibnamefont {{Kurki-Suonio}}}, \bibinfo {author} {\bibfnamefont {G.}~\bibnamefont {{Lagache}}}, \bibinfo {author} {\bibfnamefont {A.}~\bibnamefont {{L{\"a}hteenm{\"a}ki}}}, \bibinfo {author} {\bibfnamefont {J.~M.}\ \bibnamefont {{Lamarre}}}, \bibinfo {author} {\bibfnamefont {A.}~\bibnamefont {{Lasenby}}}, \bibinfo {author} {\bibfnamefont {R.~J.}\ \bibnamefont {{Laureijs}}}, \bibinfo {author} {\bibfnamefont {C.~R.}\ \bibnamefont {{Lawrence}}}, \bibinfo {author} {\bibfnamefont {S.}~\bibnamefont {{Leach}}}, \bibinfo {author} {\bibfnamefont {J.~P.}\ \bibnamefont {{Leahy}}}, \bibinfo {author} {\bibfnamefont {R.}~\bibnamefont {{Leonardi}}}, \bibinfo {author} {\bibfnamefont {J.}~\bibnamefont {{Lesgourgues}}}, \bibinfo {author} {\bibfnamefont {A.}~\bibnamefont {{Lewis}}}, \bibinfo {author} {\bibfnamefont {M.}~\bibnamefont {{Liguori}}}, \bibinfo {author} {\bibfnamefont {P.~B.}\
  \bibnamefont {{Lilje}}}, \bibinfo {author} {\bibfnamefont {M.}~\bibnamefont {{Linden-V{\o}rnle}}}, \bibinfo {author} {\bibfnamefont {M.}~\bibnamefont {{L{\'o}pez-Caniego}}}, \bibinfo {author} {\bibfnamefont {P.~M.}\ \bibnamefont {{Lubin}}}, \bibinfo {author} {\bibfnamefont {J.~F.}\ \bibnamefont {{Mac{\'\i}as-P{\'e}rez}}}, \bibinfo {author} {\bibfnamefont {B.}~\bibnamefont {{Maffei}}}, \bibinfo {author} {\bibfnamefont {D.}~\bibnamefont {{Maino}}}, \bibinfo {author} {\bibfnamefont {N.}~\bibnamefont {{Mandolesi}}}, \bibinfo {author} {\bibfnamefont {M.}~\bibnamefont {{Maris}}}, \bibinfo {author} {\bibfnamefont {D.~J.}\ \bibnamefont {{Marshall}}}, \bibinfo {author} {\bibfnamefont {P.~G.}\ \bibnamefont {{Martin}}}, \bibinfo {author} {\bibfnamefont {E.}~\bibnamefont {{Mart{\'\i}nez-Gonz{\'a}lez}}}, \bibinfo {author} {\bibfnamefont {S.}~\bibnamefont {{Masi}}}, \bibinfo {author} {\bibfnamefont {M.}~\bibnamefont {{Massardi}}}, \bibinfo {author} {\bibfnamefont {S.}~\bibnamefont {{Matarrese}}}, \bibinfo {author}
  {\bibfnamefont {F.}~\bibnamefont {{Matthai}}}, \bibinfo {author} {\bibfnamefont {P.}~\bibnamefont {{Mazzotta}}}, \bibinfo {author} {\bibfnamefont {P.~R.}\ \bibnamefont {{Meinhold}}}, \bibinfo {author} {\bibfnamefont {A.}~\bibnamefont {{Melchiorri}}}, \bibinfo {author} {\bibfnamefont {L.}~\bibnamefont {{Mendes}}}, \bibinfo {author} {\bibfnamefont {A.}~\bibnamefont {{Mennella}}}, \bibinfo {author} {\bibfnamefont {M.}~\bibnamefont {{Migliaccio}}}, \bibinfo {author} {\bibfnamefont {S.}~\bibnamefont {{Mitra}}}, \bibinfo {author} {\bibfnamefont {M.~A.}\ \bibnamefont {{Miville-Desch{\^e}nes}}}, \bibinfo {author} {\bibfnamefont {A.}~\bibnamefont {{Moneti}}}, \bibinfo {author} {\bibfnamefont {L.}~\bibnamefont {{Montier}}}, \bibinfo {author} {\bibfnamefont {G.}~\bibnamefont {{Morgante}}}, \bibinfo {author} {\bibfnamefont {D.}~\bibnamefont {{Mortlock}}}, \bibinfo {author} {\bibfnamefont {A.}~\bibnamefont {{Moss}}}, \bibinfo {author} {\bibfnamefont {D.}~\bibnamefont {{Munshi}}}, \bibinfo {author} {\bibfnamefont
  {J.~A.}\ \bibnamefont {{Murphy}}}, \bibinfo {author} {\bibfnamefont {P.}~\bibnamefont {{Naselsky}}}, \bibinfo {author} {\bibfnamefont {F.}~\bibnamefont {{Nati}}}, \bibinfo {author} {\bibfnamefont {P.}~\bibnamefont {{Natoli}}}, \bibinfo {author} {\bibfnamefont {C.~B.}\ \bibnamefont {{Netterfield}}}, \bibinfo {author} {\bibfnamefont {H.~U.}\ \bibnamefont {{N{\o}rgaard-Nielsen}}}, \bibinfo {author} {\bibfnamefont {F.}~\bibnamefont {{Noviello}}}, \bibinfo {author} {\bibfnamefont {D.}~\bibnamefont {{Novikov}}}, \bibinfo {author} {\bibfnamefont {I.}~\bibnamefont {{Novikov}}}, \bibinfo {author} {\bibfnamefont {I.~J.}\ \bibnamefont {{O'Dwyer}}}, \bibinfo {author} {\bibfnamefont {S.}~\bibnamefont {{Osborne}}}, \bibinfo {author} {\bibfnamefont {C.~A.}\ \bibnamefont {{Oxborrow}}}, \bibinfo {author} {\bibfnamefont {F.}~\bibnamefont {{Paci}}}, \bibinfo {author} {\bibfnamefont {L.}~\bibnamefont {{Pagano}}}, \bibinfo {author} {\bibfnamefont {F.}~\bibnamefont {{Pajot}}}, \bibinfo {author} {\bibfnamefont {R.}~\bibnamefont
  {{Paladini}}}, \bibinfo {author} {\bibfnamefont {S.}~\bibnamefont {{Pandolfi}}}, \bibinfo {author} {\bibfnamefont {D.}~\bibnamefont {{Paoletti}}}, \bibinfo {author} {\bibfnamefont {B.}~\bibnamefont {{Partridge}}}, \bibinfo {author} {\bibfnamefont {F.}~\bibnamefont {{Pasian}}}, \bibinfo {author} {\bibfnamefont {G.}~\bibnamefont {{Patanchon}}}, \bibinfo {author} {\bibfnamefont {H.~V.}\ \bibnamefont {{Peiris}}}, \bibinfo {author} {\bibfnamefont {O.}~\bibnamefont {{Perdereau}}}, \bibinfo {author} {\bibfnamefont {L.}~\bibnamefont {{Perotto}}}, \bibinfo {author} {\bibfnamefont {F.}~\bibnamefont {{Perrotta}}}, \bibinfo {author} {\bibfnamefont {F.}~\bibnamefont {{Piacentini}}}, \bibinfo {author} {\bibfnamefont {M.}~\bibnamefont {{Piat}}}, \bibinfo {author} {\bibfnamefont {E.}~\bibnamefont {{Pierpaoli}}}, \bibinfo {author} {\bibfnamefont {D.}~\bibnamefont {{Pietrobon}}}, \bibinfo {author} {\bibfnamefont {S.}~\bibnamefont {{Plaszczynski}}}, \bibinfo {author} {\bibfnamefont {E.}~\bibnamefont {{Pointecouteau}}},
  \bibinfo {author} {\bibfnamefont {G.}~\bibnamefont {{Polenta}}}, \bibinfo {author} {\bibfnamefont {N.}~\bibnamefont {{Ponthieu}}}, \bibinfo {author} {\bibfnamefont {L.}~\bibnamefont {{Popa}}}, \bibinfo {author} {\bibfnamefont {T.}~\bibnamefont {{Poutanen}}}, \bibinfo {author} {\bibfnamefont {G.~W.}\ \bibnamefont {{Pratt}}}, \bibinfo {author} {\bibfnamefont {G.}~\bibnamefont {{Pr{\'e}zeau}}}, \bibinfo {author} {\bibfnamefont {S.}~\bibnamefont {{Prunet}}}, \bibinfo {author} {\bibfnamefont {J.~L.}\ \bibnamefont {{Puget}}}, \bibinfo {author} {\bibfnamefont {J.~P.}\ \bibnamefont {{Rachen}}}, \bibinfo {author} {\bibfnamefont {R.}~\bibnamefont {{Rebolo}}}, \bibinfo {author} {\bibfnamefont {M.}~\bibnamefont {{Reinecke}}}, \bibinfo {author} {\bibfnamefont {M.}~\bibnamefont {{Remazeilles}}}, \bibinfo {author} {\bibfnamefont {C.}~\bibnamefont {{Renault}}}, \bibinfo {author} {\bibfnamefont {S.}~\bibnamefont {{Ricciardi}}}, \bibinfo {author} {\bibfnamefont {T.}~\bibnamefont {{Riller}}}, \bibinfo {author} {\bibfnamefont
  {I.}~\bibnamefont {{Ristorcelli}}}, \bibinfo {author} {\bibfnamefont {G.}~\bibnamefont {{Rocha}}}, \bibinfo {author} {\bibfnamefont {C.}~\bibnamefont {{Rosset}}}, \ and\ \bibinfo {author} {\bibfnamefont {G.}~\bibnamefont {{Roudier}}},\ }\href {\doibase 10.1051/0004-6361/201321569} {\bibfield  {journal} {\bibinfo  {journal} {\aap}\ }\textbf {\bibinfo {volume} {571}},\ \bibinfo {eid} {A22} (\bibinfo {year} {2014})},\ \Eprint {http://arxiv.org/abs/1303.5082} {arXiv:1303.5082 [astro-ph.CO]} \BibitemShut {NoStop}%
\bibitem [{\citenamefont {{Di Valentino}}\ \emph {et~al.}(2021)\citenamefont {{Di Valentino}}, \citenamefont {{Mena}}, \citenamefont {{Pan}}, \citenamefont {{Visinelli}}, \citenamefont {{Yang}}, \citenamefont {{Melchiorri}}, \citenamefont {{Mota}}, \citenamefont {{Riess}},\ and\ \citenamefont {{Silk}}}]{Valentino:2021-In-the-realm-of-the-Hubble-tension}%
  \BibitemOpen
  \bibfield  {author} {\bibinfo {author} {\bibfnamefont {E.}~\bibnamefont {{Di Valentino}}}, \bibinfo {author} {\bibfnamefont {O.}~\bibnamefont {{Mena}}}, \bibinfo {author} {\bibfnamefont {S.}~\bibnamefont {{Pan}}}, \bibinfo {author} {\bibfnamefont {L.}~\bibnamefont {{Visinelli}}}, \bibinfo {author} {\bibfnamefont {W.}~\bibnamefont {{Yang}}}, \bibinfo {author} {\bibfnamefont {A.}~\bibnamefont {{Melchiorri}}}, \bibinfo {author} {\bibfnamefont {D.~F.}\ \bibnamefont {{Mota}}}, \bibinfo {author} {\bibfnamefont {A.~G.}\ \bibnamefont {{Riess}}}, \ and\ \bibinfo {author} {\bibfnamefont {J.}~\bibnamefont {{Silk}}},\ }\href {\doibase 10.1088/1361-6382/ac086d} {\bibfield  {journal} {\bibinfo  {journal} {Classical and Quantum Gravity}\ }\textbf {\bibinfo {volume} {38}},\ \bibinfo {eid} {153001} (\bibinfo {year} {2021})},\ \Eprint {http://arxiv.org/abs/2103.01183} {arXiv:2103.01183 [astro-ph.CO]} \BibitemShut {NoStop}%
\bibitem [{\citenamefont {Pandey}\ \emph {et~al.}(2020)\citenamefont {Pandey}, \citenamefont {Karwal},\ and\ \citenamefont {Das}}]{Pandey:2020-Alleviating-the-H0-and}%
  \BibitemOpen
  \bibfield  {author} {\bibinfo {author} {\bibfnamefont {K.~L.}\ \bibnamefont {Pandey}}, \bibinfo {author} {\bibfnamefont {T.}~\bibnamefont {Karwal}}, \ and\ \bibinfo {author} {\bibfnamefont {S.}~\bibnamefont {Das}},\ }\href {\doibase 10.1088/1475-7516/2020/07/026} {\bibfield  {journal} {\bibinfo  {journal} {Journal of Cosmology and Astroparticle Physics}\ }\textbf {\bibinfo {volume} {2020}},\ \bibinfo {pages} {026} (\bibinfo {year} {2020})}\BibitemShut {NoStop}%
\bibitem [{\citenamefont {Chen}\ \emph {et~al.}(2021)\citenamefont {Chen}, \citenamefont {Huterer}, \citenamefont {Lee}, \citenamefont {Fert\'e}, \citenamefont {Weaverdyck}, \citenamefont {Alves}, \citenamefont {Leonard}, \citenamefont {MacCrann}, \citenamefont {Raveri}, \citenamefont {Porredon}, \citenamefont {Di~Valentino}, \citenamefont {Muir}, \citenamefont {Lemos}, \citenamefont {Liddle}, \citenamefont {Blazek}, \citenamefont {Campos}, \citenamefont {Cawthon}, \citenamefont {Choi}, \citenamefont {Dodelson}, \citenamefont {Elvin-Poole}, \citenamefont {Gruen}, \citenamefont {Ross}, \citenamefont {Secco}, \citenamefont {Sevilla-Noarbe}, \citenamefont {Sheldon}, \citenamefont {Troxel}, \citenamefont {Zuntz}, \citenamefont {Abbott}, \citenamefont {Aguena}, \citenamefont {Allam}, \citenamefont {Annis}, \citenamefont {Avila}, \citenamefont {Bertin}, \citenamefont {Bhargava}, \citenamefont {Bridle}, \citenamefont {Brooks}, \citenamefont {Carnero~Rosell}, \citenamefont {Carrasco~Kind}, \citenamefont {Carretero},
  \citenamefont {Costanzi}, \citenamefont {Crocce}, \citenamefont {da~Costa}, \citenamefont {Pereira}, \citenamefont {Davis}, \citenamefont {Doel}, \citenamefont {Eifler}, \citenamefont {Ferrero}, \citenamefont {Fosalba}, \citenamefont {Frieman}, \citenamefont {Garc\'{\i}a-Bellido}, \citenamefont {Gaztanaga}, \citenamefont {Gerdes}, \citenamefont {Gruendl}, \citenamefont {Gschwend}, \citenamefont {Gutierrez}, \citenamefont {Hinton}, \citenamefont {Hollowood}, \citenamefont {Honscheid}, \citenamefont {Hoyle}, \citenamefont {James}, \citenamefont {Jarvis}, \citenamefont {Kuehn}, \citenamefont {Lahav}, \citenamefont {Maia}, \citenamefont {Marshall}, \citenamefont {Menanteau}, \citenamefont {Miquel}, \citenamefont {Morgan}, \citenamefont {Palmese}, \citenamefont {Paz-Chinch\'on}, \citenamefont {Plazas}, \citenamefont {Roodman}, \citenamefont {Sanchez}, \citenamefont {Scarpine}, \citenamefont {Schubnell}, \citenamefont {Serrano}, \citenamefont {Smith}, \citenamefont {Suchyta}, \citenamefont {Tarle}, \citenamefont
  {Thomas}, \citenamefont {To}, \citenamefont {Varga}, \citenamefont {Weller},\ and\ \citenamefont {Wilkinson}}]{Chen-Constraints-on-dark-matter-to-dark-radiation}%
  \BibitemOpen
  \bibfield  {author} {\bibinfo {author} {\bibfnamefont {A.}~\bibnamefont {Chen}}, \bibinfo {author} {\bibfnamefont {D.}~\bibnamefont {Huterer}}, \bibinfo {author} {\bibfnamefont {S.}~\bibnamefont {Lee}}, \bibinfo {author} {\bibfnamefont {A.}~\bibnamefont {Fert\'e}}, \bibinfo {author} {\bibfnamefont {N.}~\bibnamefont {Weaverdyck}}, \bibinfo {author} {\bibfnamefont {O.}~\bibnamefont {Alves}}, \bibinfo {author} {\bibfnamefont {C.~D.}\ \bibnamefont {Leonard}}, \bibinfo {author} {\bibfnamefont {N.}~\bibnamefont {MacCrann}}, \bibinfo {author} {\bibfnamefont {M.}~\bibnamefont {Raveri}}, \bibinfo {author} {\bibfnamefont {A.}~\bibnamefont {Porredon}}, \bibinfo {author} {\bibfnamefont {E.}~\bibnamefont {Di~Valentino}}, \bibinfo {author} {\bibfnamefont {J.}~\bibnamefont {Muir}}, \bibinfo {author} {\bibfnamefont {P.}~\bibnamefont {Lemos}}, \bibinfo {author} {\bibfnamefont {A.~R.}\ \bibnamefont {Liddle}}, \bibinfo {author} {\bibfnamefont {J.}~\bibnamefont {Blazek}}, \bibinfo {author} {\bibfnamefont {A.}~\bibnamefont
  {Campos}}, \bibinfo {author} {\bibfnamefont {R.}~\bibnamefont {Cawthon}}, \bibinfo {author} {\bibfnamefont {A.}~\bibnamefont {Choi}}, \bibinfo {author} {\bibfnamefont {S.}~\bibnamefont {Dodelson}}, \bibinfo {author} {\bibfnamefont {J.}~\bibnamefont {Elvin-Poole}}, \bibinfo {author} {\bibfnamefont {D.}~\bibnamefont {Gruen}}, \bibinfo {author} {\bibfnamefont {A.~J.}\ \bibnamefont {Ross}}, \bibinfo {author} {\bibfnamefont {L.~F.}\ \bibnamefont {Secco}}, \bibinfo {author} {\bibfnamefont {I.}~\bibnamefont {Sevilla-Noarbe}}, \bibinfo {author} {\bibfnamefont {E.}~\bibnamefont {Sheldon}}, \bibinfo {author} {\bibfnamefont {M.~A.}\ \bibnamefont {Troxel}}, \bibinfo {author} {\bibfnamefont {J.}~\bibnamefont {Zuntz}}, \bibinfo {author} {\bibfnamefont {T.~M.~C.}\ \bibnamefont {Abbott}}, \bibinfo {author} {\bibfnamefont {M.}~\bibnamefont {Aguena}}, \bibinfo {author} {\bibfnamefont {S.}~\bibnamefont {Allam}}, \bibinfo {author} {\bibfnamefont {J.}~\bibnamefont {Annis}}, \bibinfo {author} {\bibfnamefont {S.}~\bibnamefont
  {Avila}}, \bibinfo {author} {\bibfnamefont {E.}~\bibnamefont {Bertin}}, \bibinfo {author} {\bibfnamefont {S.}~\bibnamefont {Bhargava}}, \bibinfo {author} {\bibfnamefont {S.~L.}\ \bibnamefont {Bridle}}, \bibinfo {author} {\bibfnamefont {D.}~\bibnamefont {Brooks}}, \bibinfo {author} {\bibfnamefont {A.}~\bibnamefont {Carnero~Rosell}}, \bibinfo {author} {\bibfnamefont {M.}~\bibnamefont {Carrasco~Kind}}, \bibinfo {author} {\bibfnamefont {J.}~\bibnamefont {Carretero}}, \bibinfo {author} {\bibfnamefont {M.}~\bibnamefont {Costanzi}}, \bibinfo {author} {\bibfnamefont {M.}~\bibnamefont {Crocce}}, \bibinfo {author} {\bibfnamefont {L.~N.}\ \bibnamefont {da~Costa}}, \bibinfo {author} {\bibfnamefont {M.~E.~S.}\ \bibnamefont {Pereira}}, \bibinfo {author} {\bibfnamefont {T.~M.}\ \bibnamefont {Davis}}, \bibinfo {author} {\bibfnamefont {P.}~\bibnamefont {Doel}}, \bibinfo {author} {\bibfnamefont {T.~F.}\ \bibnamefont {Eifler}}, \bibinfo {author} {\bibfnamefont {I.}~\bibnamefont {Ferrero}}, \bibinfo {author} {\bibfnamefont
  {P.}~\bibnamefont {Fosalba}}, \bibinfo {author} {\bibfnamefont {J.}~\bibnamefont {Frieman}}, \bibinfo {author} {\bibfnamefont {J.}~\bibnamefont {Garc\'{\i}a-Bellido}}, \bibinfo {author} {\bibfnamefont {E.}~\bibnamefont {Gaztanaga}}, \bibinfo {author} {\bibfnamefont {D.~W.}\ \bibnamefont {Gerdes}}, \bibinfo {author} {\bibfnamefont {R.~A.}\ \bibnamefont {Gruendl}}, \bibinfo {author} {\bibfnamefont {J.}~\bibnamefont {Gschwend}}, \bibinfo {author} {\bibfnamefont {G.}~\bibnamefont {Gutierrez}}, \bibinfo {author} {\bibfnamefont {S.~R.}\ \bibnamefont {Hinton}}, \bibinfo {author} {\bibfnamefont {D.~L.}\ \bibnamefont {Hollowood}}, \bibinfo {author} {\bibfnamefont {K.}~\bibnamefont {Honscheid}}, \bibinfo {author} {\bibfnamefont {B.}~\bibnamefont {Hoyle}}, \bibinfo {author} {\bibfnamefont {D.~J.}\ \bibnamefont {James}}, \bibinfo {author} {\bibfnamefont {M.}~\bibnamefont {Jarvis}}, \bibinfo {author} {\bibfnamefont {K.}~\bibnamefont {Kuehn}}, \bibinfo {author} {\bibfnamefont {O.}~\bibnamefont {Lahav}}, \bibinfo {author}
  {\bibfnamefont {M.~A.~G.}\ \bibnamefont {Maia}}, \bibinfo {author} {\bibfnamefont {J.~L.}\ \bibnamefont {Marshall}}, \bibinfo {author} {\bibfnamefont {F.}~\bibnamefont {Menanteau}}, \bibinfo {author} {\bibfnamefont {R.}~\bibnamefont {Miquel}}, \bibinfo {author} {\bibfnamefont {R.}~\bibnamefont {Morgan}}, \bibinfo {author} {\bibfnamefont {A.}~\bibnamefont {Palmese}}, \bibinfo {author} {\bibfnamefont {F.}~\bibnamefont {Paz-Chinch\'on}}, \bibinfo {author} {\bibfnamefont {A.~A.}\ \bibnamefont {Plazas}}, \bibinfo {author} {\bibfnamefont {A.}~\bibnamefont {Roodman}}, \bibinfo {author} {\bibfnamefont {E.}~\bibnamefont {Sanchez}}, \bibinfo {author} {\bibfnamefont {V.}~\bibnamefont {Scarpine}}, \bibinfo {author} {\bibfnamefont {M.}~\bibnamefont {Schubnell}}, \bibinfo {author} {\bibfnamefont {S.}~\bibnamefont {Serrano}}, \bibinfo {author} {\bibfnamefont {M.}~\bibnamefont {Smith}}, \bibinfo {author} {\bibfnamefont {E.}~\bibnamefont {Suchyta}}, \bibinfo {author} {\bibfnamefont {G.}~\bibnamefont {Tarle}}, \bibinfo
  {author} {\bibfnamefont {D.}~\bibnamefont {Thomas}}, \bibinfo {author} {\bibfnamefont {C.}~\bibnamefont {To}}, \bibinfo {author} {\bibfnamefont {T.~N.}\ \bibnamefont {Varga}}, \bibinfo {author} {\bibfnamefont {J.}~\bibnamefont {Weller}}, \ and\ \bibinfo {author} {\bibfnamefont {R.~D.}\ \bibnamefont {Wilkinson}} (\bibinfo {collaboration} {DES Collaboration}),\ }\href {\doibase 10.1103/PhysRevD.103.123528} {\bibfield  {journal} {\bibinfo  {journal} {Phys. Rev. D}\ }\textbf {\bibinfo {volume} {103}},\ \bibinfo {pages} {123528} (\bibinfo {year} {2021})}\BibitemShut {NoStop}%
\bibitem [{\citenamefont {Seto}\ and\ \citenamefont {Toda}(2021)}]{seto:2021-Comparing-early-dark-energy-and-extra-radiation}%
  \BibitemOpen
  \bibfield  {author} {\bibinfo {author} {\bibfnamefont {O.}~\bibnamefont {Seto}}\ and\ \bibinfo {author} {\bibfnamefont {Y.}~\bibnamefont {Toda}},\ }\href {\doibase 10.1103/PhysRevD.103.123501} {\bibfield  {journal} {\bibinfo  {journal} {Phys. Rev. D}\ }\textbf {\bibinfo {volume} {103}},\ \bibinfo {pages} {123501} (\bibinfo {year} {2021})}\BibitemShut {NoStop}%
\bibitem [{\citenamefont {Anchordoqui}(2019)}]{Anchordoqui2019:Ultra-high-energy-cosmic-rays}%
  \BibitemOpen
  \bibfield  {author} {\bibinfo {author} {\bibfnamefont {L.~A.}\ \bibnamefont {Anchordoqui}},\ }\href {\doibase https://doi.org/10.1016/j.physrep.2019.01.002} {\bibfield  {journal} {\bibinfo  {journal} {Physics Reports}\ }\textbf {\bibinfo {volume} {801}},\ \bibinfo {pages} {1} (\bibinfo {year} {2019})},\ \bibinfo {note} {ultra-high-energy cosmic rays}\BibitemShut {NoStop}%
\bibitem [{\citenamefont {Deur}\ \emph {et~al.}(2016)\citenamefont {Deur}, \citenamefont {Brodsky},\ and\ \citenamefont {{de Téramond}}}]{DEUR:2016-The-QCD-running-coupling}%
  \BibitemOpen
  \bibfield  {author} {\bibinfo {author} {\bibfnamefont {A.}~\bibnamefont {Deur}}, \bibinfo {author} {\bibfnamefont {S.~J.}\ \bibnamefont {Brodsky}}, \ and\ \bibinfo {author} {\bibfnamefont {G.~F.}\ \bibnamefont {{de Téramond}}},\ }\href {\doibase https://doi.org/10.1016/j.ppnp.2016.04.003} {\bibfield  {journal} {\bibinfo  {journal} {Progress in Particle and Nuclear Physics}\ }\textbf {\bibinfo {volume} {90}},\ \bibinfo {pages} {1} (\bibinfo {year} {2016})}\BibitemShut {NoStop}%
\bibitem [{\citenamefont {Xu}(2022{\natexlab{a}})}]{Xu:2022-Dark_matter-flow-dataset-part1}%
  \BibitemOpen
  \bibfield  {author} {\bibinfo {author} {\bibfnamefont {Z.}~\bibnamefont {Xu}},\ }\href {\doibase 10.5281/zenodo.6541230} {\enquote {\bibinfo {title} {Dark matter flow dataset part i: Halo-based statistics from cosmological n-body simulation},}\ } (\bibinfo {year} {2022}{\natexlab{a}})\BibitemShut {NoStop}%
\bibitem [{\citenamefont {Xu}(2022{\natexlab{b}})}]{Xu:2022-Dark_matter-flow-dataset-part2}%
  \BibitemOpen
  \bibfield  {author} {\bibinfo {author} {\bibfnamefont {Z.}~\bibnamefont {Xu}},\ }\href {\doibase 10.5281/zenodo.6569898} {\enquote {\bibinfo {title} {Dark matter flow dataset part ii: Correlation-based statistics from cosmological n-body simulation},}\ } (\bibinfo {year} {2022}{\natexlab{b}})\BibitemShut {NoStop}%
\bibitem [{\citenamefont {Xu}(2022{\natexlab{c}})}]{Xu:2022-Dark_matter-flow-and-hydrodynamic-turbulence-presentation}%
  \BibitemOpen
  \bibfield  {author} {\bibinfo {author} {\bibfnamefont {Z.}~\bibnamefont {Xu}},\ }\href {\doibase 10.5281/zenodo.6569901} {\enquote {\bibinfo {title} {A comparative study of dark matter flow \& hydrodynamic turbulence and its applications},}\ } (\bibinfo {year} {2022}{\natexlab{c}})\BibitemShut {NoStop}%
\bibitem [{\citenamefont {Xu}(2022{\natexlab{d}})}]{Xu:Dark_matter_flow_dataset_2022_all_files}%
  \BibitemOpen
  \bibfield  {author} {\bibinfo {author} {\bibfnamefont {Z.}~\bibnamefont {Xu}},\ }\href {\doibase 10.5281/zenodo.6586212} {\enquote {\bibinfo {title} {Dark matter flow dataset},}\ } (\bibinfo {year} {2022}{\natexlab{d}})\BibitemShut {NoStop}%
\bibitem [{\citenamefont {Blanco}\ \emph {et~al.}(2019)\citenamefont {Blanco}, \citenamefont {Delos}, \citenamefont {Erickcek},\ and\ \citenamefont {Hooper}}]{Blanco:2019-Annihilation-Signature-of-Hidden-Sector-Dark-Matter}%
  \BibitemOpen
  \bibfield  {author} {\bibinfo {author} {\bibfnamefont {C.}~\bibnamefont {Blanco}}, \bibinfo {author} {\bibfnamefont {M.~S.}\ \bibnamefont {Delos}}, \bibinfo {author} {\bibfnamefont {A.~L.}\ \bibnamefont {Erickcek}}, \ and\ \bibinfo {author} {\bibfnamefont {D.}~\bibnamefont {Hooper}},\ }\href {\doibase 10.1103/PhysRevD.100.103010} {\bibfield  {journal} {\bibinfo  {journal} {Phys. Rev. D}\ }\textbf {\bibinfo {volume} {100}},\ \bibinfo {pages} {103010} (\bibinfo {year} {2019})},\ \Eprint {http://arxiv.org/abs/1906.00010} {arXiv:1906.00010 [astro-ph.CO]} \BibitemShut {NoStop}%
\bibitem [{\citenamefont {Xu}(2022{\natexlab{e}})}]{Xu:2022-Postulating-dark-matter-partic}%
  \BibitemOpen
  \bibfield  {author} {\bibinfo {author} {\bibfnamefont {Z.}~\bibnamefont {Xu}},\ }\href {\doibase 10.48550/ARXIV.2202.07240} {\bibfield  {journal} {\bibinfo  {journal} {arXiv e-prints}\ ,\ \bibinfo {pages} {arXiv:2202.07240}} (\bibinfo {year} {2022}{\natexlab{e}})}\BibitemShut {NoStop}%
\bibitem [{\citenamefont {Feng}(2010)}]{Feng:2010-Dark-Matter-Candidates-from-Particle-Physics}%
  \BibitemOpen
  \bibfield  {author} {\bibinfo {author} {\bibfnamefont {J.~L.}\ \bibnamefont {Feng}},\ }\href {\doibase 10.1146/annurev-astro-082708-101659} {\bibfield  {journal} {\bibinfo  {journal} {Ann. Rev. Astron. Astrophys.}\ }\textbf {\bibinfo {volume} {48}},\ \bibinfo {pages} {495} (\bibinfo {year} {2010})},\ \Eprint {http://arxiv.org/abs/1003.0904} {arXiv:1003.0904 [astro-ph.CO]} \BibitemShut {NoStop}%
\bibitem [{\citenamefont {Castellano}\ \emph {et~al.}(2022)\citenamefont {Castellano}, \citenamefont {Fontana}, \citenamefont {Treu}, \citenamefont {Santini}, \citenamefont {Merlin}, \citenamefont {Leethochawalit}, \citenamefont {Trenti}, \citenamefont {Vanzella}, \citenamefont {Mestric}, \citenamefont {Bonchi}, \citenamefont {Belfiori}, \citenamefont {Nonino}, \citenamefont {Paris}, \citenamefont {Polenta}, \citenamefont {Roberts-Borsani}, \citenamefont {Boyett}, \citenamefont {Bradač}, \citenamefont {Calabrò}, \citenamefont {Glazebrook}, \citenamefont {Grillo}, \citenamefont {Mascia}, \citenamefont {Mason}, \citenamefont {Mercurio}, \citenamefont {Morishita}, \citenamefont {Nanayakkara}, \citenamefont {Pentericci}, \citenamefont {Rosati}, \citenamefont {Vulcani}, \citenamefont {Wang},\ and\ \citenamefont {Yang}}]{Castellano:2022-Early-Results-from-GLASS-JWST}%
  \BibitemOpen
  \bibfield  {author} {\bibinfo {author} {\bibfnamefont {M.}~\bibnamefont {Castellano}}, \bibinfo {author} {\bibfnamefont {A.}~\bibnamefont {Fontana}}, \bibinfo {author} {\bibfnamefont {T.}~\bibnamefont {Treu}}, \bibinfo {author} {\bibfnamefont {P.}~\bibnamefont {Santini}}, \bibinfo {author} {\bibfnamefont {E.}~\bibnamefont {Merlin}}, \bibinfo {author} {\bibfnamefont {N.}~\bibnamefont {Leethochawalit}}, \bibinfo {author} {\bibfnamefont {M.}~\bibnamefont {Trenti}}, \bibinfo {author} {\bibfnamefont {E.}~\bibnamefont {Vanzella}}, \bibinfo {author} {\bibfnamefont {U.}~\bibnamefont {Mestric}}, \bibinfo {author} {\bibfnamefont {A.}~\bibnamefont {Bonchi}}, \bibinfo {author} {\bibfnamefont {D.}~\bibnamefont {Belfiori}}, \bibinfo {author} {\bibfnamefont {M.}~\bibnamefont {Nonino}}, \bibinfo {author} {\bibfnamefont {D.}~\bibnamefont {Paris}}, \bibinfo {author} {\bibfnamefont {G.}~\bibnamefont {Polenta}}, \bibinfo {author} {\bibfnamefont {G.}~\bibnamefont {Roberts-Borsani}}, \bibinfo {author} {\bibfnamefont
  {K.}~\bibnamefont {Boyett}}, \bibinfo {author} {\bibfnamefont {M.}~\bibnamefont {Bradač}}, \bibinfo {author} {\bibfnamefont {A.}~\bibnamefont {Calabrò}}, \bibinfo {author} {\bibfnamefont {K.}~\bibnamefont {Glazebrook}}, \bibinfo {author} {\bibfnamefont {C.}~\bibnamefont {Grillo}}, \bibinfo {author} {\bibfnamefont {S.}~\bibnamefont {Mascia}}, \bibinfo {author} {\bibfnamefont {C.}~\bibnamefont {Mason}}, \bibinfo {author} {\bibfnamefont {A.}~\bibnamefont {Mercurio}}, \bibinfo {author} {\bibfnamefont {T.}~\bibnamefont {Morishita}}, \bibinfo {author} {\bibfnamefont {T.}~\bibnamefont {Nanayakkara}}, \bibinfo {author} {\bibfnamefont {L.}~\bibnamefont {Pentericci}}, \bibinfo {author} {\bibfnamefont {P.}~\bibnamefont {Rosati}}, \bibinfo {author} {\bibfnamefont {B.}~\bibnamefont {Vulcani}}, \bibinfo {author} {\bibfnamefont {X.}~\bibnamefont {Wang}}, \ and\ \bibinfo {author} {\bibfnamefont {L.}~\bibnamefont {Yang}},\ }\href {\doibase 10.3847/2041-8213/ac94d0} {\bibfield  {journal} {\bibinfo  {journal} {The
  Astrophysical Journal Letters}\ }\textbf {\bibinfo {volume} {938}},\ \bibinfo {pages} {L15} (\bibinfo {year} {2022})}\BibitemShut {NoStop}%
\bibitem [{\citenamefont {Neyman}\ and\ \citenamefont {Scott}(1952)}]{Neyman:1952-A-Theory-of-the-Spatial-Distri}%
  \BibitemOpen
  \bibfield  {author} {\bibinfo {author} {\bibfnamefont {J.}~\bibnamefont {Neyman}}\ and\ \bibinfo {author} {\bibfnamefont {E.~L.}\ \bibnamefont {Scott}},\ }\href {\doibase 10.1086/145599} {\bibfield  {journal} {\bibinfo  {journal} {Astrophysical Journal}\ }\textbf {\bibinfo {volume} {116}},\ \bibinfo {pages} {144} (\bibinfo {year} {1952})}\BibitemShut {NoStop}%
\bibitem [{\citenamefont {Cooray}\ and\ \citenamefont {Sheth}(2002)}]{Cooray:2002-Halo-models-of-large-scale-str}%
  \BibitemOpen
  \bibfield  {author} {\bibinfo {author} {\bibfnamefont {A.}~\bibnamefont {Cooray}}\ and\ \bibinfo {author} {\bibfnamefont {R.}~\bibnamefont {Sheth}},\ }\href {\doibase 10.1016/S0370-1573(02)00276-4} {\bibfield  {journal} {\bibinfo  {journal} {Physics Reports-Review Section of Physics Letters}\ }\textbf {\bibinfo {volume} {372}},\ \bibinfo {pages} {1} (\bibinfo {year} {2002})}\BibitemShut {NoStop}%
\bibitem [{\citenamefont {Nelson}\ \emph {et~al.}(2015)\citenamefont {Nelson}, \citenamefont {Pillepich}, \citenamefont {Genel}, \citenamefont {Vogelsberger}, \citenamefont {Springel}, \citenamefont {Torrey}, \citenamefont {Rodriguez-Gomez}, \citenamefont {Sijacki}, \citenamefont {Snyder}, \citenamefont {Griffen}, \citenamefont {Marinacci}, \citenamefont {Blecha}, \citenamefont {Sales}, \citenamefont {Xu},\ and\ \citenamefont {Hernquist}}]{NELSON:2015-The-illustris-simulation}%
  \BibitemOpen
  \bibfield  {author} {\bibinfo {author} {\bibfnamefont {D.}~\bibnamefont {Nelson}}, \bibinfo {author} {\bibfnamefont {A.}~\bibnamefont {Pillepich}}, \bibinfo {author} {\bibfnamefont {S.}~\bibnamefont {Genel}}, \bibinfo {author} {\bibfnamefont {M.}~\bibnamefont {Vogelsberger}}, \bibinfo {author} {\bibfnamefont {V.}~\bibnamefont {Springel}}, \bibinfo {author} {\bibfnamefont {P.}~\bibnamefont {Torrey}}, \bibinfo {author} {\bibfnamefont {V.}~\bibnamefont {Rodriguez-Gomez}}, \bibinfo {author} {\bibfnamefont {D.}~\bibnamefont {Sijacki}}, \bibinfo {author} {\bibfnamefont {G.}~\bibnamefont {Snyder}}, \bibinfo {author} {\bibfnamefont {B.}~\bibnamefont {Griffen}}, \bibinfo {author} {\bibfnamefont {F.}~\bibnamefont {Marinacci}}, \bibinfo {author} {\bibfnamefont {L.}~\bibnamefont {Blecha}}, \bibinfo {author} {\bibfnamefont {L.}~\bibnamefont {Sales}}, \bibinfo {author} {\bibfnamefont {D.}~\bibnamefont {Xu}}, \ and\ \bibinfo {author} {\bibfnamefont {L.}~\bibnamefont {Hernquist}},\ }\href {\doibase
  https://doi.org/10.1016/j.ascom.2015.09.003} {\bibfield  {journal} {\bibinfo  {journal} {Astronomy and Computing}\ }\textbf {\bibinfo {volume} {13}},\ \bibinfo {pages} {12} (\bibinfo {year} {2015})}\BibitemShut {NoStop}%
\bibitem [{\citenamefont {{Mo}}\ and\ \citenamefont {{White}}(2002)}]{Mo:2002-The-abundance-and-clustering-of-dark-haloes}%
  \BibitemOpen
  \bibfield  {author} {\bibinfo {author} {\bibfnamefont {H.~J.}\ \bibnamefont {{Mo}}}\ and\ \bibinfo {author} {\bibfnamefont {S.~D.~M.}\ \bibnamefont {{White}}},\ }\href {\doibase 10.1046/j.1365-8711.2002.05723.x} {\bibfield  {journal} {\bibinfo  {journal} {\mnras}\ }\textbf {\bibinfo {volume} {336}},\ \bibinfo {pages} {112} (\bibinfo {year} {2002})},\ \Eprint {http://arxiv.org/abs/astro-ph/0202393} {arXiv:astro-ph/0202393 [astro-ph]} \BibitemShut {NoStop}%
\bibitem [{\citenamefont {Peebles}(1980)}]{Peebles:1980-The-Large-Scale-Structure-of-t}%
  \BibitemOpen
  \bibfield  {author} {\bibinfo {author} {\bibfnamefont {P.~J.~E.}\ \bibnamefont {Peebles}},\ }\href@noop {} {\emph {\bibinfo {title} {The Large-Scale Structure of the Universe}}}\ (\bibinfo  {publisher} {Princeton University Press},\ \bibinfo {address} {Princeton, NJ},\ \bibinfo {year} {1980})\BibitemShut {NoStop}%
\bibitem [{\citenamefont {Xu}(2022{\natexlab{f}})}]{Xu:2022-The-evolution-of-energy--momen}%
  \BibitemOpen
  \bibfield  {author} {\bibinfo {author} {\bibfnamefont {Z.}~\bibnamefont {Xu}},\ }\href {\doibase 10.48550/ARXIV.2202.04054} {\bibfield  {journal} {\bibinfo  {journal} {arXiv e-prints}\ ,\ \bibinfo {pages} {arXiv:2202.04054}} (\bibinfo {year} {2022}{\natexlab{f}})}\BibitemShut {NoStop}%
\bibitem [{\citenamefont {Irvine}(1961)}]{Irvine:1961-Local-Irregularities-in-a-Univ}%
  \BibitemOpen
  \bibfield  {author} {\bibinfo {author} {\bibfnamefont {W.~M.}\ \bibnamefont {Irvine}},\ }\emph {\bibinfo {title} {Local Irregularities in a Universe Satisfying the Cosmological Principle}},\ \href@noop {} {\bibinfo {type} {Thesis}},\ \bibinfo  {school} {HARVARD UNIVERSITY} (\bibinfo {year} {1961})\BibitemShut {NoStop}%
\bibitem [{\citenamefont {Layzer}(1963)}]{Layzer:1963-A-Preface-to-Cosmogony--I--The}%
  \BibitemOpen
  \bibfield  {author} {\bibinfo {author} {\bibfnamefont {D.}~\bibnamefont {Layzer}},\ }\href {\doibase 10.1086/147625} {\bibfield  {journal} {\bibinfo  {journal} {Astrophysical Journal}\ }\textbf {\bibinfo {volume} {138}},\ \bibinfo {pages} {174} (\bibinfo {year} {1963})}\BibitemShut {NoStop}%
\bibitem [{\citenamefont {Richardson}(1922)}]{Richardson:1922-Weather-Prediction-by-Numerica}%
  \BibitemOpen
  \bibfield  {author} {\bibinfo {author} {\bibfnamefont {L.~F.}\ \bibnamefont {Richardson}},\ }\href@noop {} {\emph {\bibinfo {title} {Weather Prediction by Numerical Process}}}\ (\bibinfo  {publisher} {Cambridge University Press},\ \bibinfo {address} {Cambridge, UK},\ \bibinfo {year} {1922})\BibitemShut {NoStop}%
\bibitem [{\citenamefont {Kolmogoroff}(1941)}]{Kolmogoroff:1941-Dissipation-of-energy-in-the-l}%
  \BibitemOpen
  \bibfield  {author} {\bibinfo {author} {\bibfnamefont {A.~N.}\ \bibnamefont {Kolmogoroff}},\ }\href {<Go to ISI>://WOS:000201918500005} {\bibfield  {journal} {\bibinfo  {journal} {Comptes Rendus De L Academie Des Sciences De L Urss}\ }\textbf {\bibinfo {volume} {32}},\ \bibinfo {pages} {16} (\bibinfo {year} {1941})}\BibitemShut {NoStop}%
\bibitem [{\citenamefont {Navarro}\ \emph {et~al.}(1997)\citenamefont {Navarro}, \citenamefont {Frenk},\ and\ \citenamefont {White}}]{Navarro:1997-A-universal-density-profile-fr}%
  \BibitemOpen
  \bibfield  {author} {\bibinfo {author} {\bibfnamefont {J.~F.}\ \bibnamefont {Navarro}}, \bibinfo {author} {\bibfnamefont {C.~S.}\ \bibnamefont {Frenk}}, \ and\ \bibinfo {author} {\bibfnamefont {S.~D.~M.}\ \bibnamefont {White}},\ }\href {\doibase 10.1086/304888} {\bibfield  {journal} {\bibinfo  {journal} {Astrophysical Journal}\ }\textbf {\bibinfo {volume} {490}},\ \bibinfo {pages} {493} (\bibinfo {year} {1997})}\BibitemShut {NoStop}%
\bibitem [{\citenamefont {Navarro}\ \emph {et~al.}(2010)\citenamefont {Navarro}, \citenamefont {Ludlow}, \citenamefont {Springel}, \citenamefont {Wang}, \citenamefont {Vogelsberger}, \citenamefont {White}, \citenamefont {Jenkins}, \citenamefont {Frenk},\ and\ \citenamefont {Helmi}}]{Navarro:2010-The-diversity-and-similarity-of-simulated}%
  \BibitemOpen
  \bibfield  {author} {\bibinfo {author} {\bibfnamefont {J.~F.}\ \bibnamefont {Navarro}}, \bibinfo {author} {\bibfnamefont {A.}~\bibnamefont {Ludlow}}, \bibinfo {author} {\bibfnamefont {V.}~\bibnamefont {Springel}}, \bibinfo {author} {\bibfnamefont {J.}~\bibnamefont {Wang}}, \bibinfo {author} {\bibfnamefont {M.}~\bibnamefont {Vogelsberger}}, \bibinfo {author} {\bibfnamefont {S.~D.~M.}\ \bibnamefont {White}}, \bibinfo {author} {\bibfnamefont {A.}~\bibnamefont {Jenkins}}, \bibinfo {author} {\bibfnamefont {C.~S.}\ \bibnamefont {Frenk}}, \ and\ \bibinfo {author} {\bibfnamefont {A.}~\bibnamefont {Helmi}},\ }\href {\doibase 10.1111/j.1365-2966.2009.15878.x} {\bibfield  {journal} {\bibinfo  {journal} {Monthly Notices of the Royal Astronomical Society}\ }\textbf {\bibinfo {volume} {402}},\ \bibinfo {pages} {21} (\bibinfo {year} {2010})},\ \Eprint {http://arxiv.org/abs/https://academic.oup.com/mnras/article-pdf/402/1/21/18573804/mnras0402-0021.pdf}
  {https://academic.oup.com/mnras/article-pdf/402/1/21/18573804/mnras0402-0021.pdf} \BibitemShut {NoStop}%
\bibitem [{\citenamefont {{Diemand}}\ and\ \citenamefont {{Moore}}(2011)}]{Diemand:2011-The-Structure-and-Evolution-of-Cold-Dark}%
  \BibitemOpen
  \bibfield  {author} {\bibinfo {author} {\bibfnamefont {J.}~\bibnamefont {{Diemand}}}\ and\ \bibinfo {author} {\bibfnamefont {B.}~\bibnamefont {{Moore}}},\ }\href {\doibase 10.1166/asl.2011.1211} {\bibfield  {journal} {\bibinfo  {journal} {Advanced Science Letters}\ }\textbf {\bibinfo {volume} {4}},\ \bibinfo {pages} {297} (\bibinfo {year} {2011})},\ \Eprint {http://arxiv.org/abs/0906.4340} {arXiv:0906.4340 [astro-ph.CO]} \BibitemShut {NoStop}%
\bibitem [{\citenamefont {Governato}\ \emph {et~al.}(2010)\citenamefont {Governato}, \citenamefont {Brook}, \citenamefont {Mayer}, \citenamefont {Brooks}, \citenamefont {Rhee}, \citenamefont {Wadsley}, \citenamefont {Jonsson}, \citenamefont {Willman}, \citenamefont {Stinson}, \citenamefont {Quinn},\ and\ \citenamefont {Madau}}]{Governato:2010-Bulgeless-dwarf-galaxies-and-dark-matter-cores}%
  \BibitemOpen
  \bibfield  {author} {\bibinfo {author} {\bibfnamefont {F.}~\bibnamefont {Governato}}, \bibinfo {author} {\bibfnamefont {C.}~\bibnamefont {Brook}}, \bibinfo {author} {\bibfnamefont {L.}~\bibnamefont {Mayer}}, \bibinfo {author} {\bibfnamefont {A.}~\bibnamefont {Brooks}}, \bibinfo {author} {\bibfnamefont {G.}~\bibnamefont {Rhee}}, \bibinfo {author} {\bibfnamefont {J.}~\bibnamefont {Wadsley}}, \bibinfo {author} {\bibfnamefont {P.}~\bibnamefont {Jonsson}}, \bibinfo {author} {\bibfnamefont {B.}~\bibnamefont {Willman}}, \bibinfo {author} {\bibfnamefont {G.}~\bibnamefont {Stinson}}, \bibinfo {author} {\bibfnamefont {T.}~\bibnamefont {Quinn}}, \ and\ \bibinfo {author} {\bibfnamefont {P.}~\bibnamefont {Madau}},\ }\href {\doibase 10.1038/nature08640} {\bibfield  {journal} {\bibinfo  {journal} {Nature}\ }\textbf {\bibinfo {volume} {463}},\ \bibinfo {pages} {203} (\bibinfo {year} {2010})}\BibitemShut {NoStop}%
\bibitem [{\citenamefont {{McKeown}}\ \emph {et~al.}(2022)\citenamefont {{McKeown}}, \citenamefont {{Bullock}}, \citenamefont {{Mercado}}, \citenamefont {{Hafen}}, \citenamefont {{Boylan-Kolchin}}, \citenamefont {{Wetzel}}, \citenamefont {{Necib}}, \citenamefont {{Hopkins}},\ and\ \citenamefont {{Yu}}}]{McKeown:2022-Amplified-J-factors-in-the-Galactic-Centre}%
  \BibitemOpen
  \bibfield  {author} {\bibinfo {author} {\bibfnamefont {D.}~\bibnamefont {{McKeown}}}, \bibinfo {author} {\bibfnamefont {J.~S.}\ \bibnamefont {{Bullock}}}, \bibinfo {author} {\bibfnamefont {F.~J.}\ \bibnamefont {{Mercado}}}, \bibinfo {author} {\bibfnamefont {Z.}~\bibnamefont {{Hafen}}}, \bibinfo {author} {\bibfnamefont {M.}~\bibnamefont {{Boylan-Kolchin}}}, \bibinfo {author} {\bibfnamefont {A.}~\bibnamefont {{Wetzel}}}, \bibinfo {author} {\bibfnamefont {L.}~\bibnamefont {{Necib}}}, \bibinfo {author} {\bibfnamefont {P.~F.}\ \bibnamefont {{Hopkins}}}, \ and\ \bibinfo {author} {\bibfnamefont {S.}~\bibnamefont {{Yu}}},\ }\href {\doibase 10.1093/mnras/stac966} {\bibfield  {journal} {\bibinfo  {journal} {\mnras}\ }\textbf {\bibinfo {volume} {513}},\ \bibinfo {pages} {55} (\bibinfo {year} {2022})},\ \Eprint {http://arxiv.org/abs/2111.03076} {arXiv:2111.03076 [astro-ph.GA]} \BibitemShut {NoStop}%
\bibitem [{\citenamefont {{Wang}}\ \emph {et~al.}(2020)\citenamefont {{Wang}}, \citenamefont {{Bose}}, \citenamefont {{Frenk}}, \citenamefont {{Gao}}, \citenamefont {{Jenkins}}, \citenamefont {{Springel}},\ and\ \citenamefont {{White}}}]{Wang:2020-Universal-structure-of-dark-matter-haloes-over-a-mass-range}%
  \BibitemOpen
  \bibfield  {author} {\bibinfo {author} {\bibfnamefont {J.}~\bibnamefont {{Wang}}}, \bibinfo {author} {\bibfnamefont {S.}~\bibnamefont {{Bose}}}, \bibinfo {author} {\bibfnamefont {C.~S.}\ \bibnamefont {{Frenk}}}, \bibinfo {author} {\bibfnamefont {L.}~\bibnamefont {{Gao}}}, \bibinfo {author} {\bibfnamefont {A.}~\bibnamefont {{Jenkins}}}, \bibinfo {author} {\bibfnamefont {V.}~\bibnamefont {{Springel}}}, \ and\ \bibinfo {author} {\bibfnamefont {S.~D.~M.}\ \bibnamefont {{White}}},\ }\href {\doibase 10.1038/s41586-020-2642-9} {\bibfield  {journal} {\bibinfo  {journal} {\nat}\ }\textbf {\bibinfo {volume} {585}},\ \bibinfo {pages} {39} (\bibinfo {year} {2020})},\ \Eprint {http://arxiv.org/abs/1911.09720} {arXiv:1911.09720 [astro-ph.CO]} \BibitemShut {NoStop}%
\bibitem [{\citenamefont {Mo}\ \emph {et~al.}(2010)\citenamefont {Mo}, \citenamefont {van~den Bosch},\ and\ \citenamefont {White}}]{Mo:2010-Galaxy-formation-and-evolution}%
  \BibitemOpen
  \bibfield  {author} {\bibinfo {author} {\bibfnamefont {H.}~\bibnamefont {Mo}}, \bibinfo {author} {\bibfnamefont {F.}~\bibnamefont {van~den Bosch}}, \ and\ \bibinfo {author} {\bibfnamefont {S.}~\bibnamefont {White}},\ }\href@noop {} {\emph {\bibinfo {title} {Galaxy formation and evolution}}}\ (\bibinfo  {publisher} {Cambridge University Press},\ \bibinfo {address} {Cambridge},\ \bibinfo {year} {2010})\BibitemShut {NoStop}%
\bibitem [{\citenamefont {{Adams}}\ \emph {et~al.}(2014)\citenamefont {{Adams}}, \citenamefont {{Simon}}, \citenamefont {{Fabricius}}, \citenamefont {{van den Bosch}}, \citenamefont {{Barentine}}, \citenamefont {{Bender}}, \citenamefont {{Gebhardt}}, \citenamefont {{Hill}}, \citenamefont {{Murphy}}, \citenamefont {{Swaters}}, \citenamefont {{Thomas}},\ and\ \citenamefont {{van de Ven}}}]{Adams:2014-Dwarf-Galaxy-Dark-Matter-Density}%
  \BibitemOpen
  \bibfield  {author} {\bibinfo {author} {\bibfnamefont {J.~J.}\ \bibnamefont {{Adams}}}, \bibinfo {author} {\bibfnamefont {J.~D.}\ \bibnamefont {{Simon}}}, \bibinfo {author} {\bibfnamefont {M.~H.}\ \bibnamefont {{Fabricius}}}, \bibinfo {author} {\bibfnamefont {R.~C.~E.}\ \bibnamefont {{van den Bosch}}}, \bibinfo {author} {\bibfnamefont {J.~C.}\ \bibnamefont {{Barentine}}}, \bibinfo {author} {\bibfnamefont {R.}~\bibnamefont {{Bender}}}, \bibinfo {author} {\bibfnamefont {K.}~\bibnamefont {{Gebhardt}}}, \bibinfo {author} {\bibfnamefont {G.~J.}\ \bibnamefont {{Hill}}}, \bibinfo {author} {\bibfnamefont {J.~D.}\ \bibnamefont {{Murphy}}}, \bibinfo {author} {\bibfnamefont {R.~A.}\ \bibnamefont {{Swaters}}}, \bibinfo {author} {\bibfnamefont {J.}~\bibnamefont {{Thomas}}}, \ and\ \bibinfo {author} {\bibfnamefont {G.}~\bibnamefont {{van de Ven}}},\ }\href {\doibase 10.1088/0004-637X/789/1/63} {\bibfield  {journal} {\bibinfo  {journal} {\apj}\ }\textbf {\bibinfo {volume} {789}},\ \bibinfo {eid} {63} (\bibinfo {year}
  {2014})},\ \Eprint {http://arxiv.org/abs/1405.4854} {arXiv:1405.4854 [astro-ph.GA]} \BibitemShut {NoStop}%
\bibitem [{\citenamefont {{Lelli}}\ \emph {et~al.}(2016)\citenamefont {{Lelli}}, \citenamefont {{McGaugh}},\ and\ \citenamefont {{Schombert}}}]{Lelli:2016-SPARC-Mass-Models-for-175-Disk-Galaxies}%
  \BibitemOpen
  \bibfield  {author} {\bibinfo {author} {\bibfnamefont {F.}~\bibnamefont {{Lelli}}}, \bibinfo {author} {\bibfnamefont {S.~S.}\ \bibnamefont {{McGaugh}}}, \ and\ \bibinfo {author} {\bibfnamefont {J.~M.}\ \bibnamefont {{Schombert}}},\ }\href {\doibase 10.3847/0004-6256/152/6/157} {\bibfield  {journal} {\bibinfo  {journal} {\aj}\ }\textbf {\bibinfo {volume} {152}},\ \bibinfo {eid} {157} (\bibinfo {year} {2016})},\ \Eprint {http://arxiv.org/abs/1606.09251} {arXiv:1606.09251 [astro-ph.GA]} \BibitemShut {NoStop}%
\bibitem [{\citenamefont {{Li}}\ \emph {et~al.}(2020)\citenamefont {{Li}}, \citenamefont {{Lelli}}, \citenamefont {{McGaugh}},\ and\ \citenamefont {{Schombert}}}]{Li:2020-A-Comprehensive-Catalog-of-Dark-Matter-Halo-Models}%
  \BibitemOpen
  \bibfield  {author} {\bibinfo {author} {\bibfnamefont {P.}~\bibnamefont {{Li}}}, \bibinfo {author} {\bibfnamefont {F.}~\bibnamefont {{Lelli}}}, \bibinfo {author} {\bibfnamefont {S.}~\bibnamefont {{McGaugh}}}, \ and\ \bibinfo {author} {\bibfnamefont {J.}~\bibnamefont {{Schombert}}},\ }\href {\doibase 10.3847/1538-4365/ab700e} {\bibfield  {journal} {\bibinfo  {journal} {\apjs}\ }\textbf {\bibinfo {volume} {247}},\ \bibinfo {eid} {31} (\bibinfo {year} {2020})},\ \Eprint {http://arxiv.org/abs/2001.10538} {arXiv:2001.10538 [astro-ph.GA]} \BibitemShut {NoStop}%
\bibitem [{\citenamefont {{Martinsson}}\ \emph {et~al.}(2013)\citenamefont {{Martinsson}}, \citenamefont {{Verheijen}}, \citenamefont {{Westfall}}, \citenamefont {{Bershady}}, \citenamefont {{Andersen}},\ and\ \citenamefont {{Swaters}}}]{Martinsson:2014-The-DiskMass-Survey}%
  \BibitemOpen
  \bibfield  {author} {\bibinfo {author} {\bibfnamefont {T.~P.~K.}\ \bibnamefont {{Martinsson}}}, \bibinfo {author} {\bibfnamefont {M.~A.~W.}\ \bibnamefont {{Verheijen}}}, \bibinfo {author} {\bibfnamefont {K.~B.}\ \bibnamefont {{Westfall}}}, \bibinfo {author} {\bibfnamefont {M.~A.}\ \bibnamefont {{Bershady}}}, \bibinfo {author} {\bibfnamefont {D.~R.}\ \bibnamefont {{Andersen}}}, \ and\ \bibinfo {author} {\bibfnamefont {R.~A.}\ \bibnamefont {{Swaters}}},\ }\href {\doibase 10.1051/0004-6361/201321390} {\bibfield  {journal} {\bibinfo  {journal} {\aap}\ }\textbf {\bibinfo {volume} {557}},\ \bibinfo {eid} {A131} (\bibinfo {year} {2013})},\ \Eprint {http://arxiv.org/abs/1308.0336} {arXiv:1308.0336 [astro-ph.CO]} \BibitemShut {NoStop}%
\bibitem [{\citenamefont {Sofue}(2016)}]{Sofue:2016-Rotation-curve-decomposition-f}%
  \BibitemOpen
  \bibfield  {author} {\bibinfo {author} {\bibfnamefont {Y.}~\bibnamefont {Sofue}},\ }\href {\doibase 10.1093/pasj/psv103} {\bibfield  {journal} {\bibinfo  {journal} {Publications of the Astronomical Society of Japan}\ }\textbf {\bibinfo {volume} {68}} (\bibinfo {year} {2016}),\ 10.1093/pasj/psv103}\BibitemShut {NoStop}%
\bibitem [{\citenamefont {Spergel}\ and\ \citenamefont {Steinhardt}(2000)}]{Spergel:2000-Observational-Evidence-for-Self-Interacting-Cold-Dark-Matter}%
  \BibitemOpen
  \bibfield  {author} {\bibinfo {author} {\bibfnamefont {D.~N.}\ \bibnamefont {Spergel}}\ and\ \bibinfo {author} {\bibfnamefont {P.~J.}\ \bibnamefont {Steinhardt}},\ }\href {\doibase 10.1103/PhysRevLett.84.3760} {\bibfield  {journal} {\bibinfo  {journal} {Phys. Rev. Lett.}\ }\textbf {\bibinfo {volume} {84}},\ \bibinfo {pages} {3760} (\bibinfo {year} {2000})}\BibitemShut {NoStop}%
\bibitem [{\citenamefont {Rocha}\ \emph {et~al.}(2013)\citenamefont {Rocha}, \citenamefont {Peter}, \citenamefont {Bullock}, \citenamefont {Kaplinghat}, \citenamefont {Garrison-Kimmel}, \citenamefont {Oñorbe},\ and\ \citenamefont {Moustakas}}]{Rocha:2013-Cosmological-simulations-with-self-interacting-dark-matter}%
  \BibitemOpen
  \bibfield  {author} {\bibinfo {author} {\bibfnamefont {M.}~\bibnamefont {Rocha}}, \bibinfo {author} {\bibfnamefont {A.~H.~G.}\ \bibnamefont {Peter}}, \bibinfo {author} {\bibfnamefont {J.~S.}\ \bibnamefont {Bullock}}, \bibinfo {author} {\bibfnamefont {M.}~\bibnamefont {Kaplinghat}}, \bibinfo {author} {\bibfnamefont {S.}~\bibnamefont {Garrison-Kimmel}}, \bibinfo {author} {\bibfnamefont {J.}~\bibnamefont {Oñorbe}}, \ and\ \bibinfo {author} {\bibfnamefont {L.~A.}\ \bibnamefont {Moustakas}},\ }\href {\doibase 10.1093/mnras/sts514} {\bibfield  {journal} {\bibinfo  {journal} {Monthly Notices of the Royal Astronomical Society}\ }\textbf {\bibinfo {volume} {430}},\ \bibinfo {pages} {81} (\bibinfo {year} {2013})},\ \Eprint {http://arxiv.org/abs/https://academic.oup.com/mnras/article-pdf/430/1/81/3064615/sts514.pdf} {https://academic.oup.com/mnras/article-pdf/430/1/81/3064615/sts514.pdf} \BibitemShut {NoStop}%
\end{thebibliography}%
\appendix
\addtocontents{toc}{\protect\setcounter{tocdepth}{-1}}

\section{Spherical collapse in radiation and matter eras}
\label{sec:2-2-1}
The properties of dark matter particles can be highly relevant to nonlinear structure formation on small scales with an overdensity $\delta\gg1$, which is beyond the linear theory. The structure evolution in the nonlinear regime is very complicated, and, in general, numerical approaches such as N-body simulations are required. Very few analytical tools are available. The spherical collapse model (SCM) is a useful approximate \citep{Gunn:1972-Infall-of-Matter-into-Clusters}, where gravity is the only dominant force driving the nonlinear structure evolution. Therefore, it would be very instructive to follow the spherical collapse model and extend it to the radiation era to identify the necessary conditions for nonlinear structure formation and evolution on small scales.

For the spherical collapse model, we consider a spherical overdensity with a fixed mass $M$ and a radius $R(t)$ that initially reaches a maximum size $R_{\ rm ta}$ at a turnaround moment and then collapses due to its own gravity. We denote the scale factor $a_{ta}$ at turn-around radius $R_{ta}$ and introduce dimensionless variables:
\begin{equation} 
\label{2-2-1} 
\begin{split}
\tau\equiv H_{ta}t, \quad x\equiv \frac{a}{a_{ta}}, \quad y\equiv\frac{R}{R_{ta}},
\end{split}
\end{equation}
where $H_{ta}$ is the Hubble parameter at turn-around. The equation of motion for non-relativistic spherical overdensity reads \citep{Gunn:1972-Infall-of-Matter-into-Clusters}
\begin{equation} 
\label{2-2-2} 
\begin{split}
&\frac{d^2R}{dt^2}=-\frac{GM}{R^2}=-\frac{4\pi}{3}\rho_{ta}R_{ta}^3\frac{G}{R^2}, \\
&\rho_{ta} = \xi_{ta}\bar\rho_{ta}=\xi_{ta}\frac{3H_{ta}^2}{8\pi G},
\end{split}
\end{equation} 
where $\xi_{ta}$ is the ratio between sphere density at turnaround ($\rho_{ta}$) to the background density $\bar\rho_{ta}$ at that moment. 

Integrating Eq. \eqref{2-2-2} leads to a negative constant $E$ for the specific energy of the gravitationally bounded sphere
\begin{equation} 
\label{2-2-3-2-2} 
\begin{split}
E=\frac{1}{2}\left(\frac{dR}{dt}\right)^2-\frac{GM}{R}<0.
\end{split}
\end{equation} 
Using the dimensionless coordinates in Eq. \eqref{2-2-1}, the equation of motion can be rewritten as (along with the boundary conditions)
\begin{equation} 
\label{2-2-3} 
\begin{split}
&\frac{d^2y}{d\tau^2}=-\frac{\xi_{ta}}{2y^2},\\
&y|_{x=0}=0 \quad \textrm{and} \quad \left. \frac{dy}{d\tau}\right|_{x=1}=0,
\end{split}
\end{equation} 
where the spherical overdensity has an initial size $R(t=0)=0$ and a vanishing velocity at its maximum size. 

The relation between $x$ and $\tau$ or (scale factor $a$ and physical time $t$) depends on the cosmology, i.e. the content of matter $\Omega_m$, radiation $\Omega_{rad}$, dark energy $\Omega_\Lambda$, and the evolution of Hubble parameter $H$ and background density $\bar\rho$, such that
\begin{equation} 
\label{2-2-4} 
\begin{split}
H=H_0\sqrt{\Omega_m a^{-3}+\Omega_{rad} a^{-4}+\Omega_{\Lambda}} \quad \textrm{and} \quad \bar\rho = \frac{3H^2}{8\pi G}.
\end{split}
\end{equation}
In $\Lambda$CDM cosmology, $x$ and $\tau$ are related as
\begin{equation} 
\label{2-2-3-2} 
\begin{split}
\frac{dx}{d\tau}=x\sqrt{\frac{\Omega_m a_{ta}^{-3}x^{-3}+\Omega_{rad} a_{ta}^{-4}x^{-4}+\Omega_{\Lambda}}{\Omega_m a_{ta}^{-3}+\Omega_{rad} a_{ta}^{-4}+\Omega_{\Lambda}}}.
\end{split}
\end{equation} 

The analytical solution of Eq. \eqref{2-2-3} can be obtained as
\begin{equation} 
\label{2-2-5} 
\begin{split}
\tau&=\frac{1}{\sqrt{\xi_{ta}}}acot\left(\sqrt{\frac{1}{y}-1}\right)-\frac{y}{\sqrt{\xi_{ta}}}\sqrt{\frac{1}{y}-1}\\
&=\int^x_0\frac{1}{x}\sqrt{\frac{\Omega_m a_{ta}^{-3}+\Omega_{rad} a_{ta}^{-4}+\Omega_{\Lambda}}{\Omega_m a_{ta}^{-3}x^{-3}+\Omega_{rad} a_{ta}^{-4}x^{-4}+\Omega_{\Lambda}}}dx.\\
\end{split}
\end{equation}
At turnaround $y\equiv x\equiv 1$, we should have the dimensionless time
\begin{equation} 
\label{2-2-6} 
\begin{split}
\tau(y=1)=\frac{\pi}{2\sqrt{\xi_{ta}}}\equiv \tau(x=1). 
\end{split}
\end{equation}
From the analytical solution in Eq. \eqref{2-2-5}, the time derivative (or the velocity term) can be obtained as
\begin{equation} 
\label{2-2-5-3} 
\begin{split}
\frac{dy}{d\tau}=\sqrt{\xi_{ta}}\sqrt{\frac{1}{y}-1}.
\end{split}
\end{equation}
Substituting the time derivative into the equation for the specific energy $E$ (Eq. \eqref{2-2-3-2-2}), the turn-around time and turn-around size are
\begin{equation} 
\label{2-2-5-2} 
\begin{split}
t_{ta}=\pi\frac{GM} {(2|E|)^{3/2}} \quad\textrm{and} \quad R_{ta}=\frac{GM} {|E|}, 
\end{split}
\end{equation}
where both the turn-around time $t_{ta}$ and sphere size $R_{ta}$ are proportional to the shell mass $M$, i.e., the larger mass scales collapse at a later time that is consistent with the hierarchical structure formation.

We first consider the matter-dominant universe with $\Omega_{rad}=\Omega_{\Lambda}=0$. From Eqs. \eqref{2-2-5} and \eqref{2-2-6}, the critical density ratio at turn-around 
\begin{equation} 
\label{2-2-8} 
\begin{split}
\tau_{ta}=\tau(x=1)=\frac{2}{3}\quad \textrm{and} \quad \xi_{ta} = \left(\frac{3\pi}{4}\right)^2\approx 5.55.
\end{split}
\end{equation}
Now, we can find the corresponding linear overdensity $\delta=\rho/\bar\rho-1$. The ratio between sphere density $\rho(t)$ and turnaround density $\rho_{ta}$ is (the evolution of background density $\bar\rho\propto a^{-3}$)
\begin{equation} 
\label{2-2-7} 
\begin{split}
\frac{\rho}{\rho_{ta}}=\frac{1}{y^3}, \quad \frac{\bar\rho}{\bar \rho_{ta}}=\frac{1}{x^3}, \quad \frac{\rho/\bar\rho}{\rho_{ta}/\bar\rho_{ta}}=\frac{1+\delta}{\xi_{ta}}=\frac{x^3}{y^3}.
\end{split}
\end{equation}
From this, the exact relation between variables $y$ and $x$ is:
\begin{equation} 
\label{2-2-7-2} 
\begin{split}
y^3 = \left(\frac{3\pi}{4}\right)^2\frac{x^3}{1+\delta}. 
\end{split}
\end{equation}
Taylor expansion of the solution in Eq. \eqref{2-2-5} leads to the linear approximation for small $x$ and $y$:  
\begin{equation} 
\label{2-2-9} 
\begin{split}
\frac{2}{3}x^{3/2}\approx \frac{4}{3\pi}\left(\frac{2}{3}y^{3/2}+\frac{1}{5}y^{5/2}\right).
\end{split}
\end{equation}
Substitute Eq. \eqref{2-2-7-2} into Eq. \eqref{2-2-9}, we obtain the evolution of the linear overdensity $\delta^{lin}\propto a$ and the linear overdensity $\delta_{ta}^{lin}$ at turn-around,
\begin{equation} 
\label{2-2-10} 
\begin{split}
\delta^{lin}=\frac{3}{5}\left(\frac{3}{4}\pi\right)^{2/3}\frac{a}{a_{ta}} \quad \textrm{and} \quad \delta^{lin}_{ta}=\frac{3}{5}\left(\frac{3}{4}\pi\right)^{2/3}\approx 1.06.
\end{split}
\end{equation}
For fully collapsed overdensity that reaches virial equilibrium, the required time should be twice the turn-around time $t_{vir}=2t_{ta}$ ($t\propto a^{3/2}$ in matter-dominant universe). Energy conservation requires the radius of virialized overdensity to be half the turnaround radius $R_{vir}=R_{ta}/2$, which leads to the density ratio
\begin{equation} 
\label{2-2-11} 
\begin{split}
\frac{\rho_{vir}/\bar\rho_{vir}}{\rho_{ta}/\bar\rho_{ta}}=\frac{\xi_{vir}}{\xi_{ta}}=\frac{(2^{2/3})^3}{(1/2)^3}=32.
\end{split}
\end{equation}
The density ratio at virialization $\xi_{vir}$ and the corresponding linear overdensity $\delta_{vir}^{lin}$ are
\begin{equation} 
\label{2-2-12} 
\begin{split}
\xi_{vir}=32\xi_{ta}=18\pi^2 \quad\textrm{and}\quad \delta_{vir}^{lin} = 2^{2/3}\delta^{lin}_{ta}\approx 1.69. 
\end{split}
\end{equation}


\begin{figure}
\includegraphics*[width=\columnwidth]{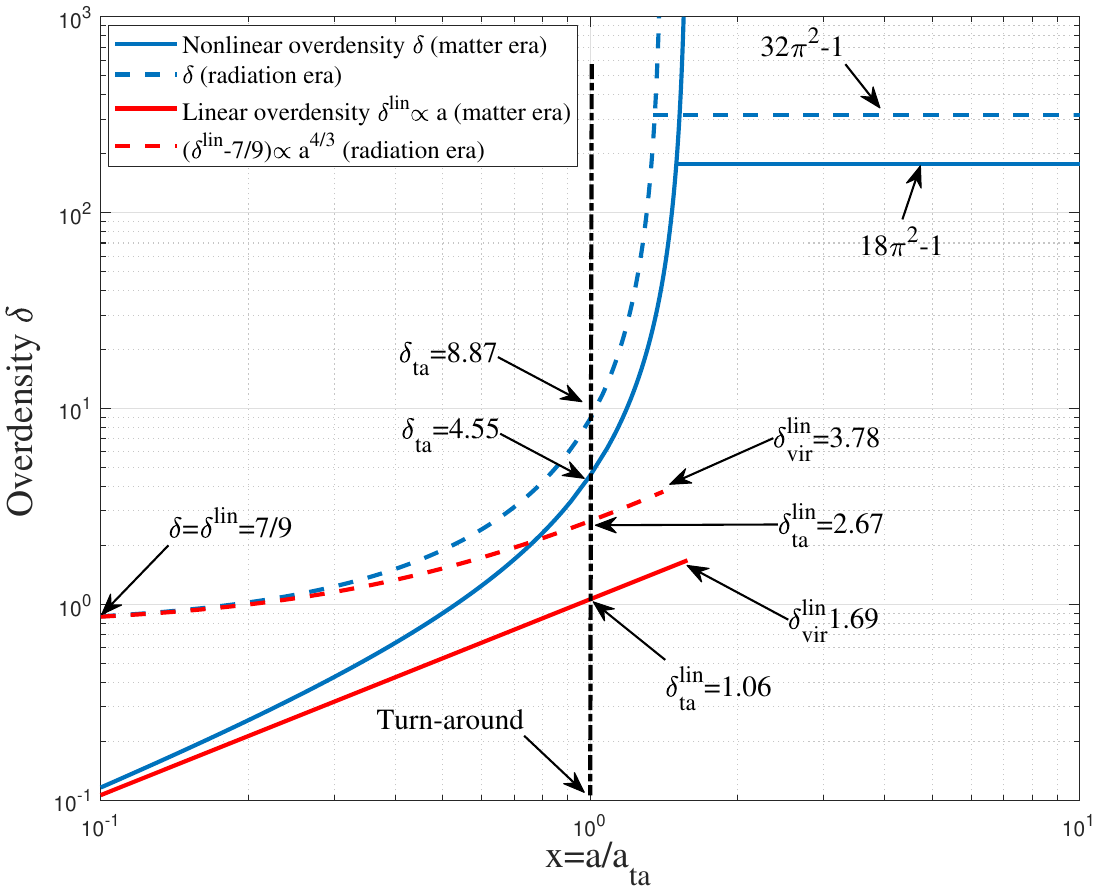}
\caption{The evolution of overdensity (relative to the total background density) with scale factor $x$ from a spherical collapse model for matter era and radiation era, respectively. The solid blue line plots the overdensity evolution in the matter era, which starts from $\delta=0$ and reaches a maximum value of $18\pi^2-1$. The corresponding linear density evolves as $\delta^{lin} \propto a$ (Eq. \eqref{2-2-10}). The dashed blue line plots the overdensity evolution in the radiation era that starts from a nonzero $\delta=7/9$ and reaches a maximum of $32\pi^2-1$. The corresponding linear overdensity evolves as $(\delta^{lin}-7/9)\propto a^{4/3}$ (Eq. \eqref{2-2-17}). In the radiation era, overdensity requires a high initial overdensity of 7/9 and grows into a nonlinear region much later than the same overdensity in the matter era.} 
\label{fig:110}
\end{figure}

Figure \ref{fig:110} summarizes the evolution of a linear overdensity $\delta^{lin}$ (solid red) in the matter era that starts from $\delta=0$, evolves as $\delta\propto a$ in the linear regime until it reaches a value of 1.06 at the turn-around (entering the nonlinear region), and a value of 1.69 when fully virialized. The actual nonlinear overdensity $\delta$ (solid blue) can grow much faster and reach a much higher maximum ($18\pi^2$-1). 

Next, we will extend the same model to a radiation-dominant universe with $\Omega_m=\Omega_{\Lambda}=0$. In the radiation era, the energy density of radiation significantly counteracts gravity, thereby preventing the formation of large-scale structures. Gravitational collapse (Eq. \eqref{2-2-2}) is only possible on very small scales where gravity exceeds expansion. This might occur because of a special phase of universe evolution before the radiation era (early matter dominant era, etc. \citep{Blanco:2019-Annihilation-Signature-of-Hidden-Sector-Dark-Matter}) or because of some special mechanisms leading to gravitational collapse on small scales that are discussed Section \ref{sec:5-2}. At this moment, we assume that the spherical collapse can occur and seek its corresponding solutions in the radiation era.

In the radiation era, the density ratio $\xi_{ta}$ at the turnaround can be obtained from Eqs. \eqref{2-2-3-2} and \eqref{2-2-6} 
\begin{equation} 
\label{2-2-13} 
\begin{split}
\tau_{ta}=\tau(x=1)={1}/{2}\quad \textrm{and} \quad \xi_{ta} = {\pi}^2\approx 9.87.
\end{split}
\end{equation}
The ratio between sphere density $\rho(t)$, turnaround density $\rho_{ta}$, and background density $\bar\rho\propto a^{-4}$ are
\begin{equation} 
\label{2-2-14} 
\begin{split}
\frac{\rho}{\rho_{ta}}=\frac{1}{y^3}, \quad \frac{\bar\rho}{\bar \rho_{ta}}=\frac{1}{x^4}, \quad \frac{\rho/\bar\rho}{\rho_{ta}/\bar\rho_{ta}}=\frac{1+\delta}{\xi_{ta}}=\frac{x^4}{y^3}.
\end{split}
\end{equation}
The exact relation between $y$ and $x$ now becomes
\begin{equation} 
\label{2-2-15} 
\begin{split}
y^3 = \pi^2x^4/(1+\delta). 
\end{split}
\end{equation}
Taylor expansion of Eq. \eqref{2-2-5} leads to the linear approximation:  
\begin{equation} 
\label{2-2-9-2} 
\begin{split}
\frac{1}{2}x^{2}\approx \frac{1}{\pi}\left(\frac{2}{3}y^{3/2}+\frac{1}{5}y^{5/2}\right).
\end{split}
\end{equation}
Using Eq. \eqref{2-2-9-2} and Eq. \eqref{2-2-15}, we obtain the evolution of the linear overdensity and the linear overdensity at the turnaround,
\begin{equation} 
\label{2-2-17} 
\begin{split}
&\delta^{lin}=\frac{7}{9}+\frac{3}{5}\left(\frac{4}{3}\right)^{\frac{4}{3}}\pi^{\frac{2}{3}}\left(\frac{a}{a_{ta}}\right)^{\frac{4}{3}},\\
&\delta^{lin}_{ta}=\frac{7}{9}+\frac{3}{5}\left(\frac{4}{3}\right)^{\frac{4}{3}}\pi^{\frac{2}{3}}\approx 2.67.
\end{split}
\end{equation}
For an over-density that is fully collapsed at time $t_{vir}=2t_{ta}$ ($t\propto a^2$ for radiation era) with $R_{vir}=R_{ta}/2$, the overdensity becomes
\begin{equation} 
\label{2-2-18} 
\begin{split}
&\frac{\rho_{vir}/\bar\rho_{vir}}{\rho_{ta}/\bar\rho_{ta}}=\frac{\xi_{vir}}{\xi_{ta}}=\frac{(2^{1/2})^4}{(1/2)^3}, \\
&\xi_{vir}=32\xi_{ta}=32\pi^2,\\
&\delta^{lin}_{vir}=\frac{7}{9}+\frac{3}{5}\left(\frac{4}{3}\right)^{\frac{4}{3}}\pi^{\frac{2}{3}}2^{2/3}\approx 3.78.
\end{split}
\end{equation}

Note that the density ratio $\xi$ refers to the ratio of the sphere density to the background density (Eq. \eqref{2-2-2}). Figure \ref{fig:110} also presents the evolution of a linear overdensity $\delta^{lin}$ in the radiation era that starts from an initial value $\delta=7/9$, evolves as $\delta-7/9 \propto a^{4/3}$ in the linear regime (Eq. \eqref{2-2-17}), until it reaches a critical value of 2.67 at the turnaround (entering the nonlinear region), and a value of 3.78 for the fully collapsed overdensity. Compared with the evolution in the matter era, we found that:
\begin{enumerate}
\item \noindent Overdensity in the radiation era evolves into a nonlinear region much later than that in the matter era with higher linear overdensity at the turnaround ($\delta^{lin}_{ta}$). Density ratio at turnaround $\xi_{ta}$ and at virial equilibrium $\xi_{vir}$ are also higher than those in the matter era.
\item \noindent On small scales, the linear perturbation theory does not apply. Our solution suggests a linear overdensity $\delta^{lin}\propto 7/9+Const\cdot a^{4/3}$.   
\item \noindent Since the universe expansion is extremely fast in the radiation era and if the gravitational collapse occurs, a special mechanism that does not involve gravity is required to give rise to the initial overdensity $\delta=7/9$ that is comparable to the density of the radiation background. The gravitational collapse can subsequently take over. In the next section, we discuss a mechanism via direct collisions. 
\end{enumerate}

\section{Structure formation by direct collisions}
\label{sec:5-2}
\begin{figure}
\includegraphics*[width=\columnwidth]{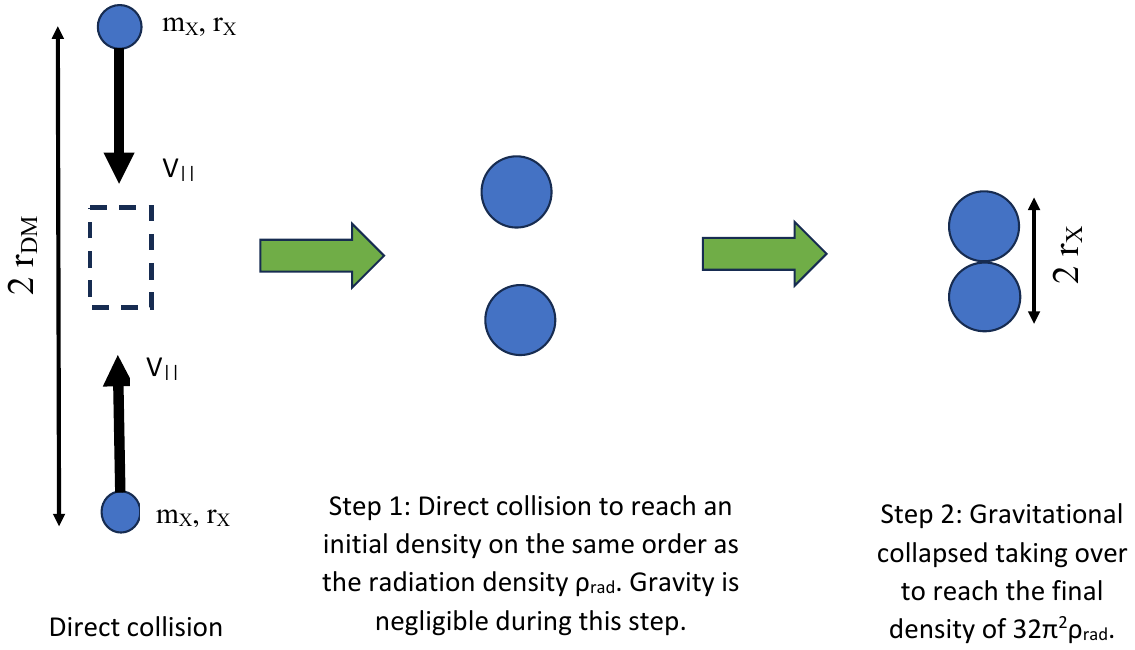}
\caption{Two-step formation of nonlinear halo structures on small scales in the radiation era. Step 1: direct collision of two X particles of mass $m_X$, size $r_X$ and a mean separation 2$r_{DM}$ to reach an initial overdensity on the same order as the radiation density $\rho_{rad}$ (Fig. \ref{fig:110} and Eq. \eqref{2-2-17}). The gravity is relatively weak during this step. Step 2: the gravitational collapse starts to take over to reach the final density of 32$\pi^2\rho_{rad}$ (Fig. \ref{fig:110} and Eq. \eqref{2-2-18}).} 
\label{fig:S2-2}
\end{figure}

In the early universe, the hierarchical formation of structures can be described as a sequential cascade process as $X+X\rightarrow XX $, $X + XX\rightarrow XXX $, $X + XXX\rightarrow XXXX $, and so on. The merging of smaller structures gives rise to larger structures. This hierarchical structure formation may reach a statistically steady state that can be described by a mass and energy cascade concept with scale-independent rates \citep{Xu:2023-Dark-matter-halo-mass-functions-and,Xu:2021-Inverse-mass-cascade-mass-function}. Appendix sections \ref{sec:5-2} and \ref{sec:5-3} provide more details on the structure evolution in the early universe. 

In the radiation era, linear theory predicts a suppressed density perturbation on large scales due to fast expansion, i.e., the Meszaros effect \citep{Meszaros:1974-The-behaviour-of-point-masses-in-an-expanding-cosmological} (see Eq. \eqref{eq:5-2-3}). The Hubble expansion suppresses perturbations on large scales. Therefore, large-scale structures smaller than the horizon ($r_t<r<r_H$ in Fig. \ref{fig:104}) remain essentially frozen as a result of radiation impeding growth. However, on small scales, nonlinear structure formation and evolution are still possible in the radiation era. 
In this section, we discuss the formation of bound structures on small scales $r<r_t$ in the radiation era, while structures on large scales $r>r_t$ are still suppressed (Fig. \ref{fig:104}). 

We begin our discussion with Fig. \ref{fig:S2-2}. Consider a collection of X particles with mass $m_X$, size $r_X$, and a mean separation of 2$r_{DM}$. The first step is direct collisions of dark matter particles to reach an initial matter density comparable to the radiation density $\rho_{rad}$. Gravity is not directly involved during this step, except through the large-scale collision velocity $V_{||}$ (Eq. \eqref{eq:5-2-3}). The second step involves the gravitational collapse of the overdensity to form virialized haloes with a final density of $32\pi^2\rho_{rad}$, as described by the spherical collapse model in Fig. \ref{fig:110}. During the first step of direct collision, gravity is neglected, and we focus on the direct collision of particles and direct collision time. For particles of size $r_X$, the cross-section is $\sigma_d=\pi r_X^2$. With the mean separation $r_{DM}=(3m_X/4\pi\rho_{DM})^{1/3}$, the mean free path $\lambda_{DM}$ and the direct collision time $\tau_{dc}$ are:
\begin{equation}
\begin{split}
&\lambda_{DM}=\frac{m_X}{\rho_{DM}\sigma_d}=\frac{4}{3}\left(\frac{r_{DM}}{r_X}\right)^2r_{DM},\\
&\tau_{dc}=\frac{\lambda_{DM}}{V_{||}}=\frac{4}{3}\left(\frac{r_{DM}}{r_X}\right)^2\frac{r_{DM}}{V_{||}}.
\end{split}
\label{eq:5-2-1}
\end{equation}
Collision time $\tau_{dc}\propto a^3$ rapidly increases with time due to the expansion $r_{DM}\propto a$ for a fixed particle size $r_X$. 

\begin{figure}
\includegraphics*[width=\columnwidth]{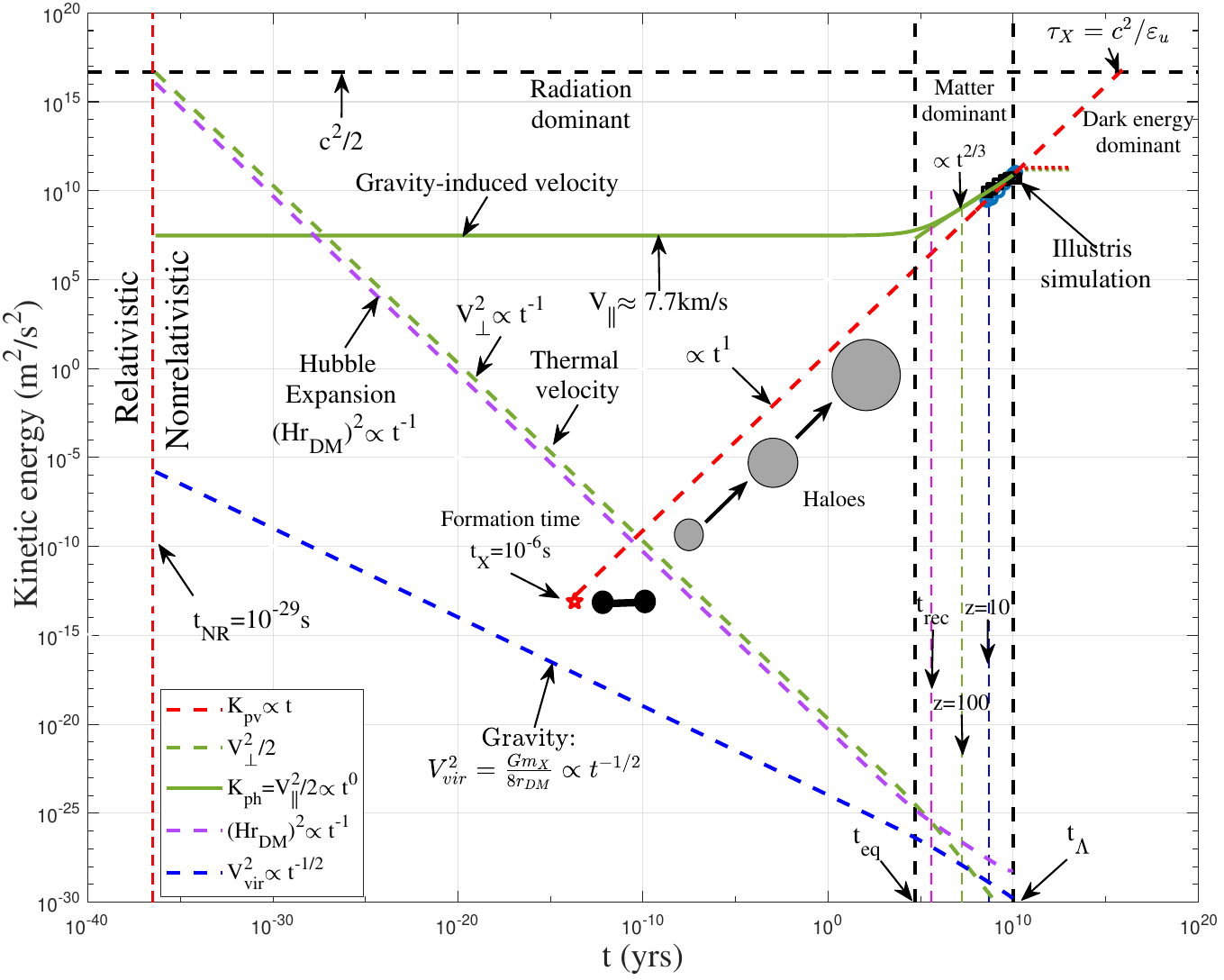}
\caption{The evolution of kinetic energy of different velocities in the radiation era for X particles with a critical mass of $m_{Xc}=10^{12}$GeV. $K_{ph}$ and $K_{pv}$ represent the kinetic energy of haloes on large scales $r>r_t$ and the kinetic energy of particles in haloes on small scales $r<r_t$, respectively. The effect of gravity is weak on large scales in the radiation era such that the virial velocity $V_{vir}$ is much smaller than other velocities (dashed blue). The perpendicular component $V_{\perp}$ (thermal velocity) and $V_H$ (due to expansion) decreases rapidly with time (Eq. \eqref{eq:5-2-6}). The large-scale parallel velocity $K_{ph}=V_{||}^2/2$ is due to gravity on large scales and is independent of particle mass and cosmic time. Direct collision due to velocity $V_{||}$ leads to the formation of the smallest two-particle bound structure at around $10^{-6}$s (red star from Eq. \eqref{eq:5-22-17}) and halo structure evolution in the radiation era (Section \ref{sec:5-3}). The kinetic energy $K_{pv}\propto t$ on small scales is due to virialization in halos and is plotted as a red dashed line. The evolution of kinetic energy in the matter era is plotted as symbols from Illustris simulation \citep{Xu:2022-Postulating-dark-matter-partic}.} 
\label{fig:111}
\end{figure}

For comparison, we also calculate the perpendicular velocity $V_{\perp}$ (thermal velocity), Hubble flow $V_{H}$, and virial velocity $V_{vir}$, 
\begin{equation}
\begin{split}
&V_{\perp} = \frac{a_{NR}}{a}c=\frac{3k_BT_{\gamma0}}{c m_{X}}a^{-1}\propto m_X^{-1}a^{-1}, \\
&V_{H} = H r_{DM} \propto m_X^{1/3}a^{-1}, \\
&V_{vir}^2=\frac{Gm_X}{8r_{DM}} \propto m_X^{2/3}a^{-1}.
\end{split}
\label{eq:5-2-6}
\end{equation}
All of these velocities depend on both particle mass and time except the gravity-induced (parallel) velocity $V_{||}$ (Section \ref{sec:5-1-2-2-2}). Figure \ref{fig:111} plots the variation of different velocities in the radiation era for X particles with a critical mass $m_{Xc}=10^{12}$GeV. The virial velocity $V_{vir}$ due to gravity is small in the radiation era when compared to the Hubble flow from expansion. The parallel peculiar velocity on large scales ($V_{||}$) is constant and independent of the particle mass $m_X$ (solid green line). Direct collisions, driven by the velocity $V_{||}$, led to the formation of the first non-linear structure (red star) and initiated the evolution of small-scale structures during the radiation era. The particle kinetic energy in small-scale structures increases linearly with time $K_{pv}\propto t$ (Eq. \eqref{eq:5-3-4-3}) due to the structure formation and virialization (red dashed line). 

The formation time $t_X$ for the smallest bound structure strongly decreases with particle mass $t_X\propto m_X^{-5}$. Heavier particles (larger $m_X$) should form smaller structures (smaller $r_X$) at an earlier time (smaller $t_X$) (Eq. \eqref{eq:5-22-17}). However, this formation time $t_X$ should also be comparable to the direct collision time scale $\tau_{dc}$ to allow a sufficient amount of time for direct collisions to occur to form these structures. With particle size $r_X$ in Eq. \eqref{eq:5-2-10-2} and velocity $V_{||}$ in Eq. \eqref{eq:5-2-5}, we are able to calculate the direct collision time $\tau_{dc}$ in Eq. \eqref{eq:5-2-1} at the time of formation $t_X$ for particles of mass $m_X$,
\begin{equation}
\begin{split}
&\tau_{dc}(m_X,t_X)=\frac{4}{3}\left(\frac{r_{DM}}{r_X}\right)^2\frac{r_{DM}}{V_{||}}.
\end{split}
\label{eq:5-2-12}
\end{equation}
With all relevant quantities, 
\begin{equation}
\begin{split}
&r_{DM}(m_X,t_X)=\left(\frac{3m_X}{4\pi\rho_{DM}}\right)^{1/3}, \quad r_X(m_X)=\frac{4\hbar^2}{G(m_{X})^3}, \\
&\rho_{DM}(m_X,t_X)=\frac{3H_0^2}{8\pi G}\Omega_{DM}a_X^{-3}, \quad t_X = \frac{a_X^2}{2H_0\sqrt{\Omega_{rad}}}.
\end{split}
\label{eq:5-2-13}
\end{equation}

The direct collision time scale $\tau_{dc}$ at time $t_X$ can be computed as a function of particle mass $m_X$,
\begin{equation}
\begin{split}
&\tau_{dc}(m_X)=\frac{2^8\pi^{3/2}}{3}\left(\frac{\Omega_{rad}^{3/2}}{H_0\Omega_{DM}^2}\right)^{1/2}\left(\frac{\hbar}{m_XV_{||}^2}\right)^{1/2}, 
\end{split}
\label{eq:5-2-14}
\end{equation}
With the formation time comparable to the direct collision time or $t_X(m_X)=\tau_{dc}(m_X)$ (Eqs. \eqref{eq:5-2-11} and \eqref{eq:5-2-14}), the only possible particle mass $m_X$  should read
\begin{equation}
\begin{split}
&m_{X}=\left(\frac{9}{64\pi}\frac{H_0\Omega_{DM}^2}{\Omega_{rad}^{3/2}}\frac{V_{||}^2\hbar^5}{G^4}\right)^{1/9}\approx m_{Xc}=10^{12} \textrm{GeV}. 
\end{split}
\label{eq:5-2-15}
\end{equation}

Therefore, the direct collision time $\tau_{dc}$ is comparable to the formation time $t_X$ only for particles with the critical mass $m_{Xc}$. This allows the formation of the smallest two-particle bound structure, which is fully consistent with the particle mass obtained from the X miracle (Eq. \eqref{eq:5-22-15}). Particles heavier than $m_{Xc}$ will have $\tau_{dc}\gg t_X$ such that there is no sufficient time to form these structures at time $t_X$. Particles lighter than $m_{Xc}$ have $\tau_{dc}\ll t_X$. However, the thermal velocity will prevent the formation of these two-particle structures due to the free streaming mass greater than the particle mass (Fig. \ref{fig:109}). Finally, only particles with a mass $m_{Xc}=10^{12}$ GeV can form the smallest bound structure as described here. This is also the earliest possible structure that particles of any mass can form. 

\section{Revisiting the WIMP miracle}
\label{sec:5-1-2-2}
The WIMP miracle, assuming gravity is negligible at freeze-out due to fast streaming and a weak-scale particle interaction, predicts weak-scale particles of 100GeV. In this Section, we first review the WIMP miracle, followed by a discussion on how to adapt the WIMP miracle for nonthermal dark matter.

In the early Universe, WIMP particles were in thermal equilibrium and interacting with standard model (SM) particles \citep{Feng:2010-Dark-Matter-Candidates-from-Particle-Physics}, 
\begin{equation}
\begin{split}
&W+W \rightleftarrows SM+SM,
\end{split}
\label{eq:5-22-112}
\end{equation}
where WIMP particles ($W$) annihilate into and are produced by SM particles. This process can be described by the Boltzmann equation that governs the evolution of particle number density $n$:
\begin{equation}
\begin{split}
&\frac{dn}{dt}=-3Hn-\langle\sigma_W v\rangle \left(n^2-n_{eq}^2\right),
\end{split}
\label{eq:5-33-1}
\end{equation}
where $\langle\sigma_W v\rangle$ represents the thermally averaged cross section of WIMPs. The first term accounts for the dilution from expansion. The $n^2$ term arises from the annihilation, while the equilibrium term $n_{eq}^2$ is from the reverse process to produce WIMPs. 

The equilibrium number density $n_{eq}$ can be obtained by integrating the Boltzmann distribution over all possible momenta $p$. Since the reaction involves creating WIMP particles, the particle energy $E$ and equilibrium number density $n_{eq}$ can be written as
\begin{equation}
\begin{split}
& E = m_Wc^2 + \frac{p^2}{2m_W}, \\
& n_{eq} \propto \int \frac{d^3p}{(2\pi)^3}\exp\left(-\frac{E}{T}\right)= g \left(\frac{m_Wk_BT}{2\pi\hbar^2}\right)^\frac{3}{2}\exp\left(-\frac{m_Wc^2}{k_BT}\right),
\end{split}
\label{eq:5-33-4-5}
\end{equation}
where $g\approx 2$ is a numerical factor. The equilibrium density is exponentially suppressed $n_{eq}\propto e^{-m/T}$ by particle mass and temperature. Heavier particles are extremely rare at a low temperature. 

Due to the expansion of the Universe, at some point, the dark matter particles become too dilute to find each other and annihilate. At that moment, particles freeze out with their comoving relic density approaching a constant. We define the freeze-out at the moment when the interaction rate is comparable to the expansion rate $H$,
\begin{equation}
\begin{split}
&n_f\langle\sigma_W v\rangle \sim H_f, \quad n_f = \frac{\bar\rho_{0}\Omega_{DM}}{m_W a_f^3},
\end{split}
\label{eq:5-33-2}
\end{equation}
where the subscript $f$ denotes the quantity at freeze-out. Here, $m_W$ is the mass of WIMPs, and $a_f$ is the scale factor. Equation \eqref{eq:5-33-2} requires the cross section to be (Hubble parameter $H_f\propto a_f^{-2}\propto T_{\gamma f}^2$)
\begin{equation}
\begin{split}
&\langle\sigma_W v\rangle \sim \frac{H_f}{n_f}\sim \frac{m_W a_f^3 H_f}{\bar\rho_{0}\Omega_{DM}}\sim \frac{m_W}{\bar\rho_{0}\Omega_{DM}} \frac{T_{\gamma 0}}{T_{\gamma f}}\frac{(k_BT_{\gamma0})^2}{\hbar c^2 M_{pl}},
\end{split}
\label{eq:5-33-3}
\end{equation}
where $T_{\gamma 0}$ and $T_{\gamma f}$ are the radiation temperature at present and freeze-out, respectively. Rearranging Eq. \eqref{eq:5-33-3} leads to
\begin{equation}
\begin{split}
&\langle\sigma_W v\rangle \sim \underbrace{\frac{k_B T_{\gamma0}}{c^2M_{pl}}}_{1} \cdot \underbrace{\frac{x_fk_B T_{\gamma0}}{c^2\bar\rho_{0}\Omega_{DM}}\cdot\frac{k_B T_{\gamma 0}}{\hbar}}_{2},\quad x_f = \frac{m_Wc^2}{k_BT_{\gamma f}},
\end{split}
\label{eq:5-33-4-4}
\end{equation}
where $x_f$ is a dimensionless constant on the order of 20 for thermal relics of different masses \citep{Feng:2010-Dark-Matter-Candidates-from-Particle-Physics}. However, $x_f$ could be very different and mass-dependent for nonthermal relics (Eq. \eqref{eq:5-22-13-2}).

With CMB temperature $T_{\gamma0}$=2.7K and Planck mass $M_{pl}=2.2\times 10^{8}$kg, the first term on the right-hand side of Eq. \eqref{eq:5-33-4-4} is dimensionless and on the order of $10^{-32}$ that sets the magnitude of $\langle\sigma_W v\rangle$. With critical density $\bar\rho_0=10^{-26}$kg/m$^3$ and the relic DM density $\Omega_{DM}$=0.24, the second term is of order unity and sets the unit of $m^3/s$. The cross section $\langle\sigma_W v\rangle$ finally reads
\begin{equation}
\begin{split}
&\langle\sigma_W v\rangle \approx 3\times 10^{-32}\frac{m^3}{s},
\end{split}
\label{eq:5-33-5-2}
\end{equation}
which is independent of the particle mass $m_W$ because $x_f$ is mass-independent. Therefore, the dark matter relic density requires a weak-scale cross section. 

To determine the particle mass, we need to assume the nature of particle interactions. For weak interaction, particle mass $m_W$ determines the S-wave annihilation cross section 
\begin{equation}
\begin{split}
&\langle\sigma_W v\rangle \propto \left(\frac{\hbar}{m_Wc}\right)^2\frac{g_w^4}{16\pi^2}c=\left(\frac{\hbar}{m_Wc}\right)^2\alpha_w^2c,
\end{split}
\label{eq:5-33-5-333}
\end{equation}
where $g_w\approx 0.65$ is the weak gauge coupling constant, $\alpha_w$ is the weak coupling constant. Therefore, for cross section in Eq. \eqref{eq:5-33-5-2}, weak-scale particles of 100 GeV are an excellent DM candidate. Note that WIMPs have a fast free streaming close to $c$ at freeze-out, and gravity can be neglected.

However, we should not exclude other possible interactions and particle masses that may also meet the same condition. We will explore the "X miracle" for nonthermal relics, in contrast to the WIMP miracle for thermal relics. To adapt the WIMP miracle to the X miracle, it is essential to understand the distinct velocity evolution of thermal and nonthermal relics.  

\section{Structure formation in radiation and matter eras}
\label{sec:5-3}
\begin{figure}
\includegraphics*[width=\columnwidth]{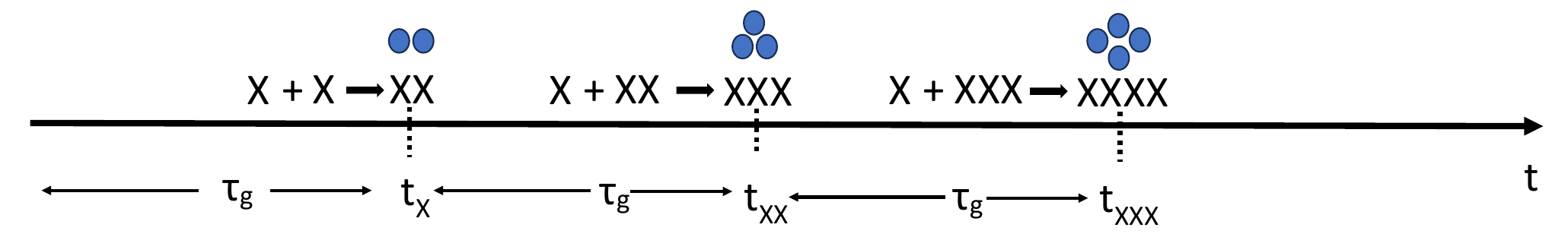}
\caption{Hierarchical structure formation via direct collisions between particles and structures. The smallest bound structure $XX$ was first formed at time $t_X$. The bound structure $XXX$ was formed at time $t_{XX}$ by merging between particle $X$ and structure $XX$, and so on for bigger structures. The average waiting time $\tau_g$ between the formation of structures of different sizes is essentially the direct collision time between substructures (Eq. \eqref{eq:5-3-1}). Through direct collision, the characteristic mass of halo structures continuously increases with time as $m_h^*(t)\propto t^{5/2}$ (Eq. \eqref{eq:5-3-4}).}  
\label{fig:S55}
\end{figure}

This section discusses the structure formation and evolution right after the freeze-out of superheavy X particles. In the radiation era, density perturbations are suppressed on large scales (linear regime) greater than the characteristic halo mass $m_h^*$ (or $r_t$ in Fig. \ref{fig:104}). On small scales (nonlinear regime), halo structures can form by direct collision on large scales, followed by gravitational collapse on small scales (Fig. \ref{fig:S2-2}). After the formation of the smallest two-particle bound structure at $t_X=10^{-6}$s with $m_h^*=2m_{Xc}$ (Eq. \eqref{eq:5-22-16}), the hierarchical structure formation takes over with the characteristic halo mass $m_h^*$ continuously increasing over time. On small scales in the nonlinear regime, halo structures grow by merging with X particles to form larger and larger halos, i.e., $X+X\rightarrow XX$, $X+XX\rightarrow XXX$, $X+XXX\rightarrow XXXX$ and so on (Fig. \ref{fig:S55}). The previous section \ref{sec:5-2} focuses on the first step $X+X\rightarrow XX$. In this section, we extend the discussion to the hierarchical structure formation with increasing structure mass and size (Fig. \ref{fig:S55}).

The hierarchical merging of structures is fundamental and complex. 
Similarly to the formation of a two-particle structure $XX$, structure evolution in the radiation era involves a direct collision between a particle and a halo of characteristic mass ${m_h}^*$. The cross-section for direct collision is $\sigma_d=\pi {r_h^*}^2$, where the halo size ${r_h}^*$ depends on the halo mass $r_h^*\propto {m_h^*}^{1/3}$. We consider the average waiting time $\tau_g^*$ as the time between two successive merges between the halo structure and dark matter particles (Fig. \ref{fig:S55}). This waiting time is just the direct collision time ($\tau_g\equiv \tau_{dc}$),
\begin{equation}
\begin{split}
&\tau_{g}^*=\tau_{dc}=\frac{4}{3}\left(\frac{r_{DM}}{r_h^*}\right)^2\frac{r_{DM}}{V_{||}}\propto \frac{a^3}{{r_h^*}^2},
\end{split}
\label{eq:5-3-1}
\end{equation}
where $r_{DM}\propto a$ is the mean separation. After every merge, the halo mass increases by the particle mass $m_X$. The evolution of $m_h^*$ reads 
\begin{equation}
\begin{split}
&\frac{dm_h^*}{dt}=\frac{m_h^*}{t}=\frac{m_X}{\tau_g^*}.
\end{split}
\label{eq:5-3-2}
\end{equation}
Haloes should have a constant density ratio to the background density (32$\pi^2$ from the spherical collapse model in Fig. \ref{fig:110}),
\begin{equation}
\begin{split}
&\rho_h^*=\frac{m_h^*}{\frac{4}{3}\pi \left(2r_h^*\right)^3}=32\pi^2\rho_{rad} \propto a^{-4}.
\end{split}
\label{eq:5-3-3}
\end{equation}
Solving Eqs. \eqref{eq:5-3-1}, \eqref{eq:5-3-2}, and \eqref{eq:5-3-3} leads to the scaling
\begin{equation}
\begin{split}
&\tau_g^*\propto a^{-3}\propto t^{-3/2}, \quad r_h^*\propto a^3\propto t^{3/2}, \quad m_h^*\propto a^5\propto t^{5/2}.
\end{split}
\label{eq:5-3-4}
\end{equation}
The waiting time $\tau_g^*$ due to the direct collision rapidly decreases due to the fast growth of the halo size $r_h^*$ and the rapid increase in the cross section $\sigma^*=\pi{r_h^*}^2$. Similar to the smallest structure (Eqs. \eqref{eq:5-22-16} and \eqref{eq:5-22-17}), the characteristic particle velocity $v_h^*$ can be obtained from virial theorem, as well as the halo formation time
\begin{equation}
\begin{split}
&{{v_h^*}^2}\propto \frac{Gm_h^*}{r_h^*} \propto a^2\propto t, \quad t_h^*\propto \frac{r_h^*}{v_h^*}\propto t.
\end{split}
\label{eq:5-3-4-3}
\end{equation}

The energy cascade parameter $\varepsilon_u$ in Eq. \eqref{eq:5-22-9} is defined as
\begin{equation}
\begin{split}
&\varepsilon_u \propto \frac{{v_h^*}^2}{t^*} \propto t^0,
\end{split}
\label{eq:5-3-4-2}
\end{equation}
which is a constant of time. The parameter $\varepsilon_u$ reflects the rate of change of halo kinetic energy per unit mass that impacts the structure formation and evolution on large scales \citep{Xu:2023-Universal-scaling-laws-and-density-slope,Xu:2023-Dark-matter-halo-mass-functions-and}, with its origin from the micro-physics of dark matter in Eq. \eqref{eq:5-22-9}. The evolution of particle velocity $v_h^*$ is presented in Fig. \ref{fig:111} (dashed red line).

\begin{figure}
\includegraphics*[width=\columnwidth]{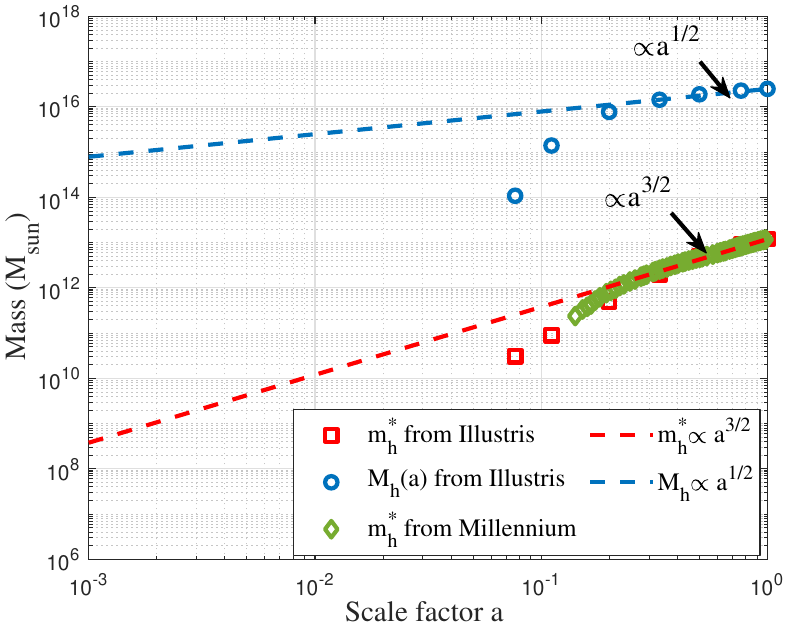}
\caption{The evolution of the characteristic halo mass $m_h^*(t)$ (red dashed) and total halo mass $M_h$ (blue dashed) in the matter era, where $m_h^*\propto a^{3/2}\propto t$ and $M_h\propto a^{1/2}\propto t^{1/3}$. Results from different cosmological N-body simulations (symbols) are also presented that are in agreement with the prediction of Eq. \eqref{eq:5-3-5-3}. The deviation at high redshift can be due to the limited mass resolution in N-body simulations.}  
\label{fig:113}
\end{figure}

For comparison, the evolution of structure in the matter era is dominated by gravity. The waiting time $\tau_g$ is inversely proportional to the halo potential $\tau_g \propto (Gm_h/r_h )^{-1}$ \citep{Xu:2023-Dark-matter-halo-mass-functions-and,Xu:2021-Inverse-mass-cascade-mass-function}. 
Larger haloes accrete mass more quickly with a shorter waiting time, where $\tau_g \propto m_h^{-2/3}$ with $r_h\propto m_h^{1/3}$. Haloes of characteristic mass $m_h^*$ have the shortest waiting time $\tau_g^*$. Therefore, equations in the matter era are
\begin{equation}
\begin{split}
&\frac{dm_h}{dt}=\frac{t_0}{10^{13}M_{\odot}}=\frac{m_X}{\tau_g},\\
&\tau_g \propto (Gm_h/r_h )^{-1}, \\
&\rho_h=\frac{m_h}{\frac{4}{3}\pi {r_h}^3}=18\pi^2\rho_{DM} \propto a^{-3}.
\end{split}
\label{eq:5-3-5}
\end{equation}
Solving these equations leads to the scalings for haloes of any mass:
\begin{equation}
\begin{split}
&\tau_g\propto m_h^{-2/3}a  \quad \textrm{and} \quad r_h\propto m_h^{1/3}a.
\end{split}
\label{eq:5-3-6}
\end{equation}
For haloes of characteristic mass $m_h^*$, the scaling reads
\begin{equation}
\begin{split}
&\tau_g^*\propto a^{0}\propto t^{0}, \quad r_h^*\propto a^{3/2}\propto t, \quad m_h^*\propto a^{3/2}\propto t,\\
&{{v_h^*}^2}\propto \frac{Gm_h^*}{r_h^*} \propto a^0, \quad t_h^*\propto \frac{r_h^*}{v_h^*}\propto t.
\end{split}
\label{eq:5-3-5-3}
\end{equation}
The total mass in all haloes $M_h\propto {m_h^*}^{1/3}\propto t^{1/3}$ \citep{Xu:2023-Dark-matter-halo-mass-functions-and}. 

\begin{table}
    \begin{center}
    \caption{Structure evolution in the radiation era and matter era}
    \label{tab:5}
    \begin{tabular}{lccccc} 
    \hline
    Quantity         & Symbol   & {Radiation era} & {Matter era}     \\
    \hline
    Time             & {t}   & $\propto a^2$   & $\propto a^{3/2}$ \\
    Hubble parameter & {Ht}  & {1/2}           & {2/3}              \\
   Large-scale overdensity    & {$\delta$} & {$a^0$}    & {$a^1$}            \\
   Large-scale KE & {$K_{ph}$}  & {$ a^0$}     &  {$a^1$}   \\
   Small-scale KE & {$K_{pv}$} & {$ t$}   &  $ t$       \\
   Total mass in all haloes & {$M_h$}     &  {$t^{5/6}$}   &  {$t^{1/3}$}  \\
   Characteristic halo mass & {$m_h^*$}   &  {$t^{5/2}$}   &  {$t$}        \\
   Characteristic halo size & {$r_h^*$}   &  {$t^{3/2}$}   &  {$t$}        \\
   Waiting time         & {$\tau_g^*$}  &  {$t^{-3/2}$} & {$t^0$} \\
   Characteristic halo velocity  & {$v_h^*$}  &  {$t^{1/2}$} & {$t^0$} \\
    \hline
    \end{tabular}
  \end{center}
\end{table}

Table \ref{tab:5} summarizes the evolution of mass and energy in both the radiation and matter eras. Figure \ref{fig:113} presents the evolution of characteristic halo mass $m_h^*$ and total mass in all haloes $M_h$ in the matter era. Different cosmological N-body simulations confirm the linear growth of the halo mass predicted by Eq. \eqref{eq:5-3-5-3}. The characteristic halo mass $m_h^*\propto t$ and the total halo mass $M_h^*\propto t^{1/3}$ in the matter era. With $m_h^*(z=0)\approx 10^{13}M_{\odot}$ and $M_h(z=0)\approx 10^{16}M_{\odot}$ from Illustris simulation, the evolution of $m_h^*$ and $M_h$ can be fully determined for the radiation era (left equations) and matter era (right equations) as
\begin{equation}
\begin{split}
& m_h^* = m_h^*(t_{eq}) \left(\frac{t}{t_{eq}}\right)^{5/2}  \quad \textrm{and} \quad m_h^* = m_h^*(t_{eq})\left(\frac{t}{t_{eq}}\right), \\
& M_h = M_h(t_{eq}) \left(\frac{t}{t_{eq}}\right)^{5/6}  \quad \textrm{and} \quad M_h = M_h(t_{eq})\left(\frac{t}{t_{eq}}\right)^{1/3}, \\
\end{split}
\label{eq:5-3-7}
\end{equation}
where the characteristic halo mass and total halo mass at matter-radiation equality are $m_h^*(t_{eq})=5\times10^{7}M_{\odot}$ and $M_h(t_{eq})=4.3\times10^{14}M_{\odot}$, respectively, where $t_{eq}=50000$yrs. 

Using Eq. \eqref{eq:5-3-2} and \eqref{eq:5-3-5-3}, the waiting time $\tau_g^*$ is:
\begin{equation}
\begin{split}
& \tau_g^* = \tau_g^*(t_{eq}) \left({t}/{t_{eq}}\right)^{-3/2} , \quad \textrm{(radiation era),} \\
& \tau_g^* = \tau_g^*(t_{eq}) =\frac{m_Xt_{eq}}{m_h^*(t_{eq})}, \quad \textrm{(matter era).} 
\end{split}
\label{eq:5-3-8}
\end{equation}
The average waiting time decreases rapidly in the radiation era ($\tau_g\propto t^{-3/2}$) and reaches a constant value in the matter era due to characteristic mass $m_h^*\propto t$ (Eq. \eqref{eq:5-3-5-3}). That constant value is dependent on the particle mass $m_X$,
\begin{equation}
\begin{split}
&\tau_g^*(t>t_{eq})=\frac{m_X}{m_h^*(t)}t=\frac{m_X}{m_h^*(t=t_0)}t_0. \\
\end{split}
\label{eq:5-3-10-1}
\end{equation}
With present characteristic halo mass of $10^{13}M_{\odot}$ and $t_0$ of the age of the universe, for particles of critical mass $m_X=m_{Xc}=10^{12}$GeV, the waiting time is (from Eq. \eqref{eq:5-3-10-1})
\begin{equation}
\begin{split}
\tau_g^*\approx 10^{-41}s\approx 18\pi^2t_p,
\end{split}
\label{eq:5-3-9}
\end{equation}
where $t_p=5.4\times 10^{-44}$s is the Planck time. Here, $\xi=18\pi^2$ is the critical density ratio in the matter era. For fixed mass density of dark matter haloes, a higher density ratio $\xi$ means a lower background matter density and, of course, a longer waiting time for X particles to merge with haloes. Therefore, $\tau_g^*$ is expected to be proportional to the density ratio $\xi$ \citep{Xu:2021-Inverse-mass-cascade-mass-function}. Since the Planck time is the smallest unit of time for any physical processes, particles with a critical mass $m_{Xc}$ have the shortest possible waiting time $18\pi^2t_p$ \citep{Xu:2021-Inverse-mass-cascade-mass-function}. The structure formation in matter era requires a critical particle mass $m_{Xc}=10^{12}$ GeV. 

Since Eq. \eqref{eq:5-3-10-1} is independent of the dark matter particle model, we write the waiting time for any particle mass $m_X$
\begin{equation}
\begin{split}
\frac{\tau_g^*}{18\pi^2t_p} = \frac{m_X}{m_{Xc}} \quad \textrm{and} \quad {\tau_g^*} \propto {m_X},
\end{split}
\label{eq:5-3-10}
\end{equation}
where the waiting time in the matter era is proportional to the particle mass. For particles with a mass $m_X\ll m_{Xc}$, the waiting time $\tau_g^*$ can be much shorter than the Planck time, which seems unfavorable. Or equivalently, there is no sufficient time for particles with mass $m_X\ll m_{Xc}$ to form haloes of $10^{13}M_{\odot}$ as observed in the current epoch unless different mechanisms exist to accelerate the mass accretion in the matter era for particles with a mass smaller than $m_{Xc}$. However, any faster super-linear halo mass evolution ($m_h^*(t)\propto t^\alpha$ with $\alpha>1$) will contradict the theory $m_h^*(t)\propto t$ and results of the N-body simulations in Fig. \ref{fig:113}. On the other hand, particles with a mass $m_X\gg m_{Xc}$ will not be able to form the smallest and earliest structure due to the collision time much larger than the formation time (Eq. \eqref{eq:5-2-14}). Again, structure formation in radiation and matter eras suggests the only viable particle mass $m_{Xc}=10^{12}$ GeV. 

Due to rapid growth in the radiation era, the characteristic halo mass $m_h^*$ can reach around $10^8M_{\odot}$ at the matter-radiation equality. This suggests that the dark matter haloes can reach an immense size earlier than we expected, which might be helpful to explain the JWST’s discovery of big and bright galaxies as early as $z=15$ \citep{Castellano:2022-Early-Results-from-GLASS-JWST}.

\section{Mass and energy cascade}
\label{sec:2-1}
While the spherical collapse model is a powerful tool for nonlinear structure evolution on small scales (Section \ref{sec:2-2-1}), it models the evolution of overdensity at a single fixed mass scale. It neglects the interactions between different scales (or haloes of different masses). In this section, we introduce a new type of analytical tool that focuses on the mass and energy flow across various scales, i.e., the mass and energy cascade in dark matter flow \citep{Xu:2021-Inverse-mass-cascade-mass-function}. This provides a "top-down" approach to postulate dark matter properties from the large-scale behavior of dark matter haloes. We will first introduce the concepts of cascade. These concepts will be validated and confirmed by N-body simulation results in Sections \ref{sec:2}, \ref{sec:2-2}, and \ref{sec:3-1}. 

First, long-range gravity requires the formation of a broad spectrum of haloes to maximize the entropy of the self-gravitating collisionless system \citep{Xu:2023-Maximum-entropy-distributions-of-dark-matter}. We note that highly localized haloes are a major manifestation of non-linear gravitational collapse \citep{Neyman:1952-A-Theory-of-the-Spatial-Distri,Cooray:2002-Halo-models-of-large-scale-str}. As the building blocks of dark matter flow (counterpart to the "eddies" in turbulence), haloes facilitate an inverse mass cascade from small to large mass scales that are absent in turbulence. The "inverse" stands for the direction of cascade from small to large scales, in contrast to the "direct" cascade from large to small scales. The halo-mediated inverse mass cascade is fully consistent with hierarchical structure formation, where haloes grow by a series of sequential merging. 
In a more realistic picture, individual haloes accrete mass irregularly on short-time scales with discrete jumps through minor and major mergers. However, if we focus on the averaged mass accretion for an ensemble of haloes of the same mass, the averaged halo growth can be much smoother on a larger time scale. On average, this can be equivalently described by continuous merging with minor mergers.   

\begin{figure}
\includegraphics*[width=\columnwidth]{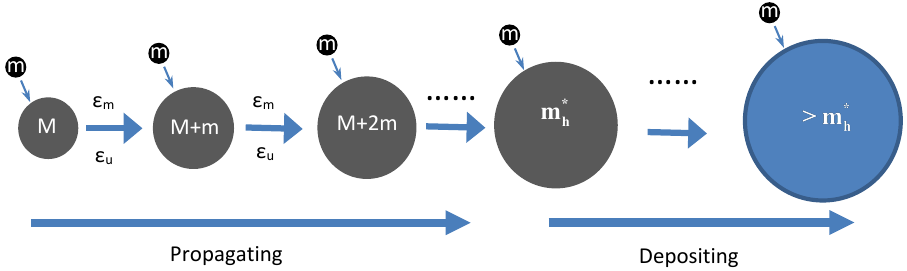}
\caption{Schematic plot for the inverse mass and energy cascade for hierarchical structure formation of dark matter haloes. Individual haloes accrete mass irregularly on short time scales with discrete jumps through minor and major mergers. However, by averaging over time and the halo ensemble, the equivalent description can be: haloes of mass $M$ merging with a single merger (free DM particles of mass $m$) leads to a mass and energy flux to a larger scale $M+m$, that is, the halo of mass $M$ moving into the next mass sale $M+m$ after merging. Because haloes have finite mass and kinetic energy, this facilitates a continuous mass and energy cascade from small to large scales across haloes of different masses. Scale-independent mass and energy flux ($\varepsilon_m$ and $\varepsilon_u$ are independent of mass scale $M$) are expected in the propagation range ($M<m_h^*$). The potential energy is directly cascaded from large to small scales at a rate of $-7/5\varepsilon_u$ (see Eq. \eqref{eq:4-1}). The cascade rate becomes scale-dependent in the deposition range ($M>m_h^*$). This concept is quantitatively demonstrated by Illustris simulations in Figs. \ref{fig:S1-2} and \ref{fig:S1-1-3}.} 
\label{fig:S4}
\end{figure}

The averaged formation of the halo structure is shown in Fig. \ref{fig:S4}. Haloes pass their mass onto larger and larger haloes until the growth of the halo mass becomes dominant over the propagation of the mass through haloes of different scales. Consequently, there is a continuous cascade of mass from the smaller to the larger mass scales with mass flux $\varepsilon_m$ independent of the mass scale in a certain range of scales (the propagation range). In that range, the mass flux into any mass scale balances the mass flux out of the same scales so that the total mass of haloes at that scale does not vary with time; i.e., these haloes propagate mass to large scales (small-scale permanence in Fig. \ref{fig:S1-5}). The mass cascaded from small scales is finally consumed to grow haloes in the deposition range at scales with $M>m_h^*$. From this description, the mass cascade can be described as follows. 

\smallbreak
\centerline{"Little haloes have big haloes, That feed on their mass;} 
\centerline{And big haloes have greater haloes, And so on to growth."} 
\smallbreak

Second, haloes have finite kinetic and potential energy. Therefore, accompanied by the mass cascade, there exists also a simultaneous energy flux (energy cascade at a rate of $\varepsilon_u$) across haloes on different scales. The kinetic energy is also cascaded from small to large scales (the same as the mass cascade), whereas the potential energy is cascaded in the opposite direction. This happens because the potential energy is defined to be negative, while the kinetic energy is positive. 
When dark matter flow reaches a statistically steady state, the rate of mass and energy cascade must be scale-independent (i.e., $\varepsilon_m$ and $\varepsilon_u$ in Fig. \ref{fig:S4} are independent of mass scale $M$). If this is not the case, there would be a net accumulation of mass and energy on some intermediate-mass scale below $m_{h}^{*}$. We exclude this possibility and require that the statistical structures of haloes be self-similar and scale-free for haloes smaller than $m_{h}^{*}$. This leads to a scale-independent cascade up to a critical mass $m_h^*$. The entire concept can be demonstrated by Illustris simulations in Section \ref{sec:2} (see Figs. \ref{fig:S1-2} and \ref{fig:S1-1-3} ). The value of $\varepsilon_u\approx 10^{-7}m^2/s^3$ for the energy cascade is estimated in Eq. \eqref{eq:5}. The scaling laws associated with the energy cascade will be derived in Section \ref{sec:3}. 

Finally, the mass and energy cascades are only relevant on small scales in the nonlinear regime, where haloes of different sizes are the dominant structures ($r_{fs}<r<r_t$ in Fig. \ref{fig:104}). The structure evolution on small scales is highly nonlinear. It can only be studied by N-body simulations with the Newtonian approximation and simplified analytical tools (spherical collapse model, etc., in Section \ref{sec:2-2-1}). The mass and energy cascade in this section provides another useful approach to understanding the nonlinear structure evolution.

\section{Inverse mass cascade in mass space}
\label{sec:2}
In this section, we present a quantitative description of the mass cascade. To validate the concepts, we used results from the large-scale cosmological Illustris simulation (Illustris-1-Dark) \citep{NELSON:2015-The-illustris-simulation}. Illustris is a suite of large-volume cosmological dark matter only and hydrodynamical simulations. The selected Illustris-1-Dark is a dark matter only simulation of a 106.5Mpc$^3$ cosmological volume with 1820$^3$ DM particles for the highest resolution. Each DM particle has a mass around $m_p=7.6\times 10^6 M_{\odot}$. The gravitational softening length is around 1.4 kpc. The simulation has cosmological parameters of dark matter density $\Omega_{DM}=0.2726$, dark energy density $\Omega_{DE}=0.7274$ at $z=0$, and Hubble constant $h=0.704$. The haloes in the simulation were identified using a standard Friends of Friends (FoF) algorithm with a link length parameter b = 0.2. Then, all dark matter particles were divided into halo particles with a total mass $M_h$ and out-of-halo particles that do not belong to any halo. Therefore, $M_h$ is the total mass of all haloes. We will focus on the evolution of mass and energy in haloes of different mass $m_h$. 

\begin{figure}
\includegraphics*[width=\columnwidth]{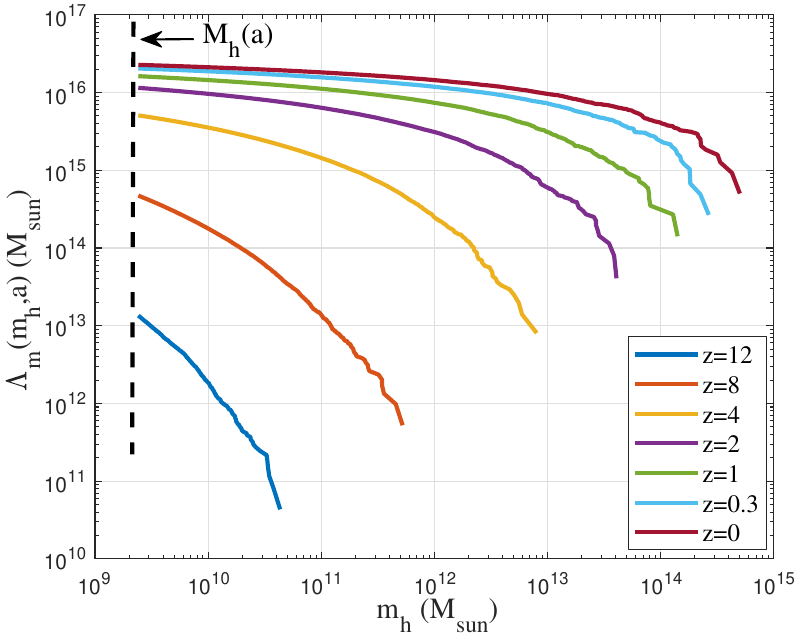}
\caption{The variation of cumulative mass function $\Lambda_m(m_h,a)$ with halo mass scale $m_h$ at different redshifts $z$ from Illustris-1-Dark simulation. The total mass $M_h(a)$ in all haloes of all sizes is computed with $m_h\rightarrow 0$, i.e. $M_h(a)=\Lambda_m(m_h=0,a)$. Using function $\Lambda(m_h,a)$, the rate of mass cascade $\Pi_m(m_h,a)$ in Eq. \eqref{ZEqnNum986318} was computed and presented in Fig. \ref{fig:S1-2}.} 
\label{fig:S1-1}
\end{figure}

First, we can mathematically express the mass flux ($\Pi_m$) across haloes of different sizes as
\begin{equation} 
\label{ZEqnNum986318} 
\begin{split}
\Pi _{m} \left(m_{h} ,a\right) &=-\int _{m_{h} }^{\infty }\frac{\partial }{\partial t} \left[M_{h} \left(a\right)f_{M} \left(m,m_{h}^{*} \right)\right] dm, \\
&=-\frac{\partial }{\partial t} \left[\int _{m_{h} }^{\infty } M_{h} \left(a\right)f_{M} \left(m,m_{h}^{*} \right) dm \right]=-\frac{\partial \Lambda_m}{\partial t},
\end{split}
\end{equation} 
where $a$ is the scale factor and $M_{h}$ is the total mass in all haloes of all sizes. Here $f_{M}(m_h,m_{h}^{*})$ is the halo mass function, that is, the probability distribution of the total mass $M_h$ in all haloes of different mass $m_h$. The total mass of all haloes with a mass between $m_h$ and $m_h+dm$ should be $M_hf_M(m_h,m_{h}^{*})dm$. The cumulative mass function $\Lambda_m(m_h, a)$ represents the total mass in all haloes greater than the scale $m_h$, that is,
\begin{equation} 
\label{ZEqnNum98631129} 
\Lambda_m(m_h,a) = \int _{m_{h}}^{\infty } M_{h} \left(a\right)f_{M} \left(m,m_{h}^{*} \right) dm.  
\end{equation} 
Figure \ref{fig:S1-1} plots the variation of the cumulative mass function $\Lambda_m(m_h, a)$ with mass scale $m_h$ and redshifts $z$ from the Illustris-1-Dark simulation. The total halo mass can be obtained by setting $m_h\rightarrow 0$ in Eq. \eqref{ZEqnNum98631129}, i.e. $M_h(a)=\Lambda_m(m_h\rightarrow 0,a)$. 

The time derivative of $\Lambda_m$ describes the mass flux from all haloes below the scale $m_h$ to all haloes above the scale $m_h$, i.e., the rate of mass cascade $\Pi_m$ in Eq. \eqref{ZEqnNum986318}. In N-body simulations, we use the difference of $\Lambda_m$ at two different redshifts $z_1$ (or $t_1$) and $z_2$ (or $t_2$) to compute the time derivative and obtain the mass cascade rate $\Pi_m$. For Illustris cosmology, the time $t$ at redshift $z$ is computed from 
\begin{equation} 
\label{ZEqnNum9863220} 
a(t)=\frac{1}{1+z(t)}=\left(\frac{\Omega_m}{\Omega_\Lambda}\right)^{1/3}\sinh^{2/3}\left(\frac{t}{2/(3H_0\sqrt{\Omega_\Lambda})}\right).
\end{equation}
\begin{figure}
\includegraphics*[width=\columnwidth]{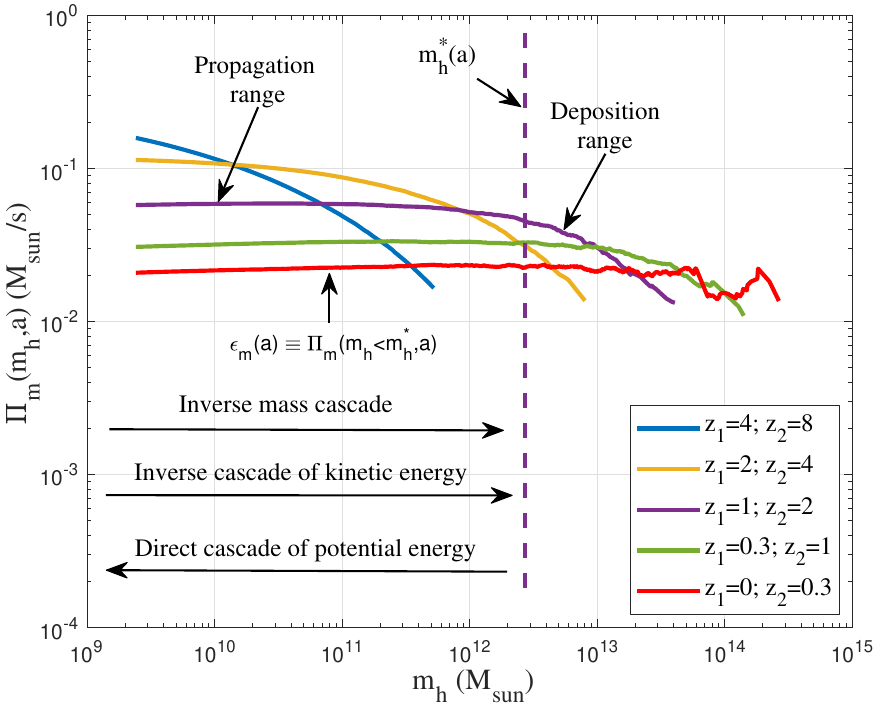}
\caption{Variation of the rate of inverse mass cascade $\Pi_m(m_h, a)$ (Eq. \eqref{ZEqnNum986318}, $<0$ for 'inverse'), calculated by the values of the cumulative mass function $\Lambda_m$ (Eq. \eqref{ZEqnNum98631129}) at two different redshifts $z_1$ and $z_2$ from Fig. \ref{fig:S1-1}. After reaching a statistically steady state, the simulation confirms the existence of a propagation range for the scales $m_h<m_h^*(a)$, a characteristic mass scale that increases with time. A scale-independent rate of the mass cascade $\varepsilon_m(a)\equiv \Pi_m(m_h, a)$ can be identified in the propagation range, which decreases with time ($\varepsilon_m(a)\propto a^{-1}$) and is around -0.02$M_{\odot}/s$ at $z=0$. Haloes in the propagation range pass their mass to larger haloes, where the group mass $m_g$ (total mass of all haloes with the same mass $m_h$) reaches a steady state (see Eq. \eqref{ZEqnNum9863159} and Fig. \ref{fig:S1-5}). There also exists a simultaneous energy cascade discussed in Section \ref{sec:2-2}.} 
\label{fig:S1-2}
\end{figure}

Figure \ref{fig:S1-2} presents the variation of the rate of the mass cascade $\Pi_m(m_h, a)$ with the mass scale $m_h$ using the values of the cumulative mass function $\Lambda_m$ at two different redshifts $z_1$ and $z_2$ in Fig. \ref{fig:S1-1}. In the schematic diagram in Fig. \ref{fig:S4}, we propose the concept that the rate of cascade $\Pi_m(m_h, a)$ is independent of the scale $m_h$ for $m_h<m_h^*$ in the propagation range since we require that the statistical structures of the haloes be self-similar and scale-free for haloes smaller than $m_h^*$. The simulation results confirm this concept, that is, the existence of a propagation range with a scale-independent cascade rate $\varepsilon_m(a)\equiv \Pi_m(m_h, a)$ for scales $m_h$ below a critical mass scale $m_h^*$. After reaching a statistically steady state, that scale-independent rate $\varepsilon_m(a)\propto a^{-1}$, that is, decreases with time and is about -0.02$M_{\odot}/s$ at $z=0$ from the Illustris simulation. The time dependence of $\varepsilon_m(a)\propto a^{-1}$ due to background expansion means a decreasing mass flux in the halo mass space. 

\begin{figure}
\includegraphics*[width=\columnwidth]{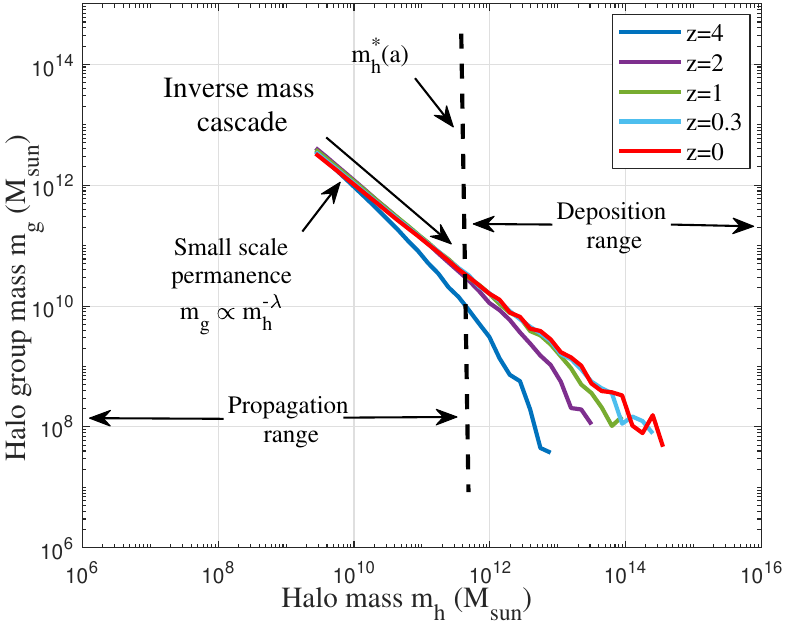}
\caption{The variation of halo group mass $m_g(m_h)$ for all haloes of the same mass $m_h$ at different redshifts $z$ from Illustris-1-Dark simulation. The figure demonstrates the small-scale permanence of the group mass $m_g$. Once the statistically steady state is established ($z\le4$), the rate of the inverse mass cascade $\varepsilon_m$ becomes scale independent (Eq. \eqref{ZEqnNum9863159}) such that the mass of the halo group $m_g$ at different redshifts $z$ collapses into a time-independent power law $m_g\propto m_h^{-\lambda}$ on small mass scales with the halo geometry parameter $\lambda\approx 0.88$ \citep{Xu:2023-Dark-matter-halo-mass-functions-and}. This is the so-called "propagation range," where the mass cascaded from the smallest (DM particle) scale is propagated to larger scales, that is, a constant group mass $m_g$. The cascaded mass is eventually consumed to grow haloes greater than $m_h^*$ (the "deposition range"). The propagation range gradually extends to large scales ($m_h^*(a)$ increases over time) due to the continuous inverse mass cascade.}
\label{fig:S1-5}
\end{figure}

As described in the schematic plot in Fig. \ref{fig:S4}, all masses cascaded from the smallest scale are propagated through the propagation range and are consumed mainly to grow haloes greater than $m_h^*$ in the deposition range. To explain this, we write the scale-independent $\varepsilon_m$ in the propagation range as
\begin{equation} 
\label{ZEqnNum986319} 
\begin{split}
&\varepsilon_m(a) \equiv \Pi _{m}(m_h,a)=-\frac{\partial M_h}{\partial t}\propto a^{-1} \quad \textrm{for}\quad m_h<m_h^*,
\end{split}
\end{equation} 
which further requires (from Eq. \eqref{ZEqnNum986318})
\begin{equation} 
\label{ZEqnNum9863159} 
\frac{\partial \Pi_m}{\partial m_h}=\frac{\partial \varepsilon_m}{\partial m_h}=\frac{\partial }{\partial t} \left[M_{h} \left(a\right)f_{M}(m_h,m_h^*)\right]=\frac{1}{m_p} \frac{\partial m_g}{\partial t}=0.
\end{equation} 
From this, we found that the total mass $m_g$ for a group of haloes with the same mass $m_h$ reaches a steady state and does not vary over time for $m_h<m_h^*$. Here, the halo group mass $m_g$ is written as 
\begin{equation} 
\label{ZEqnNum9863159-2} 
m_g(m_h) = M_{h}\left(a\right)f_{M}(m_h,m_h^*) m_p = N_h m_h, 
\end{equation} 
where $m_p$ is the mass resolution, and $N_h$ is the number of haloes in that group. Due to the steady-state group mass $m_g$, these haloes propagate the mass from small scales to grow haloes greater than $m_h^*$, that is, the "propagation range." This concept can be clearly demonstrated by the Illustris simulations in Fig. \ref{fig:S1-5} such that the halo number density follows $N_h\propto m_h^{-\lambda-1}$ in that range and is actually independent of the redshift. The mass of the halo group, $m_g$ of different redshifts, collapses onto the same time-independent power law $m_g\propto m_h^{-\lambda}$ ($\lambda=0.88$ from the Illustris simulation). This is the "so-called" small-scale permanence for the halo group mass ($m_g$) due to the scale-independent mass cascade in Fig. \ref{fig:S1-2}, i.e., the mass flux into low-mass halo group balances the mass flux out of the same group. This is also in agreement with the slow decline of the number density of low-mass haloes at low redshift \citep{Mo:2002-The-abundance-and-clustering-of-dark-haloes}. The small-scale permanence also exists for halo density profiles due to the scale-independent energy cascade (Fig. \ref{fig:S1-4}). The halo mass function and density profile can also be analytically derived based on the mass and energy cascade \citep{Xu:2023-Dark-matter-halo-mass-functions-and}.

\section{Energy cascade in mass space}
\label{sec:2-2}
Since haloes have finite energy, there is also a simultaneous energy cascade associated with the mass cascade (Fig. \ref{fig:S1-2}). This section presents a quantitative description of the energy cascade in the halo mass space. 

\begin{figure}
\includegraphics*[width=\columnwidth]{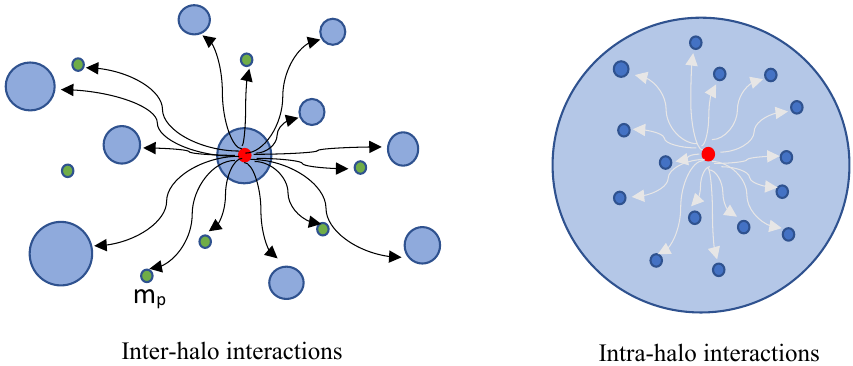}
\caption{Decomposition of kinetic energy into the contributions due to inter-halo interactions on a longer distance and intra-halo interactions on a shorter distance. The inter-halo interactions with all other particles in different haloes and all out-of-halo particles (green) are weaker on larger scales and in the linear regime ($K_{ph}$). The intra-halo interactions with all other particles in the same halo are stronger, on smaller scales, and in the nonlinear regime ($K_{pv}$). The evolution of two KEs is also presented in Fig. \ref{fig:111} and Table \ref{tab:5}.}
\label{fig:SS}
\end{figure}

To better describe the energy cascade, we start by decomposing the particle kinetic energy into two parts of different nature. In N-body simulations, every halo particle, characterized by a mass $m_{p}$, and a velocity vector $\boldsymbol{\mathrm{v}}_{\boldsymbol{\mathrm{p}}}$, should belong to one and only one parent halo. The particle velocity $\boldsymbol{\mathrm{v}}_{p}$ can be decomposed as \citep{Xu:2023-Maximum-entropy-distributions-of-dark-matter}
\begin{equation} 
\label{ZEqnNum502045} 
\boldsymbol{\mathrm{v}}_{p} =\boldsymbol{\mathrm{v}}_{h} +\boldsymbol{\mathrm{v}}_p',           
\end{equation} 
namely, the halo mean velocity, $\boldsymbol{\mathrm{v}}_{h}=\langle \boldsymbol{\mathrm{v}}_{p} \rangle_h$ , and the velocity fluctuation, $\boldsymbol{\mathrm{v}}_{p}^{'} $. Here, $\left\langle \right\rangle _{h} $ represents the average of all particles in the same halo, and $\boldsymbol{\mathrm{v}}_{h}$ represents the velocity of that halo.

Consequently, the total kinetic energy $K_p$ of a given halo particle can be divided into $K_p = K_{ph}+K_{pv}$. As shown in Fig. \ref{fig:SS}, here $K_{ph}=\boldsymbol{\mathrm{v}}_{h}^2/2$ (halo kinetic energy) is the contribution from the motion of entire haloes $\boldsymbol{\mathrm{v}}_{h}$ due to the inter-halo interaction of that particle with all other particles in different haloes and all out-of-halo particles (green). This part of the kinetic energy is related to interactions on large scales in the linear regime. While $K_{pv}={\boldsymbol{\mathrm{v}}_p'}^2/2$ (virial kinetic energy) is the contribution of the velocity fluctuation $\boldsymbol{\mathrm{v}}_p'$ due to the intra-halo interaction of that particle with all other particles in the same halo. This part of the kinetic energy is from halo virialization and is due to interactions on a shorter distance and smaller scales in the non-linear regime. Like the energy cascade associated with nonlinear interactions in turbulence, the energy cascade in this work is focused on the cascade of the virial kinetic energy $K_{pv}$ across different scales in the deeply non-linear regime.

Similar to the cumulative mass function $\Lambda_m$ in Eq. \eqref{ZEqnNum98631129}, the cumulative kinetic energies ($\Lambda_{ph}$ and $\Lambda_{pv}$) represent the total kinetic energies $K_{ph}$ and $K_{pv}$ in all haloes greater than $m_h$, such that
\begin{equation} 
\label{ZEqnNum986311299} 
\begin{split}
&\Lambda_{ph}(m_h,a) = \int _{m_{h}}^{\infty } M_{h} \left(a\right)f_{M} \left(m,m_{h}^{*} \right) K_{ph} dm,  \\
&\Lambda_{pv}(m_h,a) = \int _{m_{h}}^{\infty } M_{h} \left(a\right)f_{M} \left(m,m_{h}^{*} \right) K_{pv} dm, \\
&K_{ph} = \frac{3}{2}\sigma_h^{2}\left(m,a\right) \quad \textrm{and} \quad K_{pv} = \frac{3}{2}\sigma_v^{2}\left(m,a\right). 
\end{split}
\end{equation} 
Here, $\sigma_h^{2}$ and $\sigma_v^{2}$ are the dispersion of the one-dimensional velocity for the halo velocity $\boldsymbol{\mathrm{v}}_{h}$ and the velocity fluctuation $\boldsymbol{\mathrm{v}}_{p}^{'}$ \citep{Xu:2023-Dark-matter-halo-mass-functions-and}. 

\begin{figure}
\includegraphics*[width=\columnwidth]{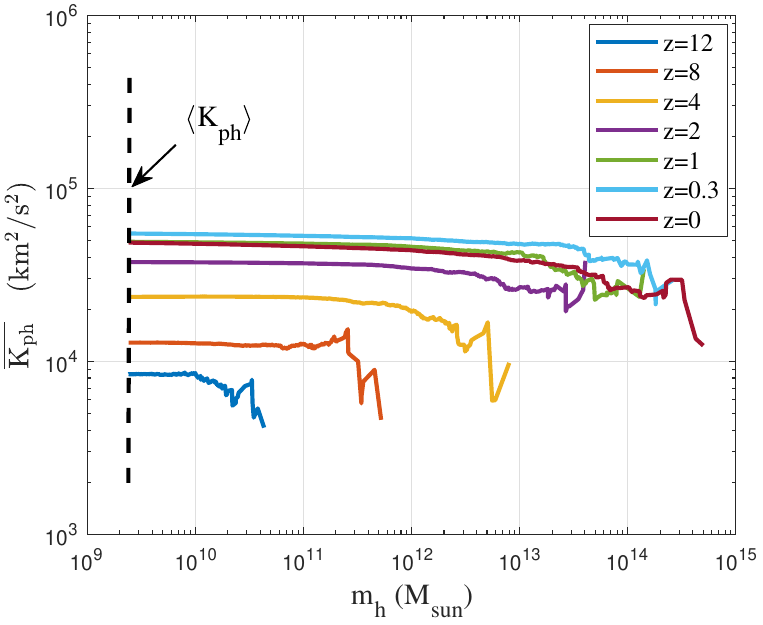}
\caption{The variation of mean halo kinetic energy $\overline {K_{ph}}$ with halo mass scale $m_h$ at different redshifts $z$ from Illustris-1-Dark simulation. The mean halo kinetic energy in all haloes of all sizes is denoted as the black dashed line with $m_h\rightarrow 0$, i.e. $\langle K_{ph} \rangle=\overline {K_{ph}}(m_h\rightarrow 0, a)$. This figure will be used to compute the evolution of mean halo kinetic energy $\langle K_{ph} \rangle$ for all halo particles in Fig. \ref{fig:S1-3}. Due to the long-range interaction with particles from different haloes, halo kinetic energy increases with time in early matter dominant universe (i.e., the linear regime with $\langle {K_{ph}}\rangle \propto t^{2/3}$ shown in Fig. \ref{fig:S1-3}). It slightly decreases at low redshift in the dark energy dominant universe due to the accelerated expansion.} 
\label{fig:S1-1-4}
\end{figure}

\begin{figure}
\includegraphics*[width=\columnwidth]{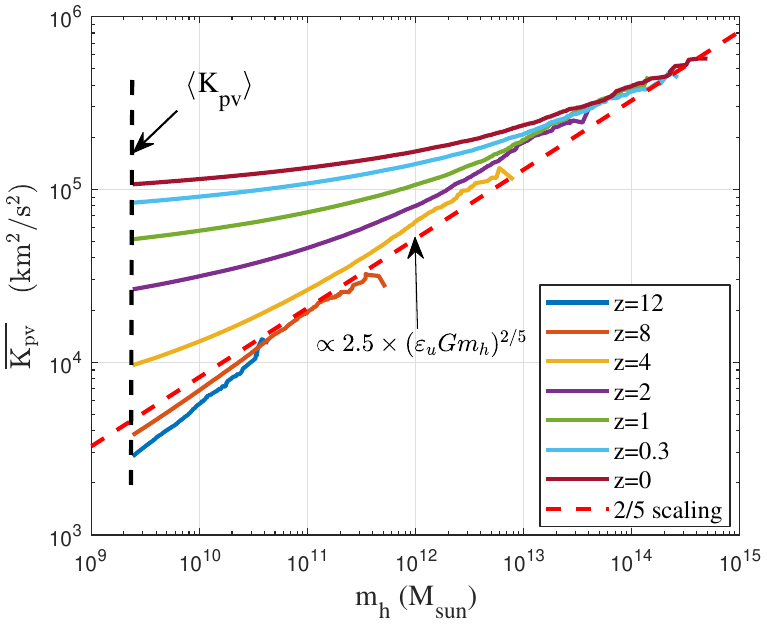}
\caption{The variation of mean virial kinetic energy $\overline {K_{pv}}$ with halo mass scale $m_h$ at different redshifts $z$ from Illustris-1-Dark simulation. The mean virial kinetic energy in all haloes of all sizes is denoted as the black dashed line with $m_h\rightarrow 0$, that is, $\langle K_{pv} \rangle=\overline {K_{pv}}(m_h\rightarrow 0, a)$ in Fig. \ref{fig:S1-3}. This figure is used to calculate the rate of the energy cascade $\varepsilon_u$ in Fig. \ref{fig:S1-1-3}. Due to the virialization of the halo and the interaction with particles in the same halo, the virial kinetic energy $K_{pv}$ increases with time (the non-linear regime with $\langle {K_{pv}}\rangle\propto t^{1}$ shown in Fig. \ref{fig:S1-3}). The dark energy has negligible effects on the virial kinetic energy due to the bounded halo structure. The 2/5 scaling ($\overline {K_{pv}}\propto m_h^{2/5}$ for $m_h\rightarrow \infty$) is also presented for comparison (Eq. \eqref{eq:11}).} 
\label{fig:S1-1-5}
\end{figure}

Next, we will use the cumulative kinetic energy and cumulative mass $\Lambda_m$ to calculate the mean specific halo kinetic energy (energy per unit mass) $\overline {K_{ph}}$ and the virial kinetic energy $\overline {K_{pv}}$ in all haloes above any mass scale $m_h$, that is,
\begin{equation} 
\label{ZEqnNum98631129911} 
\overline {K_{ph}}=\frac{\Lambda_{ph}}{\Lambda_{m}} \quad \textrm{and} \quad \overline {K_{pv}}=\frac{\Lambda_{pv}}{\Lambda_{m}}.
\end{equation}
Figures \ref{fig:S1-1-4} and \ref{fig:S1-1-5} plot the variation of the (specific) halo kinetic energy $\overline {K_{ph}}$ and the virial kinetic energy $\overline {K_{pv}}$ with the halo mass $m_h$ from Illustris simulations. The halo kinetic energy is relatively independent of $m_h$. The virial kinetic energy increases with the mass of the halo $m_h$ with $\overline {K_{pv}}\propto m_h^{2/5}$ for larger haloes. This can be explained by the scaling laws in Eq. \eqref{eq:11}. Both kinetic energies increase with time but with different scaling behavior due to the nature of interactions in the linear and nonlinear regimes (Fig. \ref{fig:S1-3}). 

Next, we will focus on the energy cascade of the specific virial kinetic energy $K_{pv}$ due to nonlinear interactions on small scales. For the inverse mass cascade in Eq. \eqref{ZEqnNum986318}, the change in total halo mass above the scale $m_h$, that is, the cumulative mass function $\Lambda_m(m_h, a)$, comes entirely from the mass cascade or the interactions between all haloes below the scale $m_h$ and all haloes above $m_h$. Similarly, the change in the (specific) virial kinetic energy $\overline {K_{pv}}$ for all haloes above the scale $m_h$ comes entirely from the energy cascade due to interactions between haloes below and above the scale $m_h$. This is because, without an energy cascade, haloes above the scale $m_h$ should be in virial equilibrium, where the specific kinetic energy $\overline {K_{pv}}$ is constant and conserved with time (see Eq. \eqref{eq:4} for an explanation), just as the total halo mass above the scale $m_h$ is conserved without a mass cascade. Therefore, similar to the mass cascade $\Pi_m$ in Eq. \eqref{ZEqnNum986318}, the rate of cascade for the virial kinetic energy $K_{pv}$ reads
\begin{equation} 
\label{ZEqnNum9863112991} 
\begin{split}
\Pi_{pv}(m_h,a) &=-\frac{\partial }{\partial t} \left( \overline {K_{pv}} \right) = -\frac{\partial }{\partial t} \left(\frac{\Lambda_{pv}}{\Lambda_m} \right) \\
&=-\frac{\partial }{\partial t}\int _{m_{h} }^{\infty } \frac{M_{h} \left(a\right)f_{M} \left(m,m_{h}^{*} \right) K_{pv}}{\int _{m_{h} }^{\infty } M_{h} \left(a\right)f_{M} \left(m,m_{h}^{*} \right) dm} dm,
\end{split}
\end{equation}
where $\overline {K_{pv}}$ is defined in Eq. \eqref{ZEqnNum98631129911}, i.e. the specific virial kinetic energy in all haloes greater than $m_h$. Similar to Eq. \eqref{ZEqnNum986318}, Eq. \eqref{ZEqnNum9863112991} describes the rate of transfer of specific virial kinetic energy ($K_{pv}$) from haloes below the scale $m_h$ to haloes above the scale $m_h$ at a rate of $\Pi_{pv}$. 

\begin{figure}
\includegraphics*[width=\columnwidth]{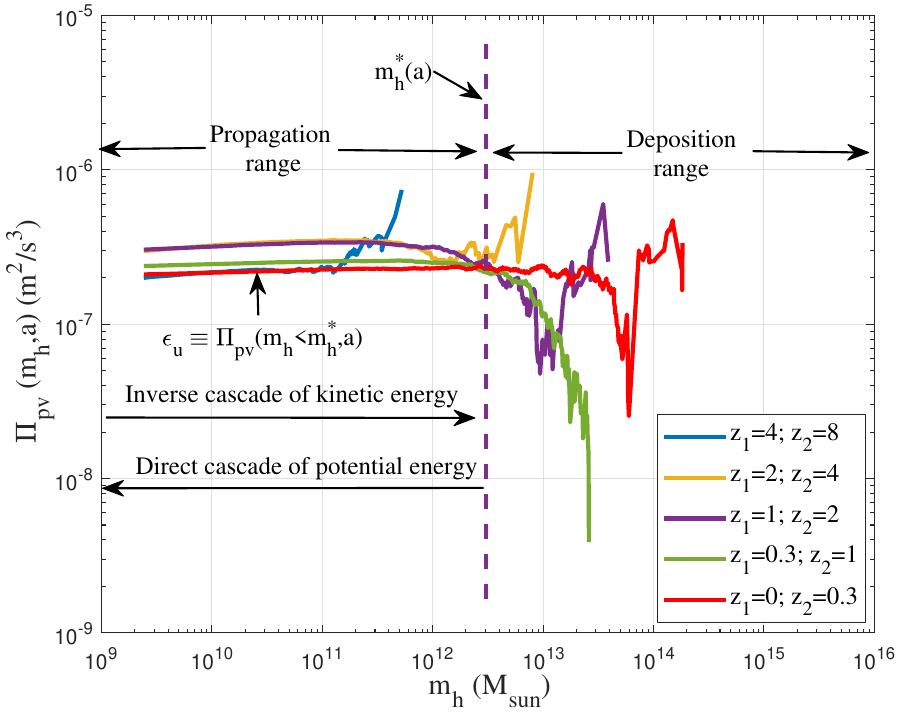}
\caption{The variation of the rate of energy cascade $\Pi_{pv}(m_h,a)$ (Eq. \eqref{ZEqnNum9863112991}) with halo mass scale $m_h$ at different redshifts $z$ from Illustris-1-Dark simulation.  A scale-independent constant rate of $\varepsilon_u$ can be identified in the propagation range for an inverse cascade of virial kinetic energy ${K_{pv}}$ from the smallest scale (single DM particle) to larger scales. That rate is also relatively independent of time and is around $\varepsilon_u=-2.5\times 10^{-7}m^2/s^3$ (also see Fig. \ref{fig:S1-3}). There also exists a simultaneous direct cascade of potential energy from large to the smallest scale at a rate of $-7/5\varepsilon_u$ (see Eq. \eqref{eq:4-1}).} 
\label{fig:S1-1-3}
\end{figure}

Figure \ref{fig:S1-1-3} plots the variation of $\Pi_{pv}$ with the halo mass $m_h$ and the redshifts $z$. The mean (specific) virial kinetic energy $\overline {K_{pv}}$ at two different redshifts $z_1$ and $z_2$ in Fig. \ref{fig:S1-1-5} was used to calculate $\Pi_{pv}$ in this figure. Similarly to the mass cascade in Fig. \ref{fig:S4}, if the statistical structures of the haloes are self-similar and scale-free for haloes smaller than the characteristic mass $m_h^*$, the rate of the energy cascade $\varepsilon_u$ should also be independent of the scale $m_h$ for $m_h<m_h^*$, i.e. the energy flux into a given group of haloes should balance the energy flux out of the same group. The simulation results confirm a scale- and time-independent rate of cascade $\varepsilon_u$. Therefore, in the propagation range ($m_h<m_h^*$),
\begin{equation} 
\label{ZEqnNum98631913} 
\begin{split}
\varepsilon_u \equiv \Pi _{pv}(m_h,a) =-\frac{\partial \langle K_{pv}\rangle}{\partial t}\propto\varepsilon_m\frac{\langle K_{pv}\rangle}{M_h},
\end{split}
\end{equation} 
where the rate of energy cascade $\varepsilon_u$ can be directly related to the rate of mass cascade $\varepsilon_m$. Since $\varepsilon_u$ is scale independent, with continuous injection of virial kinetic energy $K_{pv}$ at a constant rate of $\varepsilon_u$ on the smallest scale, we should expect the total $K_{pv}$ in all haloes of all sizes to be proportional to time $t$ or $\langle K_{pv}\rangle\propto t$. This is consistent with the solutions of the cosmic energy equation in the next section (Eq. \eqref{eq:4-1} and Fig. \ref{fig:S1-3}).

Up to this point, we have discussed the mass and energy cascade in halo mass space. The rate of the energy cascade $\varepsilon_u\approx -10^{-7}m^2/s^3$ can be estimated for galactic haloes from cosmological simulations (Fig. \ref{fig:S1-1-3}). This is consistent with $\varepsilon_u$ obtained on the smallest scale for particles of critical mass $m_{Xc}$ in Eqs. \eqref{eq:5-22-9}. This is, of course, not a mere coincidence as the key parameter $\varepsilon_u$ is a constant independent of both scale and time, as shown in Fig. \ref{fig:S1-1-3}. The value of $\varepsilon_u$ estimated for small and large haloes should be the same.

\section{Cosmic energy evolution}
\label{sec:3-1}
To better understand the energy cascade, we provide an analysis based on the energy evolution in self-gravitating collisionless dark matter flow. The equations of motion for \textit{N} collisionless particles in comoving coordinates $\boldsymbol{\mathrm{x}}$ and physical time \textit{t} read \citep{Peebles:1980-The-Large-Scale-Structure-of-t}:
\begin{equation} 
\label{ZEqnNum157264} 
\frac{d^{2} \boldsymbol{\mathrm{x}}_{i} }{dt^{2} } +2H\frac{d\boldsymbol{\mathrm{x}}_{i} }{dt} =-\frac{Gm_{p} }{a^{3} } \sum _{j\ne i}^{N}\frac{\boldsymbol{\mathrm{x}}_{i} -\boldsymbol{\mathrm{x}}_{j} }{\left|\boldsymbol{\mathrm{x}}_{i} -\boldsymbol{\mathrm{x}}_{j} \right|^{3} }  ,        
\end{equation} 
where \textit{N} particles have equal mass $m_{p}$. The Hubble parameter $H\left(t\right)={\dot{a}/a}$. Here, $H$ has a "damping" effect, which leads to the decrease in total energy of the N-body system (Eq. \eqref{eq:4} and Fig. \ref{fig:S1-3-2}). In the radiation and matter eras, 
the Hubble parameter satisfies $Ht=1/2$ and $Ht=2/3$, respectively (Table \ref{tab:5}). In the dark energy era, the Hubble parameter $H$ approaches a constant value $H_0^*=H_0\sqrt{\Omega_{\Lambda}}$, where $\Omega_{\Lambda}$ is the fraction of dark energy. 

Next, we will derive the energy evolution based on equations of motion (Eq. \eqref{ZEqnNum157264}). We first introduce a transformed time variable \textit{s} as ${ds/dt} =a^{p} $, where \textit{p} is an arbitrary exponent. In terms of the new time variable $s$, the original Eq. \eqref{ZEqnNum157264} can be transformed to 
\begin{equation} 
\label{ZEqnNum168751} 
\begin{split}
&\frac{d^{2} \boldsymbol{\mathrm{x}}_{i} }{ds^{2} } +\frac{d\boldsymbol{\mathrm{x}}_{i} }{ds} \left(p+2\right)a^{-p} H \equiv  a^{-(3+2p)}\frac{\boldsymbol{\mathrm{F}}_{i}}{m_p},\\
&\frac{\boldsymbol{\mathrm{F}}_{i}}{m_p} = -G m_p \sum _{j\ne i}^{N}\frac{\boldsymbol{\mathrm{x}}_{i} -\boldsymbol{\mathrm{x}}_{j} }{\left|\boldsymbol{\mathrm{x}}_{i} -\boldsymbol{\mathrm{x}}_{j} \right|^{3}} = -\frac{\partial P_s}{\partial \boldsymbol{\mathrm{x}}_{i}}, 
\end{split}
\end{equation} 
where $\boldsymbol{\mathrm{F}}_{i} $ is the resultant force on particle \textit{i} from all other particles, while $P_s$ is the total specific potential energy in comoving coordinates. Equation \eqref{ZEqnNum168751} reduces to the original Eq. \eqref{ZEqnNum157264} when $p=0$. With $p=-2$, the first-order derivative vanishes in Eq. \eqref{ZEqnNum168751} and \textit{s} is the time variable for integration in \textit{N}-body simulations. By setting $p=-1$, \textit{s} is the conformal time. By setting $p=-{3/2}$ along with $H_0^2=H^2a^3$ for the matter era, the equation of motion becomes 
\begin{equation} 
\label{ZEqnNum730753} 
\frac{d^{2} \boldsymbol{\mathrm{x}}_{i} }{ds^{2} } +\frac{1}{2} H_{0} \frac{d\boldsymbol{\mathrm{x}}_{i} }{ds} =\frac{\boldsymbol{\mathrm{F}}_{i} }{m_{p}}.      
\end{equation} 
In this transformed equation for the matter era, the scale factor \textit{a} is not explicitly involved. The time-dependent Hubble parameter $H$ is replaced by a Hubble constant $H_0$ (or a constant "damping"), which can offer significant convenience in analytically solving the equations of motion. 

We first identify the transformation between velocity $\boldsymbol{\mathrm{v}}_{i}$ in time variable $s$ and the peculiar velocity $\boldsymbol{\mathrm{u}}_{i}$,
\begin{equation} 
\label{ZEqnNum631619} 
\begin{split}
&\boldsymbol{\mathrm{v}}_{i} =\frac{d\boldsymbol{\mathrm{x}}_{i} }{ds} =a^{-p} \frac{d\boldsymbol{\mathrm{x}}_{i} }{dt}=a^{-p-1}\boldsymbol{\mathrm{u}}_{i}, \quad \boldsymbol{\mathrm{u}}_{i} =a\frac{d\boldsymbol{\mathrm{x}}_{i} }{dt},  \\
&K_s = K_p a^{-2p-2}, \quad P_s = aP_y,
\end{split}
\end{equation} 
where the kinetic energy $K_s$ and the potential $P_s$ in the transformed equation can now be related to the peculiar kinetic energy $K_p$ and the potential $P_y$ in the physical coordinates.

The energy evolution of the N-body system can be obtained by multiplying ${\boldsymbol{\mathrm{v}}_{i} = d\boldsymbol{\mathrm{x}}_{i} /ds}$ on both sides of Eq. \eqref{ZEqnNum168751} and adding the equation of motion for all particles together \citep{Xu:2022-The-evolution-of-energy--momen}. An exact and simple equation (in time variable $s$) for the specific kinetic energy $K_s$ and the potential energy $P_s$ can be obtained as
\begin{equation} 
\label{ZEqnNum333845} 
\frac{d K_{s} }{d s}+2HK_s(p+2)a^{-p}+a^{-(3+2p)}\frac{dP_s}{ds}=0.          
\end{equation} 
By setting $p=0$ and using the relations in Eq. \eqref{ZEqnNum631619}, the exact cosmic energy equation for energy evolution of the N-body system reads
\begin{equation} 
\label{eq:4} 
\frac{\partial E_{y} }{\partial t} +H\left(2K_{p} +P_{y} \right)=0,         
\end{equation} 
which describes the energy evolution in an expanding background. Here $K_{p}$ is the peculiar kinetic energy, $P_{y}$ is the potential energy in physical coordinates and $E_{y}=K_{p}+P_{y} $ is the total specific energy. This is also known as the Layzer-Irvine equation \citep{Irvine:1961-Local-Irregularities-in-a-Univ,Layzer:1963-A-Preface-to-Cosmogony--I--The}. Here, we derive it using a different approach.

\begin{figure}
\includegraphics*[width=\columnwidth]{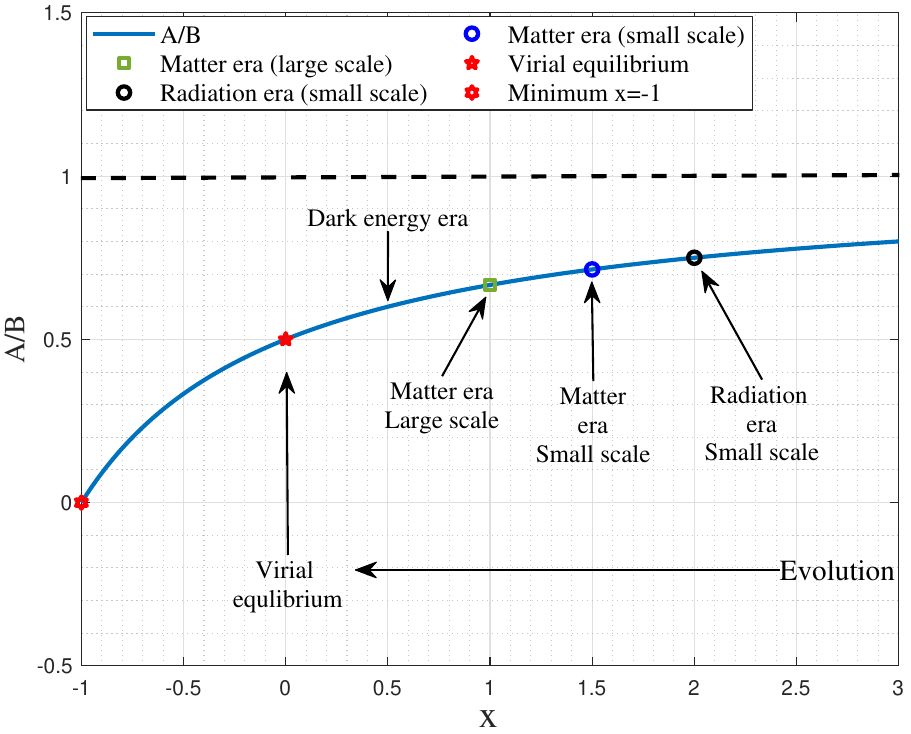}
\caption{The evolution of $A/B$ for N-body system in different eras. The solid line represents the analytical solution in Eq. \eqref{eq:4-2}. Symbols represent the solution for different eras. The initial state $x=\infty$ should have a total energy $E_y=0$ or $A=B$. In the radiation era, the kinetic energy on small scales evolves as $K_{pv}\propto a^2\propto t$ such that $A/B=3/4$ (black circle). In the matter era, the kinetic energy on large scales evolves as $K_{ph}\propto a$ such that $A/B=2/3$ (green circle), while the kinetic energy on small scales follows $K_{pv}\propto a^{3/2}\propto t$ such that $A/B=5/7$  (blue circle). For the dark energy era, $x$ decreases to $0$ and $A/B\rightarrow 1/2$ so that the system approaches the limiting virial equilibrium (red star). The minimum possible $x=-1$ is also denoted as the limiting state for the dissipative gas in the bulge where the kinetic energy of gases decreases with time as $K_p\propto a^{-1}$ \citep{Xu:2024-Cosmic-quenching-and-scaling-laws}.} 
\label{fig:105}
\end{figure}

Next, we will focus on the solutions of the cosmic energy equation in different eras. Without loss of generality, we can assume that $K_p=Aa^x$ and $P_y=-Ba^x$ and substitute them into Eq. \eqref{eq:4}, the general solution to the cosmic energy equation reads
\begin{equation} 
\label{eq:4-2} 
\begin{split}
&\frac{A}{B} = \frac{1+x}{2+x}. 
\end{split}
\end{equation} 
Solutions at different stages are identified and presented in Fig. \ref{fig:105}. First, the case of $x=\infty$ corresponds to an initial state with vanishing total energy $E_y=0$ or $A=B$. Next, we focus on the evolution in different eras.

\begin{figure}
\includegraphics*[width=\columnwidth]{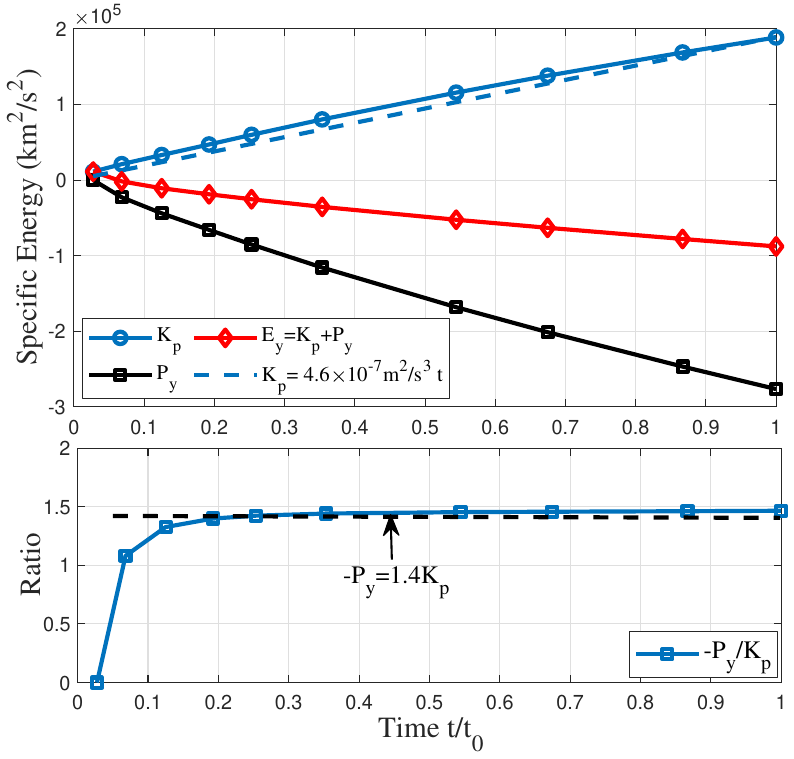}
\caption{The variation of specific (energy per unit mass) kinetic energy $K_p$, potential energy $P_y$, and total energy $E_y=K_p+P_y$ (unit: $km^2/s^2$) with time $t$ from Virgo SCDM simulation. Solution in Eq. \eqref{eq:4-1} is also presented for comparison with $\varepsilon_u=-4.6\times 10^{-7}m^2/s^3$. Simulation confirms a linear increase of $K_p=-\varepsilon_u t$ with time and negative potential energy $P_y=-1.4K_p$. The total energy $E_y=-0.4K_p$ also decreases with time, which requires a viable mechanism to dissipate the total energy $E_y$. Similar to the kinetic energy, the Figure reveals a direct cascade of potential energy at a rate of $-1.4\varepsilon_u$ from large to the smallest scale.} 
\label{fig:S1-3-2}
\end{figure}

\begin{figure}
\includegraphics*[width=\columnwidth]{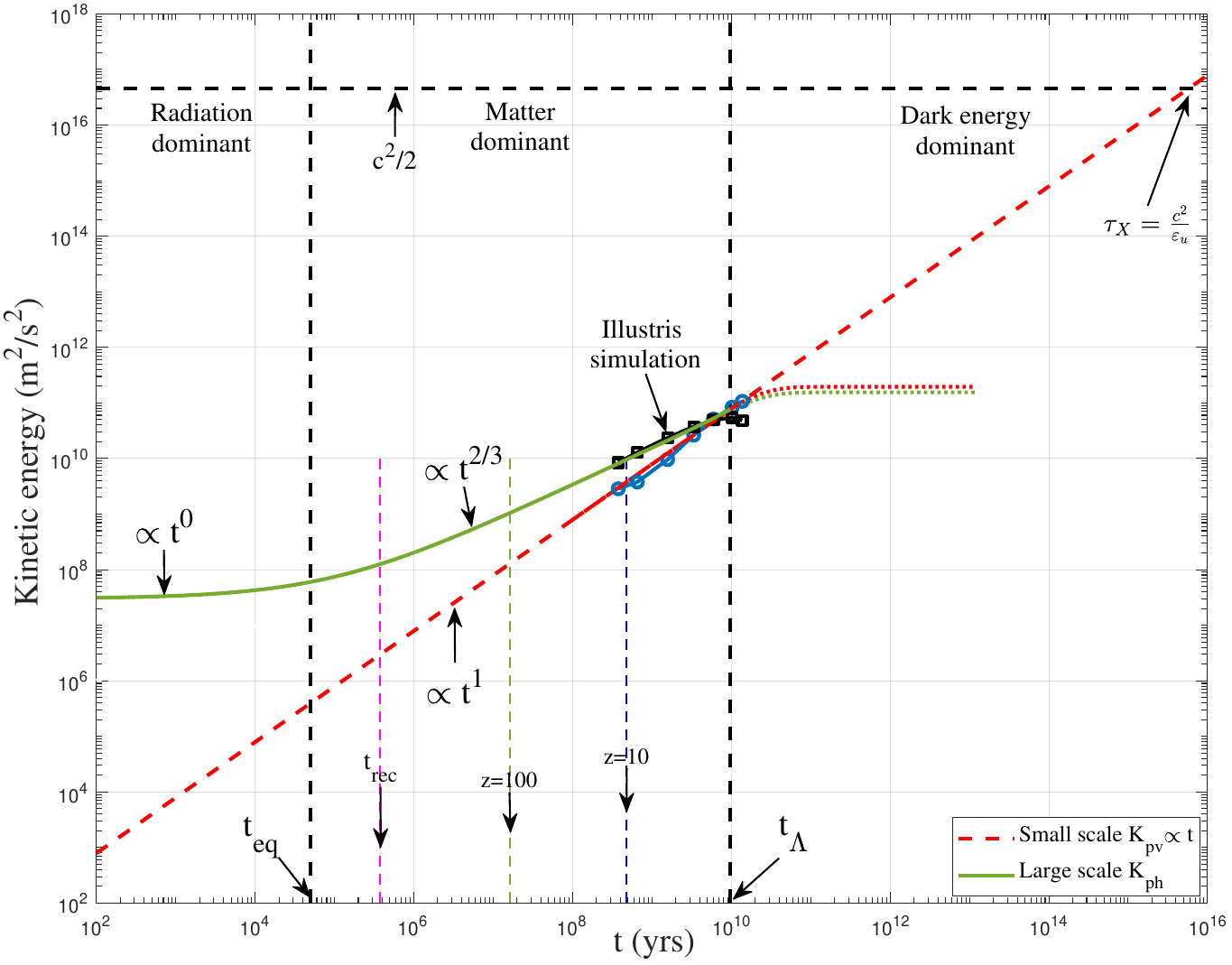}
\caption{The variation of the specific kinetic energy (energy per unit mass) $K_{ph}$ on large scales and $K_{pv}$ for haloes on small scales with time $t$ (yrs) from Illustris-1-Dark simulation. Solution in Eq. \eqref{eq:4-1} is also presented for comparison with $\varepsilon_u=-2.5\times 10^{-7}m^2/s^3$. The large-scale kinetic energy $K_{ph}$ is constant in the radiation era and increases in the matter era (i.e., the linear regime with $K_{ph}\propto t^{2/3}$). It slightly decreases at low redshift in the dark energy dominant universe due to the accelerated expansion. Due to intra-halo interactions on small scales (the non-linear regime), the small-scale kinetic energy $K_{pv}=-\varepsilon_u t$ in both radiation and matter eras. In the dark energy era, both kinetic energies level off because of the accelerated expansion and self-limiting effects of dark energy (dotted lines). By assuming that virial velocity cannot exceed the speed of light, a characteristic time $\tau_X=c^2/\varepsilon_u=10^{16}$yrs can be obtained. The true lifetime of dark matter can be greater than $\tau_X$.} 
\label{fig:S1-3}
\end{figure}

\begin{enumerate}
\item \noindent For the radiation era, the kinetic energy on large scales (halo kinetic energy $K_{ph}$) is suppressed and constant over time, similarly to the overdensity $\delta$ on large scales (Eq. \eqref{eq:5-2-5}). However, kinetic energy on small scales increases linearly with time due to the energy cascade across haloes, i.e.,
\begin{equation} 
\label{eq:4-1-1-1-1} 
\begin{split}
K_{pv}=-\varepsilon_u t\propto a^2 \quad\textrm{and}\quad K_{ph}=V_{||}^2/2\propto a^0. 
\end{split}
\end{equation} 
The exact solution is, therefore, $A/B=3/4$ with $x=2$ from Eq. \eqref{eq:4-2} (the black circle in Fig. \ref{fig:105}) for the ratio between kinetic and potential energies on small scales. The evolution of $K_{ph}$ and $K_{pv}$ in the radiation era is also presented in Figs. \ref{fig:111} and \ref{fig:S1-3}.
\newline
\item \noindent For the matter era, the kinetic energy on large scales follows $K_{ph}\propto a$ (Eq. \eqref{eq:5-2-5}), and the exact solution is $A/B=2/3$ with $x=1$ (green square in Fig. \ref{fig:105}). The kinetic energy on small scales increases linearly with time. The corresponding solution is $A/B=5/7$ with $x=3/2$ from Eq. \eqref{eq:4-2} (blue circle in Fig. \ref{fig:105}). We have solutions: 
\begin{equation} 
\label{eq:4-4-1-7} 
\begin{split}
&K_{pv}=-\varepsilon_u t \quad\textrm{and}\quad K_{ph}=-\varepsilon_u (a/a_\Lambda)^{-1/2}t.
\end{split}
\end{equation} 
Therefore, when most dark matter mass resides in haloes, the kinetic and potential energies of the entire N-body system are dominated by energy on small scales and should be 
\begin{equation} 
\label{eq:4-1} 
\begin{split}
&K_p=At=-\varepsilon_u t,\quad P_y = -Bt=\frac{7}{5} \varepsilon_u t. \\
\end{split}
\end{equation} 
\noindent This solution can be directly validated by N-body simulations. Figure \ref{fig:S1-3-2} was generated from the matter-dominant Virgo simulation (SCDM) with a matter density $\Omega_{DM}=1$ and dark energy $\Omega_{DE}=0$. A comprehensive description of this simulation can be found in \citep{Frenk:2000-Public-Release-of-N-body-simul,Jenkins:1998-Evolution-of-structure-in-cold}. The kinetic energy $K_p$ and the potential energy $P_y$ were calculated as the mean energy of all dark matter particles in the N-body system. Figure \ref{fig:S1-3-2} confirms the solution in Eq. \eqref{eq:4-1}, i.e. a linear increase of $K_p$ with time and a negative potential energy $P_y=-1.4K_p$. The total cosmic energy $E_y=-0.4K_p$ also decreases with time, as if "dissipated", even though there is no viscous force in the collisionless dark matter. This energy "dissipation" is balanced by a steady energy cascade from large to small scales when a statistically steady state is established for collisionless dark matter flow \citep{Xu:2021-Inverse-mass-cascade-mass-function}. Therefore, the rate of energy cascade ($\varepsilon_u<0$ for "inverse") equals the rate of energy "dissipation" and reads  
\begin{equation} 
\label{eq:5} 
\begin{split}
\varepsilon_{u} =-\frac{K_{p}}{t} =-\frac{3}{2} \frac{u^{2} }{t} =-\frac{3}{2} \frac{u_{0}^{2} }{t_{0} }\approx -4.6\times 10^{-7} \frac{m^{2} }{s^{3}}, 
\end{split}
\end{equation} 
where $u_{0} \equiv u\left(t=t_{0} \right) \approx 350$km/s is the one-dimensional velocity dispersion of all dark matter particles, and $t_{0}\approx$13.7 billion years is the physical time at present epoch or the age of the universe. Different simulations may have slightly different values of $u_0$ due to different cosmological parameters. However, the rate of cascade $\varepsilon_u \sim -10^{-7}m^2/s^3$ should be a good estimate. This value is also consistent with Eq. \eqref{eq:5-22-9} for particles with a critical mass of $m_{Xc}=10^{12}$GeV. 
\newline

\item \noindent For dark energy era, $A/B$ monotonically decreases with $x$ and approaches the limiting virial equilibrium with $A/B=1/2$ (red star in Fig. \ref{fig:105}). In the matter era, the total halo mass $M_h$ and the characteristic halo mass $m_h^*$ increase rapidly over time (Table \ref{tab:4}). In the dark energy era, this increase slows down due to the accelerated expansion. Their evolution can be written as (Eqs. \eqref{ZEqnNum986319} and \eqref{eq:5-3-5})
\begin{equation} 
\label{eq:4-4-1-1} 
\begin{split}
&\frac{dM_h}{dt}=-\varepsilon_m\propto a^{-1}\quad \textrm{and} \quad \frac{dm_h^*}{dt}=\frac{m_X}{\tau_g}\propto {m_h^*}^{2/3}a^{-1}.
\end{split}
\end{equation} 
The solutions for halo mass evolution in the dark energy era are
\begin{equation} 
\label{eq:4-4-1-2} 
\begin{split}
&M_h(t)=M_h(t_{\Lambda})\frac{1-C_1\exp(-H_0^*t)}{1-C_1\exp(-H_0^*t_{\Lambda})},\\
&m_h^*(t)=m_h^*(t_{\Lambda})\left(\frac{1-C_2\exp(-H_0^*t)}{1-C_2\exp(-H_0^*t_{\Lambda})}\right)^3,
\end{split}
\end{equation} 
where $C_1$ and $C_2$ are two constants that can be obtained from boundary conditions at $t_{\Lambda}$. Here, $t_{\Lambda}$ and $a_{\Lambda}$ are the time and scale factor at the equality of matter and dark energy,
\begin{equation} 
\label{eq:4-4-1-3} 
\begin{split}
&a_{\Lambda}=\left(\frac{\Omega_m}{\Omega_{\Lambda}}\right)^{1/3}, \quad t_\Lambda = \frac{2\textrm{asinh(1)}}{3H_0\sqrt{\Omega_\Lambda}},\quad \textrm{and}\quad H_0^*=H_0\sqrt{\Omega_\Lambda}.\\
\end{split}
\end{equation} 
\end{enumerate}

The virial kinetic energy on small scales increases linearly with time in both the radiation and matter eras, i.e. $K_{pv}=-\varepsilon_u t$ (see Table \ref{tab:4} for relevant quantities). In the dark energy era, the rate of the energy cascade $\varepsilon_u$ slows down. Kinetic energies on both large and small scales level off as a result of the accelerated expansion. The evolution of kinetic energy follows (from Eq. \eqref{ZEqnNum98631913}):
\begin{equation} 
\label{eq:4-4-1-4} 
\begin{split}
&\frac{dK_{pv}}{dt}=-\varepsilon_u \propto -\frac{\varepsilon_m}{M_h}K_{pv}\propto a^{-1} \quad \textrm{and} \quad \frac{dK_{ph}}{dt} \propto a^{-1},
\end{split}
\end{equation} 
with corresponding solutions (dotted lines in Fig. \ref{fig:S1-3}): 
\begin{equation} 
\label{eq:4-4-1-5} 
\begin{split}
&K_{pv}=-\varepsilon_u t_{\Lambda}-\varepsilon_u a_{\Lambda}\int^t_{t_{\Lambda}}{a}^{-1}dt,\\
&K_{ph}=-\varepsilon_u t_{\Lambda}-\frac{2}{3}\varepsilon_u a_{\Lambda}\int^t_{t_{\Lambda}}{a}^{-1}dt.
\end{split}
\end{equation}

The energy evolution of the N-body system is, of course, fully consistent with the picture of the energy cascade. With the virial kinetic energy $K_{pv}$ continuously cascaded from the smallest scale to the larger scales at a constant rate $\varepsilon_u$, $K_{pv}$ increases linearly with time $t$. Fig. \ref{fig:S1-3} presents the evolution of kinetic energy on large scales ($K_{ph}$) and small scales ($K_{pv}$) in different eras. The simulation data in Figs. \ref{fig:S1-1-4} and \ref{fig:S1-1-5} were used to plot the time evolution in Fig. \ref{fig:S1-3} (symbols). The solution in Eq. \eqref{eq:4-1} is also presented for comparison with a constant rate of the cascade $\varepsilon_u=-2.5\times 10^{-7}m^2/s^3$. As expected, the halo kinetic energy $K_{ph}$ scales as $\propto t^{2/3}$ due to inter-halo interactions in the linear regime and levels off at low redshift $z$. In the nonlinear regime, the virial kinetic energy $K_{pv}$ scales as $\propto \varepsilon_u t$ due to the inverse cascade of kinetic energy. This is true for at least up to $z=0$, as shown in the figure. Now, let us consider the following two scenarios:\\
\newline
i) Dark matter can exist independently of dark energy such that we can consider a matter-dominant universe without dark energy. In this case, the scalings for the halo mass $m_h^*\propto t$ and the kinetic energy $K_{pv} \propto -\varepsilon_u t$ can extend up to a characteristic time $\tau_X$ (red dashed line in Fig. \ref{fig:S1-3}). By assuming that the speed of particles cannot exceed the speed of light, the characteristic time should be $\tau_X \propto -c^2/\varepsilon_u = 10^{16}$yrs. Relativistic corrections might be required for a more accurate result. This scenario leads to continuous halo growth, even in the dark energy era, which is not plausible. Dark energy is required to limit and slow down the halo growth.\\
\newline
ii) Dark matter and dark energy have a related and dependent origin. In this case, dark matter must coexist with dark energy. The linear relation $K_{pv} \propto -\varepsilon_u t$ breaks down much earlier than the time $\tau_X$, as dark energy prevents the formation of large haloes. Therefore, viral velocities might level off and never reach the speed of light (dotted lines in Fig. \ref{fig:S1-3}). If this is the case, the true lifetime of dark matter can be much greater than $\tau_X=10^{16}$yrs. In this scenario, dark energy provides a self-limiting mechanism to slow down the growth of characteristic halo mass $m_h^*$ and the kinetic energy $K_{pv}$.

\begin{table}
    \begin{center}
    \caption{Cosmic mass and energy evolution in different eras}
    \label{tab:4}
    \begin{tabular}{lcccc} 
    \hline
    Quantity                &  Symbol   & \makecell{Radiation\\era} & \makecell{Matter\\era} & \makecell{Dark\\energy era} \\
    \hline
    Time                    &   {t}     & $a^2$                     &$a^{3/2}$               & $\ln a/H_0^*$\\
    Hubble parameter        &   {Ht}    & {1/2}                     & {2/3}                  & $H_0^*t$\\
    Large-scale overdensity & {$\delta$}& {$a^0$}                   & {$a^1$}                & {$a^0$}         \\
    Large-scale KE          & $K_{ph}$  & $a^0$                     &  $a^1$                 &  $a^0$\\
    Small-scale KE          & $K_{pv}$  & $t^1$                     &  $t^1$                 &  $t^0$\\
    Ratio of A to B (Eq. \eqref{eq:4-2}) &{A/B}      &  {3/4}                    & {5/7}                  &{1/2}   \\
    Characteristic halo mass &{$m_h^*$} &  {$a^5$}                  &  {$a^{3/2}$}           & {$a^0$}         \\
    Total halo mass          &{$M_h$}   & {$a^{5/3}$}               &  {$a^{1/2}$}           & {$a^0$}         \\
    Rate of mass cascade     &{$\varepsilon_m$}     & {$a^{-1/3}$}  &  {$a^{-1}$}            & {$a^{-1}$}    \\
    Rate of energy cascade   &{$\varepsilon_u$}     & {$a^{0}$}     &  {$a^0$}               & {$a^{-1}$}                \\
    \hline
    \end{tabular}
  \end{center}
\end{table}

In summary, three important observations can be obtained from the energy evolution (Table \ref{tab:4}) and the picture of the energy cascade in Figs. \ref{fig:S4}, \ref{fig:S1-2}, \ref{fig:S1-1-3}, \ref{fig:S1-3-2}, and \ref{fig:S1-3}, . 
\begin{enumerate}
\item \noindent Due to the energy cascade in halo mass space, linear evolution of kinetic energy $K_{pv}=-\varepsilon_ut$ on small scales can be obtained for both radiation era and matter era. This is also consistent with the cosmic energy equation (Eq. \eqref{eq:4}) and the N-body simulations. This is important such that the scaling laws involving $\varepsilon_u$ that we will develop in Section \ref{sec:3} are valid in both eras.  
\newline

\item \noindent Without dark energy, our universe will stuck in a statistically steady state (matter era of blue circle in Fig. \ref{fig:105}), which involves an energy cascade at a constant rate $\varepsilon_u$ and a mass cascade at a decreasing rate $\varepsilon_m$. Without dark energy, the kinetic energy $K_{pv}$ and halo mass $m_h^*$ will continuously increase until a characteristic time $\tau_X$ in Fig. \ref{fig:S1-3}. Since the total amount of dark matter in the universe is fixed, a mechanism is required to limit the growth of $m_h^*$ and $K_{pv}$. From this perspective, dark energy and dark matter must coexist with each other. Dark energy provides a self-limiting mechanism to limit the growth of dark matter structures.
\newline

\item \noindent Only with the help of dark energy is our universe able to evolve toward the limiting virial equilibrium with a decreasing rate of cascade. The mass and energy cascade only completely vanishes for the system exactly in the limiting virial equilibrium (red star in Fig. \ref{fig:105}). Similarly, haloes evolve toward the limiting virial equilibrium with a decreasing rate of cascade. The energy cascade only vanishes in completely virialized haloes. Haloes in their early stage of evolution have a higher rate of cascade than haloes in their late stage of evolution. However, both real and simulated haloes should always have a finite rate of energy cascade (next section). 
\newline
\end{enumerate}

\section{Energy cascade in haloes}
\label{sec:3-2}
The mass and energy cascade in halo mass space describes the interactions between haloes of different masses. In the radiation and matter eras, the mass and energy cascade established a statistically steady state to continuously release the system energy and maximize the entropy \citep{Xu:2023-Maximum-entropy-distributions-of-dark-matter}. In the dark energy era, the system approaches the limiting virial equilibrium with a slower mass and energy cascade (Fig. \ref{fig:105}). Similarly, there also exists an energy cascade in haloes, while these haloes evolve toward limiting equilibrium. The energy cascade only vanishes for completely virialized haloes. This section quantifies the energy cascade in haloes. The same theory can also be applied to the energy flow in dissipative gases and the associated scaling laws for bulge mass, size, and dynamics \citep{Xu:2024-Cosmic-quenching-and-scaling-laws}.  

To infer the internal structure of dark matter haloes, we examine the energy cascade in haloes at a given mass $m_h$. Similarly to the mass and energy cascade in the halo mass space (Sections \ref{sec:2} and \ref{sec:2-2}), we start by introducing the cumulative functions along the halo radial direction $r$. The cumulative mass function $\Lambda^h_m(m_h,r,z)$ (similar to $\Lambda_m$ in Eq. \eqref{ZEqnNum98631129}) represents the total mass enclosed above the scale $r$. This quantity is averaged for all haloes of the same mass $m_h$
\begin{equation} 
\label{eq:3-2-1} 
\Lambda^h_m(m_h,r,z) = \int_{r}^{\infty} \rho_h \left(m_h,r',z) \right)4\pi r'^2 dr'.  
\end{equation} 
where $\rho_h$ is the mean mass density for all haloes of the same mass. 

As usual (see Eq. \eqref{ZEqnNum502045}), next, we decompose the halo particle velocity $\boldsymbol{\mathrm{v}}_{p}$ into the halo mean velocity, $\boldsymbol{\mathrm{v}}_{h}=\langle \boldsymbol{\mathrm{v}}_{p} \rangle_h$, and the velocity fluctuation, $\boldsymbol{\mathrm{v}}_{p}^{'} $, that is, $\boldsymbol{\mathrm{v}}_{p} =\boldsymbol{\mathrm{v}}_{h} +\boldsymbol{\mathrm{v}}_p'$. Here, $\boldsymbol{\mathrm{v}}_{h}$ represents the velocity of that halo, that is, the average velocity of all particles in the same halo. Consequently, the total kinetic energy $K_p$ of a given particle can be divided into $K_p = K_{ph}+K_{pv}$. The virial kinetic energy, $K_{pv}={\boldsymbol{\mathrm{v}}_p'}^2/2$, is the contribution from the velocity fluctuation due to the intra-halo interactions on small scales in the nonlinear regime (Fig. \ref{fig:SS}). Only this part of the kinetic energy is relevant for the energy cascade in haloes. We will focus on $K_{pv}$ and introduce a cumulative kinetic energy function $\Lambda^h_{pv}$ for $K_{pv}$
\begin{equation} 
\label{eq:3-2-2} 
\begin{split}
&\Lambda^h_{pv}(m_h,r,z) = \int_{r}^{\infty} K_{pv} \rho_h \left(m_h,r',z) \right)4\pi r'^2 dr'. 
\end{split}
\end{equation} 

\begin{figure}
\includegraphics*[width=\columnwidth]{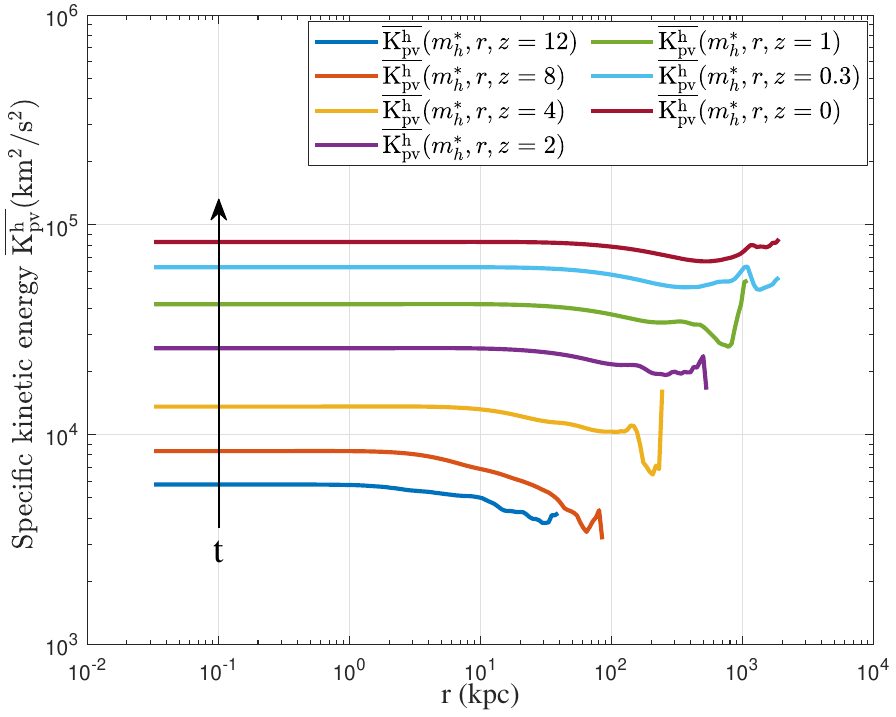}
\caption{The variation of mean specific kinetic energy $\overline{K^h_{pv}}$ with radial scale $r$ for all haloes of characteristic mass $m_h^*(z)$ at different redshift $z$. Data is calculated by Eq. \eqref{eq:3-2-3} and used to compute the rate of energy cascade $\varepsilon(m_h,z)$ in haloes in Fig. \ref{fig:107}. } 
\label{fig:106}
\end{figure}

Next, similarly to the energy cascade in the mass space (Eq. \eqref{ZEqnNum98631129911}), we introduce the specific kinetic energy on scale $r$ for all haloes of the same mass $m_h$ that reads
\begin{equation} 
\label{eq:3-2-3} 
\begin{split}
\overline {K^h_{pv}}(m_h,r,z) = \frac{\Lambda^h_{pv}}{\Lambda^h_{m}}=\frac{\int_{r}^{\infty} K_{pv} \rho_h \left(m_h,r',z) \right)4\pi r'^2 dr'}{\int_{r}^{\infty} \rho_h \left(m_h,r',z) \right)4\pi r'^2 dr'}.
\end{split}
\end{equation} 
Here, $\overline {K^h_{pv}}$ is the specific energy (energy per unit mass) contained on scales above $r$. Figure \ref{fig:106} plots the variation of the specific kinetic energy $\overline {K^h_{pv}}$ with the scale $r$ at different redshifts for haloes of characteristic mass $m_h=m_h^*$. With $\overline {K^h_{pv}}$ increasing with time, the rate of the energy cascade $\Pi^h_{pv}$ along the halo radial direction is defined as   
\begin{equation} 
\label{eq:3-2-4} 
\begin{split}
\Pi^h_{pv}(m_h,r,z) &=-\frac{\partial }{\partial t} \left( \overline {K^h_{pv}} \right) = -\frac{\partial }{\partial t} \left(\frac{\Lambda^h_{pv}}{\Lambda^h_m} \right).
\end{split}
\end{equation}

\begin{figure}
\includegraphics*[width=\columnwidth]{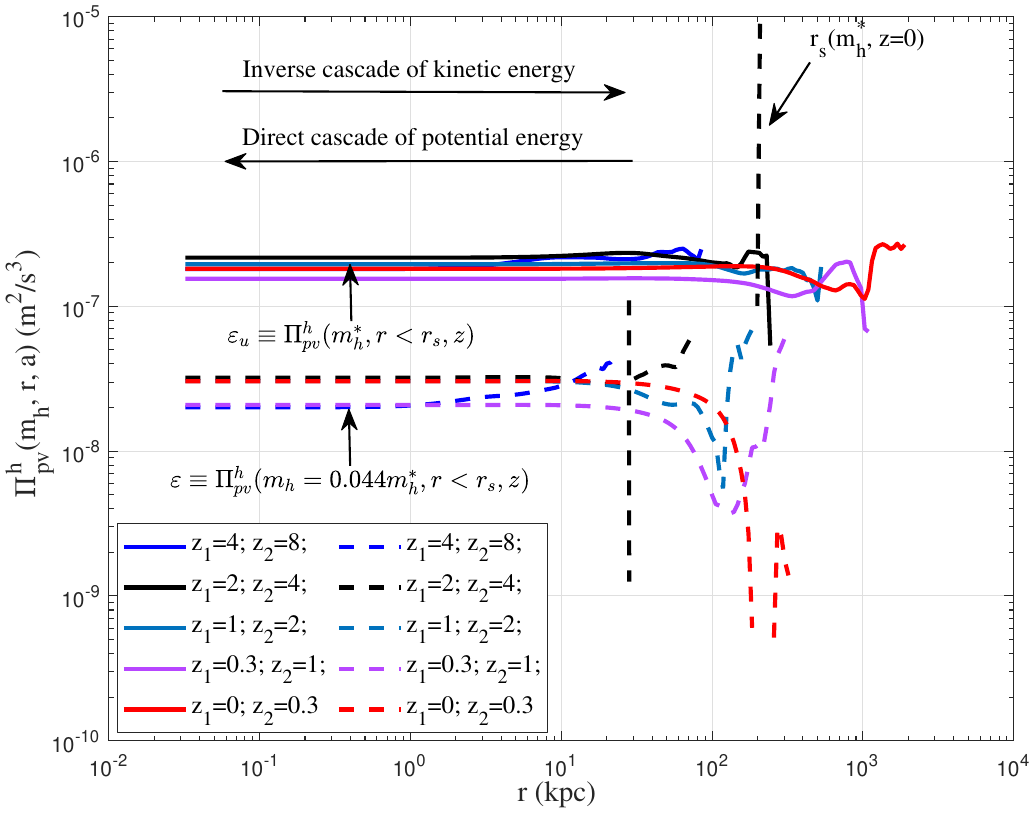}
\caption{The variation of the rate of energy cascade $\Pi^h_{pv}(m_h,r,z)$ (Eq. \eqref{ZEqnNum9863112991}) with radial scale $r$ at different redshifts $z$ for haloes of characteristic mass $m^*_h(z)$ and $m_h=0.044m^*_h(z)$. A scale-independent (independent of $r$) constant rate of $\varepsilon\equiv \Pi^h_{pv}(r<r_s)$ can be identified for an inverse cascade of virial kinetic energy ${K_{pv}}$ from halo center to larger scales. There also exists a simultaneous direct cascade of potential energy from large to the smallest scale at a rate of $-7/5\varepsilon$ (see Eq. \eqref{eq:4-1}). For haloes with characteristic mass $m^*_h$, the rate of energy cascade $\varepsilon(m_h^*,z)\equiv \varepsilon_u\approx 10^{-7}m^2/s^3$.} 
\label{fig:107}
\end{figure}

In a certain range of scales $r<r^*_s$ (inner haloes), the characteristic time scale on the small scale is very fast compared to the time scale on the large scale. The small-scale motion does not feel the slow large-scale motion directly, except through the rate of energy flux $\varepsilon$. Therefore, when a statistical equilibrium is established, similarly to the energy cascade in the halo mass space, we expect a constant $\varepsilon(m_h,z)$ that is independent of scale $r$ in the range $r<r^*_s$, i.e., 
\begin{equation} 
\label{eq:3-2-5} 
\begin{split}
\varepsilon((m_h,z)) \equiv \Pi^h _{pv}(m_h,r,z) =-\frac{\partial \overline {K^h_{pv}}}{\partial t} \quad \textrm{for}\quad r<r_s^*.
\end{split}
\end{equation} 
Figure \ref{fig:107} plots the variation of the parameter $\varepsilon(m_h,z)$ using Eq. \eqref{eq:3-2-3} and the kinetic energy $\overline {K^h_{pv}}$ in Fig. \ref{fig:106}. In this figure, a scale-independent parameter $\varepsilon$ can be clearly observed below a characteristic size $r^*_s$, usually the scale radius. The key parameter $\varepsilon$ increases with the halo mass $m_h$ and the redshift $z$. For haloes with characteristic mass $m^*_h$, the rate of energy cascade $\varepsilon(m_h^*,z)\equiv \varepsilon_u$. 

\begin{figure}
\includegraphics*[width=\columnwidth]{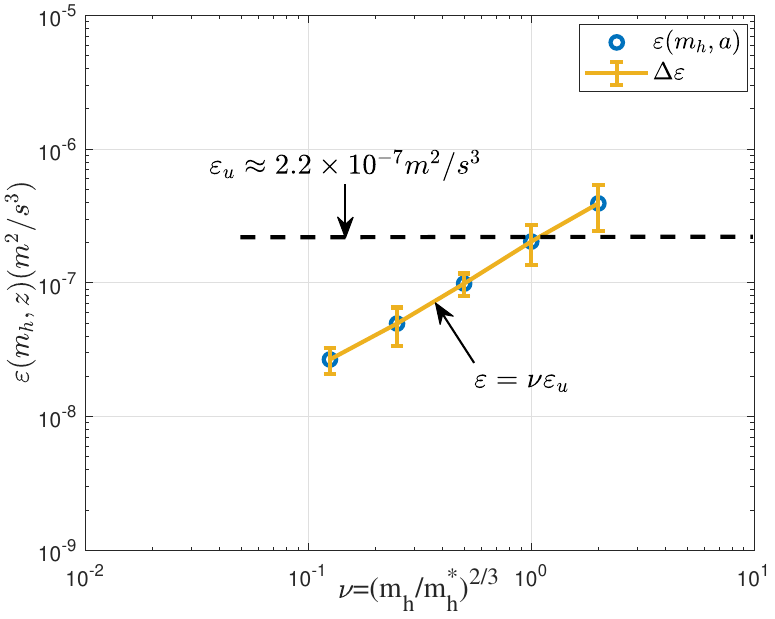}
\caption{The variation of the rate of energy cascade $\varepsilon(m_h,z)$ (Eq. \eqref{ZEqnNum9863112991}) with halo mass $m_h$ at different redshifts $z$ in terms of the dimensionless peak height parameter $\nu=(m_h/m_h^*)^{2/3}$. A scale-independent constant rate of $\varepsilon_u$ can be identified at $\nu=1$ or $m_h=m^*_h$. That rate is also relatively independent of time and is around -$2.5\times 10^{-7}m^2/s^3$ (also see Fig. \ref{fig:S1-3}).} 
\label{fig:108}
\end{figure}

Similarly, we can calculate the rate of the energy cascade for haloes of different masses. Using the data in Fig. \ref{fig:107}, Fig. \ref{fig:108} plots the variation of $\varepsilon(m_h,z)$ with the halo mass $m_h$ and the redshift $z$. The figure shows that $\varepsilon \propto m_h^{2/3}$ and increases with redshift $z$, 
\begin{equation} 
\label{ZEqnNum9863112993} 
\varepsilon(m_h,z) = \varepsilon_u \nu = \varepsilon_u \left({m_h}/{m_h^*}\right)^{2/3}\propto m_h^{2/3}a^{-1}, 
\end{equation} 
where $\varepsilon_u\equiv \varepsilon(m_h^*,z)$ is the rate of energy flow in haloes of characteristic mass $m_h^*$. The peak height parameter of dark matter haloes is defined as $\nu = ({m_h}/{m_h^*})^{2/3}$ \citep{Xu:2023-Dark-matter-halo-mass-functions-and}. The rate of energy cascade decreases with the halo mass and time as $\varepsilon\propto m_h^{2/3}a^{-1}$ while evolving towards the limiting equilibrium. We have $\varepsilon=0$ only in fully virialized haloes, a limiting state that real haloes can never reach.

\section{Scaling laws and density profiles}
\label{sec:3}
Scaling laws are well known to be associated with the energy cascade phenomenon in turbulence \citep{Richardson:1922-Weather-Prediction-by-Numerica, Kolmogoroff:1941-Dissipation-of-energy-in-the-l}. In this section, we focus on the scaling laws related to the energy cascade in spherical haloes and their effects on halo density profiles, which can be used to infer the dark matter mass and properties. 

We first consider the relevant scaling laws. Here, two perspectives are presented that may lead to these scaling laws. In the first perspective, similar to the smallest two-particle haloes (Eq. \eqref{eq:5-22-9}), the rate of the energy cascade in haloes of mass $m_h^*$ can be written as:
\begin{equation} 
\label{eq:3-1-1} 
-\varepsilon_u \propto v_r^3/r=v_r^2/(r/v_r) \quad \textrm{or} \quad v_r^2 \propto (\varepsilon_u r)^{2/3},
\end{equation} 
where $\varepsilon_u$ is the rate of cascade in these haloes (Fig. \ref{fig:107}), $v_r$ is the characteristic velocity on scale $r$, and $t_r=r/v_r$ is the time scale. This states that the kinetic energy $v_r^2$ on the scale $r$ is cascaded to large scales during a turnaround time $t_r$. The two-thirds law ($v_r^2 \propto (\varepsilon_u r)^{2/3}$) can be directly validated by N-body simulations \citep{Xu:2023-Universal-scaling-laws-and-density-slope}. On scale $r$, the virial theorem reads
\begin{equation} 
\label{eq:3-1-2} 
v_r^2 \propto \frac{Gm_r}{r},
\end{equation} 
where $m_r$ is the mass enclosed within the scale $r$. Combining Eq. \eqref{eq:3-1-1} with the virial theorem in Eq. \eqref{eq:3-1-2}, we obtain a five-thirds law for halo mass $m_r \propto \varepsilon_u^{2/3}G^{-1}r^{5/3}$ that can also be confirmed by Illustris simulations \citep{Xu:2023-Universal-scaling-laws-and-density-slope}. Similarly, the density enclosed within scale $r$ should follow $\rho_r \propto m_r/r^3 \propto \varepsilon_u^{2/3}G^{-1}r^{-4/3}$, that is, a four-thirds law for the halo density. 
We start with the two-thirds law, 

\begin{equation} 
\label{eq:10} 
-\lambda _{u} \varepsilon _{u}=\frac{2v_{r}^{2}}{r} v_{r} =\frac{2v_{r}^{2}}{{r/v_{r} }} = \frac{2v_r^2}{t_r},
\end{equation} 
where $\lambda_{u}$ is just a dimensionless numerical constant on the order of unity. Combining Eq. \eqref{eq:10} with the virial theorem in Eq. \eqref{eq:3-1-2}, we can easily obtain the scaling laws for mass scale $m_r$ (mass enclosed within $r$), the density scale $\rho_r$ (mean density of the halo enclosed within $r$), velocity scale $v_r$ (circular velocity at $r$), time $t_r$, and kinetic energy $v_r^2$, all determined by $\varepsilon_u$, $G$ and scale $r$:
\begin{equation} 
\label{eq:11} 
\begin{split}
&m_r = \alpha_r \varepsilon_u^{2/3}G^{-1}r^{5/3} \textrm{,} \quad \rho_r = \beta_r \varepsilon_u^{2/3}G^{-1}r^{-4/3}, \\
&v_r \propto (-\varepsilon_u r)^{1/3} \textrm{,} \quad t_r \propto (-\varepsilon_u)^{-1/3}r^{2/3}, \\
&v_r^2 \propto (\varepsilon_u G m_r)^{2/5},
\end{split}
\end{equation} 
where $\alpha_r\approx 5.28$ and $\beta_r\approx 1.26$ are two constants that can be determined by data fitting. The 2/5 scaling between kinetic energy $v_r^2$ and halo mass $m_r$ was also plotted in Fig. \ref{fig:S1-1-5} and is in good agreement with the Illustris simulation. These scaling laws are for haloes with characteristic mass $m^*_h$ and constant rate of cascade $\varepsilon_u$. For haloes with other masses, we replace $\varepsilon_u$ by a mass-dependent rate of the cascade $\varepsilon$ in Eq. \eqref{ZEqnNum9863112993}. For example, the inner density for the haloes of mass $m_h$ should read (-4/3 law)
\begin{equation} 
\label{eq:11-1} 
\begin{split}
\rho_r(r,m_h,z) = \beta_r \varepsilon^{\frac{2}{3}}G^{-1}r^{-\frac{4}{3}}=\beta_r \varepsilon_u^{\frac{2}{3}}G^{-1}r^{-\frac{4}{3}}\left(\frac{m_h}{m_h^*(z)}\right)^{\frac{4}{9}}. 
\end{split}
\end{equation} 
The model predicts the halo density $\rho_h(r)\propto m_h^{4/9}a^{-2/3}$ for halos of different masses with $m_h^*\propto a^{3/2}$ (Table \ref{tab:4}). In addition, scaling laws in Eq. \eqref{eq:11} predicts a limiting density slope of $\gamma=-4/3$ in the halo core region. Therefore, the inner structure of the haloes can be determined by the energy cascade in the haloes. 

We next focus on the halo density profile. The well-known core-cusp problem describes the discrepancy between cuspy halo density predicted by cosmological CDM-only simulations and core density from observations. The predicted halo density exhibits a cuspy inner density $\rho\propto r^{\gamma}$ with a wide range of $\gamma$ between -1.0 and -1.5. The NFW profile is a very popular model with a density slope of $\gamma=-1.0$ \citep{Navarro:1997-A-universal-density-profile-fr}. The density slopes of the simulated haloes are also found to be $\gamma>-1.0$ \citep{Navarro:2010-The-diversity-and-similarity-of-simulated}, $\gamma=-1.2$ \citep{Diemand:2011-The-Structure-and-Evolution-of-Cold-Dark}, and $\gamma=-1.3$ \citep{Governato:2010-Bulgeless-dwarf-galaxies-and-dark-matter-cores,McKeown:2022-Amplified-J-factors-in-the-Galactic-Centre}. There seems to be no consensus on the exact value of the asymptotic slope $\gamma$ and no solid theory for the density slope $\gamma$.  

From the mass continuity equation for a spherical halo \citep{Xu:2023-Universal-scaling-laws-and-density-slope}, we can demonstrate that the density slope $\gamma$ is highly dependent on the mean radial flow $u_{r} \left(r,t\right)$ (flow along the radial direction) as
\begin{equation} 
\label{eq:16-2} 
\begin{split}
&\gamma \approx \frac{\nu}{1-\mu}-3, \quad \nu=\frac{\partial \ln m_r(r_s)}{\partial \ln r_s},\quad \mu = \frac{u_rt}{r}\bigg /\frac{\partial \ln r_s}{\partial \ln t}.
\end{split}
\end{equation} 
Here, $r_s$ is the scale radius where the logarithmic slope is -2, and $m_r(r_s)$ is the total mass enclosed within the scale radius $r_s$. The parameter $\mu$ represents the effect of the radial flow ($u_r$). A higher radial flow $u_r$ or greater $\mu$ leads to a smaller $\gamma$ and a flatter density. The parameter $\nu=5/3$ represents the effect of the energy cascade, as we can find from the 5/3 law between $m_r$ and $r$ in Eq. \eqref{eq:11}.

Based on the effects of radial flow and energy cascade, we will consider the following scenarios:
\begin{enumerate}
\item \noindent Individual haloes with both non-zero radial flow and non-vanishing energy cascade ($u_r\ne 0$, $\varepsilon\ne 0$). Most simulated haloes are non-equilibrium dynamic objects that have both a nonzero radial flow and a non-vanishing energy cascade. The density slopes of individual haloes depend on both the radial flow and the mass accretion of each halo, as shown in Eq. \eqref{eq:16-2}. This may be the reason for the wide variety of density slopes $\gamma$ for simulated haloes. In previous work, we have analytically derived the halo density profiles based on the concept of a mass and energy cascade \citep{Xu:2023-Dark-matter-halo-mass-functions-and}. Simulated haloes with different density slopes can generally be modeled by a double-$\gamma$ density profile
\begin{equation}
\label{eq:27-2} 
\rho_h(r,t)= \rho_s(t) \left(\frac{r}{r_s}\right)^{\frac{\alpha}{\beta}-2} \exp\left(\frac{1}{\beta}\left(1-{\left(\frac{r}{r_s}\right)^{\alpha}}\right)\right),
\end{equation}
where $\rho_s(t)$ is the density at scale radius $r_s(t)$. This four-parameter double-$\gamma$ density profile ($\rho_s$, $r_s$, $\alpha$, and $\beta$ in Eq. \eqref{eq:27-2}) reduces to the standard three-parameter Einasto profile with $\alpha=2\beta$. 
\newline


\begin{figure}
\includegraphics*[width=\columnwidth]{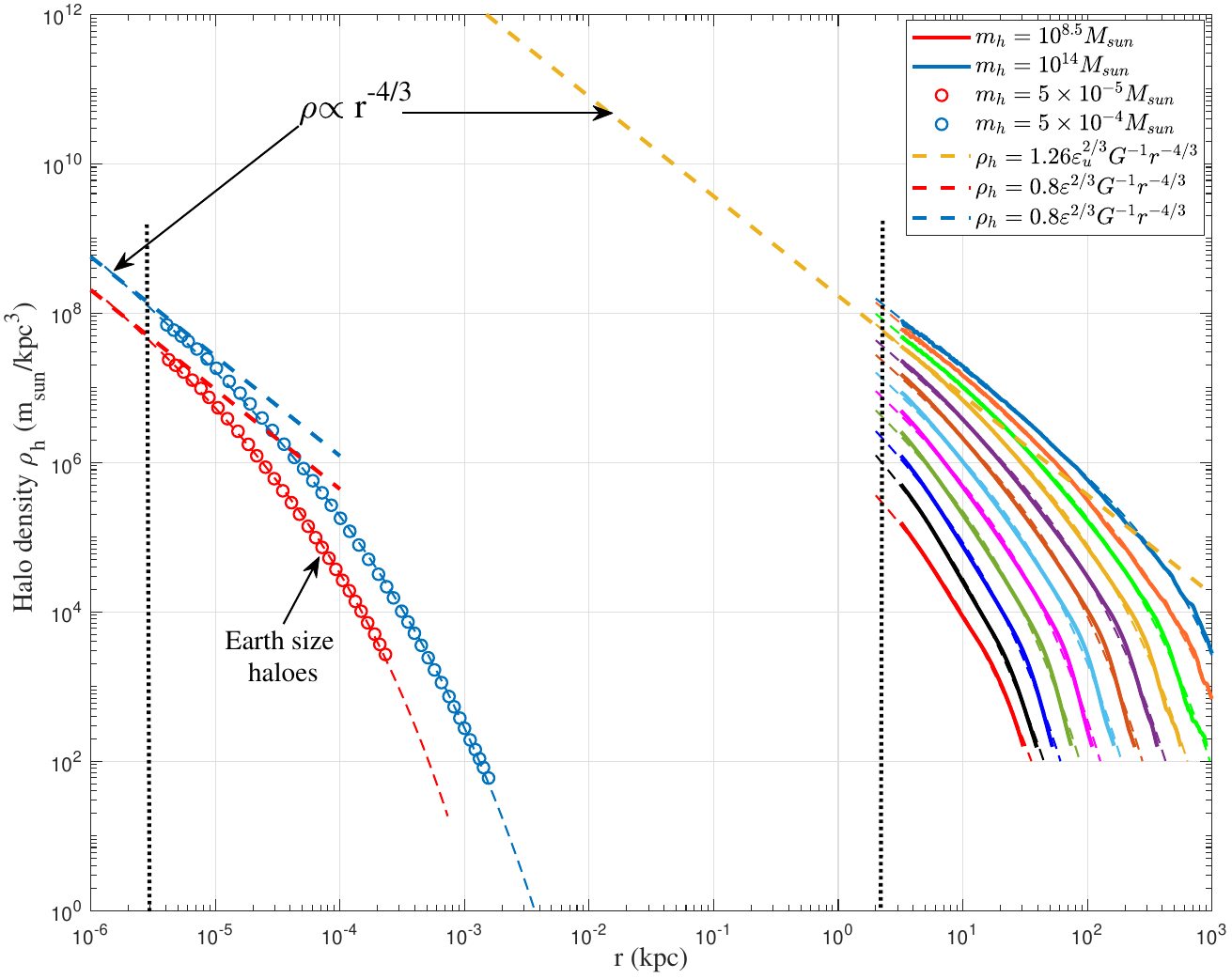}
\caption{The density profiles from cosmological simulations for haloes of different mass at $z=0$. The solid lines present the average density profiles for all haloes with a mass between $10^{\pm 0.1}m_h$ from the Illustris simulation for galactic haloes of mass $m_h=[10^{8.5} 10^{14}]M_{\odot}$. The symbols present two Earth-size simulated haloes \citep{Wang:2020-Universal-structure-of-dark-matter-haloes-over-a-mass-range}. The thin dashed lines present the double-$\lambda$ density profile from Eq. \eqref{eq:29-2}. The thick dashed (straight) lines represent the -4/3 scaling law for haloes of mass $m_h$ with the rate of energy cascade $\varepsilon\propto m_h^{2/3}$ (Eq. \eqref{eq:29-1}), and for haloes of characteristic mass $m_h^*(z=0)$ (Eq. \eqref{eq:29-3}). The dotted lines indicate the softening length. The figure demonstrates the -4/3 law from the concept of energy cascade for the density profiles of haloes of different sizes over 20 orders of magnitude. The density profiles of haloes with mass $m_h^*(z)$ at different redshifts $z$ are presented in Fig. \ref{fig:S1-4}.} 
\label{fig:103}
\end{figure}

\item \noindent Haloes with vanishing radial flow and non-vanishing energy cascade ($u_r=0$ and $\varepsilon\ne 0)$). Individual haloes have random radial flow. However, in N-body simulations, we can average out radial flow by constructing composite haloes of a given mass $m_h$ from all haloes of the same mass. This is what we did in Section \ref{sec:3-2}, where relevant quantities are averaged for all haloes with the same mass. For these composite haloes, the radial flow vanishes, and we only have the effect of the energy cascade ($\mu=0$ and $\nu=5/3$ in Eq. \eqref{eq:16-2}) such that the inner density slope $\gamma=-4/3$. The inner halo region with vanishing radial flow or $u_{r}=0$ is also expected from the stable clustering hypothesis; that is, there is no net stream motion in physical coordinate along the radial direction \citep{Mo:2010-Galaxy-formation-and-evolution, Xu:2023-On-the-statistical-theory-of-self-gravitating}. For haloes with no radial flow, we would expect an inner density with a slope of $-4/3$. The corresponding double-density profile (let $\alpha/\beta=2/3$ in Eq. \eqref{eq:27-2}) should read \citep{Xu:2023-Dark-matter-halo-mass-functions-and}
\begin{equation}
\label{eq:29-2} 
\begin{split}
&\rho_h(r,m_h,z) = \beta_r\varepsilon^{\frac{2}{3}}G^{-1}r_s^{-\frac{4}{3}}\left(\frac{r}{r_s}\right)^{-\frac{4}{3}}\exp\left[-\frac{1}{\beta}\left(\frac{r}{r_s}\right)^{\frac{2\beta}{3}}\right],\\
&\varepsilon(m_h,z) = \left({m_h}/{m_h^*(z)}\right)^{{2}/{3}}\varepsilon_u,
\end{split}
\end{equation}
where $\beta_r\approx 1$ is an amplitude parameter, $\beta$ is a shape parameter, and $r_s$ is the scale radius. Here, $\varepsilon_u=-4.6\times 10^{-7}m^2/s^3$ is the rate of the energy cascade for haloes with characteristic mass $m_h^*(z)$. While $\varepsilon(m_h,z)$ is the rate of the energy cascade in haloes of any mass $m_h$ (Eq. \eqref{ZEqnNum9863112993}). For small $r$, the inner density reduces to (recovers the scaling law in Eq. \eqref{eq:11-1})
\begin{equation}
\label{eq:29-1} 
\rho_h(r,m_h,z) = \beta_r \varepsilon^{2/3}G^{-1}r^{-4/3} \quad \textrm{for} \quad r\rightarrow 0,
\end{equation}
where the redshift dependence is incorporated into $\varepsilon$, which is dependent on the characteristic mass $m_h^*(z)$ (Eq. \eqref{eq:29-2}). For haloes of characteristic mass $m_h^*(z)$, there exists a small-scale permanence for the halo density, i.e., the density profiles at different redshifts $z$ converge to a time-unvarying scaling,
\begin{equation}
\label{eq:29-3} 
\rho_h(r,m_h^*,z) \equiv \rho_h(r) =\beta_r \varepsilon_u^{2/3}G^{-1}r^{-4/3} \quad \textrm{for} \quad r\rightarrow 0.
\end{equation}
The small-scale permanence is shown and discussed in Fig. \ref{fig:S1-4}. 
\newline

\item \noindent The last scenario is completely virialized haloes with both vanishing radial flow and vanishing energy cascade ($u_r=0$ and $\varepsilon=0$). This is the limiting virial equilibrium state (simulated or real) haloes evolve toward but can never reach. The $-4/3$ density slope should still be good in this limiting state.
\end{enumerate}

To validate the predicted scaling laws, we first present the density profiles from cosmological simulations for haloes of different masses $m_h$ at $z=0$. In Fig. \ref{fig:103}, the solid lines present the average halo density profiles for all haloes with a mass between $10^{\pm 0.1}m_h$ from the Illustris simulation, where $10^8M_{\odot}<m_h<10^{14}M_{\odot}$. The symbols present two Earth-size simulated haloes with a mass of $5\times 10^{-5}M_{\odot}$ and $5\times 10^{-4}M_{\odot}$, respectively \citep{Wang:2020-Universal-structure-of-dark-matter-haloes-over-a-mass-range}. The thin dashed lines present the double-$\lambda$ density profile from Eq. \eqref{eq:29-2}. The thick dashed (straight) lines represent the -4/3 scaling laws for haloes of mass $m_h$ that involve the rate of cascade $\varepsilon\propto m_h^{2/3}$ (Eq. \eqref{eq:29-1}). Here, $\varepsilon=\varepsilon_u$ for haloes of characteristic mass $m_h^*$ (Eq. \eqref{eq:29-3}). The figure demonstrates the -4/3 law that we obtained from the concept of energy cascade. That scaling was in agreement with the density profiles of haloes of more than 20 orders of magnitude. 

\begin{figure}
\includegraphics*[width=\columnwidth]{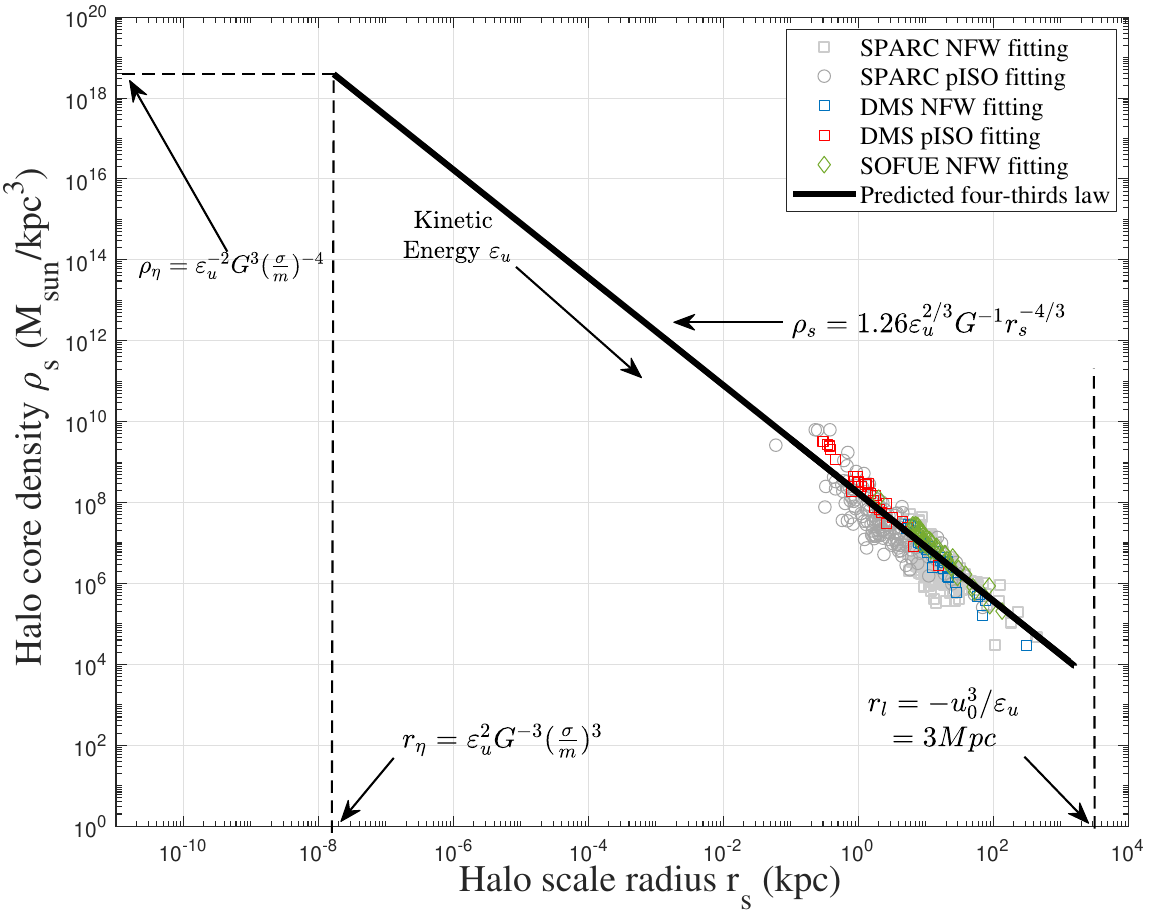}
\caption{The variation of halo core density $\rho_s$ with the halo scale radius $r_s$ from galaxy rotation curves. Each data represents the result for a given galaxy. The predicted -4/3 law in Eq. \eqref{eq:11} is also plotted for comparison (solid black line). The good agreement confirms the predicted scaling laws from the energy cascade. For self-interacting dark matter model with a cross-section $\sigma/m$, there exists the smallest structure with a size $r_{\eta}$ and a maximum density $\rho_{\eta}$ determined by $\varepsilon_u$, $G$, and $\sigma/m$ (Table \ref{tab:2}). The largest halo size $r_l=-u_0^3/\varepsilon_u$ is determined by the velocity dispersion $u_0$ and $\varepsilon_u$.}
\label{fig:5}
\end{figure}

Observational evidence of the predicted scaling laws also exists. The four-thirds law $\rho_r(r) \propto r^{-4/3}$ for the halo mass density enclosed within the scale $r$ can also be directly compared against data from galaxy rotation curves (Fig. \ref{fig:5}). Important information for dark matter haloes can be extracted from galaxy rotation curves by decomposing them into contributions from different mass components. Once the halo density model is selected, the scale radius $r_s$ and the mean density $\rho_s$ within $r_s$ can be rigorously obtained by fitting the decomposed rotation curve. In this work, for pseudo-isothermal (pISO) models \citep{Adams:2014-Dwarf-Galaxy-Dark-Matter-Density} and NFW density models \citep{Navarro:1997-A-universal-density-profile-fr}, three sources of rotation curves are used to extract $r_s$ and density $\rho_s$ within $r_s$,
\begin{enumerate}
\item \noindent SPARC (Spitzer Photometry \& Accurate Rotation Curves) including 175 late-type galaxies \citep{Lelli:2016-SPARC-Mass-Models-for-175-Disk-Galaxies,Li:2020-A-Comprehensive-Catalog-of-Dark-Matter-Halo-Models};
\item \noindent DMS (DiskMass Survey) including 30 spiral galaxies \citep{Martinsson:2014-The-DiskMass-Survey};
\item \noindent SOFUE (compiled by Sofue) with 43 galaxies \citep{Sofue:2016-Rotation-curve-decomposition-f}.
\end{enumerate}
Figure \ref{fig:5} presents the variation of the halo core density $\rho_s$ with the scale radius $r_s$ obtained from the galaxy rotation curves (square and circle symbols). Each symbol represents data from a single galaxy. The four-thirds law (Eq. \eqref{eq:11}) is also plotted (black line) with constants $\beta_r=1.26$ (equivalent to $\alpha_r=5.28$) obtained from these data. From this figure, dark matter haloes from galaxy rotation curves follow the -4/3 law across six orders of halo mass. 

\begin{figure}
\includegraphics*[width=\columnwidth]{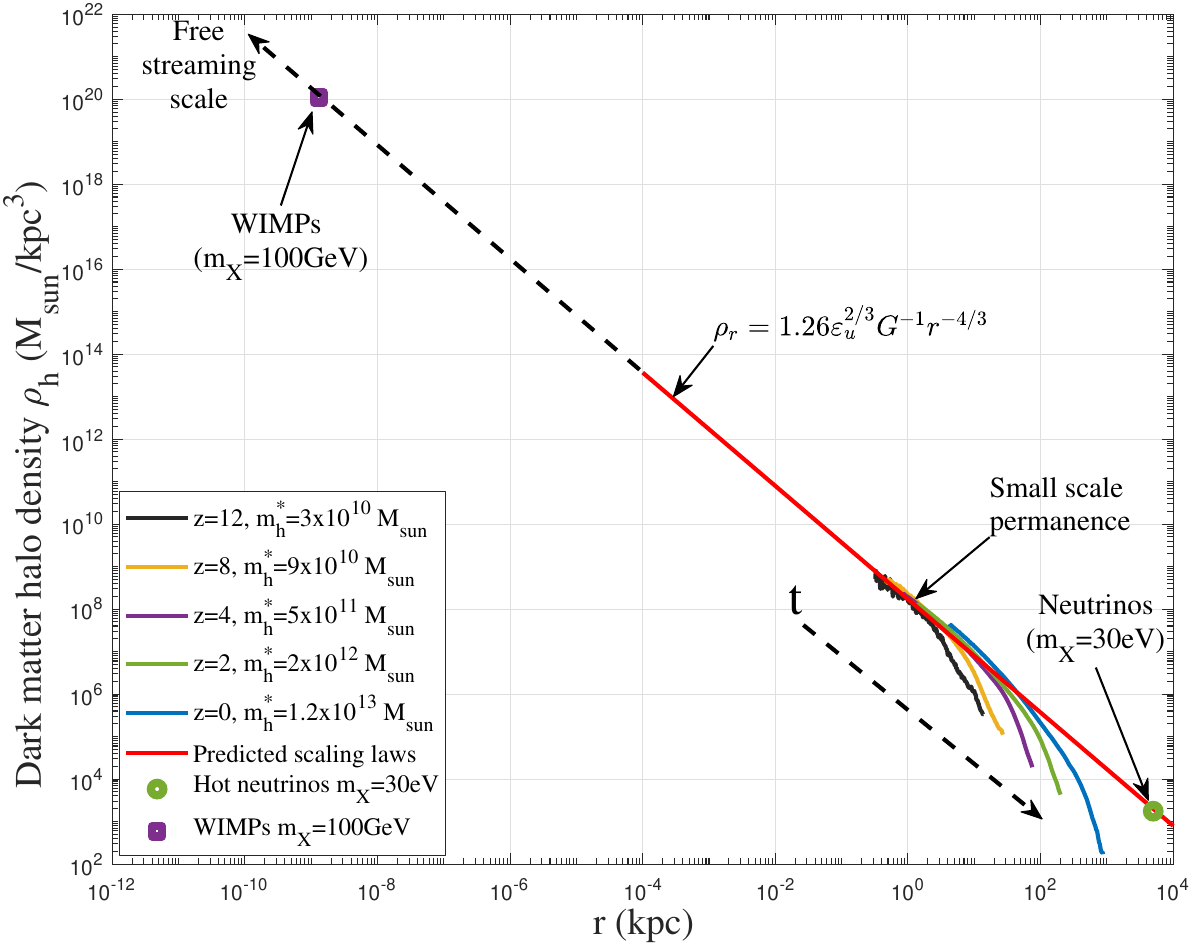}
\caption{The redshift evolution of halo density profiles for haloes with a characteristic mass $m_h^*(z)$. Density profiles for haloes with other masses (from the galaxy cluster to the Earth size) are presented in Fig. \ref{fig:103}. This figure demonstrates the small-scale permanence, i.e., the density profiles for haloes with a characteristic mass $m_h^*(z)$ at different redshifts $z$ all collapse at a small scale $r$ onto the predicted time-unvarying scaling (solid red line from Eq. \eqref{eq:29-3}). Due to the time and scale-independent rate of the cascade $\varepsilon_u$, the -4/3 scaling should extend to smaller scales (or earlier time) until reaching the smallest scale (or the formation time of the smallest structures) that depends on the nature of dark matter (Sections \ref{sec:5-1}). The halo mass, density, size, and formation time at the smallest scale are calculated in Eq. \eqref{eq:111}. Two examples, hot neutrinos (green dot) and WIMPs (purple square) are plotted in the same figure and listed in Table \ref{tab:6}.} 
\label{fig:S1-4}
\end{figure}

Finally, we are especially interested in the redshift evolution of haloes with a characteristic mass $m_h^*(z)$, which is representative. Figure \ref{fig:S1-4} shows the time evolution of the average density profiles for all haloes with characteristic masses $m_h^*(z)$ from Illustris simulations. By averaging out the radial flow, the energy cascade should give rise to a slope of $\gamma=-4/3$ (Eq. \eqref{eq:16-2}). At higher redshifts, dark matter haloes tend to be smaller with higher density. The density profiles of all dark matter haloes converge to the predicted time-unvarying scaling (solid red line) from Eq. \eqref{eq:29-3}, i.e., the small-scale permanence that can be confirmed by N-body simulations. 

The -4/3 scaling law in Fig. \ref{fig:S1-4} should extend to smaller and smaller scales until the scale of the smallest halo. That scale can be the free streaming scale, which is dependent on the nature and mass of dark matter particles (Sections \ref{sec:5-1}). The heavier particles have a smaller free streaming scale. For sufficiently heavy DM particles of mass $m_{Xc}\approx 10^{12}$GeV, the free streaming mass can be comparable to the particle mass (Fig. \ref{fig:109}) such that the -4/3 scaling can extend to the scale of the smallest structure (two-particle haloes). Since smaller structures were formed earlier (as early as the radiation era in Section \ref{sec:5-2}), extending to smaller scales is equivalent to extending to an earlier time. For dark matter candidates with known free streaming mass $M_{fs}$, scaling laws in Eq. \eqref{eq:11} can be used to calculate the relevant quantities for the smallest structure of the free streaming scale,
\begin{equation} 
\label{eq:111} 
\begin{split}
&r_{fs} = \alpha_r^{-3/5} \varepsilon_u^{-2/5}G^{3/5}M_{fs}^{3/5},\\
&\rho_{fs} = \beta_r\alpha_r^{4/5}\varepsilon_u^{6/5}G^{-9/5}M_{fs}^{-4/5},\\
&t_{fs} = \varepsilon_u^{-3/5}G^{2/5}M_{fs}^{2/5},\\
\end{split}
\end{equation}
where $r_{fs}$, $\rho_{fs}$, and $t_{fs}$ are the size, density, and formation time of the smallest haloes with a free streaming mass $M_{fs}$.

In Fig. \ref{fig:S1-4}, two examples are presented, that is, the hot neutrinos with $M_{fs}=10^{15}$M$_{\odot}$ (green dot) and standard WIMPs with $M_{fs}=10^{-6}$M$_{\odot}$ (purple square). Table \ref{tab:6} lists the relevant quantities calculated for the smallest halo structure formed by three dark matter candidates. The particle $X$ with critical mass $m_X=M_{fs}=10^{12}$ GeV was also listed, which has been extensively discussed in Sections \ref{sec:5-1} and \ref{sec:5-2}. Obviously, the cold dark matter (WIMPs and X) leads to much smaller structures formed in the earlier universe. Finally, at what scale should the -4/3 scaling stop operating strongly suggests the nature of dark matter particles (Section \ref{sec:5}).

\begin{table}
    \begin{center}
    \caption{Dark matter particle candidates and the free streaming scale}
    \label{tab:6}
    \begin{tabular}{lcccc} 
    \hline
    Quantities          &Symbol  &Hot neutrinos          & WIMPs                 & X particle     \\
    \hline
    Particle mass       &$m_X$   &30eV                   & 100GeV                & $10^{12}$GeV   \\
    Free streaming      &$M_{fs}$ &$10^{15}$M$_{\odot}$     & $10^{-6}$M$_{\odot}$    & $10^{12}$GeV        \\
    Size                &$r_{fs}$ &5Mpc                   & $10^{-9}$kpc          & $10^{-13}$m              \\
    Density             &$\rho_{fs}$ &$10^{-25}$kg/m$^3$     & $10^{-8}$kg/m$^3$     & $10^{23}$kg/m$^3$     \\
    Formation time   &$t_{fs}$    &$10^{11}$yrs           & 300yrs                & $10^{-6}$s \\
    \hline
    \end{tabular}
  \end{center}
\end{table}

\section{Particle mass and properties from scaling laws}
\label{sec:5}

In this section, we attempt to estimate the mass and properties of the cold dark matter particles from the established scaling laws in Section \ref{sec:3}. Since the rate of the energy cascade $\varepsilon_{u}$ in Eq. \eqref{eq:5} is both scale- and time-independent, the scaling laws should extend to the smallest scale (Fig. \ref{fig:S1-4}) and the earliest time when the smallest structure was formed, as early as in the radiation era (Section \ref{sec:5-2}). Here, we consider X particles with a mass $m_X\ge m_{Xc}$, or the free streaming mass is less than the particle mass (Fig. \ref{fig:109}). In this case, the smallest structures can be formed by two and only two particles. The scaling laws may extend to a scale where quantum effects become important. For particles with a mass $m_X\ll m_{Xc}$, the smallest haloes are determined by the free streaming mass (Eq. \eqref{eq:111}).  

Assuming that gravity is the only interaction between unknown dark matter particles (traditionally denoted by \textbf{\textit{X}}), the only dominant physical constants on the smallest scale are the reduced Planck constant $\hbar$ (the quantum effect), the gravitational constant $G$ (the gravitational interactions) and the rate of energy cascade $\varepsilon _{u}$. Without involving complex quantum field theory for a more rigorous treatment, a formal dimensional analysis often provides significant insights. Any physical quantities $Q$ on that scale can be expressed as $Q=\varepsilon_u^xG^y\hbar^z$, where $x$, $y$, and $z$ can be uniquely determined by dimensional analysis. Examples are the mass and length scales,
\begin{equation}
m_{X} =\left(-{\varepsilon _{u} \hbar ^{5}G^{-4} } \right)^{\frac{1}{9} } 
\label{eq:12}
\end{equation}
and
\begin{equation}
l_{X} =\left(-\varepsilon _{u}^{-1}G\hbar\right)^{\frac{1}{3} }. 
\label{eq:13}
\end{equation}

\begin{figure}
\includegraphics*[width=\columnwidth]{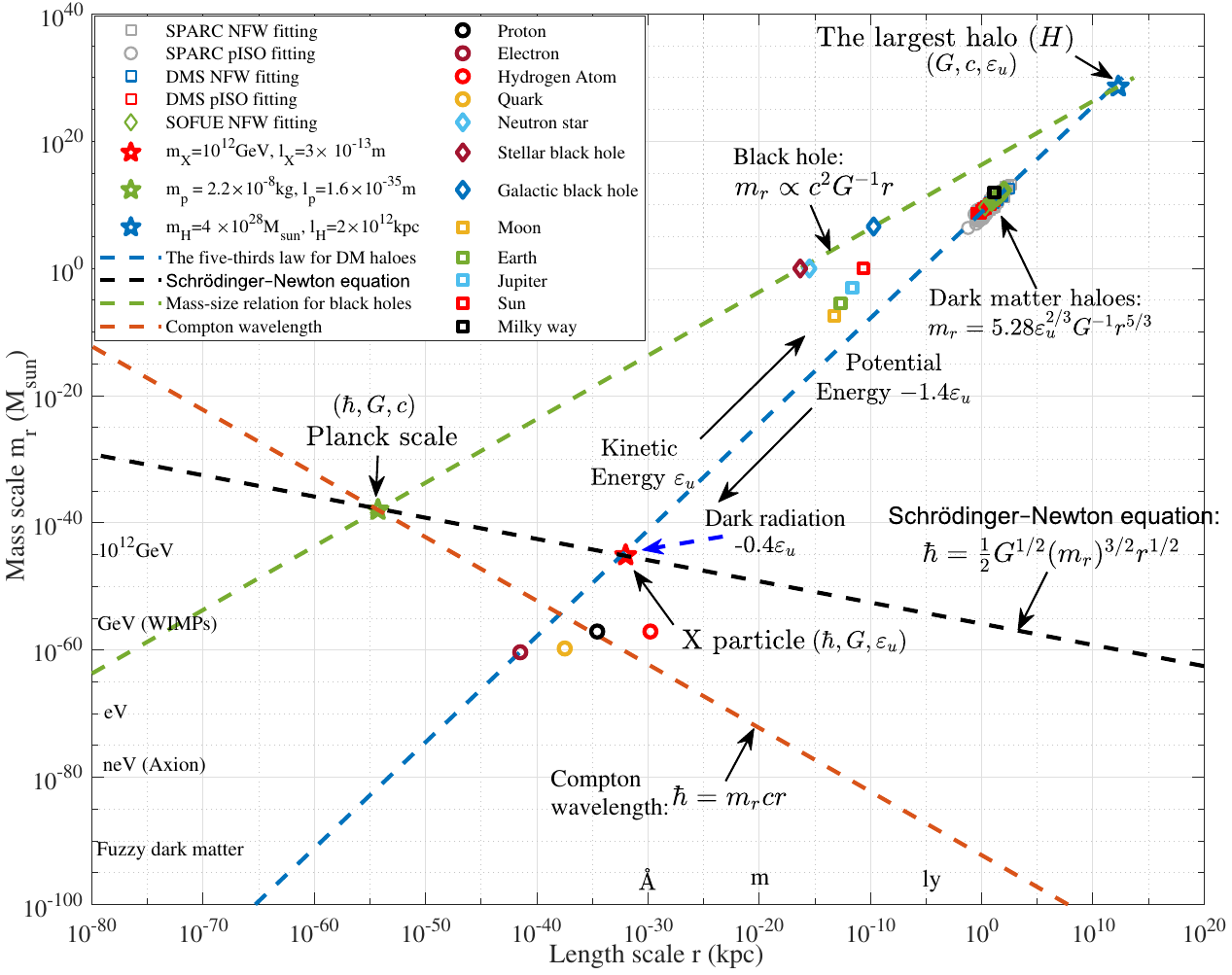}
\caption{The variation of mass scale $m_r$ and size scale $r$ for cold dark matter particle properties. This figure estimates the cold dark matter particle properties on small scales that are consistent with the dynamics of dark matter haloes on large scales. Collisionless dark matter haloes follow the scaling $m_r\propto r^{5/3}$ (the blue dashed line for the five-thirds law in Eq. \eqref{eq:11}) due to the energy cascade at a constant rate $\varepsilon_u$ for kinetic energy and $-1.4\varepsilon_u$ for potential energy (Eq. \eqref{eq:4-1} and Fig. \ref{fig:S1-3-2}). 
Dark matter haloes from different surveys of galaxy rotation curves in Fig. \ref{fig:5} follow the five-thirds law. The black dashed line represents the combined quantum and gravitational effects of Eqs. \eqref{eq:14} and \eqref{eq:16}. The green dashed line gives the mass of black holes (BHs) at a given size $r$ (Schwarzschild radius) from stellar to galactic BHs, i.e., the maximum mass at any given $r$. By extending the five-thirds law (blue dashed line) for dark matter haloes to the smallest scale (intersecting the black dashed line), the red star represents the cold dark matter particles ($X$) with predicted mass $m_X$ and size $l_X$ from Eqs. \eqref{eq:12} and \eqref{eq:13} (also see Table \ref{tab:3}). By extending the five-thirds law (blue dashed line) for dark matter haloes to the largest scale to intersect the green dashed line (the blue star), we will obtain the largest scales (the $H$ scale) of halo size ($l_H$), mass ($m_H$), and time ($t_H$ or DM lifetime $\tau_X$ in Eq. \eqref{eq:24-1-2}) (see table \ref{tab:3}). The green star represents the Planck scale. For comparison, the mass and size of some particles and astronomical objects are also shown.}
\label{fig:102}
\end{figure}

Alternatively, the same results can be obtained from a refined treatment to couple relevant physical laws on the smallest scale. This may offer a complete view than a simple dimensional analysis. Let us consider two \textbf{\textit{X}} particles on the smallest scale with separation $2r_X=l_{X}$ in the rest of the center of mass of two particles. On that scale, three relevant physics are as follows:
\begin{equation} 
\label{eq:14} 
m_{X} V_{X} \cdot {l_{X} /2} =\hbar ,          
\end{equation} 
\vspace*{-15pt}
\begin{equation} 
\label{eq:16} 
{Gm_{X} /l_{X}^{} } =2V_{X}^{2} ,          
\end{equation} 
\vspace*{-15pt}
\begin{equation} 
\label{eq:15} 
{2V_{X}^3/l_X } = a_X \cdot v_X = -\lambda_u \varepsilon_u ,           
\end{equation} 
where Eq. \eqref{eq:14} is from the uncertainty principle for momentum and position if \textbf{\textit{X}} particles exhibit wave-particle duality. Equivalently, we can also treat $l_{X}/2$ as the de Broglie wavelength of the matter wave of \textbf{\textit{X}} particle. Equation \eqref{eq:16} is from the virial theorem for the potential energy (V) and the kinetic energy (T), that is, $2T=-V$. Here, $V_X$ is the relative velocity between two particles due to their interactions and contributes to the small-scale virial kinetic energy $K_{pv}$. The last equation \eqref{eq:15} is from the two-thirds (2/3) law in Eq. \eqref{eq:10} associated with the energy cascade in dark matter reflecting the "uncertainty" principle between particle acceleration and velocity \citep{Xu:2023-Universal-scaling-laws-and-density-slope}. Since the energy cascade rate $\varepsilon_{u}$ is both scale- and time-independent, we may extend the 2/3 law down to the smallest scale for dark matter particle properties. By this approach, the dark matter particle properties on small scales are consistent with the established scaling laws for dynamics of dark matter haloes on large scales. With the following values for three constants, 
\begin{equation} 
\label{eq:17} 
\begin{split}
&\varepsilon _{u} =-4.6\times 10^{-7} {m^{2} /s^{3} },\\
&\hbar =1.05\times 10^{-34} {kg\cdot m^{2} /s},\\ 
&G=6.67\times 10^{-11} {m^{3} /(kg\cdot s^{2})} , 
\end{split}
\end{equation} 
Complete solutions of Eqs. \eqref{eq:14}, \eqref{eq:16}, and \eqref{eq:15} can be obtained (with $\lambda_{u} =1$), relevant quantities on the $X$ scale are (Fig. \ref{fig:102}), 
\begin{equation} 
\label{eq:19} 
\begin{split}
m_{X} =\left(-\frac{256\lambda _{u} \varepsilon _{u} \hbar ^{5} }{G^{4} } \right)^{\frac{1}{9} }=1.62\times 10^{-15} kg\approx10^{12} GeV,
\end{split}
\end{equation} 

\begin{equation} 
\label{eq:18} 
\begin{split}
&l_{X} =\left(-\frac{2G\hbar }{\lambda _{u} \varepsilon _{u} } \right)^{\frac{1}{3} } =3.12\times 10^{-13} m,\\
&t_{X} =\frac{l_{X} }{V_{X} } =\left(-\frac{32G^{2} \hbar ^{2} }{\lambda _{u}^{5} \varepsilon _{u}^{5} } \right)^{\frac{1}{9} } =7.51\times 10^{-7} s,
\end{split}
\end{equation} 

\begin{equation} 
\label{eq:20} 
\begin{split}
&V_{X} =\left(\frac{\lambda _{u}^{2} \varepsilon _{u}^{2} \hbar G}{4} \right)^{\frac{1}{9} } =4.16\times 10^{-7} {m/s},\\
&a_{X} =\left(-\frac{4\lambda _{u}^{7} \varepsilon _{u}^{7} }{\hbar G} \right)^{\frac{1}{9} } =1.11{m/s^{2} } .
\end{split}
\end{equation} 

Note that the solutions of mass $m_X$ and length scale $l_X$ agree with those obtained from a formal dimensional analysis in Eqs. \eqref{eq:12} and \eqref{eq:13}. The predicted particle mass is much heavier than standard WIMPs and is of the same order as the critical mass $m_{Xc}$ in Eq. \eqref{eq:12-9}. Therefore, the corresponding free streaming mass for particles of this mass should be comparable to the particle mass, allowing us to extend the scaling laws to the particle mass scale. 

The time scale $t_{X}$ (formation time of the smallest structures) is close to the characteristic time for weak interactions ($10^{-6}\sim10^{-10} s$). In contrast, the length scale $l_{X}$ is greater than the characteristic range of strong interactions ($\sim 10^{-15} m$) and weak interactions ($\sim 10^{-18} m$). The "thermally averaged cross section" of \textbf{\textit{X}} particle is around $l_{X}^{2}V_{X}=4\times 10^{-32} {m^{3}/s}$. This is on the same order as the cross-section required for the correct abundance of today via thermal production ("WIMP miracle"), where $\langle \sigma v \rangle \approx 3 \times10^{-32}m^3s^{-1}$. The "cross section $\sigma/m$" for \textbf{\textit{X}} particle is extremely small, i.e. $l_{X}^{2}/m_{X} = 6 \times 10^{-11} m^2/kg$, i.e., a fully collisionless dark matter. In addition, a new constant $\mu _{X}$ (the scale for the rate of energy dissipation) can be introduced,
\begin{equation} 
\label{eq:21} 
\begin{split}
\mu _{X}&=m_{X} a_{X} \cdot V_{X}=F_{X} \cdot V_{X} =-m_{X} \varepsilon _{u} \\
&=\left(-\frac{256\varepsilon _{u}^{10} \hbar ^{5} }{G^{4} } \right)^{\frac{1}{9} } =7.44\times 10^{-22} {kg\cdot m^{2} /s^{3} }  
\end{split}
\end{equation} 
which is a different representation of $\varepsilon _{u}$. In other words, the fundamental physical constants on the smallest scale can be $\hbar$, $G$, and the constant $\mu_{X}$. Consequently, an important energy scale is set by 
\begin{equation}
\label{eq:21-2} 
E_X={\mu _{X} t_{X} /4} ={\hbar /t_{X} } = \sqrt{\hbar\mu_{X}}/2 =0.87\times 10^{-9} \textrm{eV}.
\end{equation}
This energy scale is the binding energy of two X particles in the ground state (Eq. \eqref{eq:5-22-17}). It is equivalent to a Compton wavelength of 1.4 km or a frequency of 0.2 MHz. This can be relevant to the possible dark "radiation" (Section \ref{sec:6-1}). 

\begin{figure}
\includegraphics*[width=\columnwidth]{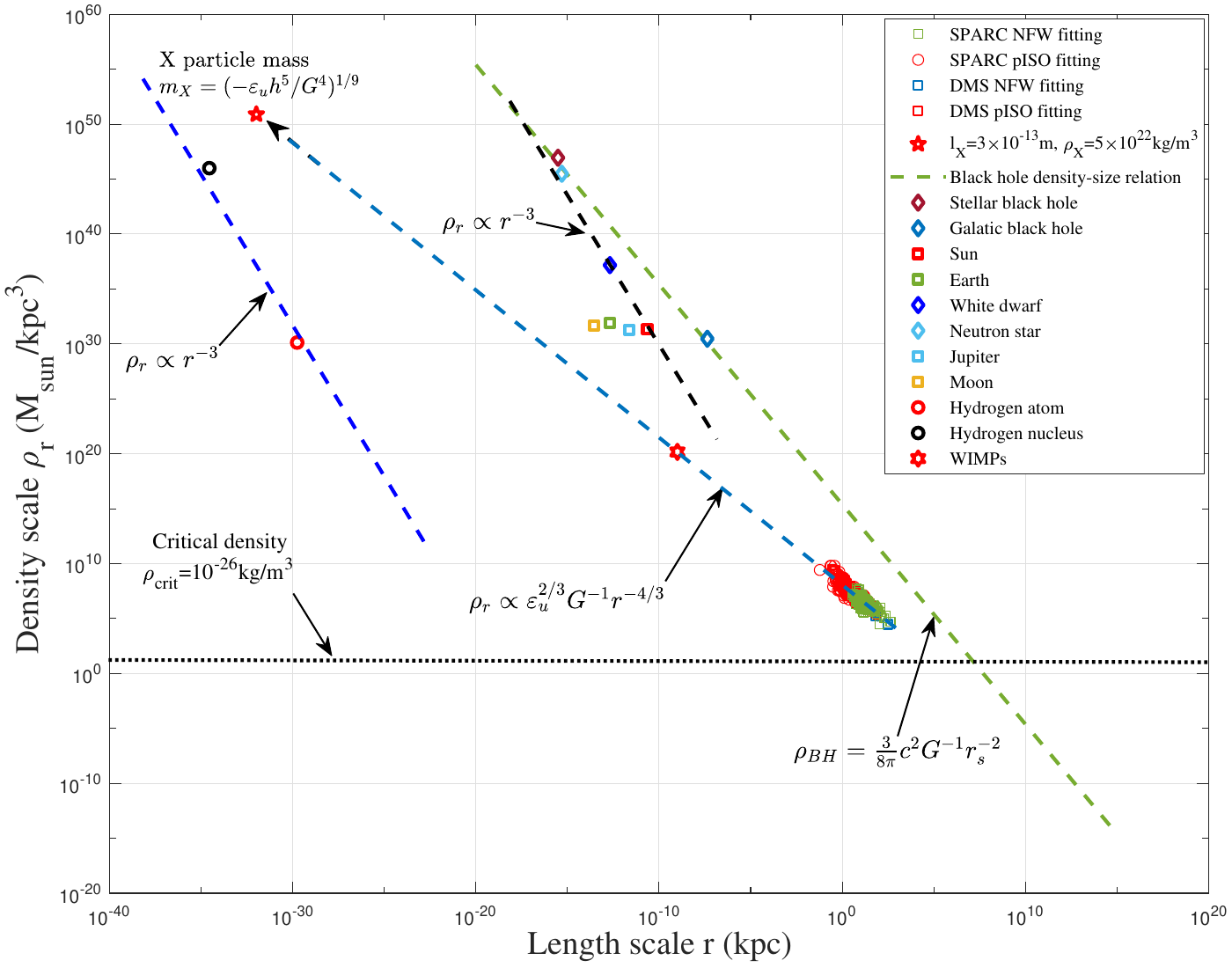}
\caption{The variation of density scale $\rho_r$ with size scale $r$ for cold dark matter particle properties. By extending the four-thirds law in Fig. \ref{fig:5} for dark matter haloes to the smallest scale where quantum effects are dominant, the red star (the $X$ scale) represents the cold DM particles with predicted density $\rho_X$ and size $l_X$ from Eqs. \eqref{eq:12} and \eqref{eq:13}. Collisionless dark matter haloes follow the scaling $\rho_r\propto r^{-4/3}$ due to the energy cascade at a constant rate $\varepsilon_u$. The densities of sun, dwarf, neutron star, and stellar black holes follow the scaling $\rho_r \propto r^{-3}$ reflecting the formation of stellar black holes from the collapse of massive stars. The density of black holes ($\rho_{BH}$) as a function of Schwarzschild radius ($r_s$) follows $\rho_{BH} \propto r_s^{-2}$, extending from stellar to galactic black holes. WIMPS from Table \ref{tab:6} is also reported (red hexagram).}
\label{fig:55}
\end{figure}

The relevant mass density is around ${m_{X} /l_{X}^{3}} \approx 5.33\times 10^{22} {kg/m^{3}}$, much higher than the nuclear density of the order of $10^{17} {kg/m^{3}}$ (Fig. \ref{fig:55}). This density is about $32\pi^2$ times the background density of the universe at the formation time $t_X$, in agreement with the spherical collapse model. The relevant pressure scale is
\begin{equation}
\label{eq:22-22}
P_X = \frac{m_{X} a_{X}}{l_{X}^{2}} = \frac{8\hbar^2}{m_X}\rho_{nX}^{5/3}=1.84\times 10^{10} P_{a},
\end{equation}
which sets the highest pressure or the possible "degeneracy" pressure of dark matter that stops further gravitational collapse. Equation \eqref{eq:22} is an analog of the degeneracy pressure of the ideal Fermi gas, where $\rho_{nX}=l_X^{-3}$ is the number density of the particle $X$. With today's dark matter density around $2.2\times 10^{-27}kg/m^3$ and local density $7.2\times 10^{-22}kg/m^3$, the mean separation between \textbf{\textit{X}} particles is about $l_u \approx 10^4m$ in the entire universe and $l_c \approx 130m$ locally. 


In summary, Fig. \ref{fig:102} presents a detailed interpretation. Different combinations of three relevant physics leads to some interesting findings in Fig. \ref{fig:102}. Combining Eq. \eqref{eq:14} with Eq. \eqref{eq:16} leads to the black dashed line in Fig. \ref{fig:102}, a relation between the mass and size of the particles ($m_r\propto r^{-1/3}$). A similar relation can also be obtained from the Schrödinger-Newton equation, a nonlinear modification of the Schrödinger equation with Newtonian gravity for self-interaction. This line represents the boundary below which quantum effects are important and dominant over gravity. 

Combining Eq. \eqref{eq:15} with Eq. \eqref{eq:16} leads to the five-thirds ($m_r\propto r^{3/5}$) law between the halo mass and the halo size (see Eq. \eqref{eq:11} for the 5/3 law), the blue dashed line in Fig. \ref{fig:102}. Dark matter haloes from galaxy rotation curves also fall on the same line. This line represents the dynamics of dark matter haloes on large scales. To find the dark matter particle properties that are consistent with the dynamics of dark matter haloes on large scales, extending the five-thirds law (blue dashed) to the smallest scale will intersect the black dashed line, i.e., the red star in Fig. \ref{fig:102}, where quantum effect becomes relevant, and scaling laws might stop operating. This should be the scale of the X particle (the $X$ scale). We show that the free streaming mass is comparable to the particle mass at this scale (Eq. \eqref{eq:12-9}).

Finally, the mass of black holes at a given size $r$ (Schwarzschild radius) is plotted as a green dashed line ($m_r\propto r$). This is the maximum mass $m_r$ on a given scale $r$. Extending the green dashed line to intersect the blue dashed line, i.e., the blue star leads to the largest scale (the $H$ scale) of dark matter haloes. This is the largest possible halo scale that can be reached during the lifetime of dark matter. The black dashed line intersects the green dashed line at the Planck scale. 

\begin{table}
    \begin{center}
    \caption{Physical quantities on the $X$ scale, $H$ scale and Planck scale}
    \label{tab:3}
    \begin{tabular}{llll} 
    \hline
    Scales &  The $X$ scale  & The $H$ scale  & The Planck scale\\
    \hline
    {Length}   &{$l_{X} =\left(-\frac{G\hbar }{\varepsilon_{u}}\right)^{{1}/{3}}$}   & {$l_{H} =-\frac{c^3}{\varepsilon_u}$} & {$l_{p} =\sqrt{\frac{\hbar G}{c^3}}$}\\
    {Time}     &{$t_{X} =\left(-\frac{G^{2}\hbar^{2}}{\varepsilon _{u}^{5}}\right)^{{1}/{9}}$}  & {$t_H=-\frac{c^2}{\varepsilon_u}$} & {$t_p=\sqrt{\frac{\hbar G}{c^5}}$}\\
    {Mass}     &{$m_{X} =\left(-\frac{\varepsilon_{u}\hbar^{5}}{G^{4}}\right)^{{1}/{9}}$}    & {$m_H=-\frac{c^5}{\varepsilon_uG}$} & {$m_p=\sqrt{\frac{\hbar c}{G}}$}\\
    \hline
    \end{tabular}
  \end{center}
\end{table}

In this plot, three key scales (Table \ref{tab:3}) are determined by different constants, i.e., the green star for the Planck scale ($\hbar$, $G$, and $c$), the red star for the smallest halo scale or the DM particle scale $X$ ($\hbar$, $G$, and $\varepsilon_u$), and the blue star for the largest halo scale $H$ ($G$, $c$, and $\varepsilon_u$). Table \ref{tab:3} provides relevant quantities on three key scales. For comparison, the mass and actual size of some typical particles and astronomical objects are also presented in the same figure.

Similarly, Figure \ref{fig:55} illustrates the variation of the density scale $\rho_r$ with size $r$. By extending the four-thirds law in Fig. \ref{fig:5} to the smallest scale where the quantum effect becomes dominant, we obtain the red star in Fig. \ref{fig:55} for cold DM particles with predicted mass $m_X$ and size $l_X$ from Eqs. \eqref{eq:12} and \eqref{eq:13}. The density and size of some typical particles and astronomical objects are also included for comparison. As expected, the density of the sun, dwarf, neutron star, and stellar black hole follows the scaling $\rho_r \propto r^{-3}$, while the density of black holes ($\rho_{BH}$) computed using the Schwarzschild radius ($r_s$) follows a simple scaling $\rho_{BH} \propto r_s^{-2}$. Thus, Fig. \ref{fig:55} displays three distinct scaling laws for density-size relations of different objects reflecting different operating physics, i.e. $\rho_r\propto r^{-4/3}$ for dark matter haloes, $\rho_r\propto r^{-3}$ for the formation of stellar black holes, and $\rho_r\propto r^{-2}$ for black holes (BH) of different sizes from stellar to galactic.

\section{Self-interacting dark matter}
\label{sec:6}
Note that the particle mass $m_{X}$ is only weakly dependent on $\varepsilon _{u}$ as $m_X \propto \varepsilon _{u}^{{1/9}}$ (Eq. \eqref{eq:19}) such that the estimation of $m_{X}$ should be fairly robust for a wide range of possible values of $\varepsilon _{u}$. A small change in $m_{X}$ requires a large change in $\varepsilon_{u}$. Unless gravity is not the only interaction, the uncertainty in the predicted $m_{X}$ should be small. In other words, if our estimate ($\varepsilon_u$ in Eq. \eqref{eq:5}) is accurate and gravity is the only interaction on the smallest scale, it seems implausible for dark matter particles with any mass far above $10^{12} \textrm{GeV}$ to produce the given value of the energy cascade rate $\varepsilon_u\approx10^{-7}m^2/s^3$. If the mass of dark matter particles has a different value, there might be some new interactions beyond gravity. This can be the self-interacting dark matter (SIDM) model as a solution for the "cusp-core" problem \citep{Spergel:2000-Observational-Evidence-for-Self-Interacting-Cold-Dark-Matter}. 

\begin{table}
    \begin{center}
    \caption{Physical scales for collisionless and self-interacting dark matter}
    \label{tab:2}
    \begin{tabular}{lll} 
    \hline
    Scales &  Fully collisionless         & Self-interacting   \\
    \hline
    {Length}   &{$l_{X} =\left(-{G\hbar }/{\varepsilon_{u}}\right)^{{1}/{3}}$}   & {$r_{\eta} =\varepsilon_{u}^{2}G^{-3}(\sigma/m)^{3}$}\\
    {Time}     &{$t_{X} =\left(-{G^{2}\hbar^{2}}/{\varepsilon _{u}^{5}}\right)^{{1}/{9}}$}       & {$t_{\eta}                 =-\varepsilon_{u}G^{-2}(\sigma/m)^{2}$}\\
    {Mass}     &{$m_{X} =\left(-{\varepsilon_{u}\hbar^{5}}/{G^{4}}\right)^{{1}/{9}}$}    & {$m_{\eta}                 =\varepsilon_{u}^4G^{-6}(\sigma/m)^{5}$}\\
    {Density}     &{$\rho_{X} =\left({\varepsilon_{u}^{10}\hbar^{-4}}/{G^{13}}\right)^{{1}/{9}}$}    & {$\rho_{\eta}                 =\varepsilon_{u}^{-2}G^{3}(\sigma/m)^{-4}$}\\
    \hline
    \end{tabular}
  \end{center}
\end{table}

For self-interacting dark matter, a key parameter is the cross-section $\sigma/m$ (in the unit: $m^2/kg$) of self-interaction that can be constrained by various astrophysical observations. Self-interaction introduces an additional scale, below which self-interaction is dominant over gravity to suppress all small-scale structures, and the scaling laws for dark matter haloes in Section \ref{sec:3} are no longer valid. In this case, the dark matter particle properties can be obtained only if the nature and dominant constants of self-interaction are known. Three constants determine the smallest scale for the existence of a halo structure, that is, the rate of the energy cascade $\varepsilon_u$, the gravitational constant $G$, and the cross-section $\sigma/m$. In other words, the cross-section might be estimated if the scale of the smallest structure is known. Table \ref{tab:2} lists the relevant scales for both collisionless and self-interacting dark matter. Taking the value of $\sigma/m=0.01m^2/kg$ used for the cosmological SIDM simulation to reproduce the right halo core size and central density \citep{Rocha:2013-Cosmological-simulations-with-self-interacting-dark-matter}, these scales are also plotted in Fig. \ref{fig:5} for illustration. 

\label{lastpage}
\end{document}